\documentclass[a4paper,11pt]{article}
\pdfoutput=1 % if your are submitting a pdflatex (i.e. if you have
\usepackage{jcappub} % for details on the use of the package, please
\usepackage{graphics}
\usepackage{graphicx}
\usepackage[T1]{fontenc} % if needed
\usepackage{footnote}
\usepackage{hyperref}
\usepackage{placeins}
\usepackage{wrapfig}
\usepackage{pgfplots}
\usepackage[utf8]{inputenc}
\usepackage{afterpage}

\usepackage{booktabs}
\usepackage[english]{babel}
\usepackage{amsmath,amssymb,amsbsy,amstext, amsthm, simplewick}
\usepackage{hyperref}
\usepackage{graphicx}
\usepackage{amsfonts}
\usepackage{amssymb}
\usepackage[small]{caption}
\usepackage{upgreek}
\usepackage{siunitx}
\usepackage{sidecap}
\usepackage{graphics}
\usepackage{graphicx}
\usepackage[T1]{fontenc} % if needed
\usepackage{footnote}
\usepackage{subfig}

\usepackage{tikz}
\usepackage{booktabs}
\usetikzlibrary{intersections}
\usetikzlibrary{calc}

\newcommand{\PPS}{\mathcal{P}}

\usepackage{layouts}

\def\be{\begin{equation}}
\def\ee{\end{equation}}
\def\ba{\begin{eqnarray}}
\def\ea{\end{eqnarray}}

\newcommand{\Fig}[1]{Fig.~{\ref{#1}}}
\newcommand{\Figs}[1]{Figs.~{\ref{#1}}}
\newcommand{\Ref}[1]{Ref.~{\cite{#1}}}
\newcommand{\Refs}[1]{Refs.~{\cite{#1}}}
\newcommand{\Sec}[1]{Sec.~\ref{#1}}

\def\reff@jnl#1{{\rm#1\/}}
\def\apj{\reff@jnl{ApJ}}       % Astrophysical Journal
\def\apjs{\reff@jnl{ApJS}}     % Astrophysical Journal, Supplement
\def\aap{\reff@jnl{A\&A}}    % Astronomy and Astrophysics, Supplement
\def\aaps{\reff@jnl{A\&AS}}    % Astronomy and Astrophysics, Supplement
\def\mnras{\reff@jnl{MNRAS}}   % Monthly Notices of the RAS
\def\prd{\reff@jnl{Phys.\ Rev.\ D}}    % Physical Review D
\def\jcap{\reff@jnl{JCAP}}    % JCAP

\newcommand{\Mp}{M_{{}_\mathrm{Pl}}}

\newcommand{\GeV}{\mathrm{GeV}}

\newcommand{\fd}{\ensuremath{f_{10}}}

\newcommand{\Nknots}{N}

\newcommand{\coreplus}{\negthinspace COrE+\/}
\newcommand{\planck}{\textit{\negthinspace Planck\/}}
\newcommand{\Planck}{\planck}

\newcommand{\fnl}{f_\text{NL}}

\newcommand{\lmin}{\ell_\text{min}}
\newcommand{\lmax}{\ell_\text{max}}
\newcommand{\CIB}{{\mathtt{CIB}}}

\def\lsim{~\rlap{$<$}{\lower 1.0ex\hbox{$\sim$}}}
\def\bsim{~\rlap{$>$}{\lower 1.0ex\hbox{$\sim$}}}

\def\gtorder{\mathrel{\raise.3ex\hbox{$>$}\mkern-14mu
             \lower0.6ex\hbox{$\sim$}}}
\def\ltorder{\mathrel{\raise.3ex\hbox{$<$}\mkern-14mu
             \lower0.6ex\hbox{$\sim$}}}

\usepackage{colortbl}
\definecolor{lightgreen}{cmyk}{0.2, 0, 0.2, 0.2}
\definecolor{lightgray}{cmyk}{0.1,0.2,0,0.1}
\definecolor{lightgray2}{cmyk}{0.1,0.1,0,0.1}

\makeatletter
\newlength{\apb@width}
\newcommand{\autoparbox}[2][c]{\settowidth{\apb@width}{#2}\parbox[#1]{\apb@width}{#2}}

\makeatother

\numberwithin{equation}{section}

\def\beq{\begin{equation}}
\def\eeq{\end{equation}}

\def\bea{\begin{eqnarray}}
\def\eea{\end{eqnarray}}

\def\beq{\begin{equation}}
\def\eeq{\end{equation}}
\def\bea{\begin{eqnarray}}
\def\eea{\end{eqnarray}}

\newcommand\limit[1]{#1\%\,\mathrm{CL}}
\newcommand{\ns}{n_{\rm s}}
\newcommand{\dif}{\mathrm{d}}

\newcommand{\sci}[2]{$#1 \times 10^{#2}$}
\newcommand{\scib}[2]{${\bf #1 \times 10^{#2}}$}

\providecommand{\sorthelp}[1]{}

\title{Exploring Cosmic Origins with CORE: Inflation}

\author[1,2]{Fabio Finelli,}
\author[3,4]{Martin Bucher,}
\author[5,6]{Ana Ach\'ucarro,}
\author[7,1,2]{Mario Ballardini,}
\author[8,9,10]{Nicola Bartolo,}
\author[11,12]{Daniel Baumann,}
\author[13]{S\'ebastien Clesse,}
\author[14]{Josquin Errard,}
\author[15,16]{Will Handley,}
\author[17,18,19]{Mark Hindmarsh,}
% ...

\affiliation[1]{INAF/IASF Bologna, via Gobetti 101, I-40129 Bologna, Italy}
\affiliation[2]{INFN, Sezione di Bologna, via Irnerio 46, I-40127 Bologna, Italy}
\affiliation[3]{APC, Astroparticule et Cosmologie, Universit\'e Paris Diderot, CNRS/IN2P3, 
CEA/lrfu, Observatoire de Paris Sorbonne Paris Cit\'e, 10, rue Alice Domon et L\'eonie Duquet, 75205 Paris Cedex 13, France}
\affiliation[4]{Astrophysics and Cosmology Research Unit, School of Mathematics, Statistics and Computer Science,
University of KwaZulu-Natal, Durban 4041, South Africa}
\affiliation[5]{Instituut-Lorentz for Theoretical Physics, Universiteit Leiden, 2333 CA, Leiden, The Netherlands}
\affiliation[6]{Department of Theoretical Physics, University of the Basque Country UPV/EHU, 48040 Bilbao, Spain}
\affiliation[7]{Dipartimento di Fisica e Astronomia, Universit\`a di Bologna, Viale Berti Pichat, 6/2, I-40127 Bologna, Italy}
\affiliation[8]{Dipartimento di Fisica e Astronomia ``Galileo Galilei'', Universit\`a degli Studi di Padova, Via Marzolo 8, I-35131, Padova, Italy}
\affiliation[9]{INFN, Sezione di Padova, Via Marzolo 8, I-35131 Padova, Italy}
\affiliation[10]{INAF, Osservatorio Astronomico di Padova, Vicolo dell'Osservatorio 5, I-35122 Padova, Italy}

\affiliation[11]{DAMTP, Centre for Mathematical Sciences, University of Cambrige, Wilberforce Road, Cambridge, CB3 0WA, UK}
\affiliation[12]{Institute of Physics, University of Amsterdam, Science Park, Amsterdam, 1090 GL, The Netherlands}
\affiliation[13]{Institute for Theoretical Particle Physics and Cosmology (TTK), RWTH Aachen University, D-52056 Aachen, Germany}
\affiliation[14]{Institut Lagrange, LPNHE, place Jussieu 4, 75005 Paris, France.}
\affiliation[15]{Astrophysics Group, Cavendish Laboratory, Cambridge, CB3 0HE, UK}
\affiliation[16]{Kavli Institute for Cosmology, Cambridge, CB3 0HA, UK}
\affiliation[17]{Department of Physics and Astronomy, University of Sussex, Falmer, Brighton, BN1 9QH, UK}
\affiliation[18]{Department of Physics, Gustaf Hallstromin katu 2a, University of Helsinki, Helsinki, Finland}
\affiliation[19]{Helsinki Institute of Physics, Gustaf Hallstromin katu 2, University of Helsinki, Helsinki, Finland}
% ...

\emailAdd{finelli@iasfbo.inaf.it,bucher@apc.univ-paris7.fr}

\abstract{We forecast the scientific capabilities 
to improve our understanding of cosmic inflation 
of CORE, a proposed CMB space satellite submitted in
response to the ESA fifth call for a medium-size mission opportunity.
The CORE satellite will map the CMB anisotropies in temperature and
polarization in 19 frequency channels spanning the range 60-600 GHz. CORE will have an aggregate noise
sensitivity of $1.7 \mu$K$\cdot \,$arcmin and an angular resolution of 5' at 200 GHz. We explore the
impact of telescope size and noise sensitivity on the inflation science return by making
forecasts for several instrumental configurations. This study assumes that the lower and higher
frequency channels suffice to remove foreground contaminations and complements other
related studies of component separation and systematic effects, which will be reported in other
papers of the series ``Exploring Cosmic Origins with CORE.'' 
We forecast the capability to determine key inflationary parameters,
to lower the detection limit for the tensor-to-scalar ratio down to the $10^{-3}$ level, to chart
the landscape of single field slow-roll inflationary models, to constrain the epoch of
reheating, thus connecting inflation to the standard radiation-matter dominated Big Bang era, to
reconstruct the primordial power spectrum, to constrain the contribution from isocurvature
perturbations to the $10^{-3}$ level, to improve constraints on the cosmic string tension to a
level below the presumptive GUT scale, and to improve the current measurements of primordial
non-Gaussianities down to the $\fnl^{\rm local} < 1$ level. For all the models explored, CORE {\em
alone} will improve significantly on the present constraints on the physics
of inflation. Its capabilities will be further enhanced by combining
with complementary future cosmological observations.}

\begin{document}
\maketitle
\flushbottom

\section{Introduction}
\label{sec:intro}

Starting with the COBE detection of a cosmic microwave background (CMB) anisotropy in 
1992 \cite{Smoot:1992td}, the precision mapping of 
the primordial CMB anisotropies in temperature and polarization has allowed us to 
characterize the initial cosmological perturbations at about the percent level 
\cite{Ade:2013zuv,Planck:2013jfk,Ade:2015xua,Ade:2015lrj}.
On the one hand, these observations serve as initial conditions to be used to understand how the 
highly clumpy and nonlinear universe at late times emerged. On the other hand, these 
observations also allow us to probe the physics of the very early universe, governed by 
unknown new physics at energy scales far beyond those scales that can be probed even with 
the most ambitious future accelerator experiments. With better observations of the CMB, we 
will be able to probe new physics at scales just below the Planck scale and establish 
meaningful constraints on theories regarding how gravity becomes unified 
with the other three fundamental interactions (i.e., strong, weak, and electromagnetic),
presumably at an energy scale around the Planck scale.

This paper describes what we may expect to learn about cosmic inflation from 
future CMB experiments, in particular from the CORE mission, which is a dedicated 
microwave polarization satellite proposed to the European Space Agency (ESA) in 
October 2016 in response to 
a Call for proposals for a future medium-sized space 
mission for the ``M5'' launch opportunity of the ESA Cosmic Vision programme. This article, which 
is part of the ``Exploring Cosmic Origins (ECO)'' collection of articles 
\cite{ecoMission,ecoInstrument,ecoSystematics,ecoParams,ecoCompSep,DeZotti:2016qfg,ecoCluster,ecoPeculiar}, 
each describing a different aspect of the CORE mission, 
%each describing a different aspect the new science that will be made possible with CORE, deals 
deals with forecasts of how the CORE data 
will improve our knowledge of the physics of cosmic inflation.
Closely related papers include the ECO paper on cosmological parameters \cite{ecoParams}
and the ECO paper on B-mode component separation \cite{ecoCompSep}.

Before the recent CORE proposal, several related proposals for a post-Planck dedicated 
microwave polarization satellite had been submitted to ESA: 
B-Pol in 2007\footnote{See www.b-pol.org for a copy of the proposal and more details.}
\cite{deBernardis:2008bf},
COrE in 2011\footnote{www.core-mission.org} \cite{Bouchet:2011ck}, 
PRISM\footnote{www.prism-mission.org} \cite{Andre:2013nfa} in 2013 (which was a higher budget  ``L" (large) class mission addressing
a broader, more ambitious science 
case), and COrE+ in 2014. 
These proposals were highly rated but none made the final cut
to selection. Similarly, in the United States, 
there have been several studies and proposals for similar missions. The CMB-Pol mission concept study 
produced a number of detailed white papers, one of which deals with inflation 
\cite{Baumann:2008aq}
and thus has much overlap with this work,
and also specialized white papers on foregrounds \cite{Dunkley:2008am,Fraisse:2008ar}
and gravitational lensing \cite{Smith:2008an},
as well as a general overview paper \cite{Baumann:2008aj}.
EPIC was proposed to NASA in 2008.
The EPIC study \cite{Bock:2009xw,Bock:2008ww}
presents a detailed conceptual design in the form of three options: a 
low-cost option, an intermediate option, and a comprehensive science option. Another US initiative
is the proposed NASA PIXIE mission \cite{Kogut:2011xw},
which would map the microwave sky using a Fourier Transform Spectrometer (FTS) 
much like the COBE FIRAS instrument 
\cite{Fixsen:1996nj} but two orders of magnitude more sensitive and with 
sensitivity to polarization. A modified version of PIXIE \cite{2016SPIE.9904E..0WK} was proposed for the 
NASA MIDEX call 2016 and could potentially be launched in 2023.
In Japan a CMB polarization space mission called LiteBIRD 
\cite{2014JLTP..176..733M,Matsumura:2016sri} is 
presently undergoing a Phase A study together with NASA. Compared to CORE, LiteBIRD 
has a smaller aperture, thus limiting its reach toward small angular scales. It is a 
lower cost mission with an earlier planned launch, according to the present schedule 
around 2025.

When inflationary cosmology was introduced in the early 1980s 
\cite{Brout:1977ix,Starobinsky:1980te,Kazanas:1980tx,Sato:1980yn,Guth:1980zm,Linde:1981mu,Linde:1983gd,Albrecht:1982wi}, it was initially greeted 
with great interest accompanied by a healthy dose of skepticism. In the few years 
following the COBE discovery, it was not at all obvious that inflation would survive a confrontation with 
forthcoming data. However, with the first clear observations of the first acoustic peaks 
\cite{deBernardis:2000sbo,Hanany:2000qf},
followed by the mapping of the three acoustic peaks by WMAP \cite{Spergel:2003cb}, and subsequently by the 
precision mapping of the five peaks by the ESA Planck mission \cite{Ade:2013zuv}, many of the competing 
models of structure formation fell by the wayside, and it turned out that rather simple
models of inflation could account for the data \cite{Planck:2013jfk,Ade:2015lrj}.

The plethora of disparate competing cosmological models existing at the time of the COBE 
discovery gradually became replaced with 
what has now become known as the `concordance' model of cosmology. More precisely, this 
is the six-parameter $\Lambda$CDM model described in detail in the WMAP and Planck papers dedicated to 
making the connection between the CMB observations and cosmological models 
\cite{Ade:2013zuv,Ade:2015xua}.
For the purposes of the present paper, it suffices to highlight the following key points:

(1) WMAP and Planck found that the data can be explained by a six-parameter $\Lambda$CDM cosmological model under 
which the scalar power spectrum takes the following simple power law form
\begin{equation}
\mathcal{P}_{\mathcal{R}}(k)=A_{\rm s}~\left( \frac{k}{k_*}\right) ^{n_{\rm s}-1}
\label{PPSeq}
\end{equation}
where $k_*$ is a pivot scale (unless otherwise stated fixed to $0.05$ Mpc$^{-1}$). 
This model also includes four additional non-inflationary parameters 
$H_0,$ $\omega_{\rm b}=\Omega_{\rm b} h^2,$ $\omega_{\rm c}=\Omega_{\rm c} h^2,$ and $\tau .$ 
The model provides a good fit to 
the data: there is no statistically significant evidence compelling us to extend this 
model despite the many extensions that have been explored 
\cite{Ade:2013zuv,Planck:2013jfk,Ade:2015xua,Ade:2015lrj}.
The $TT$, $TE$, and $EE$ CMB power 
spectra, at present most tightly constrained by the \Planck\ data, and on smaller 
angular scales by data from ACT \cite{Dunkley:2010ge} and SPT 
\cite{Keisler:2011aw}, are well accounted for by this model, which 
moreover is broadly consistent with other probes such as Baryon Acoustic Oscillations 
(BAO) \cite{Eisenstein:2005su,Percival:2009xn} 
and constraints on $\omega_{\rm b}$ derived from the observed light element abundances 
interpreted using the theory of primordial big bang nucleosynthesis. 
\Planck\ data alone have been able to set limits at the percent level on the curvature \cite{Ade:2013tyw,Ade:2015zua}, verifying one of 
the basic predictions of the simplest inflationary models. % such as the flatness of the spatial sections.
Among the caveats to an interpretation based on $\Lambda$CDM are 
disagreements at modest statistical significance, sometimes euphemistically dubbed 
`tensions' 
with determinations of $H_0$  \cite{Riess:2016jrr} and with cosmological parameters determined using cluster abundances 
\cite{Ade:2013lmv,Ade:2015fva} or galaxy shear measurements \cite{Joudaki:2016mvz,Hildebrandt:2016iqg}. 
Importantly, the model includes only the adiabatic growing mode for primordial fluctuations as predicted 
by inflation driven by a single scalar field.
No statistically significant evidence was uncovered showing that isocurvature modes were excited
\cite{Planck:2013jfk,Ade:2015lrj}, which is possible in multi-field inflationary models.
%and assumes that the
%primordial fluctuations were precisely Gaussian (as predicted to high accuracy by the
%simplest inflationary models). No statistically significant evidence was uncovered
%showing that isocurvature modes were excited
%\cite{Planck:2013jfk,Ade:2015lrj}.
%Importantly, the model 
%includes only adiabatic growing mode primordial fluctuations and assumes that the 
%primordial fluctuations were precisely Gaussian (as predicted to high accuracy by the 
%simplest inflationary models). No statistically significant evidence was uncovered 
%showing that isocurvature modes were excited 
%\cite{Planck:2013jfk,Ade:2015lrj}.

(2) One of the most significant findings, first made by WMAP at modest statistical 
significance and later by Planck at much higher statistical significance, was that the 
primordial power spectrum is not exactly scale invariant: in other words, $n_{\rm s}\ne 1$ 
\cite{Planck:2013jfk,Ade:2015lrj}. This finding is consistent with those inflationary models which have a natural exit from inflation.
%with the expectation of inflationary 
%theories, which would require unnatural fine tuning to obtain $n_{\rm s}$ very close to unity. 

(3) Another far reaching result of the ESA Planck mission was the tight constraints
established
%was the failure to find any statistically significant sign of
on primordial non-Gaussianity \cite{Ade:2013ydc,Ade:2015ava}.
% down to the level of other general relativistic signals such 
% the ISW-lensing contribution, which is now firmly detected. 
These upper bounds rule 
out at high statistical significance many of the non-standard inflationary models 
predicting a level non-Gaussianity allowed by WMAP \cite{Bennett:2012zja}.
%that theorists had developed hoping that
%hints of a signal from WMAP would be confirmed by Planck.

(4) With the presently available CMB data, the scalar power spectrum as given in 
Eq.~(\ref{PPSeq}) has been mapped out over approximately three decades in wavenumber. 
But at present, apart from upper bounds \cite{Ade:2015tva,Array:2015xqh}, almost nothing is known about the tensor mode 
power spectrum. That tensor modes should be excited is one of the most remarkable and 
surprising predictions of cosmic inflation. Yet this prediction has not yet been tested. 
Discovering primordial B modes from inflation is the primary goal of almost all future CMB 
experiments, as we detail below.

\begin{figure}
\begin{center}
                \includegraphics[width=8cm]{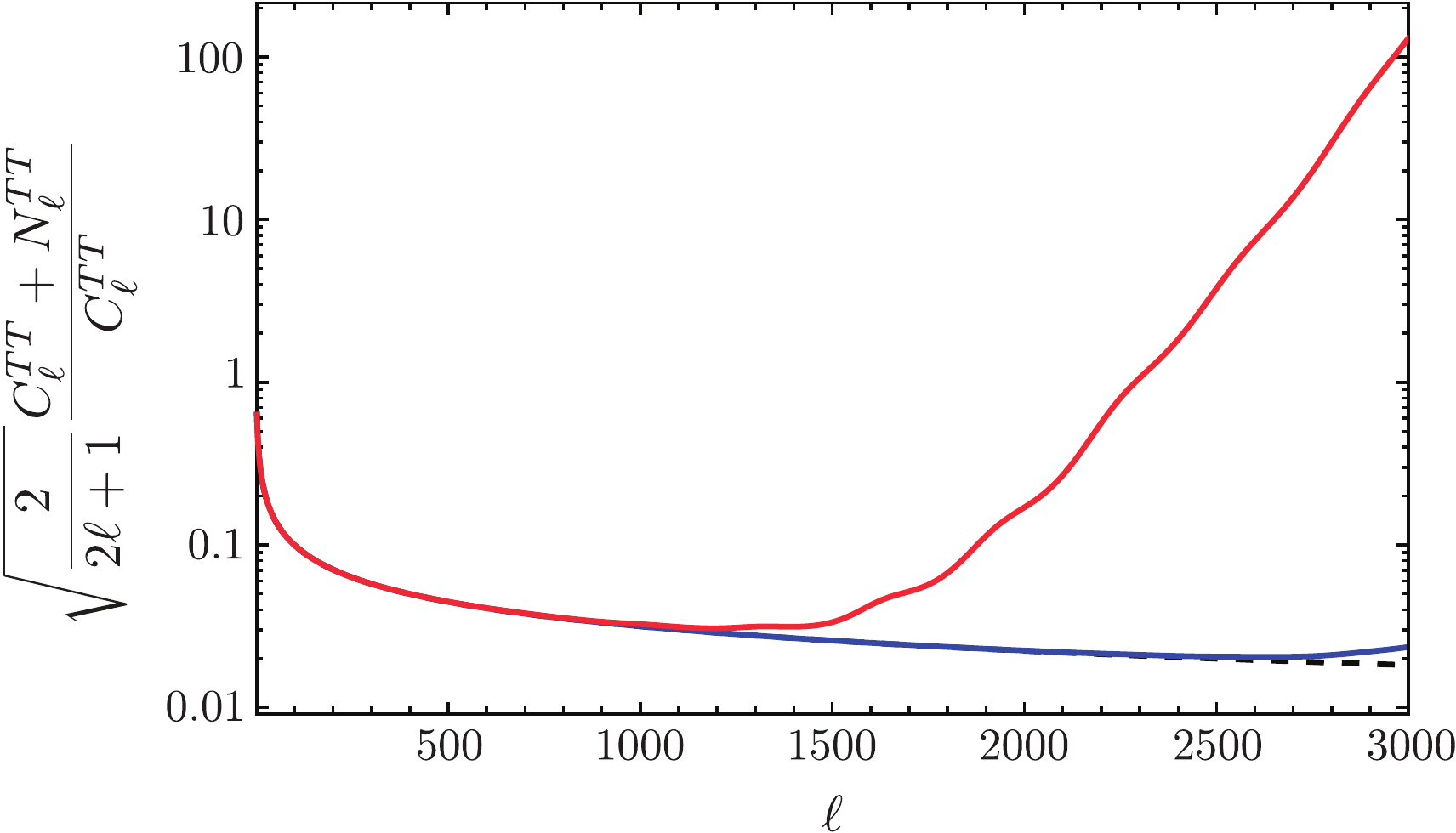}
                \includegraphics[width=8cm]{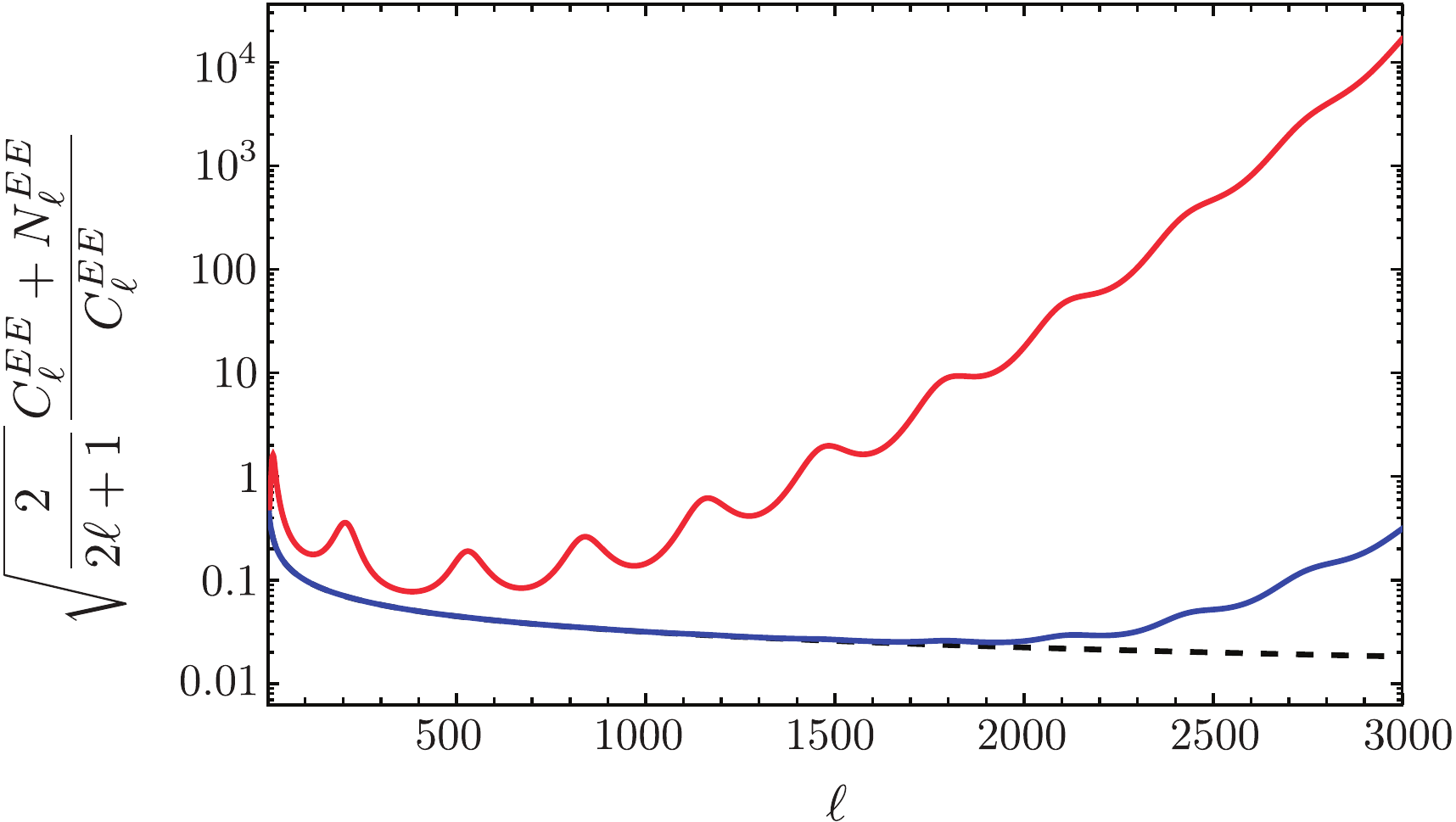}
\vspace{-1\baselineskip}
\end{center}
\caption{%We plot 
The improvement in $TT$ (left panel) and $EE$ (right) power spectra as a function of the multipole
number $\ell$ for Planck (red line) and CORE (blue line) 
up to $\ell \approx 3000$ compared to the cosmic variance limit with $f_{sky}=1$ (dashed black line).}
\label{coreVSplanck0}
\end{figure}

In order to provide a very approximate idea of how much more cosmologically relevant data 
the CORE satellite will collect compared to the data already available from Planck, we 
examine how much the error bars on the underlying theoretical power spectra will 
shrink as the result of the addition of future CORE data. In this analysis we suppose that 
the underlying stochastic process is nearly Gaussian, an assumption 
consistent with the failure of \Planck\ to turn up any statistically significant 
evidence for primordial non-Gaussianity \cite{Ade:2013ydc,Ade:2015ava}.
%so. This assumption is 
%consistent with the null results of the Planck searches for various patterns of primordial non-Gaussianity \cite{Ade:2013ydc,Ade:2015ava}. 
The precise likelihood for a given theoretical power spectrum 
for a realistic survey is complicated, but the following analytic order of magnitude 
estimate suffices \cite{Knox:1995dq}: 
%\begin{equation} 
%\left(\dfrac{\delta C_\ell }{C_\ell } \right) _{rms} \approx 
%\sqrt{\frac{2}{f_{sky}(2\ell 
%+1)}} \left( \dfrac{C_\ell +N_\ell } {C_\ell } \right) ^{1/2} \end{equation} 
\begin{equation}
\left(\dfrac{\delta C_\ell }{C_\ell } \right)_{\rm rms} \approx
\sqrt{\frac{2}{f_{\rm sky}(2\ell
+1)}} \dfrac{C_\ell +N_\ell }{C_\ell }  \,,
\label{estimate}
\end{equation}
where $f_{\rm sky}$ is the sky fraction surveyed, $C_\ell $ refers to the %expected or theoretical 
power spectrum, and $N_\ell $ is the measurement noise. % ---or more precisely, the noise in 
%the final foreground-cleaned primordial CMB map. 
%(We do not specify whether we are talking about $TT,$ $TE,$ or $EE,$ because the formula above applies to all three cases.) 
There are two regimes to consider. When $N_\ell \ltorder C_\ell ,$ which is very much the 
case at low $\ell $ for $C_\ell^{TT},$ the uncertainty is dominated by ``cosmic variance,'' 
and in terms of fixing the power spectrum, reducing $N_\ell $ further 
is of marginal added value.  Our inability to fix the power spectrum of the underlying 
stochastic process arises primarily because we can observe only one sky. In the regime 
$N_\ell \ll C_\ell ,$ the microwave sky has been mapped sufficiently well, so there is 
little motivation to construct a less noisy map. On the other hand, in the multipole 
range where $N_\ell \gtorder C_\ell ,$ there remains significant 
new information to be gained, and the 
error bars on the underlying CMB power spectra can be shrunk down further by producing 
better maps based on new data. Fig. \ref{coreVSplanck0} shows the improvement that can be reached 
with CORE over \Planck\ in temperature and polarization according to the  analytical estimate 
in Eq. (\ref{estimate}), showing that CORE will make cosmic variance limited measurements 
of the $TT$ power spectrum for $\ell \lesssim 2500$ and for $\ell \lesssim 2000$ for $EE$.

\begin{figure}
\begin{center}
%PLACE FIGURE HERE. 
                \includegraphics[width=5.3cm]{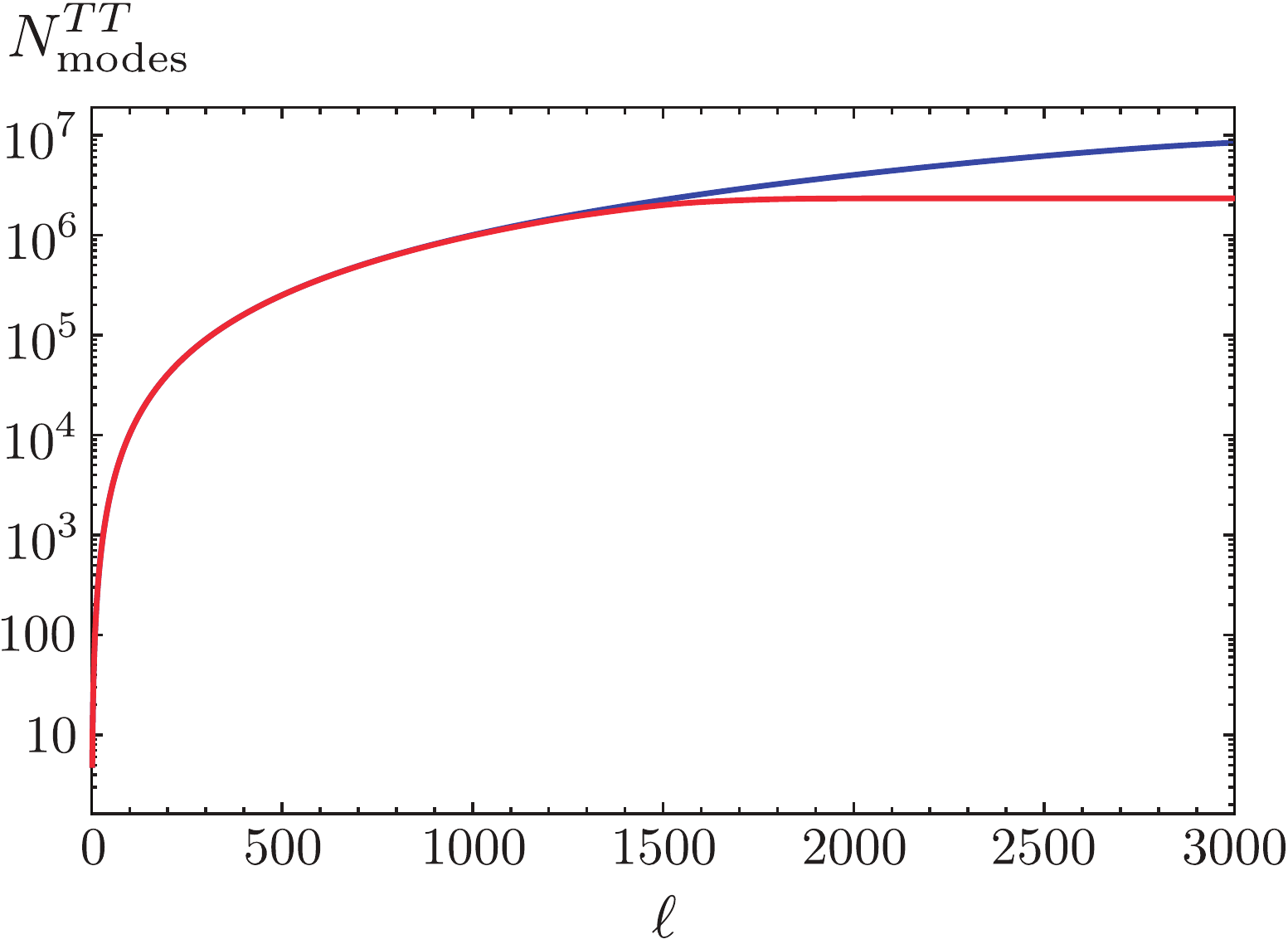} 
                \includegraphics[width=5.3cm]{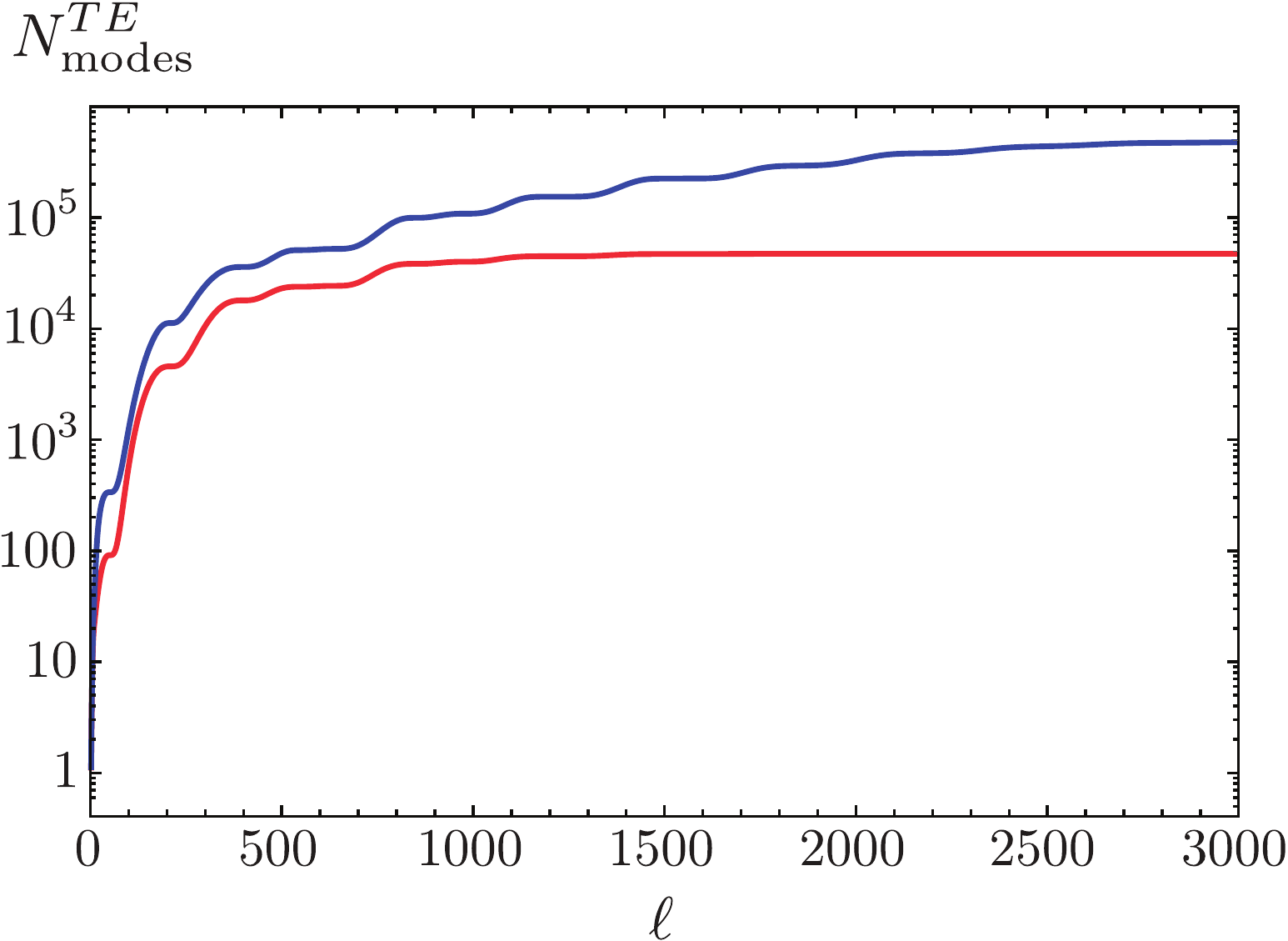} 
                \includegraphics[width=5.3cm]{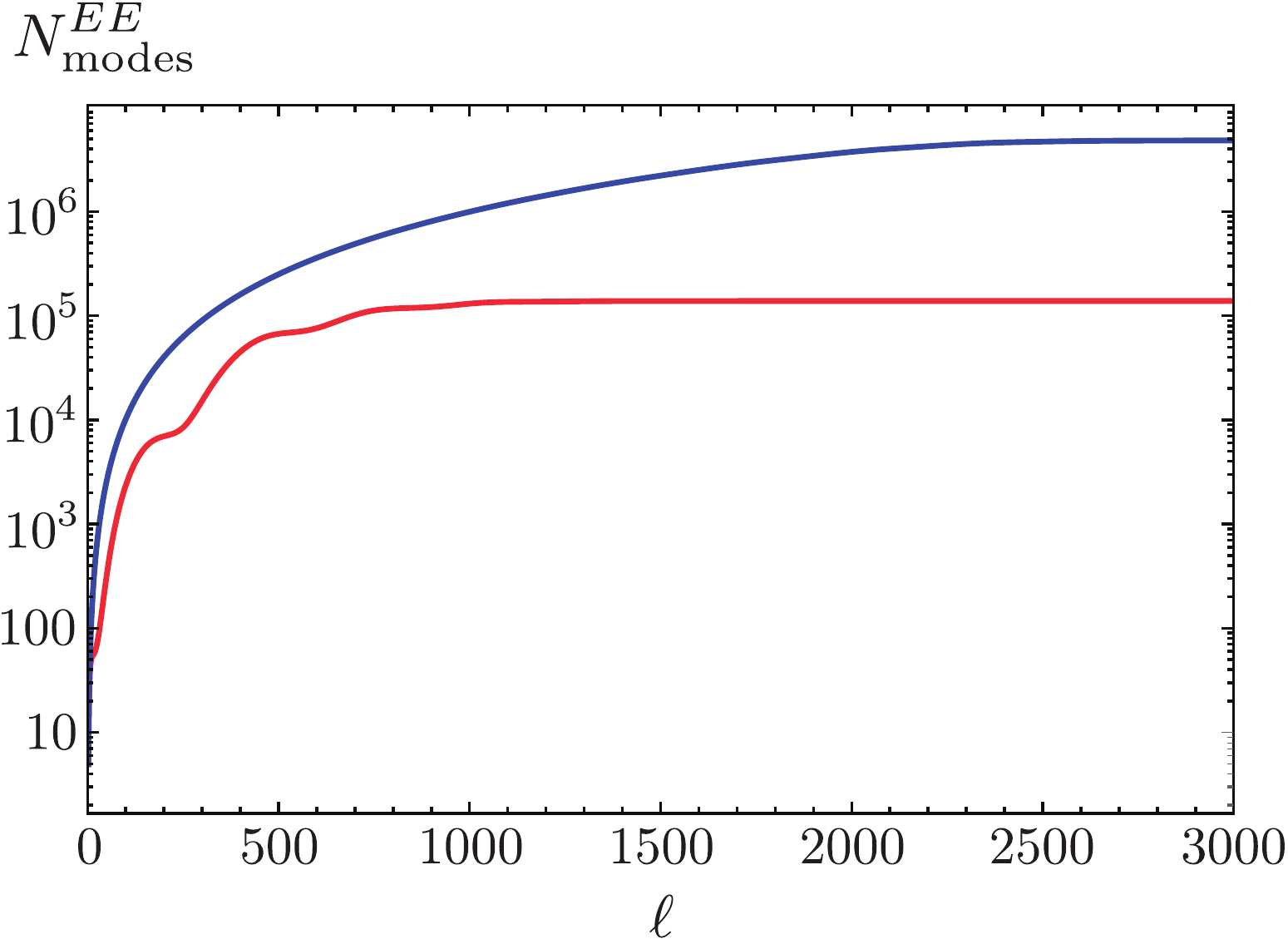}
\vspace{-1\baselineskip}
\end{center}
\caption{%We plot 
The effective number of modes defined in Eq. (\ref{numModes}) for $TT$ (left panel), $TE$ (middle panel), $EE$ (right panel) as a function of the multipole
number $\ell $ for Planck (red line) and CORE (blue line) up to $\ell \approx 3000.$
%We plot $[(c_\ell ^{AB}+n_\ell ^{AB})/c_\ell ^{AB}]^2$ as a function of the multipole 
%number $\ell $ for Planck (dashed) and CORE (solid) for $AB=TT,$ $TE,$ and $EE$ up to 
%$\ell \approx 3000.$ Below this approximate multipole number, the foreground-cleaned CMB sky is 
%dominated by the primary CMB anisotropies imprinted during inflation, as opposed to 
%secondaries anisotropies such as gravitational lensing that become dominant on very small 
%angular scales. A value near unity indicates that around the indicated multipole number the map is 
%of sufficient quality so that a less noisy map would contribute only a marginal amount of 
%additional statistically relevant information for constraining cosmological models. By 
%contrast, a value much below unity indicates that a less noisy map would contribute 
%substantial additional statistically relevant information around that multipole.
}
\label{coreVSplanck1}
\end{figure}

The effective number of modes measured by a given survey is
\begin{equation}
N_{\rm modes}=\sum _{\ell ,m} \left( \dfrac{C_\ell }{C_\ell +N_\ell }\right) ^2.
\label{numModes}
\end{equation}
For each of the power spectra $C_\ell^{TT},$ $C_\ell^{TE},$ and $C_\ell^{EE},$ we % may 
determine how many new modes 
will be measured by the various configurations of CORE compared to the existing data from 
\Planck . This analysis makes sense when we already have a good measurement of underlying power spectra. 
We therefore restrict to $C_\ell^{TT},$ $C_\ell^{TE},$ and $C_\ell^{EE},$ since the primordial $C_\ell^{BB}$ 
from inflationary gravitational waves is still unknown.
%We do not apply this analysis the primordial $C_\ell^{BB}$ from inflationary gravitational waves, for which at present the 
%theoretical target is highly uncertain---that is, uncertain
%to more than an order of magnitude. This type of 
%analysis makes sense only when we already have a good idea of the order of magnitude of 
%the underlying power spectra. For $C_\ell^{TT},$ $C_\ell^{TE},$ $C_\ell^{EE},$ thanks to Planck, we 
%already know these power spectra to within a good accuracy. The additional number of new 
%modes measured by CORE as opposed to those already measured by Planck is given by the 
%expression: 
%\begin{equation} 
%N_{modes}^{(NEW)}=\sum _{\ell ,m} \left[ \left( 
%\Delta N_{\rm modes} =\sum _{\ell ,m} \left[ \left( 
%\dfrac{C_\ell }{C_\ell +N_\ell ^{\rm CORE} }\right) ^2 - \left( \dfrac{C_\ell }{C_\ell 
%+N_\ell ^{Planck} }\right) ^2 \right]. 
%\end{equation}
%Fig.~\ref{coreVSplanck} plots the signal-to-noise per multipole, equal to $[C_\ell 
%/(C_\ell +N_\ell )]^2$ (with cosmic variance included) for the TT, TE, and EE spectra for 
%Planck and for the various configurations of CORE up to $\ell \approx 3000,$ 
%beyond where other effects start to dominate. We observe that for measuring the EE power spectrum, 
%a lot remains to be gained from a more sensitive survey, especially at high $\ell $ near 
%the threshold ($\ell \approx 3000$) where other effects such as gravitational lensing, the Sunyaev-Zeldovich effect, etc.~start to take over. 
Fig.~\ref{coreVSplanck1} plots the effective number of modes for the $TT$, $TE$, and $EE$ spectra for \Planck\ and CORE up to 
$\ell \approx 3000$. 
We observe that for measuring the $EE$ power spectrum,
a lot remains to be gained from a more sensitive survey, especially at high $\ell$.
We also note that interesting room remains % some scope still remains 
for improving our knowledge of the $TT$ power spectrum on smaller scales, but 
our ability to remove secondary anisotropies and foreground residuals rather than the noise is 
the limitation to extracting new information regarding primordial anisotropies in this region.
CORE has a great potential to enhance our knowledge of extragalactic sources \cite{DeZotti:2016qfg} 
and therefore to better characterize the CMB high-$\ell$ damping tail of temperature anisotropies.
%See the companion paper of this series \cite{DeZotti:2016qfg} on the potential of CORE to enhance
%our knowledge of extragalactic sources. 
See also Ref.~\cite{Wu:2016hej} for a recent paper studying the CORE capabilities to
advance our understanding of the anisotropies of the 
cosmic far-infrared background.
%for a recent paper dedicated to cosmic far-infrared background anisotropies with CORE.

%but it is the limitation of our ability to remove the other contaminants to the primordial signal 
%rather than cosmic variance that limits our ability to extract new information regarding 
%the primordial anisotropies.

The previous discussion has emphasized the scientific objective of measuring the 
primordial power spectrum. But this style of analysis can also be applied to forecasting 
how much various non-Gaussian analyses will improve when the data from CORE is 
used instead of the less accurate existing Planck data. As a concrete example, 
consider the constraints on local bispectral non-Gaussianity as predicted by many 
non-minimal inflationary models such as those having more than one scalar field. An 
approximate analytic expression for the information constraining the parameter 
$f^{\rm local}_{\rm NL}$ in a sky map spanning the multipole range $[\ell _{\rm min}, \ell _{\rm max}]$ 
at a signal-to-noise ratio equal to or larger than one is given by \cite{Bucher:2009nm}
\begin{equation}
O(1)~\left( f^{{\rm local} }_{\rm NL}\right) ^2~\ell^2_{\rm max} \ln \left( \dfrac{\ell _{\rm max}}{\ell _{\rm min}}\right) 
\end{equation}
where the presence of the logarithmic factor emphasizes the importance of full sky 
coverage as one typically obtains from a space-based experiment. The bottleneck for 
improving on Planck arises from the ${\ell _{\rm max}}^2$ factor, which is proportional to the 
number of modes as defined in 
Eq.~(\ref{numModes}).

The purpose of this paper is to study the capabilities of the ESA M5 mission proposal 
CORE and to compare them with those obtained from other designs for CMB space missions, such as the 
JAXA LiteBIRD configuration, the LiteCore designs, and the CORE+ proposal 
submitted for the previous ESA M4 mission call. This comparison sheds light on the role of 
different angular resolutions and raw detector sensitivities on the 
constraints on inflation to be expected from a future CMB space mission.

While space provides the most hospitable environment
for searching for primordial B modes, a number of ground based and balloon
experiments also seeking to detect B modes are now
either underway or in the planning stage. These
include Keck Array/BICEP 3/Bicep Array \cite{Grayson:2016smb}, Spider \cite{Fraisse:2011xz}, POLARBEAR-2 and the Simons Array \cite{Suzuki:2015zzg}, 
Advanced ACTpol \cite{Henderson:2015nzj}, SPT-3G \cite{Benson:2014qhw}, 
Piper \cite{Gandilo:2016sqn}, CLASS \cite{Harrington:2016jrz}, LSPE \cite{Aiola:2012sfi}, QUIJOTE
\citep{Rubino-Martin2012} and the US DOE Stage 4 (S4) experiment \cite{Abazajian:2016yjj}.
Although capable of a much finer angular resolution, 
ground based experiments must overcome a number of
limitations absent for experiments deployed in
space, such as atmospheric emission and absorption, ground pickup through
beam far sidelobes, unstable time-varying observing conditions and limitations to sky and frequency
coverage. In particular, the channels most difficult to access
from the ground are those at high frequencies where the polarized dust
emission is most intense. The role of polarized dust emission, first measured
by \Planck\ at high frequencies from space, has been shown to be of key importance not only
for the correct interpretation \cite{Ade:2015tva} of the B-mode detection by BICEP 2 \cite{Ade:2014xna}, but also for a more accurate
determination of the optical depth from E-mode polarization \cite{Ade:2013kta,Aghanim:2015xee,Aghanim:2016yuo,Adam:2016hgk}. 

We stress however the complementarity of CORE and S4. One example of such
complementarity is a more efficient delensing
of the primordial B modes. Although CORE will be able to delens using its own data
alone, combining with data from S4 hold promise to delens down to lower values of $r.$ Likewise, the maps
from CORE at frequencies inaccessible from the ground are likely to provide invaluable information for the S4
analysis.

The organization of the paper is as follows. Section 2 presents the connection 
between inflation and fundamental physics. %This Section describes the theory behind the observables both in the scalar and the tensor sectors as well as consistency tests. 
Section 3 describes the methodology used for forecasting the performance of CORE and of other alternative configurations.
%the forecasts reported in this paper. 
%In particular the models used for forecasting the performance of CORE and of other alternative 
%configurations are described in detail. 
%Sections 4 and 5 describe the constraints would 
%be obtained from measuring the scalar and tensor contributions to the CMB power spectra 
%in temperature and polarization. 
In Section 4, the forecasts for key inflationary parameters such as the scalar tilt and its scale dependence, 
the tensor-to-scalar ratio, and the spatial 
curvature are presented. The expected constraints on slow-roll parameters and a Bayesian comparison among slow-roll inflationary parameters are discussed in Section 5.
In Section 6, the forecasts for spectrum reconstruction with CORE are explored. 
Here a nonparametric analysis attempting to find statistically significant
features in the primordial power spectrum is considered.
Section 7 discusses tests of the adiabaticity of the primordial fluctuations
based on searching for primordial isocurvature modes.
The expected constraints on primordial non-Gaussianities are studied in Section 8.
%Section 8 deals with constraints beyond the 
%two-point function---in other words, non-Gaussian predictions of non-standard inflationary models. 
%Section 7 discusses tests of the adiabaticity of the primordial fluctuations 
%based on searching for primordial isocurvature modes. 
Section 9 forecasts the expected constraints on topological defect models.
%In Section 9 power spectrum reconstruction 
%is discussed. Here a non-parametric analysis attempting to find statistically significant 
%features in the primordial power spectrum is explored. 
Finally, Section 10 presents some 
conclusions.

\section{Inflation and fundamental physics}
\label{sec:theory}

At the current level of sensitivity, the initial conditions of the universe 
are described by just two numbers: the amplitude of primordial curvature perturbations
$A_{\mathrm s}$ and its spectral index $n_{\mathrm s}$. 
Moreover, the form of the power spectrum follows from the weakly broken scaling symmetry of the inflationary spacetime and is therefore rather generic. For these reasons,
it is hard to extract detailed information about the microphysical origin of inflation from current observations.
With future observations, we hope 
to detect extensions of the simple 
two-parameter description of the initial conditions (see Table~\ref{tab:parameters}).  As we will describe in this section, these observations have the potential to reveal much more about the physics of the inflationary era.  

\begin{table}[b!]
\begin{center}
 \begin{tabular}{cllc} 
  \toprule
 Parameter & Meaning & Physical Origin & Current Status	\\				
  \midrule[0.065em]
%	$A_{\rm s}$ &Scalar amplitude & $H$, $\dot H$, $c_{\rm s}$& $(2.14\pm 0.05)\times 10^{-9}$\\
%		$n_{\rm s}$ &Scalar tilt & $\dot H$, $\ddot H$, $\dot c_{\rm s}$ & $0.960\pm 0.007$\\
       $A_{\rm s}$ &Scalar amplitude & $H$, $\dot H$, $c_{\rm s}$& $(2.13\pm 0.05)\times 10^{-9}$\\ 
               $n_{\rm s}$ &Scalar tilt & $\dot H$, $\ddot H$, $\dot c_{\rm s}$ & $0.965 \pm 0.005$\\
  \midrule
	$\mathrm{d}n_\mathrm{s}/\mathrm{d}\ln k$ & Scalar running & $\dddot H$, $\ddot c_{\rm s}$  & only upper limits \\
		$A_{\rm t}$ & Tensor amplitude & $H$ & only upper limits\\
		$n_{\rm t}$ & Tensor tilt & $\dot H$ & only upper limits\\
			$r$ & Tensor-to-scalar ratio & $\dot H$, $c_{\rm s}$  & only upper limits \\
			$\Omega_{\rm k}$ &Curvature &Initial conditions & only upper limits\\
			$f_{\rm NL}$ &Non-Gaussianity &Extra fields, sound speed, $\cdots$ & only upper limits \\
			$S$ &Isocurvature &Extra fields & only upper limits \\
			$G\mu$ &Topological defects &End of inflation& only upper limits \\
	  \bottomrule 
 \end{tabular}
\caption{Summary of key parameters in inflationary cosmology, 
together with their likely physical origins and current observational constraints. 
At present, only upper limits exist for all parameters except $A_{\rm s}$ and~$n_{\rm s}$~\cite{Ade:2015lrj}. }
\label{tab:parameters}
\end{center}
\end{table}

\subsection{Physics of inflation}
\label{sec:physics}

\subsubsection{Ultraviolet sensitivity}

It is rare that our understanding of Planck-scale physics matters for the description of low-energy phenomena. This is because even large changes in the couplings to Planck-scale degrees of freedom usually have small effects on observables at much lower energies. Quantum gravity is irrelevant (in the technical sense) for experiments at the LHC.  It is therefore a remarkable feature of inflation that it is sensitive to the structure of the theory at the Planck scale.  Order-one changes in the interactions with Planck-scale degrees of freedom generically have significant effects on the inflationary dynamics. 
As we will see, this ultraviolet (UV) sensitivity is especially strong in models with observable levels of gravitational waves.
Writing down a theory of inflation therefore requires either making strong assumptions about the UV embedding, or formulating the theory in a UV-complete framework.
On the other hand, the UV sensitivity of inflation is also an opportunity to learn about the nature of quantum gravity from future cosmological observations. 

\subsubsection{Inflation in string theory}

String theory remains the most promising framework for addressing the issues raised by the UV sensitivity of the inflationary dynamics.  The question of consistency with quantum gravity is particularly pressing in models of large-field inflation with observable levels of gravitational waves (see Sec.~\ref{ssec:r}).  Effective field theory (EFT) models of large-field inflation have to assume protective internal symmetries for the inflaton field in order to forbid dangerous UV corrections.
Such symmetries are generically broken in quantum gravity, so it is unclear whether the success of an EFT realization of large-field inflation survives its embedding into a UV-complete framework. In string theory these abstract questions can 
in principle be addressed by concrete computations. One of the main advances in the field were the first semi-realistic models 
of large-field inflation in string theory~\cite{Silverstein:2008sg, McAllister:2008hb} 
(see also~\cite{Marchesano:2014mla, Dimopoulos:2005ac}).  Although work remains to be done to scrutinize the details 
of these models, they provide the first concrete attempts to study the UV sensitivity of large-field inflation directly. 

\subsubsection{Inflation in supergravity}

Another interesting question concerns the realization of inflation in supergravity, which is 
an intermediate platform between top-down string theory  and bottom-up effective field theory approaches. The possible inflationary dynamics and couplings to other fields are then restricted by local supersymmetry. At the two-derivative level, the scalar fields of ${\cal N}=1$ supergravity span a so-called K\"{a}hler manifold, while the potential energy is dictated by an underlying superpotential. 

The literature of inflation in supergravity is vast, so we here restrict ourselves to a few comments on recent developments. In recent years, various ways have been found to realize 
the Starobinsky model~\cite{Starobinsky:1980te} in supergravity~\cite{Ellis:2013xoa, Kallosh:2013xya, Buchmuller:2013zfa, Farakos:2013cqa, Ferrara:2013wka} providing a natural target of future B-mode searches. Moreover, a large class of supergravity models was shown to draw its main properties from the K\"{a}hler geometry. For instance, the stability of inflationary models is determined by specific components of the scalars curvature, see e.g.~\cite{Covi:2008ea, Covi:2008cn}. Similarly, their inflationary predictions follow from the curvature rather than by their potential. The latter type of models are referred to as $\alpha$-attractors \cite{Kallosh:2013yoa}, and give a spectral index in excellent agreement
with the latest Planck data. Moreover, the level of tensor modes is directly 
related to the curvature $R_K$ of the hyperbolic manifold (in Planck units)
\begin{align}
r = \frac{8}{(-R_K) N^2} 
\end{align}
where $N = O(60)$ indicates
the total number of $e$-folds of the observable part of the inflationary epoch.  
More generally, bounds on the curvature tensor of the K\"ahler manifold \cite{Covi:2008ea, Covi:2008cn} 
imply an interesting constraint on
the sound speed $c_s \gtrsim 0.4$ for ${\cal N}=1$ supergravity models in which  a single chiral superfield evolves during inflation, suggesting that this scenario can be constrained by a measurement of non-Gaussianity~\cite{Hetz:2016ics}.

Finally, in addition to linearly realized supersymmetry, non-linear realizations have been proposed.
Non-linear realizations offer a number of phenomenological simplifications, 
such as the absence of possibly tachyonic directions~\cite{Ferrara:2014kva}. 
They can be regarded as effective descriptions when supersymmetry is spontaneously 
broken, as occurs during inflation. Examples of supersymmetric effective field theories of inflation 
are~\cite{Allahverdi:2006iq,Allahverdi:2006we,Baumann:2011nk,Delacretaz:2016nhw}. 
%Recent examples of supersymmetric EFTs of inflation are~\cite{Baumann:2011nk,Delacretaz:2016nhw}.

\subsubsection{Inflation in the Standard Model}

To date, only one scalar field has been observed directly: the Standard Model
(SM) Higgs field. Simplicity compels us to consider this as a possible inflaton candidate.
Confined to SM interactions alone, the potential of the Higgs singlet is never flat enough to inflate. However, if the Higgs couples non-minimally to gravity, then the kinetic mixing with the graviton results in an exponentially flat potential at large enough field values~\cite{Bezrukov:2007ep}. 
With enough assumptions about running in the intermediate field regime, one then makes predictions for CMB observables in terms of SM parameters at low energies~\cite{Bezrukov:2007ep, DeSimone:2008ei}. The tensor-to-scalar ratio predicted by these models is $r \sim 10^{-3}$.

\subsubsection{Inflation in effective field theory}

The most conservative way of describing inflationary observables is in terms of an effective theory of adiabatic fluctuations~\cite{Cheung:2007st}.  Given an expansion history defined by the time-dependent Hubble rate $H(t)$, fluctuations are described  
by the Goldstone boson $\pi$ associated with the spontaneously broken time translation symmetry. In the absence of additional light degrees of freedom, the Goldstone boson is related in a simple way to the comoving curvature perturbation 
${\cal R} = -H \pi$.  At quadratic order in fluctuations and to lowest order in derivatives, the effective action for $\pi$ contains two time-dependent parameters: $\epsilon_1(t) \equiv - \dot H/H^2$ and $c_{\rm s}(t)$. 
The latter characterizes the sound speed of $\pi$ fluctuations.
The amplitude of the power spectrum of curvature perturbations then is
\begin{equation}
A_{\rm s} = \frac{1}{8\pi^2} \frac{1}{\epsilon_1 c_{\rm s}} \frac{H^2}{M_{\rm pl}^2}\, .
\end{equation}
The near scale-invariance of the power spectrum requires the time dependence of $\epsilon_1(t)$ and $c_s(t)$ to be mild.  
Interestingly, the nonlinearly realized time translation symmetry relates a small value 
of $c_{\rm s}$ to a cubic operator in the action for $\pi$, leading to enhanced non-Gaussianity of the fluctuations 
with $f_{\rm NL} \sim c_{\rm s}^{-2}$ (see Sec.~\ref{ssec:NG}).  Additional higher-order operators of the effective action for $\pi$ are associated with additional free parameters.

\subsection{Tensor observables}

The most important untested
prediction of inflation concerns the existence of tensor modes, 
arising from quantum fluctuations of the metric. A detection would have a 
tremendous impact as it would be the first experimental signature of quantum gravity. 

\subsubsection{Tensor amplitude}
\label{ssec:r}

The amplitude of inflationary tensor modes is typically expressed in terms of the
tensor-to-scalar ratio $r\equiv A_{\rm t}/A_{\rm s}$. 
The parameter $r$ provides a measure of the expansion rate during inflation
\begin{align}
H &= \SI{7.2e12}{GeV} \left(\frac{r}{0.001}\right)^{1/2}\, ,
\end{align}
which can be related to the energy scale of inflation, $\rho^{1/4} = \SI{6.1e15}{GeV}\, ({r/0.001})^{1/4}$.  The observation of primordial tensor modes at the level $r>0.001$ would therefore associate inflation with physics at the GUT scale.

\vskip 4pt 
Although there is no definitive prediction for the magnitude 
of $r$, there exist simple arguments why $r >0.001$ is a theoretically interesting observational target:
\begin{itemize}
\item Famously, for inflationary models driven by a fundamental scalar field, the value of $r$ is related to the total field excursion~\cite{lyth1997would, Baumann:2006cd}
\beq
\frac{\Delta \phi}{\Mp} \approx \frac{N}{90} \left( \frac{r}{0.001} \right)^{1/2} \, .
\eeq
An observation of tensor modes above the per thousand level would therefore imply large-field inflation with super-Planckian field excursions.

\item A well-motivated ansatz for the inflationary observables, satisfied by a large class of inflationary models, corresponds to an expansion in $1/N$ with leading terms \cite{Mukhanov:2013tua, Roest:2013fha, Garcia-Bellido:2014gna, Garcia-Bellido:2014wfa, Creminelli:2014nqa}
\begin{align}
n_{\rm s} = 1 - \frac{p}{N} \,, \quad r = \frac{r_0}{N^q} \,.
\end{align}
Interestingly, this simple scaling leads to two universality classes. The first has $q=1$ and $r_0 = 8 (p-1)$, comparable to quadratic inflation which is already under serious tension and will be probed further with ground-based experiments. The second has $p=q$, comparable to Starobinsky inflation and $\alpha$-attractors~\cite{Starobinsky:1980te, Kallosh:2013yoa} and leads to a per thousand level of~$r$. \end{itemize}
The above two arguments make the range between $10^{-3}$ and $10^{-2}$, which includes a variety of specific models \cite{Martin:2013tda}, a theoretically interesting regime for the tensor-to-scalar ratio.

\subsubsection{Tensor tilt}

In the event of a detection of primordial tensor modes, it will be interesting to probe the scale dependence of the spectrum.
In standard single-field slow-roll inflation, the tensor tilt satisfies a consistency relation, $n_{\rm t} = -r/8$. Unfortunately, this makes the expected tensor tilt too small to be detectable with future CMB experiments. Nevertheless, it remains interesting to look for larger deviations from the consistency conditions. It would be striking to find a blue tensor tilt 
$n_{\rm t} >0$, which in the context of inflation would require a violation of the 
null energy condition.\footnote{A blue tensor tilt can arise if large 
curvature corrections during inflation lead to a tensor sound speed 
with non-trivial evolution~\cite{Baumann:2015xxa}.  In Einstein frame, this effect would correspond to a stable 
violation of the null energy condition as discussed in \cite{Creminelli:2014wna, Creminelli:2006xe} (see also \cite{Cai:2015yza,Cai:2016ldn}). 
However, it is hard to make this effect large without awakening ghosts in the effective theory. 
Alternatively, a blue tensor tilt can also arise in models with a non-standard spacetime symmetry 
breaking pattern, such as solid \cite{Endlich:2012pz} or supersolid \cite{Bartolo:2015qvr} inflation. 
In these scenarios, spatial reparameterization invariance is spontaneously broken during inflation, by 
means of background fields with space-dependent vacuum expectation values. Since spatial diffeomorphism
invariance is broken, there is no symmetry preventing the  tensor modes from  acquiring an effective mass during 
inflation \cite{Cannone:2014uqa}.  The graviton  mass, if sufficiently large,  can lead to a blue spectrum 
for primordial tensor modes, as explicitly shown in concrete realizations in \cite{Endlich:2012pz,Bartolo:2015qvr}. 
After inflation ends, the field configuration can rearrange itself so as to recover space reparameterization 
symmetry and set the effective graviton mass equal to zero.}

\subsubsection{Graviton mass}

If the graviton has a mass $m_g$, the dispersion relation of gravitational waves is modified: $\omega^2 = k^2 +m_g^2$. 
For masses comparable to the Hubble rate at recombination this has a significant effect on the B-mode spectrum. In that case, the tensor mode oscillates on superhorizon scales which adds power to the B-mode spectrum on large angular scales ($\ell < 100$).
Observing primordial B-modes but not finding this excess in large-scale power 
would put a strong constraint on the graviton mass, $m_g < \SI{e-30}{eV}$ \cite{Dubovsky:2009xk} (compared to $m_g < \SI{1.2e-22}{eV}$ from LIGO~\cite{TheLIGOScientific:2016src}).

\subsubsection{Non-vacuum sources}

So far we have assumed that the primordial tensor modes are generated by vacuum fluctuations.  
In principle, there could also be tensor modes produced by non-vacuum fluctuations, 
such as the fluctuations that could arise from particle production during inflation~\cite{Senatore:2011sp, Mirbabayi:2014jqa, Cook:2011hg, Ozsoy:2014sba}. One may be concerned that this could destroy the relationship between the size of $r$ and the energy scale of inflation. However, it has been shown that non-vacuum fluctuations cannot be parametrically larger than vacuum fluctuations without violating bounds on primordial non-Gaussianity~\cite{Mirbabayi:2014jqa}. The B-mode amplitude therefore remains a good measure of the energy scale of inflation.  

\subsubsection{Non-Gaussianity}

If the amplitude of primordial tensors is large, it may become feasible to study higher-order correlators involving tensor modes. 
Of particular interest is the tensor-scalar-scalar bispectrum~$\langle h {\cal R} {\cal R} \rangle$. 
In single-field inflation, the form of the squeezed limit of this bispectrum is fixed by the fact 
that a long 
wavelength tensor fluctuation is locally equivalent to a spatially 
anisotropic coordinate transformation~\cite{Maldacena:2002vr}. 
This consistency condition is more robust that the corresponding consistency condition for the scalar correlator 
$\langle {\cal R} {\cal R}{\cal R} \rangle$ in the sense that it cannot be violated 
by the presence of additional scalar fields. 
Observing non-analytic corrections to the consistency condition for $\langle h {\cal R} {\cal R} \rangle$ 
(e.g., through a measurement of $\langle BTT \rangle$~\cite{Meerburg:2016ecv}) would be a signature of extra 
higher-spin particles during inflation~\cite{Arkani-Hamed:2015bza, Lee:2016vti} 
or of a different symmetry breaking pattern in the EFT 
of inflation~\cite{Bordin:2016ruc, Endlich:2012pz, Gruzinov:2004ty, Bartolo:2015qvr, Maleknejad:2011jw,Adshead:2012kp}.\footnote{Models with space-dependent background fields have a symmetry breaking pattern different from 
standard single-field inflation and can lead to scenarios in which inflation is not a strong isotropic 
attractor. In this case, anisotropies are not diluted exponentially fast by inflation.
This implies that tensor fluctuations are not adiabatic during inflation and 
that violations of  consistency relations in the 
tensor-scalar-scalar correlation functions are generally expected~\cite{Bordin:2016ruc}. In these models, 
the squeezed limit for the $\langle h {\cal R} {\cal R} \rangle$ three-point function can have an amplitude 
much larger than in standard inflationary scenarios \cite{Akhshik:2014bla,Bartolo:2015qvr}. 
In addition, such non-standard  behavior for the  tensor modes  leads to indirect, distinctive 
consequences for correlation functions of the scalar perturbation, namely a 
quadrupolar contribution to the scalar two-point function $\langle  {\cal R} {\cal R} \rangle$ 
and a direction-dependent contribution to the counter-collinear limit of the four-point function $\langle  {\cal R} {\cal R}  {\cal R} {\cal R} \rangle$~\cite{Bordin:2016ruc}.}

\subsection{Scalar observables}

Future CMB observations will improve constraints on primordial scalar fluctuations through precision measurements of the damping tail of the temperature anisotropies and the polarization of the anisotropies. In this Section, we will describe what can be learned from these measurements.

\subsubsection{Running}

\begin{SCfigure}
\includegraphics[width=.55\textwidth]{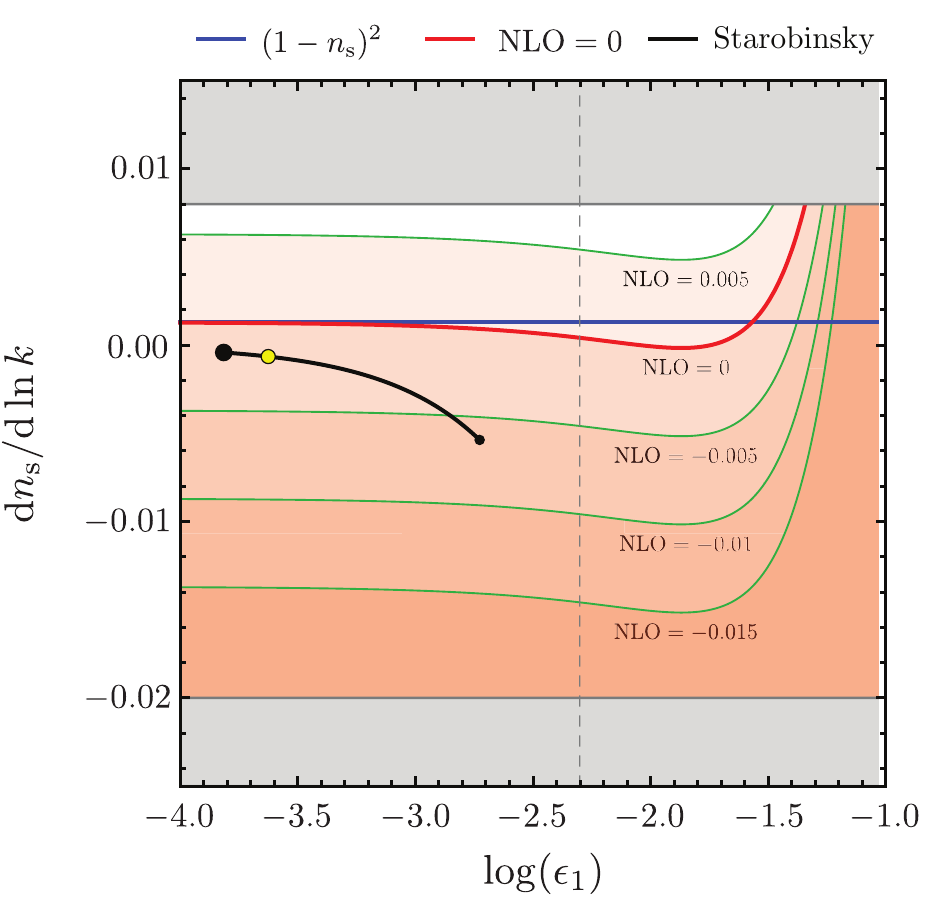}
\caption{\footnotesize
The plot \cite{Cabass:2016giw} shows the running $\mathrm{d}n_\mathrm{s}/\mathrm{d}\ln k$ as function of 
$\epsilon_1$ for different values of the NLO slow-roll parameters. Notice that the uncertainty in $\ns$ is smaller than the thickness of the lines in the plot. In red we show $\mathrm{d}n_\mathrm{s}/\mathrm{d}\ln k$ for $\mathrm{NLO} = 0$, while the blue line is its asymptotic value $(1-\ns)^{2}\approx 0.0013$. The black line  
shows the predictions of the Starobinsky model \cite{Starobinsky:1980te} (with $N$ going from $20$ to $70$), with the yellow dot being its prediction for $N = 56$ (chosen to reproduce the observed value of $\ns$). The gray bands show the values of $\mathrm{d} n_\mathrm{s}/\mathrm{d}\ln k$ excluded (at $\limit{95}$) by \emph{Planck} $TT$, $TE$, $EE$ + lowP data, while the gray dashed vertical line shows the current bound on $\epsilon_1 = r/(16c_\mathrm{s})$ assuming $c_\mathrm{s} = 1$.}
\label{fig:NLOplot}
\end{SCfigure}

Assuming that deviations of the primordial power spectrum from a perfect power law are small, they can be parameterized with the running of the spectral tilt $\mathrm{d}n_\mathrm{s}/\mathrm{d}\ln k$.
%\beq
%\nrun \equiv
% \left.\frac{d\ns}{d\ln k}\right|_{k=k_\star}\, .
%\eeq
%\begin{equation}
%\Delta^2_\zeta(k) = A_\mathrm{s} \left( \frac{k}{k_\star} \right)^{\ns-1 + \frac{1}{2}\nrun \log(k/k_\star) }\,.
%\end{equation}
Current constraints from Planck give $\mathrm{d}n_\mathrm{s}/\mathrm{d}\ln k = -0.0057^{+0.0071}_{-0.0070}\, \text{($\limit{68}$)}$~\cite{Ade:2015lrj}. The standard prediction of single-field inflation is conveniently written in terms of observables as \cite{Cabass:2016giw}
\bea
%  1 - \ns &=& 2\epsilon-\frac{\epsilon_{,N}}{\epsilon}-\frac{c_{\mathrm{s},N}}{c_\mathrm{s}} \label{eq:eom1} \\
%     &=& \frac{r}{8 c_\mathrm{s}} - \frac{r_{,N}}{r}\,\,, \label{eq:eom2} \\
 %    \nrun &=& 2\epsilon_{,N}-\frac{r_{,NN}}{r} + \bigg(\frac{r_{,N}}{r}\bigg)^{2}\,\,, 
\frac{\mathrm{d}n_\mathrm{s}}{\mathrm{d}\ln k} = (1-\ns)^{2} - 6\epsilon_1 (1-\ns) + 8\epsilon_1^{2} - \left( \frac{rs}{8c_\mathrm{s}}+\frac{r_{,NN}}{r} \right) 
\label{vicina}
\eea
where $r=16\epsilon_1 c_\mathrm{s}$, $ s\equiv c_{\mathrm{s},N}/c_\mathrm{s} $ and $\ast_{,N}$ refers to a derivative with respect to the number of $e$-foldings from the end of inflation, namely $H\dif t=-\dif N$. Barring cancellations, one expects $\mathrm{d}n_\mathrm{s}/\mathrm{d}\ln k\sim (1-\ns)^{2}$. While $ \epsilon_1$ is related to $ r $ once $ c_{s} $ is known or constrained and $ \ns $ is measured, the last two terms in (\ref{vicina}) make their first appearance in $\mathrm{d}n_\mathrm{s}/\mathrm{d}\ln k$. In this sense they are next-to-leading order (NLO) parameters
\bea
\mathrm{NLO}\equiv\frac{rs}{8c_\mathrm{s}} + \frac{r_{,NN}}{r} \ \xrightarrow{\ c_\mathrm{s} = 1\ } \ \frac{\epsilon_{1,NN}}{\epsilon_1}\,.
\eea
Since $ \ns $ is relatively well constrained, it is convenient to summarize current and future constraints in terms of $ \epsilon_1$ and $\mathrm{d}n_\mathrm{s}/\mathrm{d}\ln k$ as in Fig.~\ref{fig:NLOplot} (from \cite{Cabass:2016giw}).

%%%%%%%%%%%%%%%%%%%%%%%%%%%%%%%%%%%%%%%%%%%
\vskip 4pt
Deviations from a power-law behavior can be expanded one order further to include the running of the running $\mathrm{d}^2 n_\mathrm{s}/\mathrm{d}\ln k^2$. Current constraints
from Planck give  $\mathrm{d}^2 n_\mathrm{s}/\mathrm{d}\ln k^2=0.025\pm0.013$ ($\limit{68}$)~\cite{Ade:2015lrj}.  A potential detection of the running of running with CORE would be in conflict with the single-field, slow-roll paradigm, which generally predicts a much smaller value of order~$ (1-\ns)^{3}$.

\subsubsection{Non-Gaussianity}
\label{ssec:NG}

In standard single-field slow-roll inflation, the flatness of the inflaton potential constrains the size of interactions in the inflaton fluctuations. These inflaton fluctuations are therefore expected to be highly Gaussian. Significant non-Gaussianity in the initial conditions can nevertheless arise in simple extensions of the standard single-field slow-roll paradigm. 

\paragraph{Sound speed} 

Higher-derivative inflaton interactions have been proposed as a mechanism to slow down the inflaton evolution even 
in the presence of a steep potential~\cite{Alishahiha:2004eh}. A consequence of these interactions are a reduced 
sound speed\footnote{In Dirac-Born-Infeld (DBI) inflation~\cite{Alishahiha:2004eh} significant reductions in the speed of sound are 
possible and radiatively stable due to a nonlinearly realized Lorentz symmetry protecting the structure of the DBI action.} 
for the inflaton interactions which leads to a significant level of non-Gaussianity peaked in the equilateral 
configuration~\cite{Chen:2006nt}. In the framework of the EFT of inflation~\cite{Cheung:2007st}, the relation 
between small $c_{\rm s}$ and large $f_{\rm NL} \sim c_{\rm s}^{-2}$ is a consequence of the nonlinearly realized 
time translation symmetry. Current constraints on equilateral non-Gaussianity imply $c_{\rm s} > 0.024$~\cite{Ade:2015ava}. 
This is still an order of magnitude away from the unitarity bound derived in~\cite{Baumann:2015nta}.

CORE will also constrain the closely related class of single-field slow-roll models in which the speed of sound of the inflaton fluctuations 
is not much less than unity and slowly varying.\footnote{This is expected,
for instance from integrating out heavy degrees of freedom in a UV
completion in which the background is protected by some internal
symmetry softly broken by the inflationary potential if the
inflaton is a pseudo-Goldstone boson with derivative interactions to
these fields. 
This happens automatically in multi-scalar models if the group orbits are curved trajectories in the sigma model metric.}
 In general, these models have lower values of $r = 16
\epsilon_1 c_{\rm s}$ than their $c_{\rm s} = 1$ counterparts, but the reductions
are moderate. If the speed of sound is approximately constant,
unitarity implies a lower bound $c_s \gtrsim 0.3 $ in the absence
of protective symmetries~\cite{Baumann:2014cja,Baumann:2015nta}.  Probing $r$ down to
$10^{-3}$ will constrain, and in some cases rule out any such
deformations of the `vanilla' slow-roll models~\cite{Achucarro:2015rfa}. 

\paragraph{Extra fields} Non-Gaussianity is a powerful way to detect the presence of extra fields during inflation.  The self-interactions of these fields are not as strongly constrained as those of the inflaton. Non-Gaussianity in these hidden sectors can then be converted into observable non-Gaussianity in the inflaton sector, e.g.~\cite{Enqvist:2001zp, Lyth:2001nq, Moroi:2001ct, Chen:2009zp, Baumann:2011nk, Noumi:2012vr, Arkani-Hamed:2015bza,Lee:2016vti}. By measuring the precise momentum scaling of the squeezed bispectrum, one can in principle extract the masses and spins of any extra particles present during inflation.
Since inflation excites all degrees of freedom with masses up to the inflationary Hubble scale (which may be as high as $10^{14}$ GeV), this potentially allows us to probe the particle spectrum far above the reach of terrestrial colliders.

\paragraph{Excited initial states}

By its very nature, inflation is very efficient at diluting initial inhomogeneities. Any traces of a pre-inflationary state in the CMB would either require inflation to have lasted not too much longer than 
required to solve the horizon problem, or to 
have the initial state contain excitations beyond 
the usual Bunch-Davies vacuum at arbitrarily short distances. This is
a problematic proposition for various reasons (see e.g.~\cite{Banks:2002nv, Kaloper:2002cs}). 
The effects of excited initial states were studied in an EFT analysis in
\cite{Holman:2007na, Agarwal:2012mq}. The signal in the bispectrum peaks in the flattened momentum 
configuration $k_1 \approx k_2 \approx 2 k_3$. This is because the flattened configuration is sensitive to the presence of higher-derivative interactions that were more relevant at early times.
 
 \subsubsection{Features}
 \label{sssec:features}

The presence of any localized bumps or oscillatory features in the
power spectrum or other correlation functions provides an interesting and powerful probe of deviations from
the simplest realizations of single-field slow-roll inflation.  A variety of
physical processes can generate spectral features at any time during
or after inflation (see \cite{Chluba:2015bqa} for a review). Most interesting for us are features that originated 
during inflation, as these can probe energy scales well beyond $H$~\cite{Achucarro:2010da, Flauger:2016idt} and may reveal new mass scales and interactions 
difficult to probe in any other way. The production of features
during inflation may or may not involve violations of the slow-roll dynamics, or adiabaticity, and may or may not involve particle production.

Small amplitude periodic or localized features in the couplings of the EFT of inflation 
($\epsilon_1(t)$, $c_{\rm s}(t)$, etc.) lead to modulated oscillatory features superimposed on the almost scale-invariant 
power spectrum, bispectrum, and higher order $n$-point functions. 
Because of their common origin, any such features are correlated~\cite{Chen:2006xjb} and the specific form of the correlations can be used to identify their origin and to improve their detectability~\cite{Achucarro:2012fd, Mooij:2016dsi}, in particular with joint searches in the power spectrum and bispectrum~\cite{Meerburg:2015owa}.  Periodic variations in $\epsilon_1(t)$  (for instance due to monodromy or small-scale structure in the scalar potential) can lead to a resonant enhancement of oscillatory features in the power spectrum or in the non-Gaussianity~\cite{Chen:2008wn, Flauger:2010ja}.
Abrupt  changes in $\epsilon_1$ and $c_{\rm s}$, typically associated with steps in the potential and other theoretically well motivated interruptions of slow roll (see e.g.~\cite{Hotchkiss:2009pj} and other examples discussed in \cite{Ade:2015ava}), can also lead to oscillations and an enhancement of the bispectrum.  In all cases it is important to check that the time and energy scales associated with the generation of the features are compatible with the use of the effective single-field or low energy description. 

\begin{figure}[t!]
\centering
\includegraphics[width=0.8\textwidth]{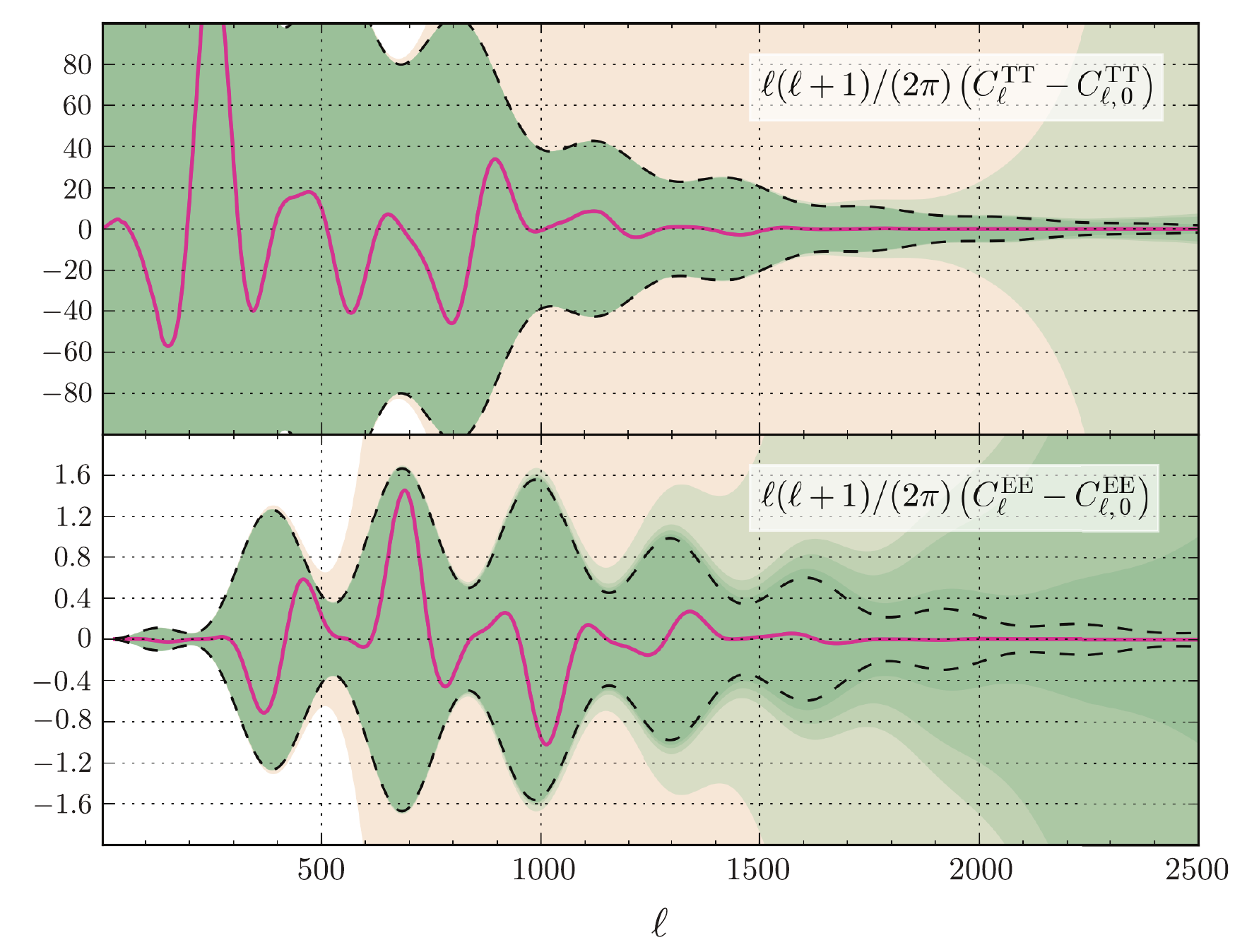}
\caption{
\label{fig:cl}
Example of a feature  in the CMB power spectrum relative to the corresponding $c_{\rm s} = 1$ 
featureless one $C_{\ell,0}$ due to a transient reduction in the speed of sound $c_{\rm s}$ 
in single-field slow-roll inflation. CORE's approximated $68\%$ error bars are represented in 
increasingly darker shades of green for the configurations in Tables~\ref{tab:CORE-bands} and \ref{tab:specifications} 
in Sec.~\ref{sec:three}. LiteBIRD's sensitivity corresponds to the orange-shaded region. 
The dashed line is the standard deviation due to cosmic variance. 
CORE's increased sensitivity in the polarization power spectrum is
cosmic variance limited up to 
$\ell\approx 1500.$ 
This power spectrum feature is accompanied by a correlated feature in the bispectrum, given in~(\ref{DeltaBispectrum}). }
\end{figure}

The superior constraining power of CORE is illustrated in Fig.~\ref{fig:cl} for the particular case of transient, moderate reductions in $c_{\rm s}$ away from $c_{\rm s}=1$ 
(a situation that is well-motivated theoretically and fully compatible with uninterrupted slow-roll and the single-field 
EFT description~\cite{Achucarro:2012fd, Achucarro:2014msa}). These features look like enveloped linear oscillations in $k$, 
both in the primordial power spectrum and the bispectrum, and approximately so in the respective CMB projections, persisting 
over a relatively large range of scales.  The power spectrum feature ${\Delta {\cal P}_{\cal R}} / {{\cal P}_{\cal R}}$ can 
be shown to be the Fourier transform of the reduction in $c_{\rm s}$:  $
{\Delta {\cal P}_{\cal R}} / {{\cal P}_{\cal R}}
=k\int_{-\infty}^0{\rm d}\tau\,(1-c_{\rm s}^{-2})\sin(2k\tau)
$.  
The primordial bispectrum is given by
\begin{align}
\Delta  B_{\cal R}(k_1, k_2, k_3) = \frac{(2\pi)^4 {\cal P}_{\cal R}^2}{(k_1k_2k_3)^2} \Bigg[ c_0(k_i) \frac{\Delta {\cal P}_{\cal R}}{{\cal P}_{\cal R}} \left(\frac{k_t}{2}\right) &+
  c_1(k_i) \frac{{\rm d}}{{\rm d} \log k_t} \frac{\Delta {\cal P}_{\cal R}}{{\cal P}_{\cal R}} \left(\frac{k_t}{2}\right)  \nonumber  \\[2pt]
  & +
  c_2(k_i) \left(\frac{{\rm d}}{{\rm d} \log k_t}\right)^2 \frac{\Delta {\cal P}_{\cal R}}{{\cal P}_{\cal R}} \left(\frac{k_t}{2}\right)  \Bigg]\,,
  \label{DeltaBispectrum}
\end{align}
where $k_t \equiv k_1+k_2+k_3$ and the coefficients $c_i$
are known functions \cite{Achucarro:2012fd} of $k_1$, $k_2$, $k_3$.

\subsubsection{Isocurvature}

The primordial seed perturbations for structure formation and CMB anisotropies could have been either
pure curvature perturbations (the adiabatic mode), cold dark matter, baryon, or neutrino density isocurvature perturbations
(i.e., spatial perturbations in the ratios of number densities of different 
particle species), neutrino velocity isocurvature perturbations or an arbitrarily
correlated mixture of some or all of these \cite{Bucher:1999re,Bucher:2000hy}. 
A detection of any type of isocurvature would be a smoking gun for multi-field inflationary
models and would rule out single-field models. However, this is 
a one way implication. Lack of an isocurvature contribution to the CMB anisotropies does not rule out multi-field inflation, since even
a large isocurvature contribution present immediately after inflation can be wiped out by later
processes \cite{Mollerach:1989hu,Weinberg:2004kf,Beltran:2005gr}. Planck has set tight upper bounds on the possible isocurvature 
contribution \cite{Planck:2013jfk,Ade:2015lrj}.
However, as we see in Section~\ref{sec:testing}, future CMB missions, in particular with CORE's sensitivity,
may improve the constraints by a factor of 5 in simple one-parameter isocurvature extensions to
the adiabatic $\Lambda$CDM model. Conversely, this means that there is a relatively large window for an
observation of a per thousand level isocurvature contribution. A CMB mission optimized for detecting or constraining
the primordial tensor-to-scalar ratio is also excellent for 
breaking the degeneracy between the large-scale isocurvature
and tensor contributions. This mitigates the possibility of misinterpreting a nearly scale-invariant isocurvature contribution
as a tensor contribution or vice versa. Finally, it is important to test how accurate and unbiased the determination of the
standard cosmological parameters is if instead of the usually assumed purely adiabatic primordial mode, we allow for more
general initial conditions of perturbations.

\subsection{Beginning and end of inflation}

\subsubsection{Spatial curvature}

False vacuum decay leads to pockets of space with negative spatial curvature~\cite{Coleman:1980aw}. 
The sign and the size of the curvature parameter $\Omega_{\rm k}$ are therefore interesting probes 
of the pre-inflationary state. Unfortunately, this will be hard to observe, given
the efficiency with which the inflationary expansion dilutes pre-existing inhomogeneities including curvature.  
Values of $|\Omega_{\rm k}| > 10^{-4}$ only survive if inflation did not last longer than the minimal duration
required to solve the horizon problem~\cite{Guth:1980zm}.

\subsubsection{Topological defects}

Topological defects can form in symmetry-breaking phase transitions in the early Universe \cite{Kibble:1976sj}. 
These transitions happen at the end of inflation in a class of models  
called hybrid inflation \cite{Linde:1993cn,Copeland:1994vg}.
Alternatively, defects can form in phase transitions during 
inflation \cite{Kamada:2014qta,Ringeval:2015ywa} or afterwards~\cite{Kibble:1976sj,Rajantie:2001ps}.
Searching for topological defects therefore yields information about the phase transition that formed them, 
the high-energy physics responsible for the phase transition, and ultimately can distinguish 
between models of inflation or even rule them out. 
The most commonly encountered defects in models of inflation-scale physics are 
cosmic strings (see \cite{Copeland:2011dx,Hindmarsh:2011qj} for reviews).

Certain topological defects, including cosmic strings,  produce a scale-invariant spectrum of density fluctuations 
and therefore are in principle compatible with the standard cosmology.  By making precise measurements of the CMB, 
we can check how large an
admixture of defects our Universe allows.  The CMB temperature and polarization power spectra are proportional 
to the fourth power of the symmetry breaking scale, and so the CMB constraints on 
the fractional contribution of defects $f_{10}$ (by convention the fraction is taken at multipole 
$\ell=10$ in the temperature power spectrum)
can be translated into constraints 
on that scale. As an example of the importance of searching for signals from defects, minimal D-term hybrid inflation is 
effectively ruled out because of its cosmic string contribution to the 
temperature power spectrum  \cite{Battye:2010hg}.\footnote{Hybrid inflation ends when a second ``waterfall" field becomes unstable,  triggering a symmetry-breaking phase 
transition. In  ${\cal N} =1$ supergravity models, both the inflaton and the waterfall field are complex scalars 
(belonging to ${\cal N} =1$ chiral supermultiplets) and therefore the defects formed are cosmic strings. 
In the D-term case, the mass per unit length of the strings is fixed by the scale of inflation, and this is enough to rule out the minimal model.}

An important feature of the perturbations coming from topological defects is that they create a B-mode polarization 
signal. This signal would be in competition with that produced by primordial gravitational waves. In order to analyze 
the B-mode polarization signal correctly, the possibility that topological defects are the seed of the B-modes must 
be included. It may also be that gravitational waves and defects can both 
seed B-modes,  and thus one has to be able to distinguish between them \cite{Urrestilla:2008jv}.

\section{Experimental configurations and forecasting methodology}
\label{sec:three}

{\small
\begin{table}[h]
\begin{center}
{\footnotesize
\begin{tabular}{|c|c|c|c|c|c|c|c|}
\hline
Channel & FWHM  &  $N_{\rm det}$  &  $\Delta T$  &  $\Delta P$  & $\Delta I$ & $\Delta I$  & $\Delta y\times 10^6$ \\ 
%&  PS  ($5\sigma$) \\
GHz     &       [arcmin] &    &  [$\mu K$ arcmin] &  [$\mu K$ arcmin] &  [$\mu K_{\rm RJ}$ arcmin]  & [kJy/sr arcmin] 
&  [$y_{\rm SZ}$ arcmin] \\
%& [mJy]  \\
\hline
\hline

60      & 17.87   & 48      & 7.5          & 10.6         & 6.81          & 0.75         & -1.5                            \\
70      & 15.39   & 48      & 7.1          & 10           & 6.23          & 0.94         & -1.5                    \\
80          & 13.52   & 48      & 6.8          & 9.6          & 5.76          & 1.13         & -1.5                 \\
90         & 12.08   & 78      & 5.1          & 7.3          & 4.19          & 1.04         & -1.2                 \\
100       & 10.92   & 78      & 5.0            & 7.1          & 3.90          & 1.2          & -1.2             \\
115       & 9.56    & 76      & 5.0            & 7.0            & 3.58          & 1.45         & -1.3            \\
130      & 8.51    & 124     & 3.9          & 5.5          & 2.55          & 1.32         & -1.2               \\
145      & 7.68    & 144     & 3.6          & 5.1          & 2.16          & 1.39         & -1.3                    \\
160      & 7.01    & 144     & 3.7          & 5.2          & 1.98            & 1.55         & -1.6                 \\
175        & 6.45    & 160     & 3.6          & 5.1          & 1.72          & 1.62         & -2.1             \\
195     & 5.84    & 192     & 3.5          & 4.9          & 1.41          & 1.65         & -3.8                 \\
220      & 5.23    & 192     & 3.8          & 5.4          & 1.24          & 1.85         & -                  \\
255      & 4.57    & 128     & 5.6          & 7.9          & 1.30          & 2.59         & 3.5                 \\
295       & 3.99    & 128     & 7.4          & 10.5         & 1.12          & 3.01         & 2.2                 \\
340      & 3.49    & 128     & 11.1         & 15.7         & 1.01            & 3.57         & 2.0                     \\
390       & 3.06    & 96      & 22.0           & 31.1         & 1.08          & 5.05         & 2.8                \\
450      & 2.65    & 96      & 45.9         & 64.9         & 1.04            & 6.48         & 4.3                 \\
520      & 2.29    & 96      & 116.6        & 164.8        & 1.03            & 8.56         & 8.3            \\
600       & 1.98    & 96      & 358.3        & 506.7        & 1.03            & 11.4         & 20.0             \\
\hline
\hline
Array         &              & 2100 & 1.2  & 1.7 &  &  & 0.41 \\
\hline
\end{tabular}
}
\end{center}
\caption{\small {\bf CORE-M5 proposed frequency channels.}
Sensitivities are calculated assuming $\Delta \nu/\nu=30\%$ bandwidth,
60\% optical efficiency, a total noise twice the expected photon noise
from the sky, and the optics of the instrument at 40K temperature.
This configuration has 2100 detectors, of which about 45\% lie 
in the CMB channels between 130 and 220 GHz. The six CMB channels yield 
an aggregated CMB sensitivity of $2\,\mu$K $\cdot $ arcmin 
($1.7\,\mu$K $\cdot $ arcmin for the full array).
}
\label{tab:CORE-bands}
\end{table}
}

The purpose of this section is to explain the 
assumptions regarding the CORE instrumental 
capabilities for arriving at 
the forecasts presented in this paper. For a more detailed description
and discussion of the CORE instrument, we refer the
reader to the companion ECO mission paper \cite{ecoMission}, 
the ECO instrument paper \cite{ecoInstrument} and the ECO systematics paper \cite{ecoSystematics}.

As part of the pre-proposal studies reported in the ``Exploring Cosmic Origin with CORE'' series,
this paper mainly deals with the impact of the telescope size and noise sensitivity on the inflation science results.
CORE will map the sky in temperature and polarization in 19 frequency channels spanning the 
$60-600$ GHz range with
noise sensitivities and angular resolution summarized in Table \ref{tab:CORE-bands}.
For our forecast we use the only the six channels in the frequency range $130-220$ GHz
under the ideal assumption that foreground contaminations in these bands are completely removed
by the lower and higher frequency channels and systematic effects are under control. In the following we will refer to this 
configuration as CORE-M5.
This same assumption that lower and higher frequencies suffice to remove completely
foreground contamination in the inverse noise weighted combination of the 
six central frequency channels is used in the companion ECO paper on cosmological parameters.
A dedicated paper of this ECO series \cite{ecoCompSep} studies in great detail the capability
to measure primordial B-mode polarization by a component separation approach that 
makes use of all the frequency channels
and takes into account the foreground contamination of anomalous microwave emission, thermal dust emission, synchrotron, and point sources.
Updating all the science forecast for cosmology with the inclusion of the main results of this component separation 
dedicated study \cite{ecoCompSep} is left for a future work, but we will however refer to the impact of foreground 
residuals as estimated in \cite{Errard:2015cxa}.

\begin{wrapfigure}{L}{0.6\textwidth}
\begin{center}
                \includegraphics[width=9cm]{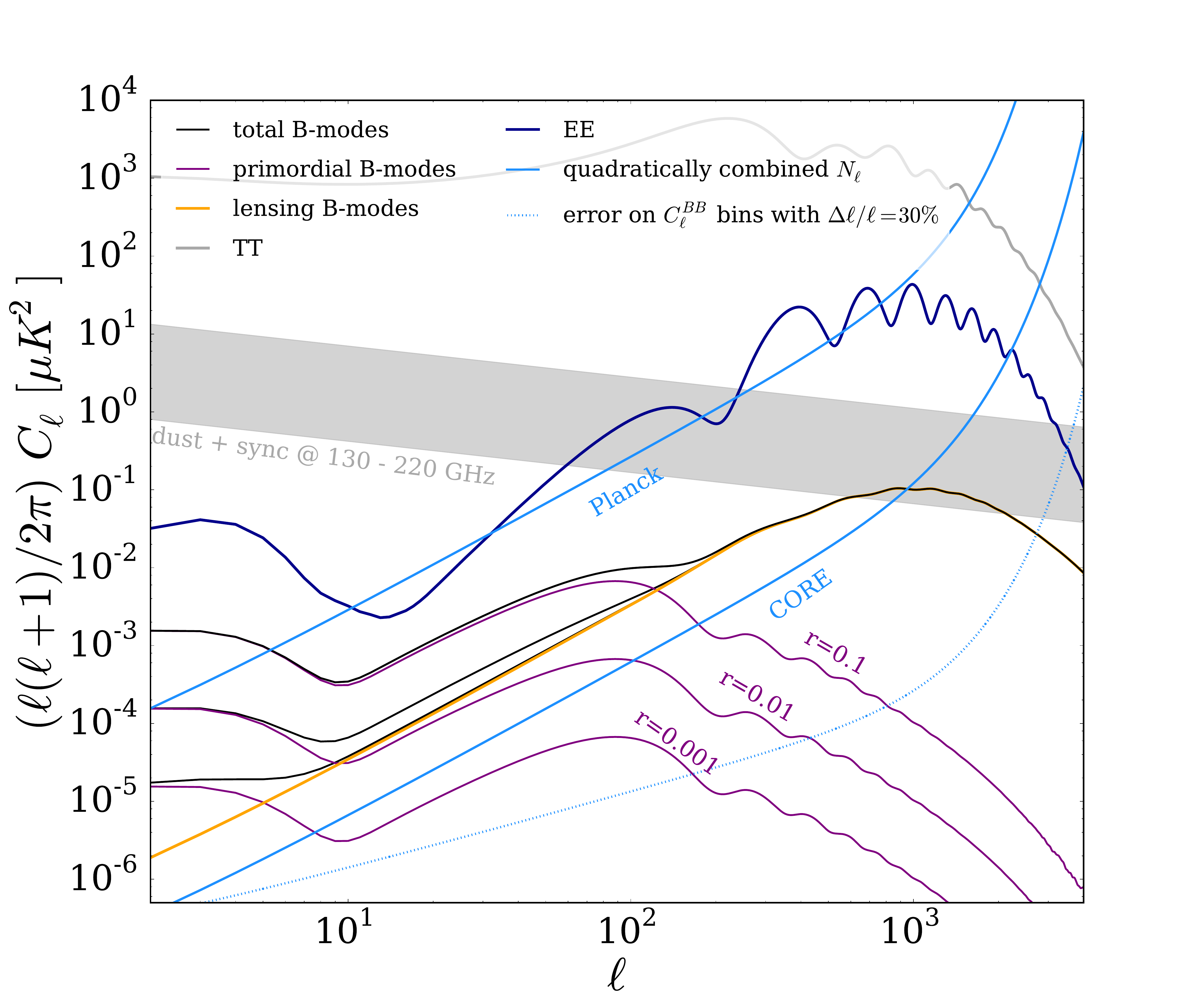}
\end{center}
\vspace{-1\baselineskip}
\caption{
Comparison of cosmological signals,
in particular the primordial and lensed $B$-modes, with diffuse astrophysical
foregrounds amplitudes. The noise levels for \Planck\ and CORE are also displayed.
%of diffuse astrophysical
%foregrounds amplitudes with cosmological signals,
%in particular the primordial and lensed $B$-modes.
%The sensitivity to the primordial 
%signal is limited by galactic emissions and gravitational lensing (from~\cite{Errard:2015cxa}).
}
\label{fig:FGspectra}
\end{wrapfigure}

%In this paper, we also include a few other 
%configurations for comparison. 

In this paper, we compare our forecasts for CORE-M5 to those obtained with experimental configurations
from other concepts for the next space missions dedicated to CMB polarization, 
such as (a) the LiteBIRD-ext configuration \cite{Errard:2015cxa} for JAXA
LiteBIRD \cite{2014JLTP..176..733M}, (b) three configurations with the same noise 
sensitivity in $\mu K\cdot $arcmin 
as LiteBIRD-ext,
but with higher angular resolution thanks to a larger telescope of 80 cm (LiteCORE-80) or
120 cm (LiteCORE-120) or
150 cm (LiteCORE-150), and (c) the COrE+ proposal previously submitted to ESA in
response to the M4 mission call and its version optCOrE+ with an extended mission duration.
Note that we will refer to the LiteCORE-80 forecasts as those representative for 
a possible downscoped version of CORE-M5, called MiniCORE \cite{ecoInstrument}, with a 80 cm telescope and covering the reduced 100-600 GHz frequency.
For all these configurations,
we consider an inverse noise weighted combination of central frequency channels, 
always under the assumption that lower and higher frequency channels remove completely 
the foreground contamination. Table \ref{tab:specifications} reports the experimental specifications for the central frequency channels of these configurations.
This comparison sheds light on the role that angular resolution and noise detector sensitivity
plays in determining what new conclusions concerning
inflation can be extracted from the data of a particular concept for a future CMB space mission. 
See also \cite{Verde:2005ff,Creminelli:2015oda,Kamionkowski:2015yta} 
for previous forecasts for inflation science comparing different future experiments dedicated to CMB polarization.  

This comparison is important since the cost of a CMB 
polarization space mission is largely determined by 
the size of the instrument, or equivalently the size 
of the primary aperture. Because the resolution is 
diffraction limited, the angular size is inversely 
proportional to the effective mirror diameter.  
LiteBIRD envisages a smaller mirror size and thus 
would have a coarser angular resolution. On the one 
hand, detecting primordial B modes from inflation 
does not require necessarily an exquisite angular resolution. 
There are the two windows for detecting such B modes: the 
so-called ``reionization'' bump, situated on very large 
angular scales not accessible from the ground (i.e., 
$\ell \ltorder 10$); and the so-called 
``recombination'' bump, for which the statistically 
exploitable information is centered around $\ell 
\approx 80.$ [See Fig.~\ref{fig:FGspectra}.]
Above $\ell \gtorder 100,$ beyond 
approximately where the modes had already entered the 
horizon at last scattering, the primordial $B$ mode 
signal plummets while the contaminant signal from 
gravitational lensing continues to rise. Given the current 
constraints on $r$ \cite{2016PhRvL.116c1302B}, there is not much exploitable information regarding the primordial 
B modes on scales with $\ell \gtorder 100.$ 
%This is the argument why exquisite angular resolution is not 
%required.

However there are two reasons other than ancillary 
science why an angular resolution extending well 
beyond $\ell \approx 100$ is highly desirable. 
Firstly, when $r\ltorder 10^{-2},$ the  
gravitational lensing signal starts to exceed any 
possible primordial $B$ mode signal beyond the reionization bump. If this gravitational lensing $B$ mode cannot be removed, it 
becomes the dominant source of noise when the 
detector noise level becomes smaller than $\approx 5 
\mu K\cdot $ arcmin. Two approaches to dealing with 
this parasitic lensing are possible. (1) One can try 
to characterize the power spectrum of this lensing 
contaminant at exquisite precision. In this case 
$C^{BB}_\ell $ can be measured after broad binning 
(i.e., with $\Delta \ell /\ell \approx 1$) to an 
accuracy of approximately $\tilde{N}_\ell ^{BB}/\ell $ in the 
broad bins, where here $\tilde{N}_\ell^{BB}$ includes both the 
gravitational lensing and the detector noise 
contributions. The success of this approach obviously 
relies on knowing $\tilde{N}_\ell^{BB}$ from lensing, or 
$C_\ell^{BB,{\rm lensing}}$ at the same accuracy. (2) 
Another more ambitious approach is what is called 
``de-lensing.'' Delensing is predicting $a_{\ell 
m}^{B,{\rm lensing}}$ based on $a_{\ell m}^{E}$ and 
$a_{\ell m}^{\phi, {\rm lensing}}$ where $\phi $ is the CMB 
lensing potential and subtracting this prediction 
from the measured B mode map. $a_{\ell m}^{B,{\rm lensing}}$ 
at low $\ell$ of interest for detecting primordial B mode signal may be though of 
as the low $\ell$ ``white noise'' tail of many small, 
weakly-correlated gravitationally lensed regions. 
Mathematically, $a_{\ell m}^{B,lensing}$ may be 
expressed as a sort of convolution of $a_{\ell 
m}^{E}$ and $a_{\ell m}^{\phi ,lensing},$ and the 
multipole numbers making the dominant contribution 
have $\ell \approx O(10^3).$ This is why both 
exquisite angular resolution and sensitivity are 
needed to be able to ``de-lens.''

%\vspace{-3cm} 
\begin{table}[t]\footnotesize
\begin{center}
\begin{tabular}{|c|c|c|c|} %\footnotesize %c|
\hline
Channel [GHz] & FWHM [arcmin] & $\Delta T $ [$\mu$K arcmin] & $\Delta P $ [$\mu$K arcmin] \\
\hline
%\multicolumn{4}{|c|}{LiteBird,  $l_{\mathrm{max}} = 1350, f_{\mathrm{sky}} = 0.7$  }    \\
\multicolumn{4}{|c|}{LiteBIRD,  $f_{\mathrm{sky}} = 0.7$  }    \\
\hline
% $ 80$ & $55$  & $8.8$ & $12.5$ \\
% $ 90$ & $49$  & $7.1$ & $10.0$ \\
% $100 $ & $43$  & $8.5$ & $12.0$ \\
% $120 $ & $36$  & $6.7$ & $9.5$ \\
% $ 140$ & $31$  & $5.3 $ & $7.5$ \\
% $ 166$ & $26$  & $5.0$ & $7.0$ \\
% $ 195$ & $22$  & $3.6$ & $5.0$ \\
 $ 78$ & $55$  & $8.8$ & $12.5$ \\
 $ 88.5$ & $49$  & $7.1$ & $10.0$ \\
 $100 $ & $43$  & $8.5$ & $12.0$ \\
 $118.9 $ & $36$  & $6.7$ & $9.5$ \\
 $ 140$ & $31$  & $5.3 $ & $7.5$ \\
 $ 166$ & $26$  & $5.0$ & $7.0$ \\
 $ 195$ & $22$  & $3.6$ & $5.0$ \\
\hline
\multicolumn{4}{|c|}{ LiteCORE-80,  $f_{\mathrm{sky}} = 0.7$  }    \\
%\multicolumn{4}{|c|}{LiteCORE-80,  $l_{\mathrm{max}} = 2400, f_{\mathrm{sky}} = 0.7$  }    \\
\hline
$ 80$ & $20.2$  & $8.8$ & $12.5$ \\
$ 90$ & $17.8$  & $7.1$ & $10.0$ \\
$100 $ & $15.8$  & $8.5$ & $12.0$ \\
$120 $ & $13.2$  & $6.7$ & $9.5$ \\
$ 140$ & $11.2$  & $5.3 $ & $7.5$ \\
$ 166$ & $8.5$  & $5.0$ & $7.0$ \\
$ $195 & $8.1$  & $3.6 $ & $5.0$ \\
\hline
%\multicolumn{4}{|c|}{LiteCORE-120,  $l_{\mathrm{max}} = 3000, f_{\mathrm{sky}} = 0.7$  }    \\
\multicolumn{4}{|c|}{ LiteCORE-120, $f_{\mathrm{sky}} = 0.7$ }    \\
\hline
 $ 80$ & $13.5$  & $8.8$ & $12.5$ \\
 $ 90$ & $11.9$  & $7.1$ & $10.0$ \\
 $ 100$ & $10.5$  & $8.5$ & $12.0$ \\
 $ 120$ & $8.8$  & $6.7$ & $9.5$ \\
 $ 140$ & $7.4$  & $5.3 $ & $7.5$ \\
 $ 166$ & $6.3 $  & $5.0$ & $7.0$ \\
 $ 195$ & $5.4 $  & $3.6 $ & $5.0$ \\
\hline
\multicolumn{4}{|c|}{LiteCORE-150, $f_{\mathrm{sky}} = 0.7$ }    \\
\hline
 $ 80$ & $10.8$  & $8.8$ & $12.5$ \\
 $ 90$ & $9.5$  & $7.1$ & $10.0$ \\
 $ 100$ & $8.4$  & $8.5$ & $12.0$ \\
 $ 120$ & $7.0$  & $6.7$ & $9.5$ \\
 $ 140$ & $5.9$  & $5.3 $ & $7.5$ \\
 $ 166$ & $5.0$  & $5.0$ & $7.0$ \\
 $ 195$ & $4.3$  & $3.6 $ & $5.0$ \\
\hline
%\multicolumn{4}{|c|}{\coremfive\, $l_\umax = 3000, f_{\mathrm{sky}} = 0.7$  }   \\
%\hline
% $ 130$ & $8.51$  & $3.9$ & $5.5$ \\
% $ 145 $ & $7.68$  & $3.6$ & $5.1$ \\
% $ 160$ & $7.01$  & $3.7$ & $5.2$ \\
% $175 $ & $6.45 $  & $3.6$ & $5.1$ \\
% $195 $ & $5.84 $  & $3.5$ & $4.9$ \\
% $220 $ & $5.23 $  & $3.8$ & $5.4$ \\
%\hline
\multicolumn{4}{|c|}{(opt) COrE+, $f_{\mathrm{sky}} = 0.7$  }   \\
\hline
% $\textcolor{gray}{80}$ & $10.5$  & $9.1 \, (6.4)$ & $12.9 \, (9.1)$ \\
% $\textcolor{gray}{90}$ & $9.3$  & $6.5 \, (4.6)$ & $9.2 \, (6.5)$  \\
% $\textcolor{gray}{80}$ & $\textcolor{gray}{10.5}$  & $\textcolor{gray}{9.1} \, \textcolor{gray}{(6.4)}$ 
%& $ \textcolor{gray}{12.9} \, \textcolor{gray}{(9.1)}$ \\
% $\textcolor{gray}{90}$ & $\textcolor{gray}{9.3}$  & $\textcolor{gray}{6.5}\, \textcolor{gray}{(4.6)}$ & $\textcolor{gray}{9.2} \, \textcolor{gray}{(6.5)}$  \\
 $ 100$ & $8.4$  & $6.0 \, (4.2)$ & $8.5 \, (6.0)$ \\
 $ 115$ & $7.3$  & $5.0 \, (3.5)$ & $7.0 \, (5.0)$ \\
 $ 130$ & $6.5$  & $4.2 \, (3.0)$ & $5.9 \, (4.2))$ \\
 $ 145 $ & $5.8$  & $3.6 \, (2.5)$ & $5.0 \, (3.6)$ \\
 $ 160$ & $5.3 $  & $3.8 \, (2.7)$ & $5.4 \, (3.8)$ \\
 $175 $ & $4.8 $  & $3.8 \, (2.7)$ & $5.3 \, (3.8)$ \\
 $195 $ & $4.3 $  & $3.8 \, (2.7)$ & $5.3 \, (3.8)$ \\
 $220 $ & $3.8 $  & $5.8 \, (4.1)$ & $8.1 \, (5.8)$ \\
\hline
\end{tabular}
\label{tab:specifications}
\caption{Experimental specifications for 
LiteBIRD-ext, LiteCORE-80, LiteCORE-120, LiteCORE-150, and (opt) COrE+. 
\label{tab:specifications}
}
\end{center} 
\end{table}\mbox{}

\subsection{Simplified likelihood for forecasts}
\label{sec:simplified_likelihood}

Assuming that CMB anisotropies are Gaussian and statistically isotropic, we use 
the following $\chi_\mathrm{eff}^2$ for our science forecasts 
\cite{Knox:1995dq,Hamimeche:2008ai}:
\be
\chi_\mathrm{eff}^2 = - 2 \ln {\cal L} = 
\sum_{\ell} (2\ell+1) f_\mathrm{sky} \left\{ \mathrm{Tr} [ \hat{{\bf C}}_\ell \bar{{\bf C}}_\ell^{-1}] +
\ln |\hat{{\bf C}}_\ell \bar{{\bf C}}_\ell^{-1}| - n \right\},
\label{like}
\end{equation}
where $\bar{{\bf C}}_\ell$ and $\hat{{\bf C}}_\ell$ denote the theoretical and observed 
data covariance matrices, respectively,
and $n$ is a normalization factor. The covariance matrix depends on the power spectra 
$C_\ell^{XY}$ where $X$ and $Y$ 
can take the values $T, E$ (with $n=2$), $T, E, B$ or $T, E, P$ (with $n=3$), 
or $T, E, B, P$ (with $n=4$), where $P$ denotes lensing potential. 
In the most general case the theoretical covariance matrix is
\begin{equation}
\label{eq:covariance_definition}
\bar{\boldsymbol{C}}_\ell \equiv
 \begin{bmatrix}
\bar{C}_\ell^{TT} + N_\ell^{TT} & \bar{C}_\ell^{TE} & 0 & \bar{C}_\ell^{TP} \\
\bar{C}_\ell^{TE} & \bar{C}_\ell^{EE} + N_\ell^{EE} & 0 & \bar{C}_\ell^{EP} \\
0 & 0 & \bar{C}_\ell^{BB} + N_\ell^{BB} & 0  \\
\bar{C}_\ell^{TP}  & \bar{C}_\ell^{EP} & 0 & \bar{C}_\ell^{PP} + N_\ell^{PP}
\end{bmatrix} ,
\end{equation}
where $N_\ell^{\rm XX}$ are the noise power spectra. These account for the 
instrumental noise $N_\ell^{\mathrm XX, \rm inst};$ when $X=B$ and 
internal delensing is possible, the residual delensing error can also be included. 
%the effective noise
%from gravitational lensing can also be included. In the case of delensing 
%this effective noise is the residual delensing error. 
For $X = T, E, B,$ for each experimental configuration studied, we 
consider an inverse-variance weighted sum of the noise sensitivity convolved with 
a Gaussian beam window function for each frequency channel $\nu$:
\be
N_\ell^{\mathrm XX, \rm inst} = \left[ \sum_\nu \frac{1}{N_{\ell \, \nu}^{\mathrm XX, \rm inst}} \right]^{-1} \,,
\ee
with 
\be
N_{\ell \, \nu}^{\mathrm XX, \rm inst} 
= w_{X \, \nu}^{-1} \exp \left[ \ell (\ell+1) \frac{\theta_{{\rm FWHM} \, \nu}}{8 \ln 2} \right] .
\ee
Here $w_{{\rm E} \, \nu}^{-1/2} = w_{{\rm B} \, \nu}^{-1/2}$ ($=w_{{\rm T} \, \nu}^{-1/2} \sqrt{2})$ is the
detector noise level on a steradian patch for polarization (temperature) and 
$\theta_{\rm FWHM}$ being the full width half maximum (FWHM) of 
the beam in radians for a given frequency channel $\nu$.
In Eq. (\ref{like}) we consider $\ell_\mathrm{max}=1350$ for LiteBIRD, $\ell_\mathrm{max}=2400$ for LiteCORE-80 and 
$\ell_\mathrm{max}=3000$ for all the other CORE configurations in Table \ref{tab:CORE-bands} and \ref{tab:specifications}.
We assume here that beam and other systematic uncertainties are much smaller than the statistical errors. 

In addition to the primary CMB temperature and polarization anisotropies, we consider the cosmological
information contained in the CMB lensing potential power spectrum $C_\ell^{PP}$.
Most of the experimental configurations studied in this paper allow a precise reconstruction
of the CMB lensing potential.
It is well known that $C_\ell^{PP}$ helps to break parameter degeneracies
and to determine the absolute 
neutrino masses \cite{2003PhRvL..91x1301K}. We therefore fold in the CMB
lensing information in our likelihood [corresponding to $n=3$
case when considering $T,E,P$ or the most general case
$n=4$ in Eq.~(\ref{eq:covariance_definition})] by neglecting the temperature/polarization-lensing cross-correlation ($C_\ell^{TP} \,,  C_\ell^{EP}$) 
and considering lensed spectra $C_\ell^{TT} \,, C_\ell^{TE} \,, C_\ell^{EE}$, as explained in the ECO paper on cosmological parameters \cite{ecoParams}. 

%with the same methodology as used
%in the companion paper \cite{2003PhRvL..91x1301K}.

We use the publicly available Einstein-Boltzmann codes {\tt CAMB}~\cite{Lewis:1999bs}
%\footnote{\tt http://camb.info/}~\cite{Lewis:1999bs}
to compute the theoretical predictions for temperature, polarization, 
and gravitational lensing deflection power spectra.
We use {\sl Recfast} to model the cosmic recombination history in the forecasts. 
For future data analyses, the differences with the detailed cosmological 
recombination codes {\sl CosmoRec} \cite{Chluba:2010ca} and {\sl HyReci} 
\cite{AliHaimoud:2010dx} 
can be avoided but have shown to be small ($\approx 
0.3 \sigma$ for $n_{\rm S}$) even at the level of precision that could be reached
with CORE \cite{ecoParams}.
Unless otherwise specified, we consider for our mock data a fiducial 
$\Lambda$CDM cosmology compatible with $Planck$ constraints \cite{planck2014-a10}.
We assume a flat Universe with a cosmological constant and three massless neutrinos, $\Omega_{\mathrm b} h^2 = 0.02214$ 
and $\Omega_{\mathrm c} h^2 = 0.1206$ as the baryon and cold dark matter
physical densities, respectively.
We choose the optical depth to reionization $\tau = 0.0581$, the 
Hubble parameter $H_0 = 66.89$ km s$^{-1}$ Mpc$^{-1}$,
$A_{\mathrm s} = 2.1179 \times 10^{-9}$ and spectral index $n_{\mathrm s} 
= 0.9625$ as amplitude and tilt for the spectrum of primordial perturbations, respectively.
We generate posterior probability
distributions for the parameters using either the Metropolis-Hastings
algorithm implemented in {\tt CosmoMC}~\mbox{\citep{Lewis:2002ah}},
the nested sampling
algorithm {\tt MultiNest} \citep{Feroz:2007kg,Feroz:2008xx,Feroz:2013hea},
or {\tt PolyChord}, which combines nested sampling with slice sampling~\citep{Handley:2015fda}.

\subsection{Dealing with gravitational lensing}

In the search to detect primordial gravitational waves,
CMB lensing can be considered as equivalent to an 
additional noise source whose power spectrum in turn 
depends on a number of 
other non-inflationary cosmological parameters.
Gravitational lensing mixes the $E$ and $B$ polarization modes \citep{2006PhR...429....1L}.
The amplitude of these lensing $B$-modes peaks at around a multipole of 
$\ell\approx 1000$ and is always larger than the
primordial signal on scales smaller than the reionization bump for the current upper bounds on $r$ 
\cite{2016PhRvL.116c1302B}.
Lensing is therefore, after the diffuse astrophysical foregrounds, the second largest 
contaminant in the search for $r,$ cf. Fig.~\ref{fig:FGspectra}.

Techniques have been proposed to \emph{delens} the CMB polarization maps 
\citep{2015PhRvD..92d3005S,2015ApJ...807..166S,2007PhRvD..76l3009M,2004PhRvD..69d3005S, 
2002PhRvL..89a1304K,2002PhRvL..89a1303K,2012JCAP...06..014S}---that is, to reduce the 
extra \emph{variance} induced by lensing. Delensing proposes to subtract the 
lensing induced B mode from the observed CMB B mode map, which requires estimating 
the unlensed CMB as well as the matter distribution $P$ (in the form of a lensing potential).  
The performance of delensing depends on the 
experimental characteristics and on the choice for the matter distribution 
estimator (for a detailed analysis of the delensing efficiency and its 
propagation to limits on $r$ see~\cite{Errard:2015cxa}). 
In this paper, we consider 
internal CMB delensing, in order to investigate the performance of each 
experimental configuration alone without the help of external data. 
Given its limited internal delensing capabilities, we consider the full signal in B-mode polarization 
(primordial plus lensing) for LiteBIRD, %~\cite{Errard:2015cxa}, 
whereas we consider the full delensed option for the various CORE configurations. 
As an estimate for the noise spectrum $N_\ell^{PP}$ of the CMB lensing potential, we consider 
the $EB$ estimator \cite{Okamoto_Hu_2003} as in \cite{Errard:2015cxa}.

\section{Probing inflationary parameters with CORE}
\label{sec:four}

In this Section we begin the presentation of our science forecasts dedicated to the physics of 
inflation for the CMB space mission configurations previously described. Different priors 
can be used to compare inflationary theoretical predictions to CMB anisotropies power spectra. In 
this Section we employ the physical parameterization of the primordial power spectra of scalar and 
tensor perturbations, extending Eq.~(\ref{PPSeq}) as:
\begin{align}
\mathcal{P}_{\cal R}(k)
&= A_\mathrm{s} \left( \frac{k}{k_*}\right)^{n_\mathrm{s}-1 +
\frac{1}{2} \, \frac{\mathrm{d}n_\mathrm{s}}{\mathrm{d}\ln k} \ln(k/k_*) + ... } \label{scalarps}\\
%\frac{1}{6} \, \frac{\mathrm{d}^2n_\mathrm{s}}{\mathrm{d}\ln k^2} \left( \ln(k/k_*) \right)^2}, \label{scalarps}\\
\mathcal{P}_\mathrm{t}(k)
&= r \, A_\mathrm{s} \left( \frac{k}{k_*}\right)^{n_\mathrm{t} + \frac{1}{2} \,
\frac{\mathrm{d}n_\mathrm{t}}{\mathrm{d}\ln k} \ln(k/k_*) + ... } \,.
%\frac{1}{6} \, \frac{\mathrm{d}^2n_\mathrm{t}}{\mathrm{d}\ln k^2} \left( \ln(k/k_*) \right)^2} \,.
\label{tensorps}
\end{align}
Physical parameters such 
as $A_\mathrm{s}$, $n_\mathrm{s}$, $\mathrm{d}n_\mathrm{s}/\mathrm{d}\ln k$, %$\mathrm{d}^2n_\mathrm{s}/\mathrm{d}\ln k^2$, 
$r$, $n_\mathrm{t}$ will be considered as the primary 
parameters, together with $\omega_{\mathrm b} $, $\omega_{\mathrm c}$, $\tau$ and 
$\theta_{\mathrm{MC}}$ (the latter being the CosmoMC variable for the angular size of sound 
horizon, i.e., $r_* / D_A$ ). In the next Section the dependence of the 
slow roll predictions for the primordial power spectra on the Hubble flow functions (HFFs 
henceforth) will be investigated and the HFFs will be sampled as the primary parameters. The consistent 
results obtained by these two approaches have been studied in 
\cite{Hamann:2008pb,Finelli:2009bs,Planck:2013jfk}.  In the next Subsections we will discuss the 
improvements on the measurement of $n_\mathrm{s}$ and its scale dependence, the expected 
constraints on the tensor-to-scalar ratio and the tensor tilt, and finally the forecasts 
for the constraints on the spatial curvature.

\subsection{Forecasts for the spectral index and its scale dependence}

One of the main \Planck\ results has been to provide an accurate measurement of $n_{\mathrm s} < 
1$ that rules out the Harrison-Zeldovich scale invariant primordial power spectrum at more than $5 
\sigma$ \cite{Planck:2013jfk,Ade:2015lrj}. This measurement suggests a preference for models 
with a natural exit from inflation and ruled out hybrid inflationary models predicting 
$n_{\mathrm s} > 1$.

%\begin{table}[h]
%\begin{center}\footnotesize
%\begin{tabular}{|c|c|c|c|c|}
%\hline
%Parameter  &  LiteBIRD,$\,$ TE  &  LiteCORE-120,$\,$ TEP  &  CORE-M5,$\,$TEP  &  \coreplus ,$\,$TEP  \\
%\hline
%$\Omega_{\mathrm{b}} h^2$  &  $0.02214\pm0.00013$ & $0.022140\pm0.000040$ & $0.022141\pm0.000037$ & $0.022141\pm0.000033$ \\
%$\Omega_{\mathrm{c}} h^2$  &  $0.12059\pm0.00099$ & $0.12059\pm0.00029$ & $0.1205857\pm0.00027$ & $0.12058\pm0.00025$ \\
%$ 100\theta_{\mathrm{MC}}$  &  $1.03922\pm0.00030$ & $1.039223\pm0.000080$ & $1.039223\pm0.000077$ & $1.039224\pm0.000073$ \\
%$ \tau$  &  $0.0582\pm0.0021$ & $0.0582\pm0.0021$ & $0.0583\pm0.0020$ & $0.0582\pm0.0020$ \\
%$ n_\mathrm{s}$  &  $0.9625\pm0.0034$ & $0.9625\pm0.0015$ & $0.9625\pm0.0014$ & $0.9626\pm0.0014$ \\
%$ \ln(10^{10} A_\mathrm{s})$  &  $3.0533\pm0.0045$ & $3.0532\pm0.0037$ & $3.0533\pm0.0035$ & $3.0533\pm0.0034$ \\
%\hline
%$H_0$  &  $66.89\pm0.47$ & $66.90\pm0.12$ & $66.89\pm0.11$ & $66.90\pm0.10$ \\
%$\sigma_8$  &  $0.8285\pm0.0039$ & $0.8285\pm0.0011$ & $0.8285\pm0.0011$ & $0.8285\pm0.0010$ \\
%\hline
%\end{tabular}
%\end{center}
%\caption{Forecast 68\% CL constraints on primary and derived cosmological parameters
%for LiteBIRD, LiteCORE-120, CORE-M5 and \coreplus  configurations for the $\Lambda$CDM model.
%}
%\label{tab:lambdacdm}
%\end{table}

\begin{table}[h]
\begin{center}\footnotesize
\begin{tabular}{|c|c|c|c|c|}
\hline
Parameter  &  LiteBIRD,$\,$ TE  &  LiteCORE-80,$\,$ TEP  &  CORE-M5,$\,$TEP  &  \coreplus ,$\,$TEP  \\
\hline
$\Omega_{\mathrm{b}} h^2$  &  $0.02214\pm0.00013$ & $0.022140\pm0.000052$ & $0.022141\pm0.000037$ & $0.022141\pm0.000033$ \\
$\Omega_{\mathrm{c}} h^2$  &  $0.12059\pm0.00099$ & $0.12058\pm0.00033$ & $0.1205857\pm0.00027$ & $0.12058\pm0.00025$ \\
$ 100\theta_{\mathrm{MC}}$  &  $1.03922\pm0.00030$ & $1.039220\pm0.000099$ & $1.039223\pm0.000077$ & $1.039224\pm0.000073$ \\
$ \tau$  &  $0.0582\pm0.0021$ & $0.0582\pm0.0020$ & $0.0583\pm0.0020$ & $0.0582\pm0.0020$ \\
$ n_\mathrm{s}$  &  $0.9625\pm0.0034$ & $0.9625\pm0.0016$ & $0.9625\pm0.0014$ & $0.9626\pm0.0014$ \\
$ \ln(10^{10} A_\mathrm{s})$  &  $3.0533\pm0.0045$ & $3.0533\pm0.0038$ & $3.0533\pm0.0035$ & $3.0533\pm0.0034$ \\
\hline
$H_0$  &  $66.89\pm0.47$ & $66.89\pm0.14$ & $66.89\pm0.11$ & $66.90\pm0.10$ \\
$\sigma_8$  &  $0.8285\pm0.0039$ & $0.8285\pm0.0013$ & $0.8285\pm0.0011$ & $0.8285\pm0.0010$ \\
\hline
\end{tabular}
\end{center}
\caption{Forecast 68\% CL constraints on primary and derived cosmological parameters
for LiteBIRD, LiteCORE-80, CORE-M5 and \coreplus  configurations for the $\Lambda$CDM model.
}
\label{tab:lambdacdm}
\end{table}

Table \ref{tab:lambdacdm} reports the forecast 68\% CL uncertainties on the cosmological 
parameters of the $\Lambda$CDM model obtained by LiteBIRD, LiteCORE-80, CORE-M5, and \coreplus. 
Note that LiteCORE-80, CORE-M5 and \coreplus  results include lensing in the mock likelihood. All 
four configurations will provide a nearly cosmic-variance limited measurement of the $EE$ power 
spectrum at low multipoles and will therefore lead to a determination of the optical depth with an
uncertainty close to the ideal case. 
%Because of a better noise sensitivity but a 
%slightly worst angular resolution, CORE-M5 results are only marginally better than those achievable by LiteCORE-120. 
Because of a better noise sensitivity and angular resolution, CORE-M5 results are better than those achievable by LiteCORE-80.
The \coreplus  configuration with a 1.5 m class telescope, leads to further 11\%, 
6\%, 4\%, and 3\% improvements in the determination of $\Omega_{\mathrm{b}} h^2$, $\Omega_{\mathrm{c}} h^2$,
$\theta_{\mathrm{MC}},$ and $\ln(10^{10} A_\mathrm{s})$, respectively, with the uncertainties in 
the other primary parameters unchanged.

CORE-M5 tightens 
the uncertainties of the primary cosmological 
parameters with respect to the \Planck\ 2015 results including high-$\ell$ polarization 
by factors between 4 and 10, and 
between 3 and 5 with respect to those including the most recent determination of the 
reionization optical depth \cite{planck2014-a10}. The LiteBIRD improvement on the \Planck\ 
results is rather different: there is a significant reduction in uncertainties on $\tau$, $\ln(10^{10} A_\mathrm{s})$. 
But no gain on parameters such 
as $\theta_{\mathrm{MC}}$ \cite{planck2014-a10} is expected due to the coarser angular resolution.
It is interesting to note that a larger reduction of uncertainty with respect to most recent \Planck\ 
results occurs on derived parameters such as the Hubble constant $H_0.$ Given the current tension 
between CMB and the most recent local measurements of $H_0$ \cite{Riess:2016jrr}, CORE-M5 will help 
to clarify whether new physics beyond the $\Lambda$CDM model is required.
See \cite{ecoParams} for an analogous discussion of the larger 
expected improvement on the uncertainty on 
$\sigma_8$, whose value obtained from the CMB currently disagrees by about 2$\sigma$ 
with the measurement obtained using galaxy shear measurements \cite{Joudaki:2016mvz,Hildebrandt:2016iqg}.

CORE-M5 will achieve an accuracy in the determination of $n_{\mathrm s}$ more than three 
times better than the current \Planck\ uncertainty. The CORE-M5 error on $n_{\mathrm s}$ will be 
comparable to the standard theoretical uncertainty arising from the entropy generation stage after inflation 
\cite{Planck:2013jfk}, which enters into the comparison of CMB data with specific inflationary 
models. The predictions for a given inflationary model will be sensitive not only to the form 
of the potential during inflation but also to the subsequent
reheating stage during which the 
inflaton decays, leading to a fully thermalized Universe. 
In the next Section this
aspect will be explored 
quantitatively for the inflationary models studied in Ref. \cite{Martin:2013tda}.

Inflation as modelled by a slowly rolling scalar field generically predicts a small running of 
the spectral index of the primordial spectra \cite{Kosowsky:1995aa}, which 
in terms of the slow-roll parameters 
is of higher order than the deviation of the scalar spectral index from unity. Although a theoretical prior based on the 
slow roll approximation makes the running of the scalar spectral index undetectably small for the 
precision of cosmological observations achieved so far \cite{Pahud:2007gi,Adshead:2010mc}, its 
value inferred by CMB data was not merely a consistency check. Negative values of the running 
could be easily accommodated by pre-\Planck\ data since the WMAP first year data release. See for example 
the WMAP 7 year data combined with SPT result $\mathrm{d} n_{\mathrm{s}}/ \mathrm{d} \ln k = - 0.024 \pm 
0.011$ \cite{Hou:2012xq}. With the most precise measurement of the CMB anisotropies in the region 
of the higher acoustic peaks, the \Planck\ data are compatible with a vanishing running 
of the spectral index. See the \Planck\ 2015 temperature and polarization (TT,TE,EE + lowP) 
constraint $\mathrm{d} n_{\mathrm{s}}/ \mathrm{d} \ln k = -0.006 \pm 0.007$ at 68\% CL 
\cite{Ade:2015lrj} or the most recent result $\mathrm{d} n_{\mathrm{s}}/ \mathrm{d} \ln k = 
-0.003 \pm 0.007$ at 68\% CL, obtained with the latest measurement of the reionization optical 
depth $\tau$ \cite{planck2014-a10}.

Table \ref{tab:running} reports the forecasts for the cosmological parameters when 
$\mathrm{d} n_{\mathrm{s}}/\mathrm{d} \ln k$ is allowed to vary in addition to the parameters of 
the standard $\Lambda$CDM model for a fiducial cosmology in which $\mathrm{d} 
n_{\mathrm{s}}/\mathrm{d} \ln k=0$. As indicated in 
Table \ref{tab:runningplusr}, the LiteBIRD forecast 
error on the running does not improve significantly on the current \Planck\ constraints.
Moreover, 
the LiteBIRD uncertainty on the spectral index increases by 44\% when the running is allowed to 
vary. By contrast, CORE-M5 will be able to reduce 
the current uncertainty 
by approximatively a factor of 
3. The forecast error on the spectral index increases by only 7\% when the running is allowed 
to vary. \coreplus  would only marginally improve on the CORE-M5 forecast uncertainties for the 
running.

\begin{figure}
        \centering
\begin{center}
                \includegraphics[width=8cm]{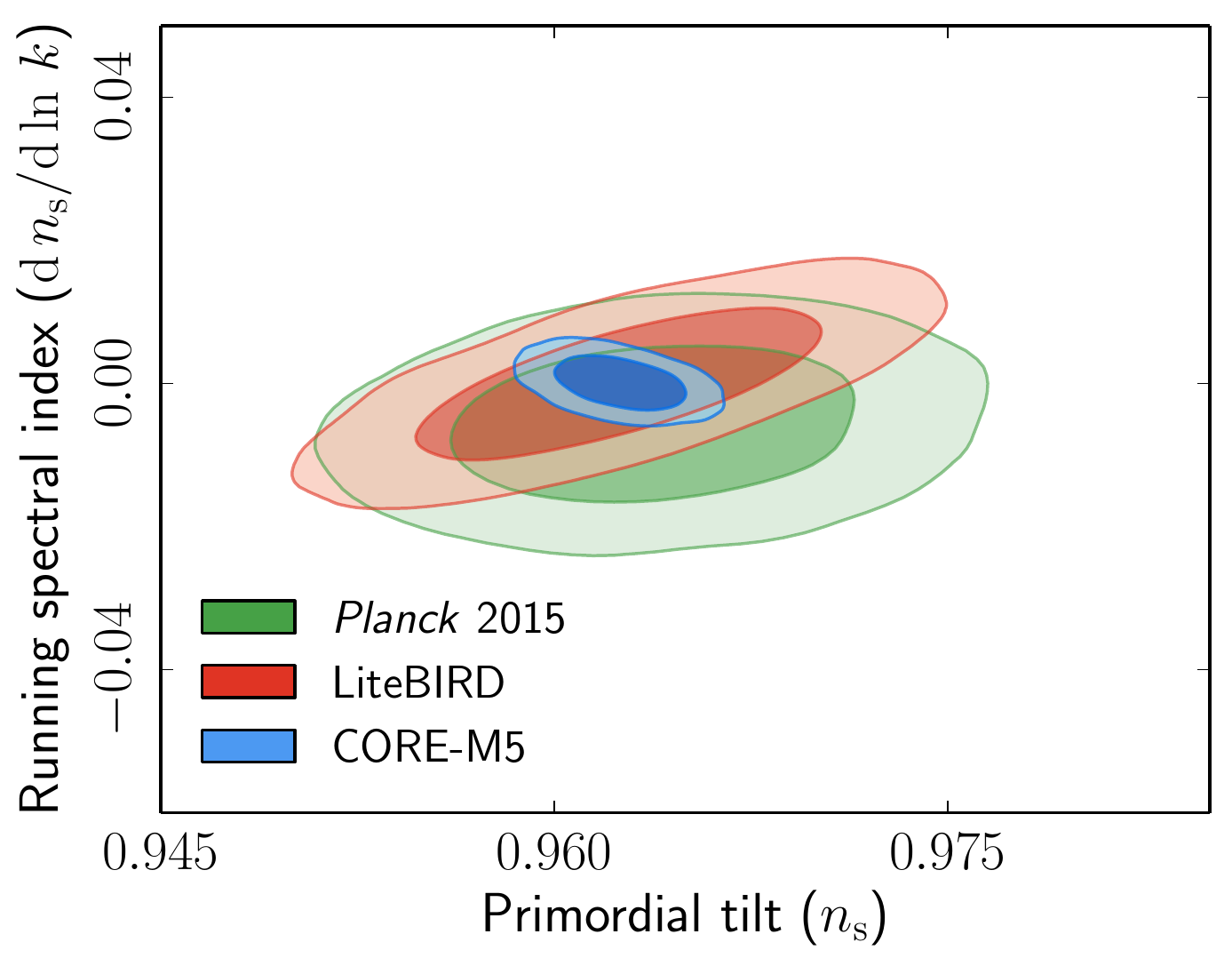}
\end{center}
\vspace{-1\baselineskip}
\caption{\footnotesize
Forecast 68\% and 95\% CL 2D marginalized regions for $(n_{\mathrm s}, d n_{\mathrm s} / d 
\ln k)$ for CORE-M5 (blue) and LiteBIRD (red). These forecasts assume $d n_{\mathrm s} / d \ln 
k=0$ as the fiducial value. The green contours showing the 68\% and 95\% CL for Planck 2015 TT + 
lowP \cite{Ade:2015lrj} are displayed for comparison. Note that the Planck 2015 marginalized 
regions are based on real data whose best fit is different from the fiducial cosmology used in 
this paper.}
\label{fig:running}
\end{figure}

It is now understood that the negative running spectral index allowed by the
pre-\Planck\ data was driven by the low amplitude of the temperature power spectrum at $\ell 
\lesssim 40$ for the WMAP data. The relatively larger impact of this anomaly at $\ell \lesssim 40$ 
for WMAP data could have then conspired through calibration uncertainties in the combination with 
complementary CMB data at higher angular resolution such as SPT to accommodate a larger negative 
running. Thanks to the measurement of $C_\ell$ from $\ell=2$ to $2500$, \Planck\ was able to show 
that the poor fit obtained by the running to the low-$\ell$ anomaly at $\ell \lesssim 40$ can be 
improved by other templates for 
deviations from a power law primordial power spectrum on large scales, although not at a significant statistical level \cite{Ade:2015lrj}. 
As an example of such extensions, we consider a scale dependence 
of the running (i.e., a running of the running $\mathrm{d}^2 n_\mathrm{s}/\mathrm{d} \ln k^2$), 
which was already discussed in Section 2.
Although there is no statistical evidence of a scale dependence of $n_\mathrm{s}$ to second order, a 
combination of positive running and positive running of the running with a lower value of the 
spectral index at $k_* = 0.05$ Mpc$^{-1}$ are allowed by \Planck\ 2015 data, e.g. $n_\mathrm{s} = 
0.9586 \pm 0.0056$, $\mathrm{d} n_\mathrm{s}/ \mathrm{d} \ln k = 0.09 \pm 0.010$, $\mathrm{d}^2 
n_\mathrm{s}/\mathrm{d} \ln k^2 = 0.025 \pm 0.013$ at 68\% CL with \Planck\ TT,TE,EE + lowP, 
leading to a $\Delta \chi^2 \sim - 5$ improvement over the $\Lambda$CDM \cite{Ade:2015lrj}. 
A conservative reconstruction of the inflationary potential including a 
fourth derivative of the inflaton potential beyond the slow-roll approximation supports similar 
considerations \cite{Ade:2015lrj}. Table \ref{tab:runrunning} reports the forecast
uncertainties of the cosmological parameters when the running and the running of the running are 
allowed to vary around a fiducial cosmology with $\mathrm{d} n_{\mathrm{s}}/ \mathrm{d} \ln k = 
\mathrm{d}^2 n_{\mathrm{s}}/ \mathrm{d}\ln k^2 = 0$. We obtain $\sigma (\mathrm{d} 
n_{\mathrm{s}}/ \mathrm{d} \ln k) \approx 0.0024 (0.0023)$ and $\sigma (\mathrm{d}^2 
n_{\mathrm{s}}/ \mathrm{d} \ln k^2) \approx 0.0046 (0.0043)$ as CORE-M5 (\coreplus ) uncertainties 
on the running and on the running of the running, respectively. The CORE-M5 forecast
uncertainties improve approximatively by a factor of 5 and 3 with respect to LiteBIRD for 
$\mathrm{d} n_{\mathrm{s}}/ \mathrm{d} \ln k$ and $\mathrm{d}^2 n_{\mathrm{s}}/ \mathrm{d} \ln 
k^2$, respectively. Note that LiteBIRD does not 
significantly improve on the \Planck\ constraints. 
There is a trade off between the better LiteBIRD noise sensitivity leading to a nearly 
cosmic variance limited measurement of the E-mode polarization at low and intermediate multipoles 
and the \Planck\ higher resolution. This trade-off between LiteBIRD and \Planck\ will be evident 
also in the different analysis presented in the next Sections.

%\begin{table}[h]
%\begin{center}\footnotesize
%\begin{tabular}{|c|c|c|c|c|}
%\hline
%Parameter  &  LiteBIRD,$\,$ TE  &  LiteCORE-120,$\,$ TEP  &  CORE-M5,$\,$ TEP  &  \coreplus ,$\,$ TEP  \\
%\hline
%$\Omega_{\mathrm{b}} h^2$  &  $0.02214\pm0.00013$ & $0.022140\pm0.000046$ & $0.022139\pm0.000044$ & $0.022140\pm0.000038$ \\
%$\Omega_{\mathrm{c}} h^2$  &  $0.1206\pm0.0011$ & $0.12058\pm0.00030$ & $0.12059\pm0.00028$ & $0.12059\pm0.00027$ \\
%$ 100\theta_{\mathrm{MC}}$  &  $1.03922\pm0.00030$ & $1.039222\pm0.000083$ & $1.039225\pm0.000078$ & $1.039223\pm0.000072$ \\
%$ \tau$  &  $0.0583\pm0.0021$ & $0.0582\pm0.0020$ & $0.0582\pm0.0020$ & $0.0582\pm0.0019$ \\
%$ n_\mathrm{s}$  &  $0.9625\pm0.0049$ & $0.9625\pm0.0016$ & $0.9625\pm0.0016$ & $0.9626\pm0.0016$ \\
%$ d n_{\mathrm{s}}/ d \ln k$  &  $0.0000\pm0.0067$ & $0.0000\pm0.0025$ & $0.0000\pm0.0024$ & $0.0000\pm0.0023$ \\
%$ \ln(10^{10} A_\mathrm{s})$  &  $3.0534\pm0.0055$ & $3.0531\pm0.0036$ & $3.0532\pm0.0036$ & $3.0531\pm0.0035$ \\
%\hline
%$H_0$  &  $66.90\pm0.51$ & $66.90\pm0.13$ & $66.90\pm0.11$ & $66.90\pm0.10$ \\
%$\sigma_8$  &  $0.8285\pm0.0040$ & $0.8285\pm0.0012$ & $0.8285\pm0.0011$ & $0.8285\pm0.0010$ \\
%\hline
%\end{tabular}
%\end{center}
%\caption{Forecast 68\% CL constraints on primary and derived cosmological parameters
%when the running of the scalar spectral index is allowed to vary for LiteBIRD, LiteCORE-120, 
%CORE-M5 and \coreplus  configurations. These forecasts assume $\mathrm{d} n_{\mathrm s} / 
%\mathrm{d} \ln k=0$ as the fiducial value.
%}
%\label{tab:running}
%\end{table}

\begin{table}[h]
\begin{center}\footnotesize
\begin{tabular}{|c|c|c|c|c|}
\hline
Parameter  &  LiteBIRD,$\,$ TE  &  LiteCORE-80,$\,$ TEP  &  CORE-M5,$\,$ TEP  &  \coreplus ,$\,$ TEP  \\
\hline
$\Omega_{\mathrm{b}} h^2$  &  $0.02214\pm0.00013$ & $0.022140\pm0.000059$ & $0.022139\pm0.000044$ & $0.022140\pm0.000038$ \\
$\Omega_{\mathrm{c}} h^2$  &  $0.1206\pm0.0011$ & $0.12058\pm0.00034$ & $0.12059\pm0.00028$ & $0.12059\pm0.00027$ \\
$ 100\theta_{\mathrm{MC}}$  &  $1.03922\pm0.00030$ & $1.039225\pm0.000099$ & $1.039225\pm0.000078$ & $1.039223\pm0.000072$ \\
$ \tau$  &  $0.0583\pm0.0021$ & $0.0583\pm 0.0021$ & $0.0582\pm0.0020$ & $0.0582\pm0.0019$ \\
$ n_\mathrm{s}$  &  $0.9625\pm0.0049$ & $0.9625\pm0.0017$ & $0.9625\pm0.0014$ & $0.9626\pm0.0016$ \\
$ d n_{\mathrm{s}}/ d \ln k$  &  $0.0000\pm0.0067$ & $0.0000\pm0.0030$ & $0.0000\pm0.0024$ & $0.0000\pm0.0023$ \\
$ \ln(10^{10} A_\mathrm{s})$  &  $3.0534\pm0.0055$ & $3.0531\pm0.0038$ & $3.0532\pm0.0036$ & $3.0531\pm0.0035$ \\
\hline
$H_0$  &  $66.90\pm0.51$ & $66.90\pm0.15$ & $66.90\pm0.11$ & $66.90\pm0.10$ \\
$\sigma_8$  &  $0.8285\pm0.0040$ & $0.8285\pm0.0014$ & $0.8285\pm0.0011$ & $0.8285\pm0.0010$ \\
\hline
\end{tabular}
\end{center}
\caption{Forecast 68\% CL constraints on primary and derived cosmological parameters
when the running of the scalar spectral index is allowed to vary for LiteBIRD, LiteCORE-80,
CORE-M5 and \coreplus  configurations. These forecasts assume $\mathrm{d} n_{\mathrm s} /
\mathrm{d} \ln k=0$ as the fiducial value.
}
\label{tab:running}
\end{table}

\begin{figure}
        \centering
\begin{center}
                \includegraphics[width=15cm]{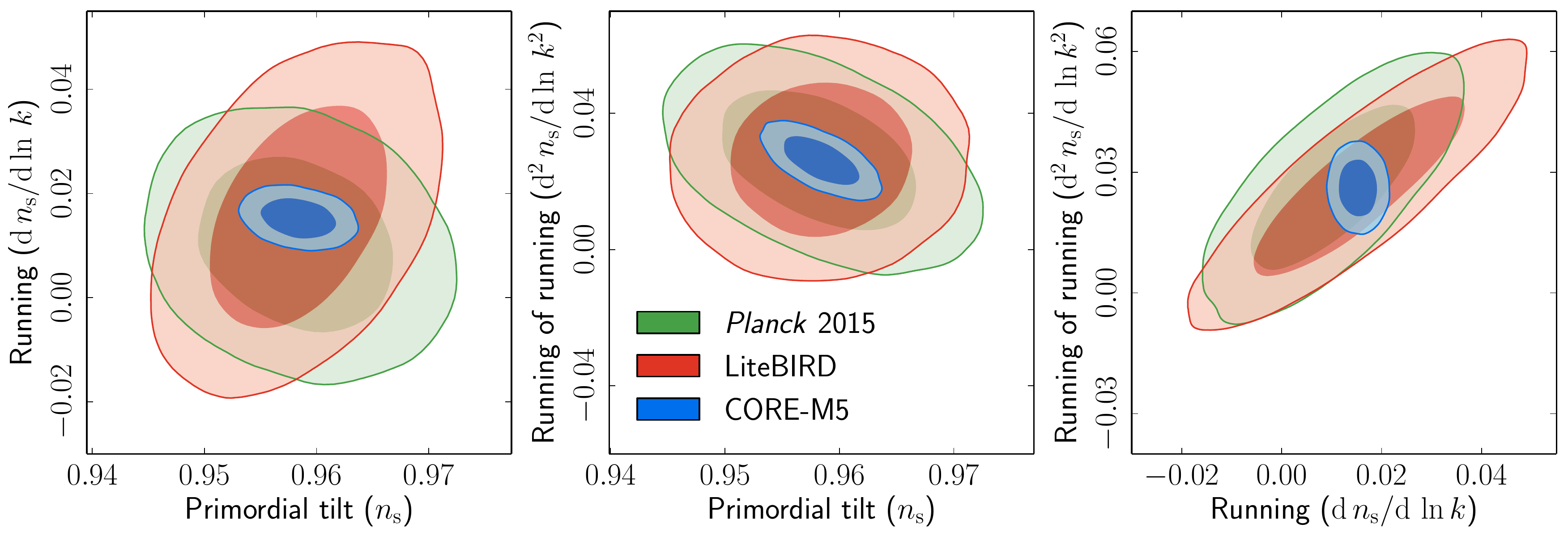}
\end{center}
\vspace{-1\baselineskip}
\caption{\footnotesize
Forecast 68\% and 95\% CL 2D marginalized regions for $(n_{\mathrm s}, \mathrm{d} n_{\mathrm 
s} / \mathrm{d} \ln k)$ (left panel), $(n_{\mathrm s}, \mathrm{d}^2 n_{\mathrm s} / \mathrm{d} 
\ln k^2)$ (middle panel) and $(\mathrm{d} n_{\mathrm s}/\mathrm{d} \ln k, \mathrm{d}^2 n_{\mathrm 
s} / \mathrm{d} \ln k^2)$ (right panel) for CORE-M5 (blue) and LiteBIRD (red). These forecasts 
assume as the fiducial values the \Planck\ 2015 best fits including the running of the running 
\cite{Ade:2015lrj}. The green contours showing the 68\% and 95\% CL for Planck 2015 TT,TE,EE + 
lowP are displayed for comparison.}
\label{fig:runrunning}
\end{figure}

Fig.~\ref{fig:runrunning} shows instead the 2D forecast uncertainties when the spectral 
index is allowed to vary up to second order for a fiducial cosmology chosen based on the 
\Planck\ 2015 TT,TE,EE + lowP best fit \cite{Ade:2015lrj}, i.e., $n_\mathrm{s} = 0.9583$, 
$\mathrm{d} n_\mathrm{s}/ \mathrm{d} \ln k = 0.015$, $\mathrm{d}^2 n_\mathrm{s}/\mathrm{d} \ln 
k^2 = 0.026$. For CORE-M5 (LiteBIRD) we obtain $\mathrm{d} n_\mathrm{s}/ \mathrm{d} \ln k = 
0.0153 \pm 0.0025$ ($\mathrm{d} n_\mathrm{s}/ \mathrm{d} \ln k = 0.015 \pm 0.013$) and 
$\mathrm{d}^2 n_\mathrm{s}/\mathrm{d} \ln k^2 = 0.0261 \pm 0.0045$ ($\mathrm{d}^2 
n_\mathrm{s}/\mathrm{d} \ln k^2 = 0.026 \pm 0.014$) at 68\% CL (the uncertainties are 
essentially unchanged from the case with a fiducial case with no scale dependence of the spectral 
index). CORE-M5 has therefore the capability to probe at a statistically significant 
level the large running of the running which currently leads 
to an improved fit to data.\footnote{Note 
that CORE-M5 forecast uncertainties on the running of the running also
improve on the capabilities of future missions dedicated to the measurements of CMB spectral
distortions as PIXIE, which can probe smaller wavelengths inaccessible using CMB anisotropies
because of Silk damping. The
forecast 68\% uncertainties for $\mathrm{d} n_\mathrm{s}/ \mathrm{d} \ln k$ and $\mathrm{d}^2
n_\mathrm{s}/ \mathrm{d} \ln k^2$ reported in Ref. \cite{Cabass:2016ldu} by combining \Planck\
and spectral distortions by PIXIE are $0.0065$ and $0.0045$, respectively, (i.e., slightly larger
than those reported for CORE-M5).} The CORE-M5 capabilities to probe the scale dependence 
of the spectral index $n_\mathrm{s}$ down to slow-roll predictions
will be further improved by future galaxy surveys \cite{Pourtsidou:2016ctq}.

\begin{table}[h]
\begin{center}\footnotesize
\begin{tabular}{|c|c|c|c|c|}
\hline
Parameter  &  LiteBIRD,$\,$ TE  &  LiteCORE-80,$\,$TEP  &  CORE-M5,$\,$TEP  &  \coreplus,$\,$TEP  \\
\hline
$\Omega_{\mathrm{b}} h^2$  &  $0.02215\pm0.00014$ & $0.022139\pm0.000061$ & $0.022141\pm0.000044$ & $0.022142\pm0.000039$ \\
$\Omega_{\mathrm{c}} h^2$  &  $0.1206\pm0.0011$ & $0.12059\pm0.00040$ & $0.12058\pm0.00032$ & $0.12057\pm0.00031$ \\
$ 100\theta_{\mathrm{MC}}$  &  $1.03922\pm0.00030$ & $1.03922\pm0.00010$ & $1.039224\pm0.000077$ & $1.039222\pm0.000074$ \\
$ \tau$  &  $0.0583\pm0.0023$ & $0.0582\pm0.0021$ & $0.0582\pm0.0021$ & $0.0582\pm0.0020$ \\
$ n_\mathrm{s}$  &  $0.9625\pm0.0050$ & $0.9625\pm0.0026$ & $0.9625\pm0.0022$ & $0.9625\pm0.0021$ \\
$ \mathrm{d} n_{\mathrm{s}}/ d \ln k$  &  $0.000\pm0.014$ & $0.0001\pm0.0031$ & $0.0000\pm0.0024$ & $0.0000\pm0.0023$ \\
$ \mathrm{d}^2 n_{\mathrm{s}}/ d \ln k^2$  &  $0.000\pm0.014$ & $-0.0003\pm0.0060$ & $0.0000\pm0.0046$ & $0.0000\pm0.0043$ \\
$ \ln(10^{10} A_\mathrm{s})$  &  $3.0534\pm0.0055$ & $3.0532\pm0.0043$ & $3.0532\pm0.0040$ & $3.0533\pm0.0038$ \\
\hline
$H_0$  &  $66.90\pm0.50$ & $66.89\pm0.17$ & $66.90\pm0.13$ & $66.89\pm0.13$ \\
$\sigma_8$  &  $0.8284\pm0.0055$ & $0.8285\pm0.0021$ & $0.8285\pm0.0016$ & $0.8285\pm0.0015$ \\
\hline
\end{tabular}
\end{center}
\caption{Forecast 68\% CL constraints on primary and derived cosmological parameters
when the scale dependence of the scalar spectral index is allowed to vary up to its second
derivative for the LiteBIRD, LiteCORE-80, CORE-M5, and \coreplus  configurations. These forecasts
assume $\mathrm{d} n_{\mathrm s} / \mathrm{d} \ln k = \mathrm{d}^2 n_{\mathrm s} / \mathrm{d} \ln
k^2 = 0$ as the fiducial values.
}
\label{tab:runrunning}
\end{table}

\subsection{Joint forecasts for $n_{\mathrm s}$ and $r$}

Only recently has the constraint on $r$ from the  B-mode polarization 
become competitive with the constraint
obtained using the temperature and E-mode polarization anisotropies alone.  
The BICEP 2/Keck Array/\Planck\ (BKP) 
joint cross-correlation provided the same upper bound as that obtained with the \Planck\ 
nominal mission temperature data in combination with the WMAP large angular scale polarization 
\cite{Planck:2013jfk}, i.e., $r < 0.12$ at 95\% CL \cite{Ade:2015tva}. The combination of 
\Planck\ 2015 temperature and E-mode polarization data with the BKP cross-correlation tightened 
this constraint to $r < 0.08$ at 95\%CL \cite{Ade:2015xua,Ade:2015lrj}, thus 
strongly disfavoring what was believed to be the simplest inflationary model, i.e., a chaotic 
model with a quadratic potential $V(\phi) \propto \phi^2$ \cite{Ade:2015lrj}. See 
\cite{2016PhRvL.116c1302B} for the most recent BICEP 2/Keck Array/\Planck\ joint 
cross-correlation including the Keck Array 95 GHz and the WMAP 23, 33 GHz bands, slightly 
improving the 95\% CL upper bound on $r$ to $0.07$.

With its mean noise level of $2 ~\mu K \cdot \,$arcmin and its angular resolution of 5' at 200 
GHz, CORE-M5 has the capability to probe several classes of large-field inflationary models, 
for which the excursion of the inflaton between the time at which the observable wavelength exited 
the Hubble radius and the end of inflation is of the order of the Planck mass. We first consider 
$r=10^{-3}$ as a fiducial value, which was discussed in Section 2 as an interesting target motivated by recent developments in 
supergravity \cite{Ferrara:2016fwe}. We sample linearly on the six cosmological parameters and on the tensor-to-scalar ratio $r$ at the 
pivot scale $k_*=0.05$ Mpc$^{-1}$, with the tensor tilt satisfying $n_\mathrm{t} = -r/8$ as 
predicted by canonical single field slow-roll models. The results for LiteBIRD, LiteCORE-80, 
CORE-M5, and \coreplus\   are presented in Table \ref{tab:r001}. Note that the LiteCORE-80, CORE-M5,
and \coreplus\ results include either internal CMB delensing for B-mode polarization and the CMB 
lensing as an additional observable in the mock likelihood. The noise sensitivity and the CMB x 
CMB delensing capability leads to $r=0.00100 \pm 0.00021$ at 68\% CL for CORE-M5. The CORE-M5 uncertainty 
is nearly half of $3.6 \times 10^{-4}$, 
%($3.8 \times 10^{-4}$), 
which can be reached by LiteBIRD %(LiteBIRD \cite{Matsumura:2016sri}) 
and improves on that obtained by 
LiteCORE-80 by 29\%. The more ambitious \coreplus  configuration improves on the CORE-M5 uncertainty 
by 10\%. 
As an approximate idea of the impact of foreground uncertainties, %According to Ref. \cite{Errard:2015cxa}, 
let us note that a Fisher matrix approach including foreground residuals and delensing \cite{Errard:2015cxa} 
leads to an uncertainty on the tensor-to-scalar ratio 
as $\sigma (r) \approx 3.2 \times 10^{-4}$ for a fiducial case with $r=10^{-3}$ for CORE-M5.  

CORE-M5 will target the Starobinsky $R^2$ model \cite{Starobinsky:1980, 
Mukhanov:1981xt,Starobinsky:1983zz}, which predicts $r \approx 12/N_*^2$ (approximatively $r \sim 
0.0042$ for $N_* \approx 53$). The Starobinsky model is currently the simplest among the 
inflationary models allowed by Planck data \cite{Planck:2013jfk,Martin:2013nzq,Ade:2015lrj} and 
has been at the center stage of several theoretical developments in supergravity 
\cite{Ketov:2010qz,Ellis:2013xoa,Farakos:2013cqa,Ferrara:2013wka}, as explained in Section 2. By 
assuming $r=0.0042$ as the fiducial value, compatible with the Starobinsky $R^2$ model for the 
fiducial value of $n_{\mathrm s}$ considered in Section 3, we obtain $n_{\mathrm s} = 0.9625 
\pm 0.0030$, $r = 0.00424 \pm 0.00057$, %($n_{\mathrm s} = 0.9625 \pm 0.0031$, $r = 0.00425 \pm 0.00062$) 
and $n_{\mathrm s} = 0.9625 \pm 0.0015$, $r = 0.00421 \pm 0.00028$ 
at 68\% CL, for LiteBIRD % (LiteBIRD \cite{Matsumura:2016sri}) 
and CORE-M5, respectively.

In the worst case scenario in which the energy scale of inflation is such that the primordial 
B-mode polarization is below the sensitivity of any experimental configuration studied here, we 
find $1.8 \times 10^{-4}$ 
and $4.3 \times 10^{-4}$ %($4.7 \times 10^{-4}$) 
as the CORE-M5 and LiteBIRD %(LiteBIRD \cite{Matsumura:2016sri}) 
95\% CL upper bounds on $r$ taking into account noise sensitivity only, respectively. CORE-M5 could improve by more 
than a factor of 2 on the 95\% CL upper bound on $r$ obtained with LiteBIRD. 

\begin{table}[h]
\begin{center}\footnotesize
\begin{tabular}{|c|c|c|c|c|}
\hline
Parameter  &  LiteBIRD,$\,$ TEB  &  LiteCORE-80,$\,$ TEBP  &  CORE-M5,$\,$ TEBP  &  \coreplus ,$\,$ TEBP  \\
\hline
$\Omega_{\mathrm{b}} h^2$  &  $0.02215\pm0.00010$ & $0.022136\pm0.000057$ & $0.022141\pm0.000037$ & $0.022140\pm0.000034$ \\
$\Omega_{\mathrm{c}} h^2$  &  $0.12056\pm0.00046$ & $0.12058\pm0.00033$ & $0.12058\pm0.00026$ & $0.12059\pm0.00025$ \\
$ 100\theta_{\mathrm{MC}}$  &  $1.03922\pm0.00028$ & $1.03923\pm0.000010$ & $1.039223\pm0.000077$ & $1.039223\pm0.000072$ \\
$ \tau$  &  $0.0582_{-0.0022}^{+0.0020}$ & $0.0582\pm0.0020$ & $0.0582\pm0.0020$ & $0.0582\pm0.0019$ \\
$ n_\mathrm{s}$  &  $0.9626\pm0.0030$ & $0.9625\pm0.0016$ & $0.9625\pm0.0015$ & $0.9625\pm0.0014$ \\
$ \ln(10^{10} A_\mathrm{s})$  &  $3.0531\pm0.0041$ & $3.0531\pm0.0034$ & $3.0532\pm0.0034$ & $3.0531\pm0.0034$ \\
$r$  &  $0.00104 \pm 0.00036$ & $0.00103 \pm0.00027$ & $0.00100 \pm 0.00021$ & $0.00101\pm0.00019$ \\
\hline
$H_0$  &  $66.91\pm0.24$ & $66.90\pm0.14$ & $66.90 \pm 0.11$ & $66.90 \pm 0.10$ \\
$\sigma_8$  &  $0.8284\pm0.0021$ & $0.8285\pm0.0014$ & $0.8285 \pm0.0011$ & $0.8285\pm0.0010$ \\
\hline
\end{tabular}
\end{center}
\caption{Forecast 68\% CL constraints on primary and derived cosmological parameters
when $r$ is allowed to vary for LiteBIRD, LiteCORE-80, CORE-M5, and \coreplus  configurations.
These forecasts assume $r=10^{-3}$ as the fiducial value.
\label{tab:r001}
}
\end{table}

\begin{figure}
\begin{center}
\includegraphics[width=8.5cm]{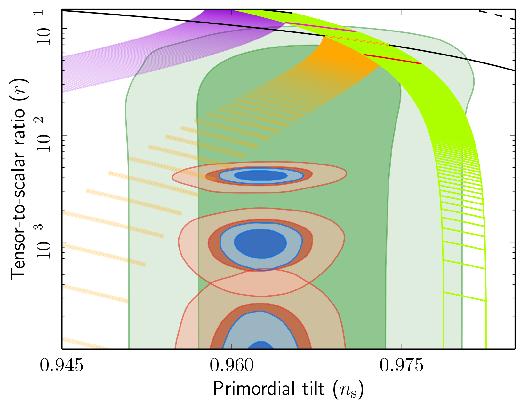} 
\end{center}
\vspace{-1\baselineskip}
\caption{\footnotesize 
68\% and 95\% CL 2D marginalized forecast regions for $(n_{\mathrm s}, r)$ for CORE-M5 (blue) 
and LiteBIRD (red). Three reference cosmologies are considered: a value for the tensor-to-scalar 
ratio consistent with the $R^2$ model ($r \approx 0.0042$), $r=0.001,$ and a third case in 
which the level of primordial gravitational waves is undetectably small (i.e., $r=0$). The green 
contours showing the 68\% and 95\% CL for Planck 2015 TT + lowP data combined with the BKP 
joint cross-correlation \cite{Ade:2015lrj} are also displayed for comparison. We show the 
predictions for natural inflation \cite{Freese:1990rb,Adams:1992bn} (purple band), the 
hilltop quartic model \cite{Boubekeur:2005zm} (orange discrete band), 
and power law chaotic \cite{Linde:1983gd} (light green discrete band) models, accounting for representative 
uncertainties in the post-inflationary era with $47 < N_* < 57$. These inflationary models 
consistent with the current data can be ruled out by CORE-M5. Note the logarithmic scale on the 
$y$-axis and that the pivot scale considered here is $k_* = 0.05$ Mpc$^{-1}$.}
\label{fignsr}
\end{figure}

The forecast CORE-M5 and LiteBIRD 68\% and 95\% CL 2D marginalized regions in 
$(n_\mathrm{s},r)$ for $r = 0.0042$, $r=0.001$ and $r=0$ are displayed in Fig.~\ref{fignsr}. 
These forecasts are compared with the corresponding regions for the real Planck 2015 data 
combined with the BKP joint cross-correlation for B-mode polarization \cite{Ade:2015lrj}. 
Theoretical predictions for a few inflationary models in agreement with 
current data are also shown: natural inflation (purple band), the hilltop quartic model 
(orange discrete band), and power law chaotic models (light green discrete band), all displayed 
with standard representative uncertainties for the post-inflationary era. Fig.~\ref{fignsr} shows 
how CORE-M5 can discriminate better than LiteBIRD among models which fit current observations at a 
similar statistical significance \cite{Ade:2015lrj}.

We now analyze the case in which either the scalar and tensor power spectra have a non-vanishing 
running of the spectral index. We restrict ourselves to the standard slow-roll case and we use 
consistency conditions for the tensor tilt and its running to the next-to-leading order, i.e.,
$n_\mathrm{t}=-r(2-r/8-n_\mathrm{s})/8$ and $\mathrm{d}n_\mathrm{t}/\mathrm{d}\ln k = 
r(r/8+n_\mathrm{s}-1)/8$ at $k_* = 0.05$ Mpc$^{-1}$. When $r$ and 
$\mathrm{d}n_\mathrm{s}/\mathrm{d}\ln k$ are allowed to vary, the following constraints have been 
obtained with \Planck\ 2015 TT + lowP + BKP \cite{Ade:2015lrj}: $n_\mathrm{s} = 0.9661 \pm 
0.0064$, $\mathrm{d}n_\mathrm{s}/\mathrm{d}\ln k = -0.012^{+0.009}_{-0.008}$ at 68\% CL and $r < 
0.09$ at 95\% CL.

Table \ref{tab:runningplusr} reports the forecasts obtained by assuming $r=0.0042$ and 
$\mathrm{d}n_\mathrm{s}/\mathrm{d}\ln k = -0.0007$ as fiducial values, which are representative 
for the Starobinsky $R^2$ model \cite{Starobinsky:1980,
Mukhanov:1981xt,Starobinsky:1983zz} including the running of the spectral index. The CORE-M5 
forecast uncertainties for the running are approximatively a factor of 3.5 smaller than the 
current uncertainties, whereas LiteBIRD will be able to provide only a 30\% improvement. Allowing a 
non-vanishing running in the fit, the forecast errors on $n_\mathrm{s}$ increase by $23\%$ for 
LiteBIRD, but just a $7\%$ for LiteCORE-80, CORE-M5 and COrE+, with respect to the case of 
power law spectra. The forecast uncertainties reported in Tables \ref{tab:running} and \ref{tab:runningplusr} show 
that none of the configurations alone will allow precise measurements of running down to the 
level of $\mathrm{d}n_\mathrm{s}/\mathrm{d}\ln k \sim (n_\mathrm{s}-1)^2/2$, but the CORE 
configurations will certainly be more powerful when combined with future large scale structure or 
21 cm observations \cite{Adshead:2010mc,Pourtsidou:2016ctq}.

\begin{table}[h]
\begin{center}\footnotesize
\begin{tabular}{|c|c|c|c|c|}
\hline
Parameter  &  LiteBIRD,$\,$ TEB  &  LiteCORE-80,$\,$ TEBP  &  CORE-M5,$\,$ TEBP  &  \coreplus ,$\,$ TEBP  \\
\hline
$\Omega_{\mathrm{b}} h^2$  &  $0.02214 \pm 0.00011$ & $0.022140 \pm 0.000060$ &
$0.022140 \pm 0.000043$ & $0.022140 \pm 0.000038$ \\
$\Omega_{\mathrm{c}} h^2$  &  $0.12057 \pm 0.00047$ & $0.12058 \pm 0.00033$ &
$0.12058 \pm 0.00027$ & $0.12059 \pm 0.00026$ \\
$ 100\theta_{\mathrm{MC}}$  &  $1.03922 \pm 0.00028$ & $1.039223 \pm 0.000099$ &
$1.039224 \pm 0.000077$ & $1.039222 \pm 0.000073$ \\
$\tau$  &  $0.0583\pm0.0021$ & $0.0582 \pm 0.0020$ & $0.0582 \pm 0.0020$ & $0.0582 \pm 0.0019$ \\
$ n_\mathrm{s}$  &  $0.9625 \pm 0.0037$ & $0.9625 \pm 0.0017$ & $0.9625 \pm 0.0016$
& $0.9625 \pm 0.0015$ \\
$ {\mathrm{d}} n_{\mathrm{s}}/ {\mathrm{d}} \ln k$  &  $-0.0007 \pm 0.0059$ & $-0.0007 \pm 0.0030$ &
$-0.0007 \pm 0.0024$ & $-0.0007 \pm 0.0023$ \\
$ \ln(10^{10} A_\mathrm{s})$  &  $3.0532 \pm 0.0045$ & $3.0532 \pm 0.0037$ & $3.0532
\pm 0.0035$ & $3.0532 \pm 0.0035$ \\
$r$  &  $0.0043 \pm 0.0006$ & $0.0042 \pm 0.0004$ & $0.00421 \pm 0.00028$ &
$0.00421 \pm 0.00026$ \\
\hline
$H_0$  &  $66.91 \pm 0.25$ & $66.90 \pm 0.15$ & $66.89 \pm 0.11$ & $66.89 \pm 0.11$ \\
$\sigma_8$  &  $0.8283 \pm 0.0025$ & $0.8283 \pm 0.0014$ & $0.8283 \pm 0.0011$ &
$0.8283 \pm 0.0010$ \\
\hline
\end{tabular}
\end{center}
\caption{
Forecast 68\% CL constraints on the primary and derived cosmological parameters when $r$ and 
the running of the spectral index are allowed to vary for the LiteBIRD, LiteCORE-80, and CORE-M5 
configurations. These forecasts assume $r=0.0042$ and $\mathrm{d} n_{\mathrm{s}}/ d \ln k = 
-0.0007$ as the fiducial values.
}
\label{tab:runningplusr}
\end{table}

\begin{figure}
\begin{center}
\includegraphics[width=15cm]{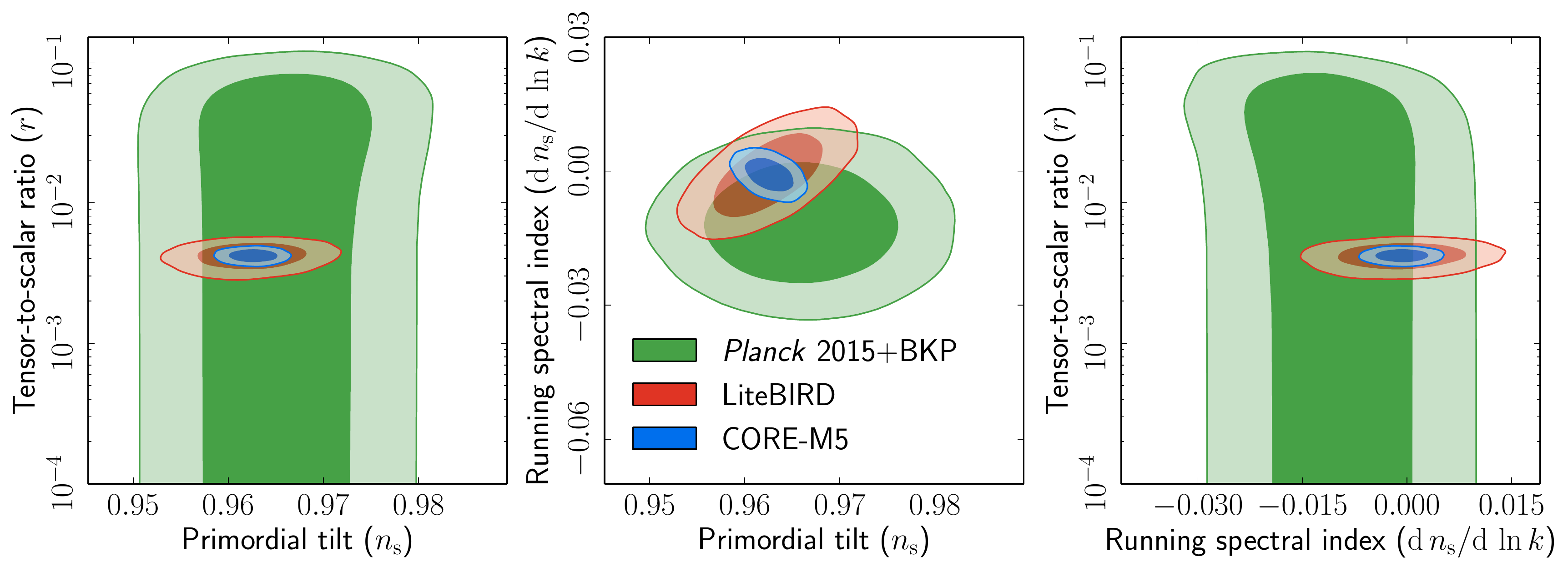}
\end{center}
\vspace{-1\baselineskip}
\caption{\footnotesize
Forecast 68\% and 95\% CL 2D marginalized regions for $(n_{\mathrm s}, r)$ (left panel), 
$(n_{\mathrm s}, {\mathrm d} n_{\mathrm s} / d \ln k)$ (middle panel) and $(r, {\mathrm d} 
n_{\mathrm s}/d \ln k)$ (right panel) for CORE-M5 (blue) and LiteBIRD (red) obtained by allowing 
the tensor-to-scalar ratio and the running to vary. These forecasts assume $r=0.0042$ and $d 
n_{\mathrm{s}}/ d \ln k = -0.0007$ as the fiducial values. The green contours show the 68\% and 95\% 
CL for Planck 2015 data combined with the BKP joint cross-correlation \cite{Ade:2015lrj} are 
shown for comparison. Note that the Planck 2015 contours are based on real data whose best fit is 
different from the fiducial cosmology used in this paper. }
\label{fignsrrun}
\end{figure}

We now study the dependence of the constraint on ($n_{\mathrm s} \,, r$) with respect to 
uncertainties in the neutrino physics. See the companion paper on cosmological parameters 
\cite{ecoParams} for an extensive study of how CORE-M5 could improve our current knowledge of 
neutrino physics. We consider as an example the case in which the number of neutrino species 
$N_{\mathrm{eff}}$ is allowed to vary around the standard value $3.046$. It is known that the 
damping tail of the power spectra of CMB anisotropies is very sensitive to the number of neutrino 
species and therefore partially degenerate with the scalar spectral index. The impact of the 
degeneracy of a varying $N_{\mathrm{eff}}$ with $n_{\mathrm s}$ on establishing the statistical 
significance of the departure from scale invariance and on the combined constraints on 
$(n_{\mathrm s}, r)$ has been studied for the \Planck\ data \cite{Planck:2013jfk,Ade:2015lrj}. When 
$N_{\mathrm{eff}}$ is allowed to vary simultaneously with the tensor amplitude, 
the following constraints are 
obtained using the \Planck\ 2015 temperature and polarization data \cite{Ade:2015lrj}: 
$n_\mathrm{s} = 0.964 \pm 0.010$, $N_{\mathrm{eff}} = 3.02^{+0.20}_{-0.21}$ at 68\% CL. The 
constraint on $r$ is unchanged by allowing $N_{\mathrm{eff}}$ to vary, i.e., $r < 0.11$ at 95\% 
CL.

Table \ref{tab:Neffplusr} reports the forecasts for primary and derived parameters when $r$ and 
$N_{\mathrm{eff}}$ are simultaneously varied in addition to the parameters of the standard 
$\Lambda$CDM model, assuming a fiducial model with $r=0.0042$ and $N_{\mathrm{eff}}=3.046$. 
LiteBIRD improves on the \Planck\ uncertainties by approximatively 40\% in 
$n_\mathrm{s}$ and 20\% in $N_{\mathrm{eff}}$. By contrast, CORE-M5 has the capability to reduce by approximatively a 
factor of 4 the \Planck\ 2015 uncertainties on $n_\mathrm{s}$ and $N_{\mathrm{eff}}$, 
respectively. LiteCORE-80 forecasts are slightly worst than CORE-M5, %in particular for $N_{\mathrm{eff}}$, 
whereas COrE+ can further improve the CORE-M5 forecasts on $n_\mathrm{s}$ and $N_{\mathrm{eff}}$ by 3.7\% and 7.5\%, respectively.
See Fig.~\ref{tab:Neffplusr} for the 2D joint 68 and 95\% CL marginalized regions for 
$(n_{\mathrm s}, r)$, $(n_{\mathrm s}, N_{\mathrm{eff}})$ and $(N_{\mathrm{eff}},r)$ expected 
from LiteBIRD and CORE-M5, compared with the current \Planck\ 2015 results.

\begin{table}[h]
\begin{center}\footnotesize
\begin{tabular}{|c|c|c|c|c|}
\hline
Parameter  &  LiteBIRD,$\,$ TEB  &  LiteCORE-80,$\,$ TEBP  &  CORE-M5,$\,$ TEBP  &  \coreplus ,$\,$ TEBP  \\
\hline
$\Omega_{\mathrm{b}} h^2$  &  $0.02215 \pm 0.00019$ & $0.022142 \pm 0.000085$ & $0.022141 \pm 0.000057$ & $0.022141 \pm 0.000049$ \\
$\Omega_{\mathrm{c}} h^2$  &  $0.1207 \pm 0.0022$ & $0.12062 \pm 0.00093$ & $0.12060 \pm 0.00061$ & $0.12060 \pm 0.00057$ \\
$ 100\theta_{\mathrm{MC}}$  &  $1.03921 \pm 0.00041$ & $1.039223 \pm 0.000099$ & $1.039221 \pm 0.000091$ & $1.039222 \pm 0.000085$ \\
$ \tau$  &  $0.0582 \pm 0.0020$ & $0.0582 \pm 0.0020$ & $0.0582 \pm 0.0019$ & $0.0581 \pm 0.0018$ \\
$ n_\mathrm{s}$  &  $0.9626 \pm 0.0060$ & $0.9626 \pm 0.0028$ & $0.9625 \pm 0.0027$ & $0.9625 \pm 0.0026$ \\
$ \ln(10^{10} A_\mathrm{s})$  &  $3.053 \pm 0.071$ & $3.0533 \pm 0.0038$ & $3.0532 \pm 0.0035$ & $3.0531 \pm 0.0035$ \\
$r$  &  $0.00424 \pm 0.00057$ & $0.00421 \pm 0.00038$ & $0.00421 \pm 0.00028$ & $0.00421 \pm 0.00026$ \\
$N_{\mathrm{eff}}$  &  $3.05 \pm 0.16$ & $3.048 \pm 0.045$ & $3.046 \pm 0.040$ & $3.046 \pm 0.037$ \\
\hline
$H_0$  &  $66.9 \pm 1.1$ & $66.90 \pm 0.34$ & $66.90 \pm 0.31$ & $66.89 \pm 0.28$ \\
$\sigma_8$  &  $0.8284 \pm 0.0058$ & $0.8284 \pm 0.0023$ & $0.8284 \pm 0.0020$ & $0.8283 \pm 0.0019$ \\
\hline
\end{tabular}
\end{center}
\caption{
Forecast 68\% CL constraints on primary and derived cosmological parameters when $r$ and
$N_{\mathrm{eff}}$ are allowed to vary for the LiteBIRD, LiteCORE-80, CORE-M5, and \coreplus\
configurations. These forecasts assume $r=0.0042$ and $N_{\mathrm{eff}}=3.046$ as fiducial
values.
}
\label{tab:Neffplusr}
\end{table}

\begin{figure}
        \centering
\begin{center}
                \includegraphics[width=15cm]{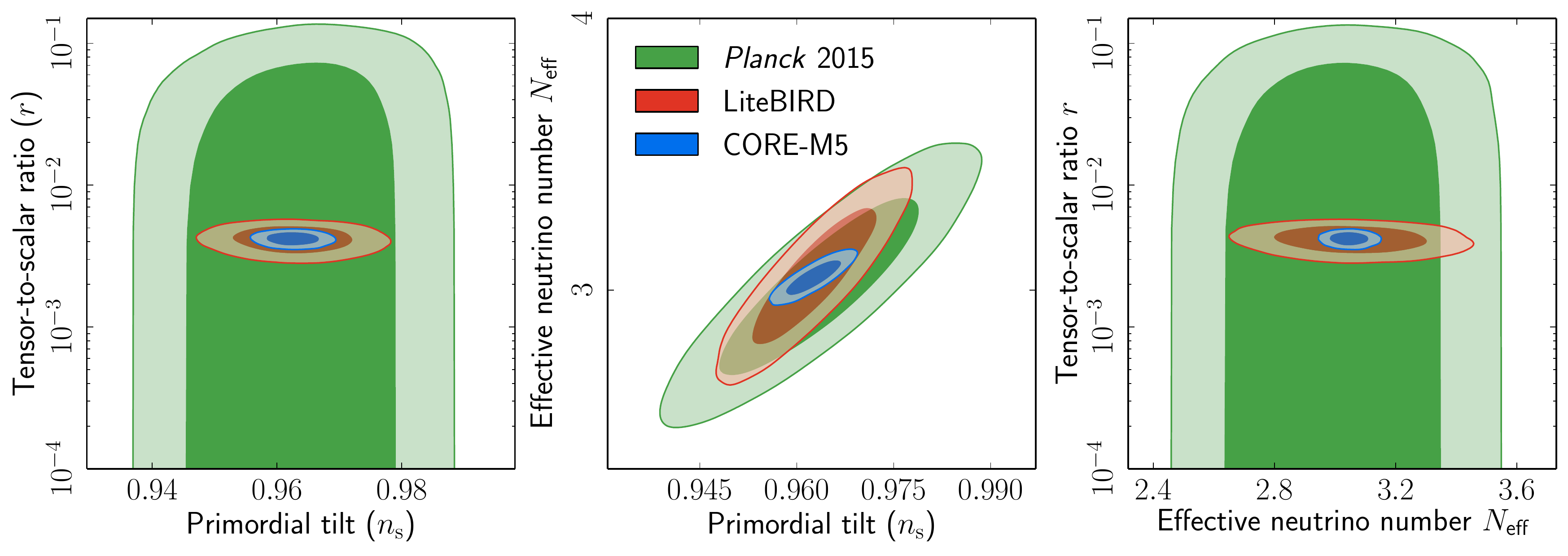}
\end{center}
\vspace{-1\baselineskip}
\caption{\footnotesize 
Forecast 68\% and 95\% CL 2D marginalized regions for the $(n_{\mathrm s}, r)$ (left panel), 
$(n_{\mathrm s}, N_{\mathrm{eff}})$ (middle panel) and $(N_{\mathrm{eff}},r)$ (right panel) for 
CORE-M5 (blue) and LiteBIRD (red) obtained by allowing the tensor-to-scalar ratio and the running 
to vary. These forecasts assume $r=0.0042$ and $N_{\mathrm{eff}}=3.046$ as fiducial values. The 
68\% and 95\% CL marginalized contours for Planck 2015 TT,TE,EE + lowP (green) are shown for 
comparison \cite{Ade:2015lrj}. Note that the Planck 2015 contours are based on real data whose 
best fit is different from the fiducial cosmology used in this paper.}
\label{fig:Neffplusr}
\end{figure}

\subsection{Beyond the consistency condition for the tensor tilt}

The consistency condition for the tensor tilt predicted by canonical single field slow-roll 
inflation is a key theoretical prior commonly assumed given the current sensitivity of CMB 
anisotropies measurements to the tensor contribution. Deviations from $n_{\mathrm t} \approx - 
r/8$ are expected for well motivated extensions of the simplest scalar field inflationary models, 
which include generalized Lagrangians \cite{1999PhLB..458..219G,Kobayashi:2010cm}, multifield 
inflation \cite{Bartolo:2001rt,Wands:2002bn,Byrnes:2006fr}, and gauge inflation 
\cite{Maleknejad:2012fw}. More radical departures from a nearly scale-invariant Gaussian spectrum 
of gravitational waves are expected in presence of non-vanishing anisotropic stresses from 
additional components during inflation \cite{Barnaby:2010vf} (see \cite{Guzzetti:2016mkm} for a recent review) or in alternative models to 
inflation \cite{Gasperini:1992em,Boyle:2003km,Brandenberger:2006xi}.

\begin{figure}
        \centering
\begin{center}
                \includegraphics[width=7.5cm]{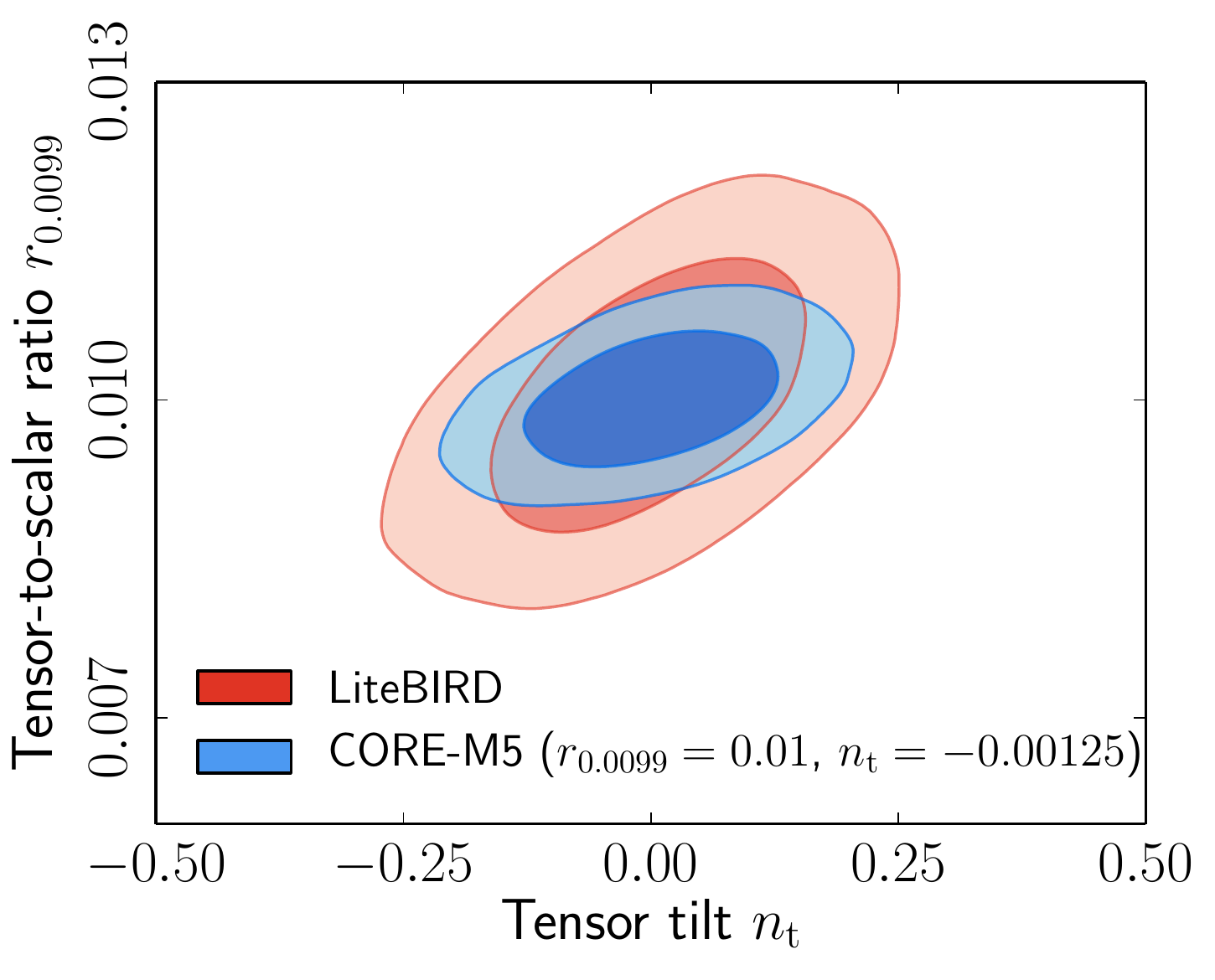}\includegraphics[width=7.5cm]{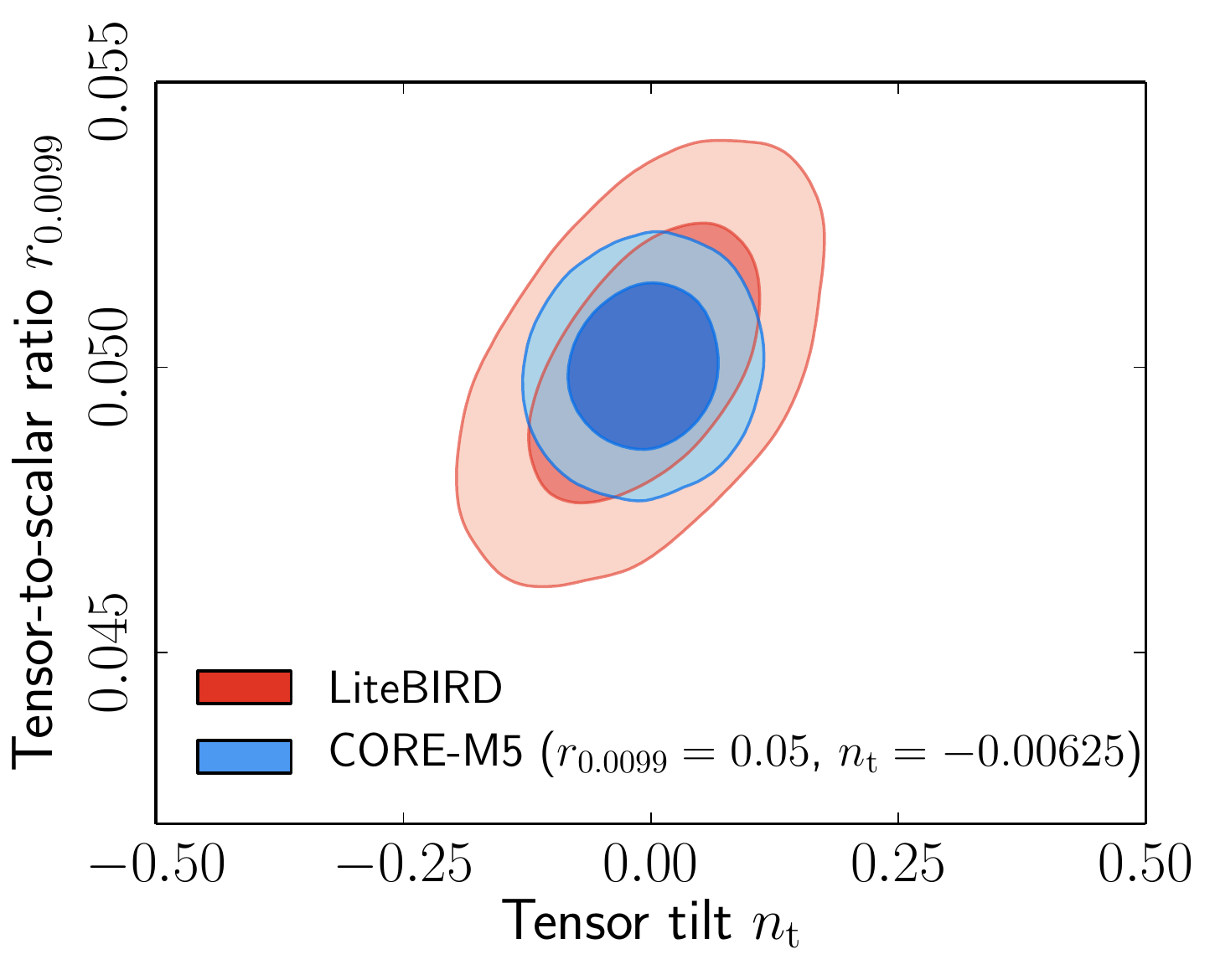}
\end{center}
\vspace{-1\baselineskip}
\caption{\footnotesize
Forecasts for the $(n_{\mathrm t}, r)$ marginalized regions at the 68\% and 95\% CL for CORE-M5 
(blue) and LiteBIRD (red) considering $r=0.01, n_{\mathrm t}=-r/8=-0.00125$ (left panel) and 
$r=0.05, n_{\mathrm t}=-r/8=-0.00625$ (right panel). The tensor pivot scale is $k_{*, {\mathrm 
t}} = 0.0099$ Mpc$^{-1}$. 
%and we use the same axis ranges as in the S4 science book 
%\cite{Abazajian:2016yjj} to allow for an easy comparison to the forecast constraints from 
%future ground based experiments.
}
\label{fig:tensortilt}
\end{figure}

In case of a statistically significant detection of primordial B-mode polarization, the next 
observational target would be the measurement of the tensor tilt $n_{\mathrm t}$. As a benchmark 
for our assessment of the expected uncertainties on the tensor tilt, we consider $(r=0.05, 
n_{\mathrm t}=-r/8=-0.00625)$ and $(r=0.01, n_{\mathrm t}=-r/8=-0.00125)$ at $k_{*, {\mathrm t}} 
= 0.0099$ Mpc$^{-1}$ \footnote{In this 
Section we consider two different pivot scales for scalar and tensor 
power spectra in Eqs.~(\ref{scalarps}) and (\ref{tensorps}). We retain 
the standard pivot scale at $0.05$ Mpc$^{-1}$ for the scalar perturbations.} 
as fiducial values compatible with current constraints, and consistent with 
those considered in the S4 science book \cite{Abazajian:2016yjj}. The two dimensional marginalized regions 
expected from CORE-M5 and LiteBIRD are shown in Fig. \ref{fig:tensortilt}.
As a forecast 68\% CL uncertainty for the tensor tilt we obtain $\sigma (n_{\mathrm t}) = 0.05$ 
($0.08$) for $r=0.05$ and $\sigma (n_{\mathrm t}) = 0.08$ ($0.1$) for $r=0.01$ with CORE-M5 
(LiteBIRD) specifications. The CORE-M5 uncertainties on $r$ remain basically unchanged when we allow for 
free $n_{\mathrm t}$ ($\sigma (r)=0.0009$ and $\sigma (r)=0.0004$ for $r=0.05$ and $r=0.01$ as fiducial values, respectively). 
Although the verification of the consistency relation for the tensor tilt 
was expected out of reach for values consistent with the current upper bounds on $r$ \cite{Huang:2015gca,Errard:2015cxa}, 
the CORE capability to probe either the recombination and the reionization peak is a key aspect for measuring the slope of the tensor power spectrum.
%will be an added value with respect to ground experiments.

\subsection{Constraints on the curvature}

Since the measurement of the first acoustic peak of the CMB temperature power spectrum 
\cite{deBernardis:2000sbo}, the consistency of cosmological observations with a flat Universe has 
been an important test in favor of inflation. \Planck\ data set tight constraints on the spatial 
curvature at the percent level. The \Planck\ 2015 result $\Omega_{\mathrm k} = - 
0.040^{+0.024}_{-0.016}$ at 68\% CL obtained by temperature and polarization data is further 
tightened by the inclusion of Planck lensing, i.e., $\Omega_{\mathrm k} = - 
0.004^{+0.008}_{-0.007}$ \cite{Ade:2015xua,Ade:2015zua}, or by the inclusion of 
BAO \cite{Ade:2015xua}. We now discuss the capabilities of the different experimental 
configurations for a future CMB space mission in constraining the spatial curvature.

As from Table \ref{tab:omegak}, the LiteBIRD forecast uncertainty for $\Omega_{\mathrm k}$ 
based on TE does not improve on the current \Planck\ constraints including lensing. The CORE-M5 
forecast uncertainty based on TE is approximatively an order of magnitude better than the 
corresponding \Planck\ constraint and six times better than what is expected from LiteBIRD. By 
including the lensing information, the CORE-M5 forecast uncertainty on $\Omega_{\mathrm k}$ 
further improves by approximatively 18\%. \coreplus  can further reduce the CORE-M5 uncertainty on 
$\Omega_{\mathrm k}$ by 5\%.

It is interesting to note that CORE-M5 will provide an accurate measurement of the $E$-mode 
polarization spectrum in the region of the acoustic peaks sufficient to remove 
almost completely the CMB 
degeneracy of $\Omega_{\mathrm k}$ with the other cosmological parameters. The amount of spatial 
curvature allowed by $Planck$ 2015 temperature and polarization data and improving on the 
$\Lambda$CDM fit with a $\Delta \chi^2 \sim - 5$ \cite{Ade:2015xua} will be therefore probed 
by CORE-M5 without the use of the lensing information. These improvements in the uncertainties of 
spatial curvature might open new avenues in the understanding of the global dynamics of inflation 
\cite{Kleban:2012ph} or could transform our understanding of inflation showing that our Universe 
is inside a large bubble embedded in the background vacuum space.

\begin{table}[h]
\begin{center}\footnotesize
\begin{tabular}{|c|c|c|c|c|}
\hline
Parameter  &  LiteBIRD,$\,$ TE  &  CORE-M5,$\,$TE  &  CORE-M5,$\,$ TEP  & \coreplus ,$\,$ TEP \\
\hline
$\Omega_{\mathrm{b}} h^2$  &  $0.02215\pm0.00014$ & $0.022140\pm0.000041$ & $0.022141\pm0.000040$ & $0.022141\pm0.000034$ \\
$\Omega_{\mathrm{c}} h^2$  &  $0.1206\pm0.0010$ & $0.12059\pm0.00068$ & $0.12057\pm0.00067$ & $0.12057\pm0.00064$ \\
$ 100\theta_{\mathrm{MC}}$  &  $1.03922\pm0.00030$ & $1.039219\pm0.000086$ & $1.039219\pm0.000086$ & $1.039220\pm0.000081$ \\
$ \tau$  &  $0.0583\pm0.0021$ & $0.0582\pm0.0020$ & $0.0583\pm0.0020$ & $0.0582\pm0.0019$ \\
$ n_\mathrm{s}$  &  $0.9625\pm0.0034$ & $0.9626\pm0.0018$ & $0.9626\pm0.0018$ & $0.9626\pm0.0018$ \\
$ \ln(10^{10} A_\mathrm{s})$  &  $3.0533\pm0.0046$ & $3.0532\pm0.0043$ & $3.0533\pm0.0043$ & $3.0532\pm0.0042$ \\
$\Omega_{\mathrm{k}}$  &  $-0.001\pm0.012$ & $0.0001\pm0.0021$ & $0.0000\pm0.0019$ & $0.0000\pm0.0018$ \\
\hline
$H_0$  &  $66.8_{-6.1}^{+4.8}$ & $66.98\pm0.81$ & $66.90\pm0.65$ & $66.90\pm0.63$ \\
$\sigma_8$  &  $0.827\pm0.010$ & $0.8286\pm0.0040$ & $0.8285\pm0.0040$ & $0.8285\pm0.0039$ \\
\hline
\end{tabular}
\end{center}
\caption{
Forecast 68\% CL constraint on primary and derived cosmological parameters when the spatial 
curvature $\Omega_{\mathrm{k}}$ is allowed to vary for LiteBIRD and CORE-M5 without and with the 
lensing information, and \coreplus . These forecasts assume $\Omega_{\mathrm{k}}=0$ as fiducial value.
}
\label{tab:omegak}
\end{table}

\begin{figure}[h!]
\centering 
\includegraphics[width=1\textwidth]{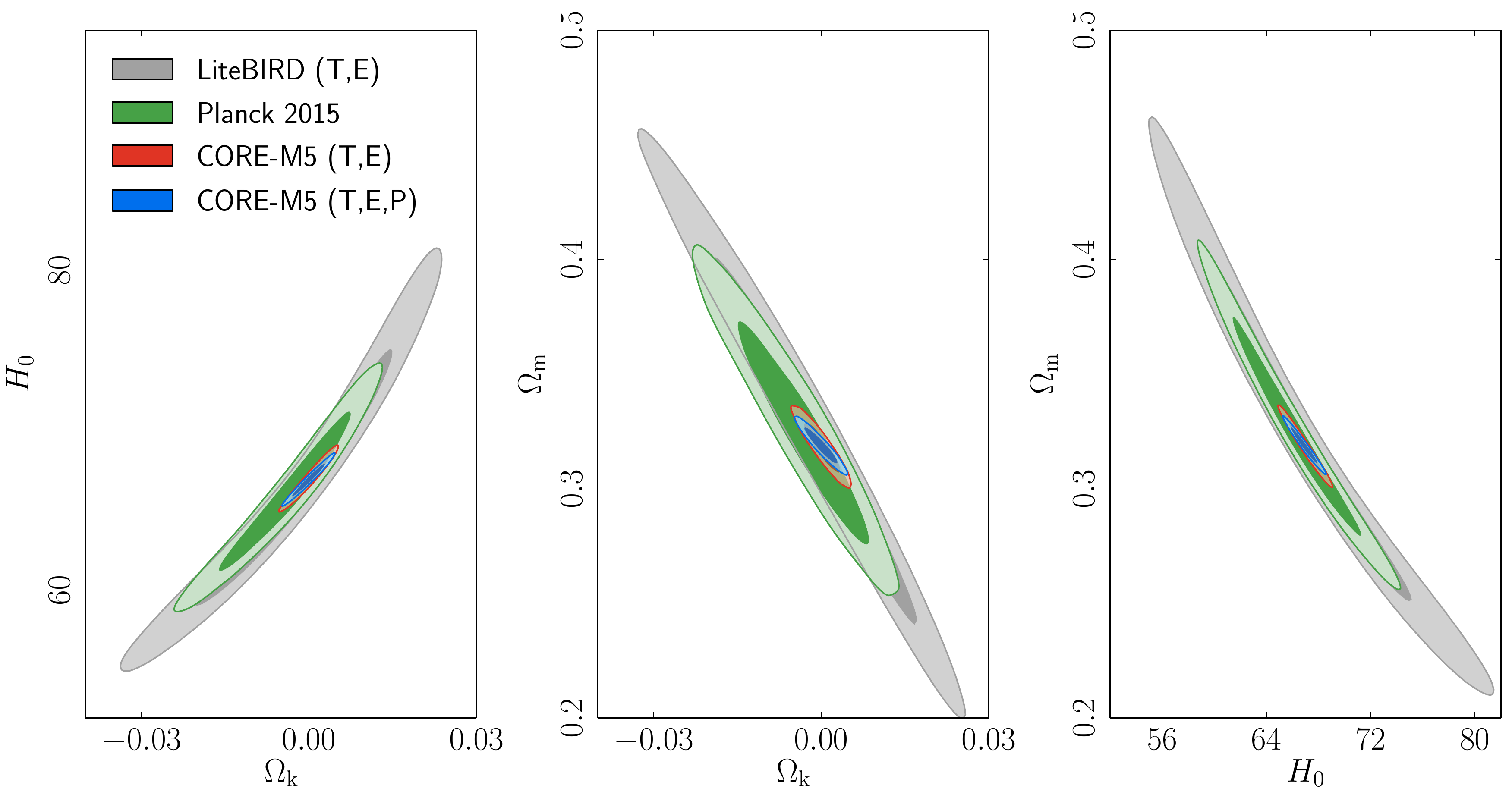}
\caption{\footnotesize 
Forecast 68\% and 95\% CL marginalized regions for $(\Omega_{\mathrm k}, H_0)$ (left panel), 
$(\Omega_{\mathrm k}, \Omega_{\mathrm m})$ (middle panel) and $(H_0, \Omega_{\mathrm m})$ (right 
panel) for LiteBIRD (grey) and CORE-M5 (blue) obtained by allowing $\Omega_{\mathrm k}$ to vary. 
These forecasts assume $\Omega_{\mathrm{k}}=0$ as fiducial value. The
68\% and 95\% CL marginalized contours for Planck 2015 TT,TE,EE + lowP + lensing (green) are shown for
comparison \cite{Ade:2015xua}. Note that the Planck 2015 contours are based on real data whose
best fit is different from the fiducial cosmology used.
} 
\label{fig:AlphaK}
\end{figure}

\subsection{Summary}

The forecasts presented in this Section show how the experimental configuration of the CORE 
mission proposed to ESA allows for a significant improvement on the uncertainties of key 
inflationary parameters with respect to the current measurements largely based on Planck. CORE-M5 
performs better than the LiteCORE-80 and LiteCORE-120 configurations (the latter with a slightly larger telescope but 
higher noise due to the smaller number of detectors than CORE-M5). Thanks to a 1.5 m telescope, COrE+ 
leads to an improvement with respect to CORE-M5, which is however modest compared to the higher 
cost connected to such larger mirror.

Our forecasts show that the LiteBIRD noise sensitivity (based on the LiteBIRD-ext configuration 
\cite{Errard:2015cxa}) 
%which is based on the most updated LiteBIRD configuration \cite{Matsumura:2016sri}) 
guarantees a measurement of the optical depth at the same level of the 
other configurations considered. Due to the smaller aperture, LiteBIRD provides an 
uncertainty on the spectral index $n_{\mathrm s}$ at least twice that of CORE-M5.  Moreover, 
LiteBIRD will not provide a significant improvement with respect to 
the most recent \Planck\ results 
\cite{planck2014-a10} in the measurement of $\Omega_{\mathrm b} h^2$, $\theta_{\mathrm{MC}}$ (and 
therefore $H_0$), the scale dependence of the spectral index, and $\Omega_{\mathrm k}$. 
Concerning the search for primordial B-mode polarization, the LiteBIRD forecast uncertainty in $r$ is 
approximatively twice as large as from CORE-M5. The lower 
noise sensitivity and higher angular resolution of CORE-M5 becomes even more important in the 
comparison with LiteBIRD when theoretical assumptions are relaxed as the case of varying 
the effective number of neutrinos $N_{\mathrm{eff}}$ shows.

\section{Constraints on slow-roll inflationary models}
\label{sec:five}

At present, the full set of CMB temperature and polarization anisotropy 
measurements~\cite{Ade:2013sjv, Adam:2015rua, Array:2015xqh, Ade:2015lrj} can be accounted for 
in a minimal setup, where inflation is driven by a single scalar field $\phi$ with 
a canonical kinetic term, minimally coupled to gravity,
and evolving in a flat potential $V(\phi)$. In this slow-roll regime, the dynamics are
often parameterized in terms of the HFFs
\begin{equation}
\epsilon_{n+1} = \frac{\mathrm{d}\ln\epsilon_n}{\mathrm{d}N}
\end{equation}
where $\epsilon_0=H_\mathrm{in}/H$ and $N=\ln a$ is the number of $e$-folds. 
The values of the $\epsilon_n$ parameters characterize the predictions of each model. 
At second order in slow roll, one has $n_\mathrm{s}=1-2\epsilon_1 - \epsilon_2 - 2\epsilon_1^2 - (2C+3)\epsilon_1 \epsilon_2 -C \epsilon_2 \epsilon_3$, 
$r= 16\epsilon_1 + 16C \epsilon_1 \epsilon_2$ and $\mathrm{d}n_\mathrm{s} / \mathrm{d} \ln k=-2 \epsilon_1 \epsilon_2 - \epsilon_2 \epsilon_3$, 
where the parameter $C \approx -0.73$ is a numerical constant. In \Sec{sec:primordialparameters}, we derive forecasts on the parameters $\epsilon_i$ 
in various experimental configurations.

Since particle physics beyond the electroweak scale remains elusive, and given that inflation can proceed at energy scales as large as $10^{16}\GeV$, 
hundreds of inflationary scenarios have been proposed. A systematic Bayesian analysis for a selection of inflationary models 
listed in \Ref{Martin:2013tda} has revealed that one third of them can now be considered as ruled out 
\cite{Martin:2013tda, Martin:2013nzq, Vennin:2015eaa}, 
while the vast majority of the preferred scenarios are of the plateau type, i.e., 
they are such that the potential $V(\phi)$ is a monotonic 
function that asymptotes to a constant value when $\phi$ goes to infinity. This still leaves us with about $50$ ``favored'' potentials of those listed in 
\Ref{Martin:2013tda}, corresponding to various extensions of the Standard Model of particle physics between which it is therefore still impossible to discriminate. 
In this Section, we quantify the ability of the proposed mission to improve this picture and to provide further insight into the physics of the very early Universe.

For reference, we will consider two fiducial models of inflation, both of the plateau type and belonging to the favored models according to the 
recent \Planck\ measurements. The first is the Starobinsky model (SI), and corresponds to a case in which $r$ 
is sufficiently large so that $B$-modes should be detected with CORE. The second fiducial model is Mutated Hilltop Inflation (MHI), where the mass 
scale $\mu$ appearing in this potential has been taken to $\mu=0.01\Mp$ (further details about these models can be found in \Ref{Martin:2013tda}), 
and corresponds to a case in which $r$ is too small for $B$-modes to be detected. In both cases, the reheating temperature has been fixed 
to $T_\mathrm{reh}=10^8\,\mathrm{GeV}$ with an average equation of state $\bar{w}_\mathrm{reh}=0$, 
while the post-inflationary evolution is 
assumed to be described by a flat $\Lambda\mathrm{CDM}$ 
model with $\Omega_\mathrm{b}h^2=0.0223$, $\Omega_\mathrm{c}h^2=0.120$, 
$\Omega_\nu h^2=6.45 \times 10^{-4}$, $\tau=0.0931$ and $h= 0.674$. 
In this Section we have not folded in $C_\ell^{\mathrm{PP}}.$ We 
have cross-checked our results with two independent numerical pipelines.
The first combines the Einstein-Boltzmann code \texttt{CLASS}~\cite{2011JCAP...07..034B} and the sampling code \texttt{MontePython}~\cite{2013JCAP...02..001A}, 
and the second relies on \texttt{CAMB}~\cite{Lewis:1999bs} combined with the sampler \texttt{CosmoMC}~\cite{Lewis:2002ah}. 

\subsection{Constraints on slow-roll parameters}
\label{sec:slowrollparameters}

\subsubsection{Impact of apparatus size and sensitivity}
\label{sec:primordialparameters}
 
\begin{figure}[h!]
\begin{center}
\includegraphics[width=0.495\textwidth]{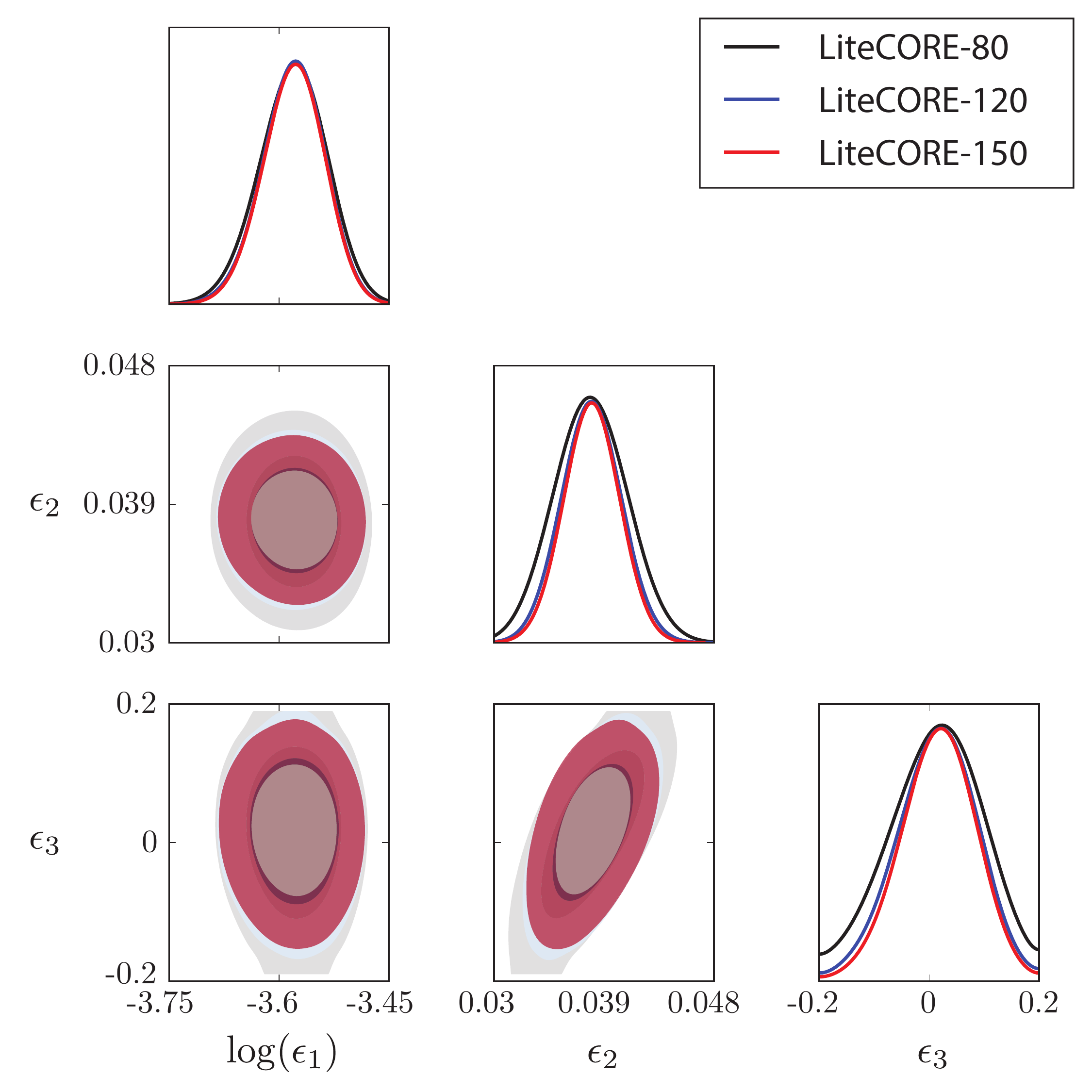}
\includegraphics[width=0.495\textwidth]{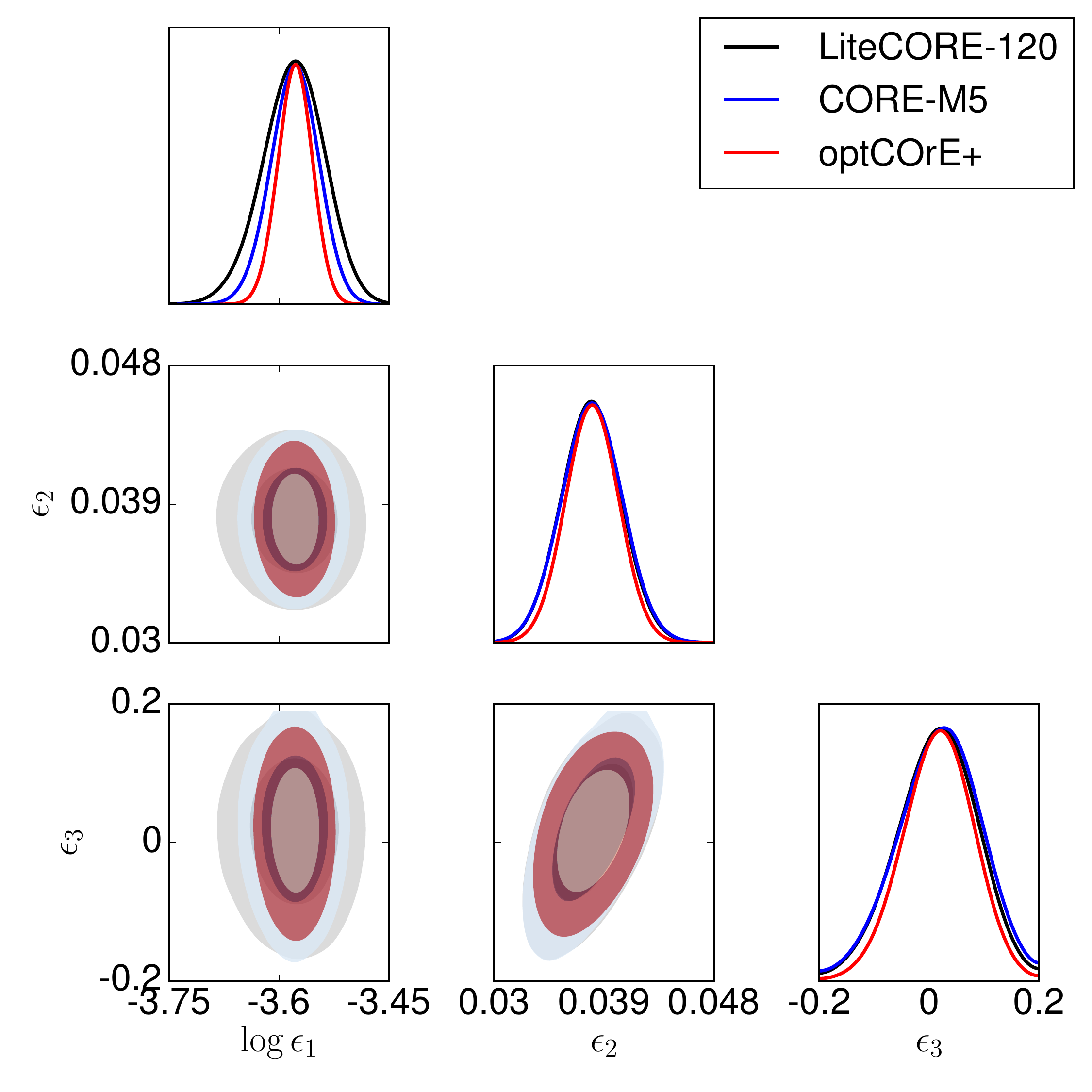}
\caption{Compared forecasts on the Hubble flow parameters (1D and 2D marginalized posterior distributions after CMBxCMB delensing) when the fiducial model is SI.}
\label{fig:eps:HI}
\end{center}
\end{figure}
 
\begin{figure}[h!]
\begin{center}
\includegraphics[width=0.495\textwidth]{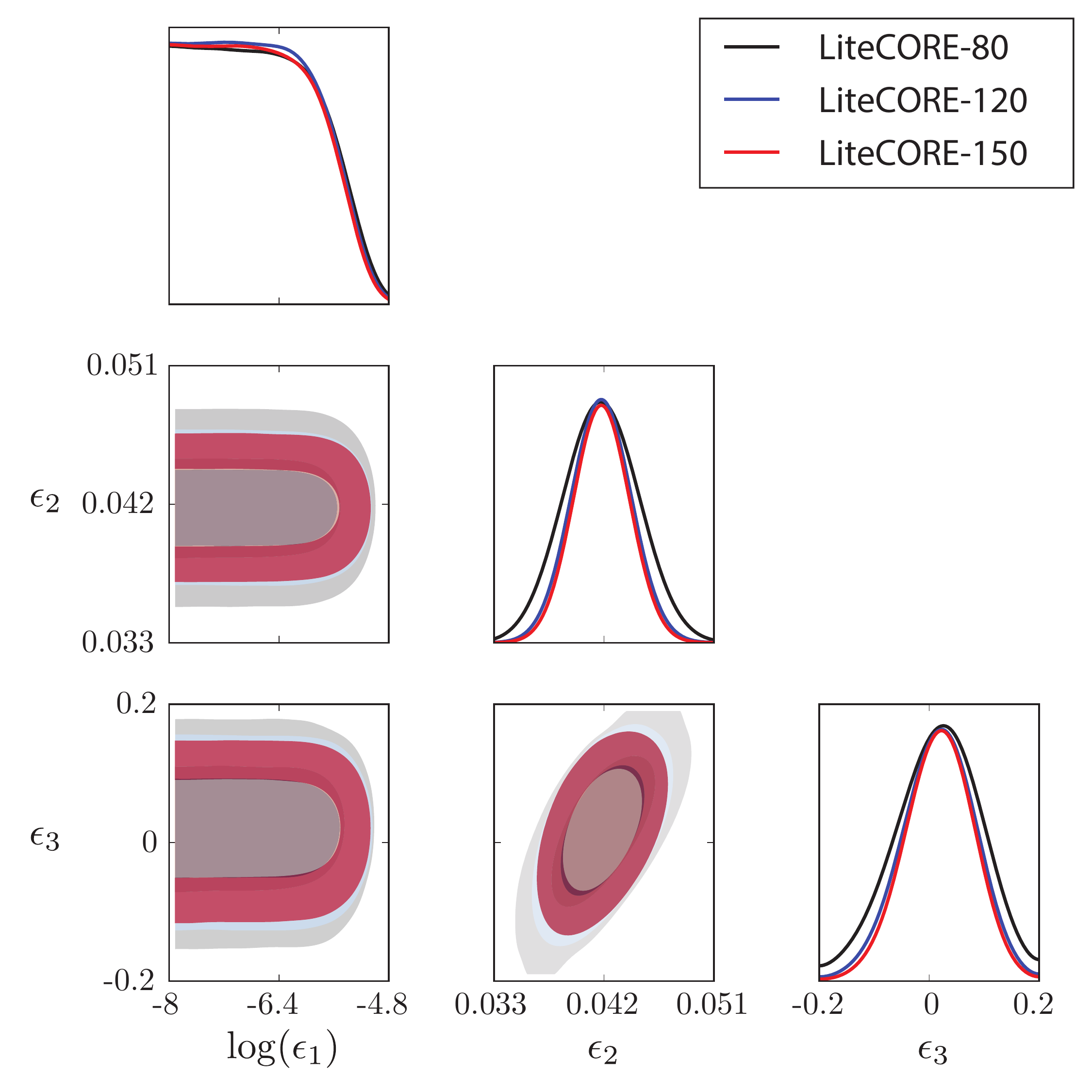}
\includegraphics[width=0.495\textwidth]{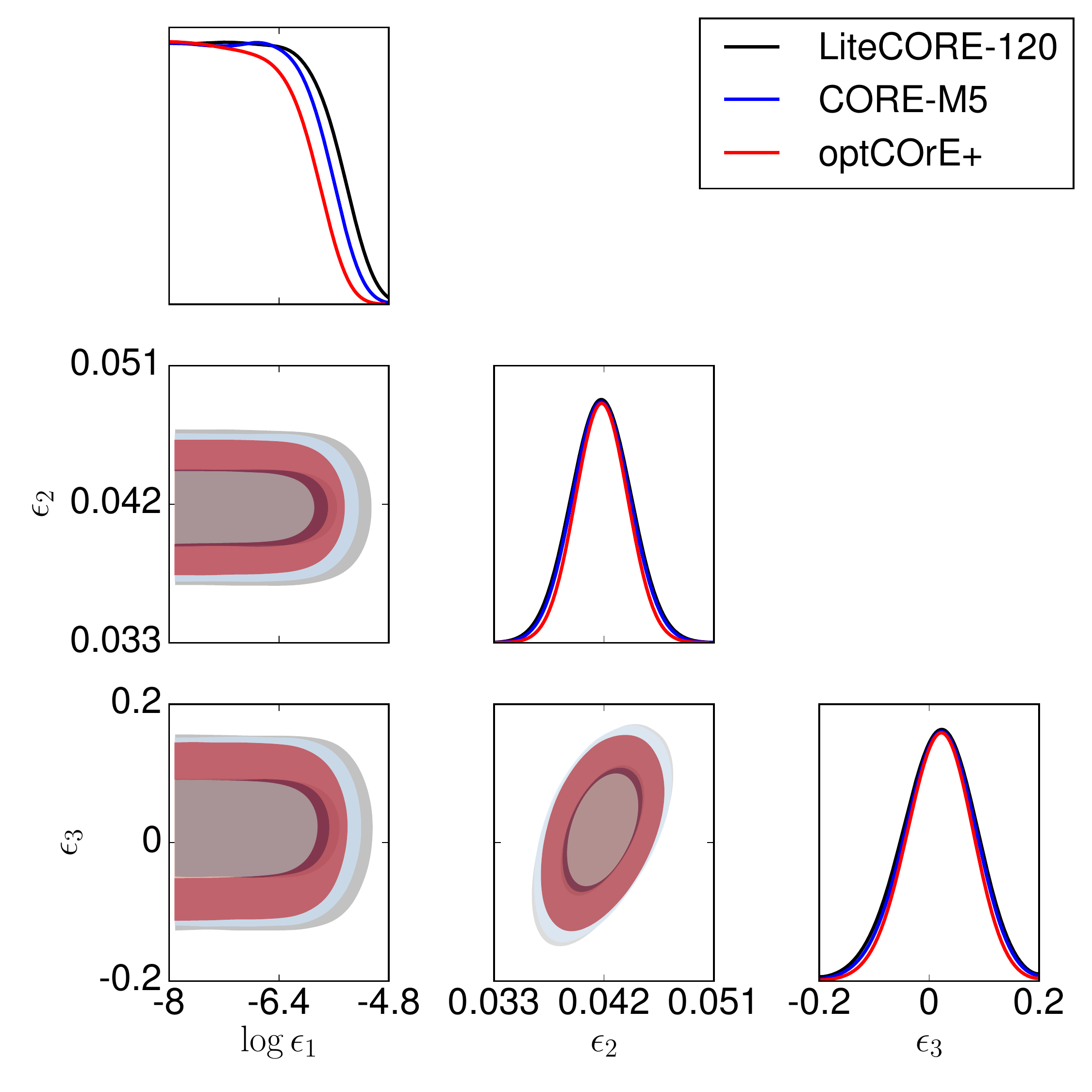}
\caption{Compared forecasts on the Hubble flow parameters (1D and 2D marginalized posterior distributions after CMBxCMB delensing) when the fiducial model is MHI.}
\label{fig:eps:MHI}
\end{center}
\end{figure}
 
The 1D and 2D marginalized posterior distributions on the $\epsilon_n$ parameters after CMBxCMB delensing are displayed 
in Figs.~\ref{fig:eps:HI} and~\ref{fig:eps:MHI}. At leading order in slow roll, $\epsilon_1$ is given by $r/16$, 
and since the order of magnitude of $r$ is unknown, a uniform log prior 
with $\log\epsilon_1\in[-7,-0.7]$ is used, 
while flat priors $\epsilon_2,\epsilon_3\in[-0.2,0.2]$ are employed. The corresponding 1$\sigma$ bounds or 2$\sigma$ upper limits are given in Table~\ref{tab:forecasts}.
\begin{table} [h!]  %\label{tab:forecasts}
\begin{footnotesize}
\begin{tabular}{c|c|c|c|c|c} 
 \multicolumn{6}{c}{\textit{\small{SI fiducial model}}}   \\ 
 \hline
\footnotesize{Parameter}  & \footnotesize{LiteCORE-80} & \footnotesize{LiteCORE-120} & \footnotesize{LiteCORE-150} & \footnotesize{CORE-M5} & \footnotesize{optCOrE+} \\ \hline 
$\log \epsilon_{1 }\, (-3.58)$  & $-3.581_{-0.040}^{+0.045}$  &$-3.580_{-0.037}^{+0.042}$  &   $-3.580_{-0.036}^{+0.040}$  &$-3.579_{-0.029}^{+0.029}$ &$-3.578_{-0.021}^{+0.022}$    \\
$\epsilon_{2 }\, (0.0381)$  & $0.0379_{-0.0029}^{+0.0028}$ & $0.0380_{-0.0023}^{+0.0023}$ &  $0.0380_{-0.0021}^{+0.0021}$ & $0.0380_{-0.0022}^{+0.0022}$&  $0.0380_{-0.0020}^{+0.0019}$\\
$\epsilon_{3 }\, (0.0193)$ & $0.013_{-0.076}^{+0.091}$ &$0.014_{-0.066}^{+0.075}$ &  $0.016_{-0.062}^{+0.069}$ &  $0.018_{-0.066}^{+0.078}$& $0.016_{-0.059}^{+0.063}$ \\
\hline
$\log r\, (-2.388)$ &  $-2.387_{-0.040}^{+0.045}$ & $-2.385_{-0.037}^{+0.042}$ & $-2.386_{-0.036}^{+0.040}$ & $-2.387_{-0.029}^{+0.029}$ &  $-2.386_{-0.021}^{+0.022}$ \\
$n_\mathrm{s}\, (0.9619)$  &$0.9621_{-0.0021}^{+0.0021}$ & $0.9620_{-0.0020}^{+0.0020}$&  $0.9620_{-0.0020}^{+0.0019}$  &  $0.9619_{-0.0022}^{+0.0022}$ &  $0.9619_{-0.0019}^{+0.0019}$\\
$\mathrm{d}n_\mathrm{s}/\mathrm{d}\ln k\, (-0.0008)$  &  $-0.0008_{-0.0030}^{+0.0031}$& $-0.0008_{-0.0026}^{+0.0026}$ &  $-0.0008_{-0.0024}^{+0.0024}$ &  $-0.0008_{-0.0028}^{+0.0028}$& $-0.0008_{-0.0023}^{+0.0023}$  \\
\hline
%\label{tab:forecasts}
 \end{tabular} \\
 \end{footnotesize}
%\end{table}
%
%\begin{table} [h!]
\begin{footnotesize}
\begin{tabular}{c|c|c|c|c|c} 
 \multicolumn{6}{c}{\textit{\small{MHI fiducial model}}}   \\ 
 \hline
\footnotesize{Parameter} &\footnotesize{LiteCORE-80} & \footnotesize{LiteCORE-120} & \footnotesize{LiteCORE-150} & \footnotesize{CORE-M5} & \footnotesize{optCOrE+} \\ \hline 
$\log \epsilon_{1 }\, (-7.66)$ &  $< -5.40$ &$< -5.44$  &  $<-5.46$ & $< -5.50$ &  $< -5.79$  \\
$\epsilon_{2 }\, (0.0419)$ & $0.0418_{-0.0029}^{+0.0029}$  & $0.0418_{-0.0023}^{+0.0023}$   & $0.0418_{-0.0021}^{+0.0021}$ & $0.0418_{-0.0023}^{+0.0022}$ & $0.0418_{-0.0020}^{+0.0020}$ \\
$\epsilon_{3 }\, (0.0209)$& $0.016_{-0.070}^{+0.083}$ & $0.017_{-0.060}^{+0.069}$ &  $0.019_{-0.057}^{+0.063}$  & $0.020_{-0.062}^{+0.071}$&  $0.018_{-0.054}^{+0.059}$\\
\hline
$\log r\, (-6.469)$ & $< -4.19$ & $< -4.24$ &$-4.25$&  $<-4.29$ &  $<-4.58$\\
$n_\mathrm{s}\, (0.9587)$ & $0.9588_{-0.0021}^{+0.0021}$  & $0.9588_{-0.0020}^{+0.0020}$ &  $0.9588_{-0.0020}^{+0.0019}$    &  $0.9588_{-0.0022}^{+0.0022}$& $0.9588_{-0.0019}^{+0.0019}$ \\
$\mathrm{d}n_\mathrm{s}/\mathrm{d}\ln k\, (-0.0009)$ & $-0.0009_{-0.0031}^{+0.0031}$ & $-0.0009_{-0.0026}^{+0.0026}$ &  $-0.0009_{-0.0024}^{+0.0024}$ &  $-0.0009_{-0.0028}^{+0.0028}$ &  $-0.0009_{-0.0023}^{+0.0023}$ \\
\hline
 \end{tabular} \\
 \end{footnotesize}
 \caption{Hubble flow parameters 1$\sigma$ bounds or 2$\sigma$ 
upper limit after CMBxCMB delensing. The corresponding constraints on the power spectra parameters $\log r$, 
$n_\mathrm{s}$ and $\mathrm{d}n_\mathrm{s}/\mathrm{d}\ln k$ are also given and have been 
obtained by importance sampling.  The numbers in parenthesis in the first column 
are the fiducial values used for the corresponding parameters.
}
\label{tab:forecasts}
\end{table}
In order to determine the optimal balance between the mirror size, the number of detectors, 
mission duration, and the total mission cost, an important task is to analyze how 
the forecasts for inflation are modified, first when varying the angular resolution at fixed 
sensitivity, and second when improving or degrading the sensitivity, 
always by using the experimental configurations presented in Section \ref{sec:three}.  

One obtains very similar results for $\epsilon_1$ with LiteCORE-80, LiteCORE-120 and LiteCORE-150, 
because these 
experimental configurations have the same 
sensitivity but different angular resolutions. For $\epsilon_2$ and $\epsilon_3$, some improvement 
occurs when going from LiteCORE-80 
to LiteCORE-120, mainly because of the degraded angular resolution, which reduces the lever arm to 
probe the spectral tilt and the running 
of the scalar power spectrum. Otherwise, very similar results are 
obtained for LiteCORE-120 and LiteCORE-150. Hence there is no 
clear gain in increasing the mirror size at fixed sensitivity for diameters beyond 120 cm.  

On the other hand, the constraints on $\epsilon_1$ are improved when going 
from LiteCORE to CORE-M5 (by about 30$\%$ for the SI fiducial), and 
from CORE-M5 to optCOrE+ (by about 25$\%$ for SI). This illustrates 
the importance of the sensitivity of the experiment for recovering 
the first Hubble flow parameter.  Nevertheless there is no significant improvement 
in the recovery of the second and third Hubble flow parameters.   

We conclude that the sensitivity plays the dominant role, and thus the number and quality 
of detectors at the focal plane. The number of detectors typically scales with the size of the focal plane, 
and thus with the size of the experiment, which itself determines the angular resolution.  Nevertheless, 
one finds that the angular resolution alone plays a minor role, as long as it suffices to probe 
the lensing spectrum and efficiently delens the B-mode. These results give strong support 
to the CORE-M5 specifications, which optimize the sensitivity through the number of detectors and the 
mission duration, for a slightly reduced mirror size compared to the previous COrE+ specifications.

%\FloatBarrier
\subsubsection{Removing low multipoles and delensing }
\label{sec:lowmultipoles}
\begin{figure}[h!]
\begin{center}
\includegraphics[width=\textwidth]{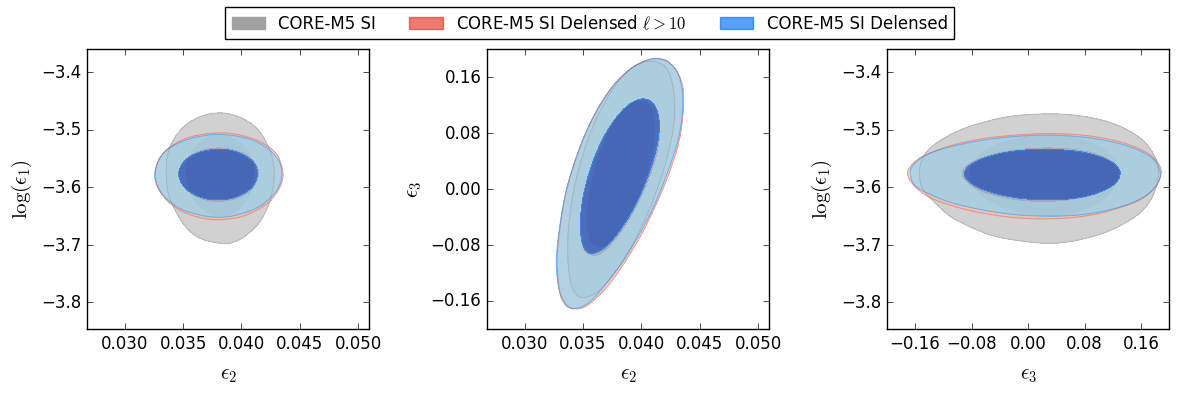}
\includegraphics[width=\textwidth]{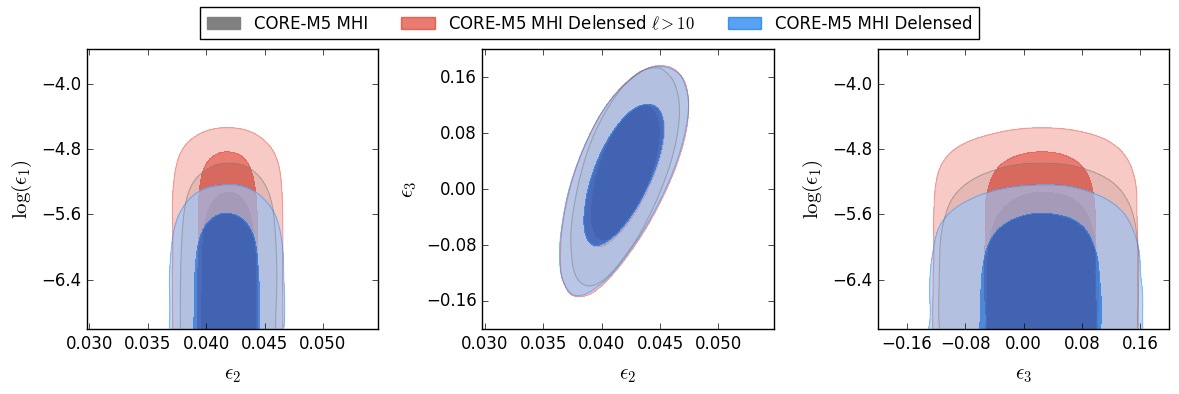}
\caption{Forecasts on the Hubble flow parameters presented in \Sec{sec:primordialparameters} 
(blue), when the low multipoles $\ell < 10$ are removed (red), and when delensing 
is not performed (grey), for CORE-M5 and SI (top panel) and MHI (bottom panel) as the fiducial model.}
\label{fig:lowl}
\end{center}
\end{figure}
 
The forecasts derived in Section~\ref{sec:primordialparameters} 
rely on the assumption that low multipoles can be correctly measured and 
that delensing can be performed successfully.
However, these two tasks 
can be challenging in practice, and this is why in this Subsection we assess 
the degradation of our results when the first multipoles $\ell <10$ are removed 
from the analysis and when delensing is not carried out. In \Fig{fig:lowl}, 
the 2D posterior contours on the Hubble flow parameters are displayed for CORE-M5 
with and without $\ell <10$ and with and without delensing when SI and MHI are the fiducial models. 

For SI, one can see that the constraints are almost unchanged when removing low multipoles, 
and slightly degraded without delensing.
We found that this translates 
into no substantial difference for the constraints in the space of inflationary models and 
reheating expansion history (see Secs.~\ref{sec:modelComparison} and~\ref{sec:reheating}). 
We therefore conclude that our analysis is robust under low multipoles removal and delensing in this case. 

For MHI, however, the constraints on $\epsilon_2$ and $\epsilon_3$ do not change much when the 
low multipoles are removed, but $\epsilon_1$ is substantially affected. If delensing is not 
performed, the constraint on $\epsilon_1$ is weaker, but the main degradation comes from 
low multipole removal. We also found 
that in this case substantial differences for model comparison and reheating constraints are obtained.

\FloatBarrier
\subsection{Inflationary model comparison}
\label{sec:modelComparison}
\begin{figure}[h!]
\begin{center}
\includegraphics[width=0.49\textwidth]{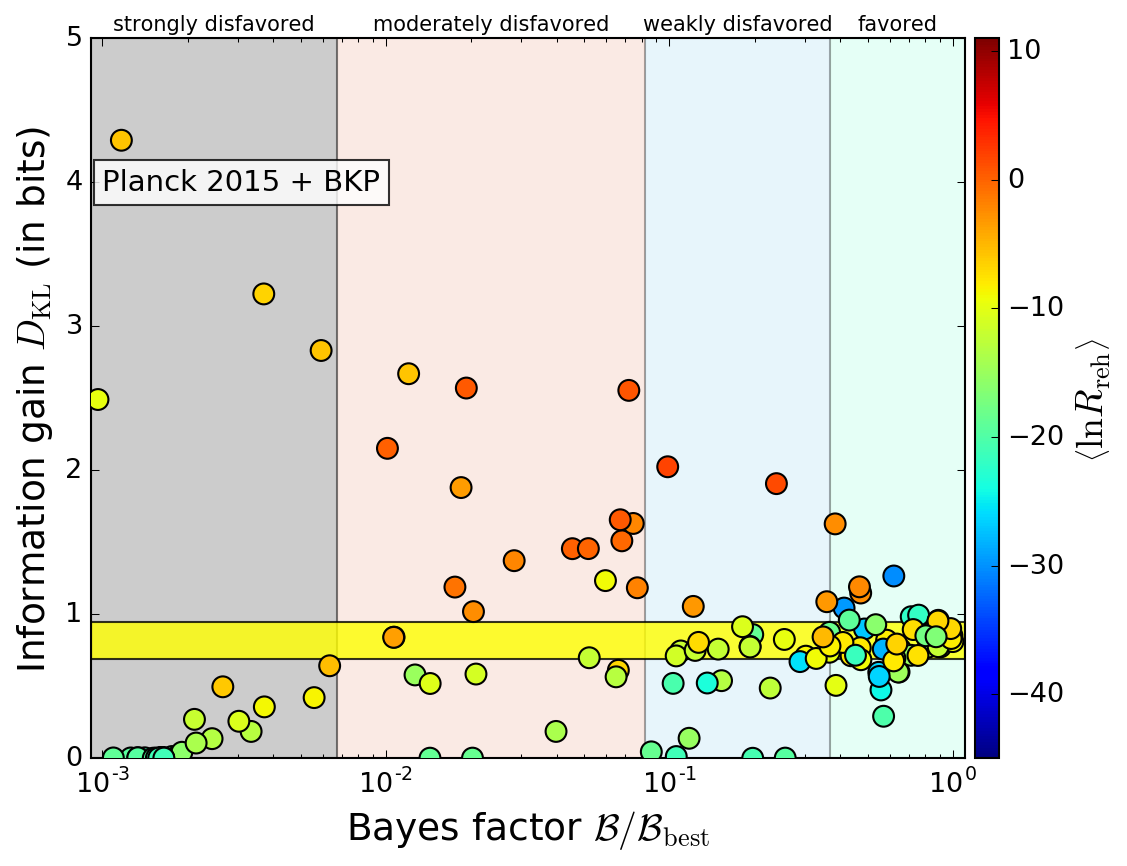}
\includegraphics[width=0.49\textwidth]{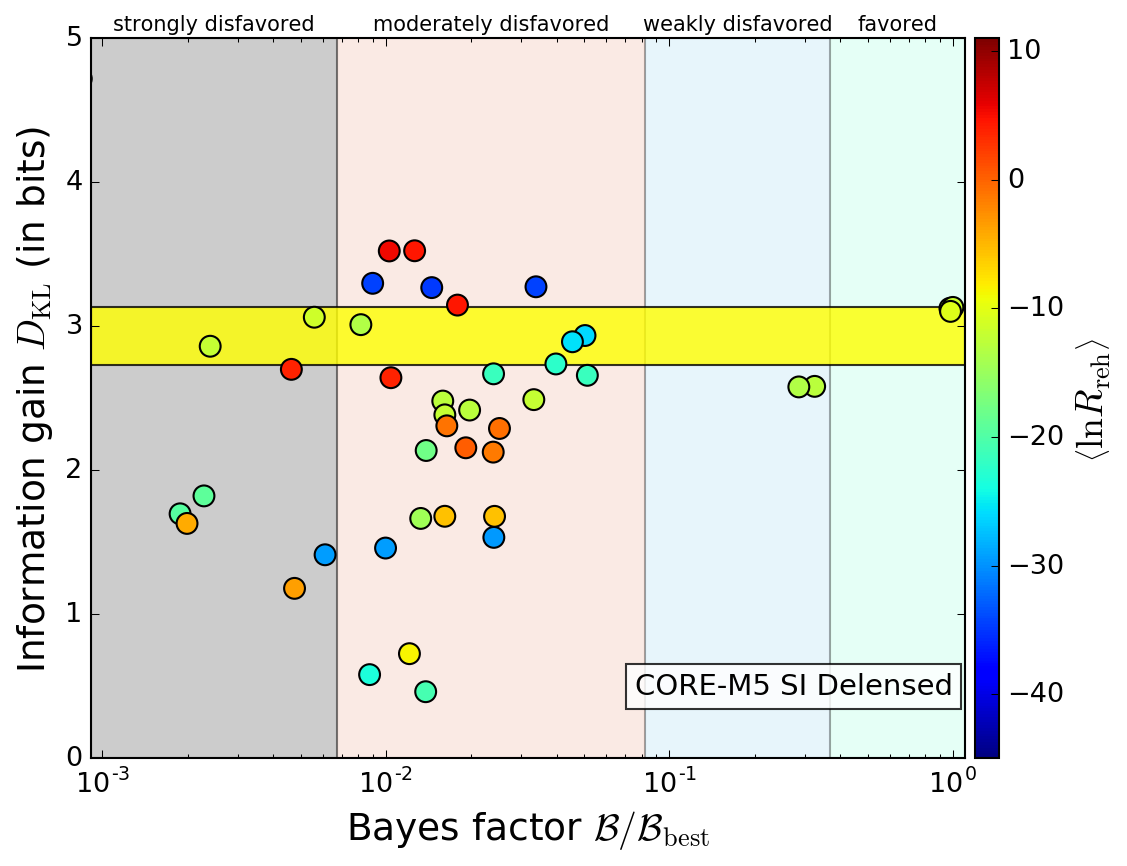}
\caption{Information gain on reheating measured by the Kullback-Leibler divergence $D_\mathrm{KL}$ (in bits) on the reheating parameter $\ln R_{\mathrm{reh}}$ for \Planck\ 2015 + BKP joint cross-correlation~\cite{Ade:2015tva} (left panel, results taken from \Ref{Martin:2016oyk}) and for CORE-M5 if SI is the fiducial model, as a function of the Bayesian evidence normalized to the best model. Each circle represents one model and the $x$-axis is divided into Jeffreys categories. The circle color traces the mean value of the reheating parameter $\ln R_{\mathrm{reh}}$. The yellow band represents the 1$\sigma$ deviation around the mean value for $D_\mathrm{KL}$.}
\label{fig:Rreh:Planck}
\end{center}
\end{figure}

We now investigate 
how the constraints of \Sec{sec:primordialparameters} translate 
into a Bayesian comparison of inflationary models. 
As mentioned above, \Planck\ has allowed us to rule out one 
third of single-field slow-roll models listed in \Ref{Martin:2013tda}, 
but still left us with about $27$\% of favored models. In this Section, we study how much this number could be decreased with a CORE-type mission.

In practice, we computed the Bayesian evidence~\cite{Cox:1946,Jeffreys:1961,deFinetti:1974,Trotta:2008qt} $\mathcal{E}$ 
of all inflationary models $\mathcal{M}_i$ listed in Encyclopaedia Inflationaris~\cite{Martin:2013tda} 
using the priors proposed in \Ref{Martin:2013nzq}. We have followed the same method as described 
in \Refs{Ringeval:2013lea, Martin:2013nzq, Martin:2014lra, Martin:2014rqa, Vennin:2015egh, Hardwick:2016whe} 
which relies on the \texttt{ASPIC} library~\cite{aspic} and 
involves the derivation of an effective marginalized likelihood function depending 
only on the primordial parameters. All evidences have then been derived using 
the \texttt{MULTINEST} nested sampling algorithm~\cite{Mukherjee:2005wg, Feroz:2007kg, Feroz:2008xx} 
with a target accuracy of $10^{-4}$ and the number of live points equal to $30000$.

Under the principle of indifference, two models $\mathcal{M}_i$ and $\mathcal{M}_j$ can be 
compared by computing the ratio $\mathcal{E}_i/\mathcal{E}_j$ of their Bayesian evidence. 
This ratio is called the Bayes factor and is displayed in \Figs{fig:Rreh:Planck} and~\ref{fig:Rreh:1} 
for various experimental setups (\Planck\, LiteCORE-120, CORE-M5, and optCOrE+) and for the two 
fiducial models SI and MHI considered in this work. Each model is represented with a circle and 
its position along the $x$-axis stands for its Bayes factor with the best model found in each case. 
The vertical colored stripes stand for Jeffreys empirical scale where if $\ln(\mathcal{E}_i/\mathcal{E}_j)>5$, 
$\mathcal{M}_j$ is said to be ``strongly disfavored'' with respect to $\mathcal{M}_i$, ``moderately disfavored'' 
if $2.5<\ln(\mathcal{E}_i/\mathcal{E}_j)<5$, ``weakly disfavored'' if $1<\ln(\mathcal{E}_i/\mathcal{E}_j)<2.5$, 
and the result is said to be ``inconclusive'' if $\vert \ln(\mathcal{E}_i/\mathcal{E}_j)\vert<1$.

Compared to \Planck\ where $52$ models are still favored, one can see that only a few models among those listed in \Ref{Martin:2013tda} would 
remain favored if SI is taken as the fiducial model, 
and around $15$ with MHI, corresponding to a large improvement of the constraints in model space. 
Interestingly though, the level of constraints 
increases 
only moderately 
when going from LiteCORE to CORE-M5 and optCOrE+. 
 
\subsection{Reheating}
\label{sec:reheating}
%
%Inflation also needs to be connected to the subsequent hot Big Bang phase through an era of reheating, 
Inflation is connected to the subsequent hot Big Bang phase through an era of reheating,
during which the energy contained in the inflationary fields eventually decays into the Standard Model 
degrees of freedom. The amount of expansion during this epoch determines the amount of expansion between 
the Hubble crossing time of the physical scales probed 
in the CMB and the end of inflation~\cite{Martin:2006rs, Martin:2010kz, Easther:2011yq, Dai:2014jja, Rehagen:2015zma}. 
As a consequence, the kinematics of reheating affects the time frame during which the fluctuations probed in cosmological 
experiments emerge, and hence the location of the observable window along the inflationary potential. 
This effect can be used to extract constraints on a certain combination $\ln R_{\mathrm{reh}}$ of the 
averaged equation-of-state parameter $\bar{w}_{\mathrm{reh}}$ during reheating and the energy density at the end of reheating $\rho_{\mathrm{reh}}$,
\begin{eqnarray}
\ln R_{\mathrm{reh}} & = 
\displaystyle
\frac{1-3\bar{w}_{\mathrm{reh}}}{12\left(1+\bar{w}_{\mathrm{reh}}\right)}
\ln\left(\frac{\rho_{\mathrm{reh}}}{\rho_{\mathrm{end}}}\right)+\frac{1}{4}\ln\left(\frac{\rho_{\mathrm{end}}}{\Mp^4}\right)
\end{eqnarray}
where $\rho_\mathrm{end}$ is the energy density at the end of inflation. 
Around $0.8$ bits of information on this quantity can be extracted~\cite{Martin:2014nya, Martin:2016oyk} 
from the latest CMB measurements. However, it is still at a level where  
if the averaged equation-of-state parameter during reheating takes a fixed value that is not too close 
to $-1/3$ (for instance, vanishes), the reheating temperature cannot be constrained very well.
\begin{figure}%[t]
\begin{center}
\includegraphics[width=0.49\textwidth]{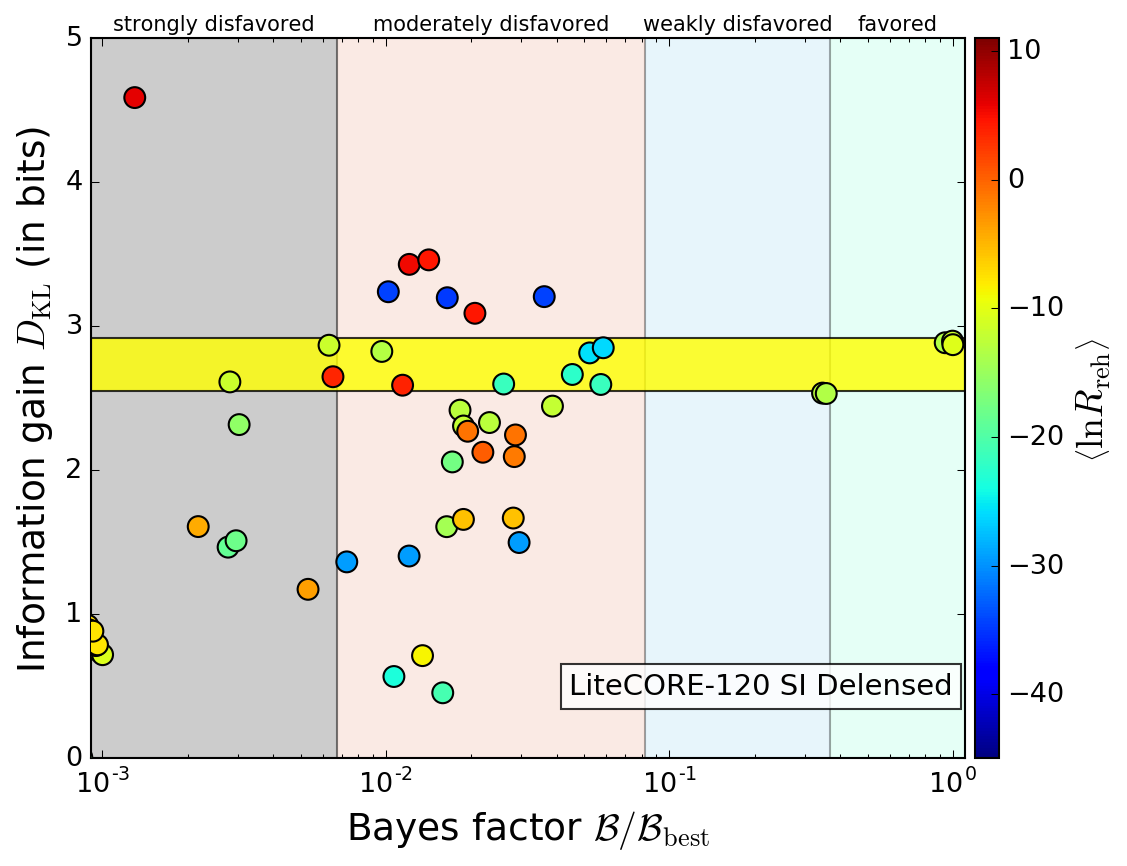}
\includegraphics[width=0.49\textwidth]{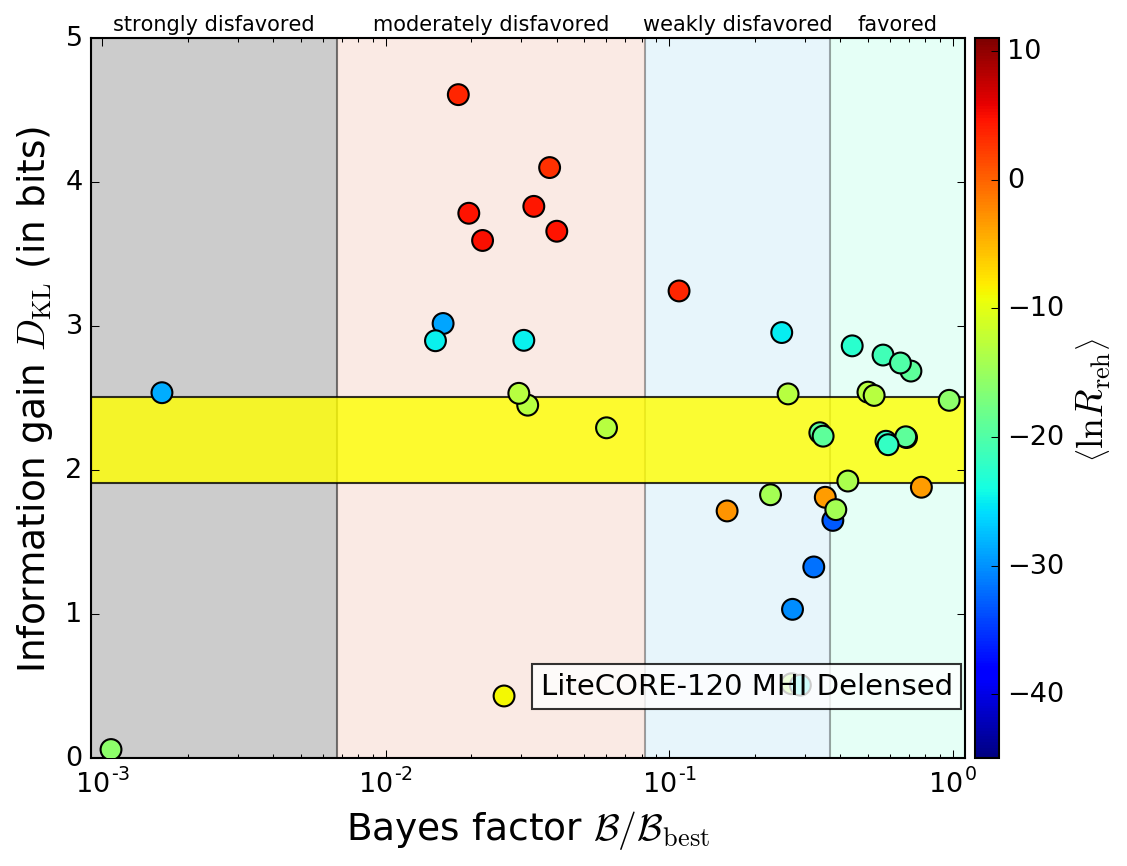}
\includegraphics[width=0.49\textwidth]{section_five/figures/corem5_hi_D}
\includegraphics[width=0.49\textwidth]{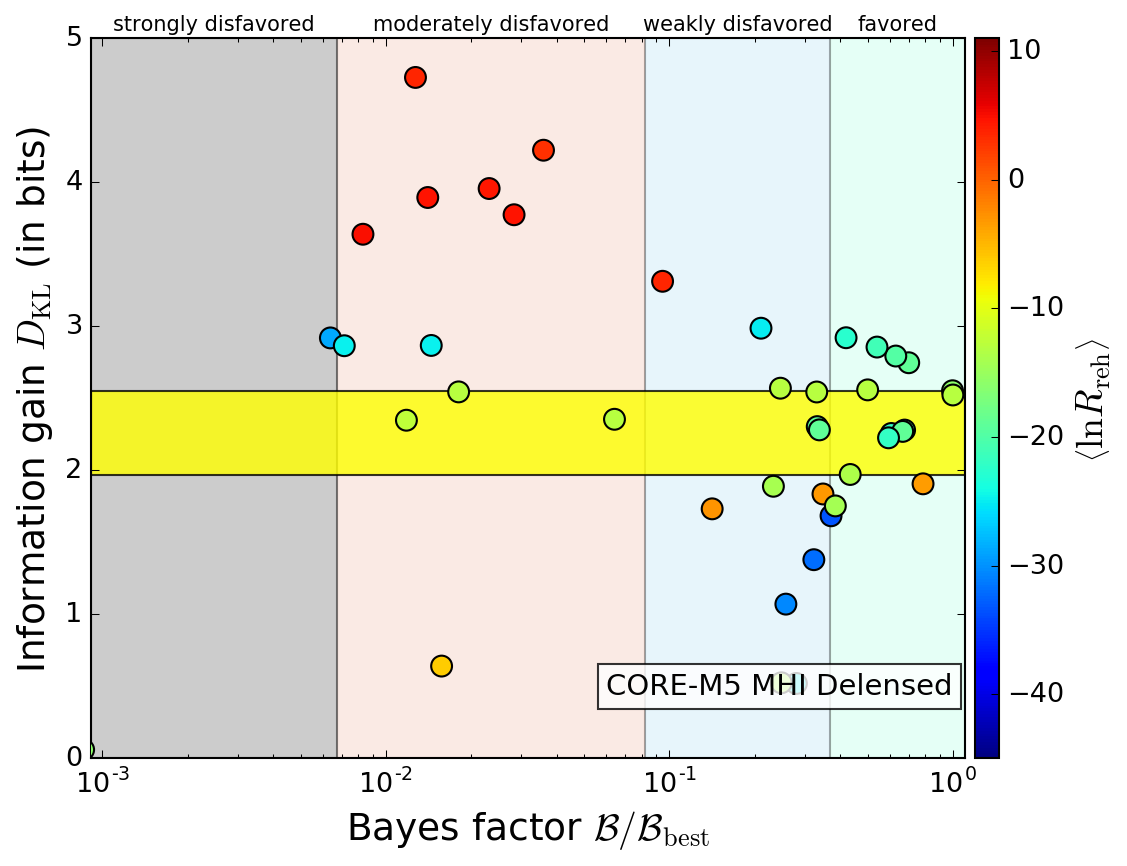}
\includegraphics[width=0.49\textwidth]{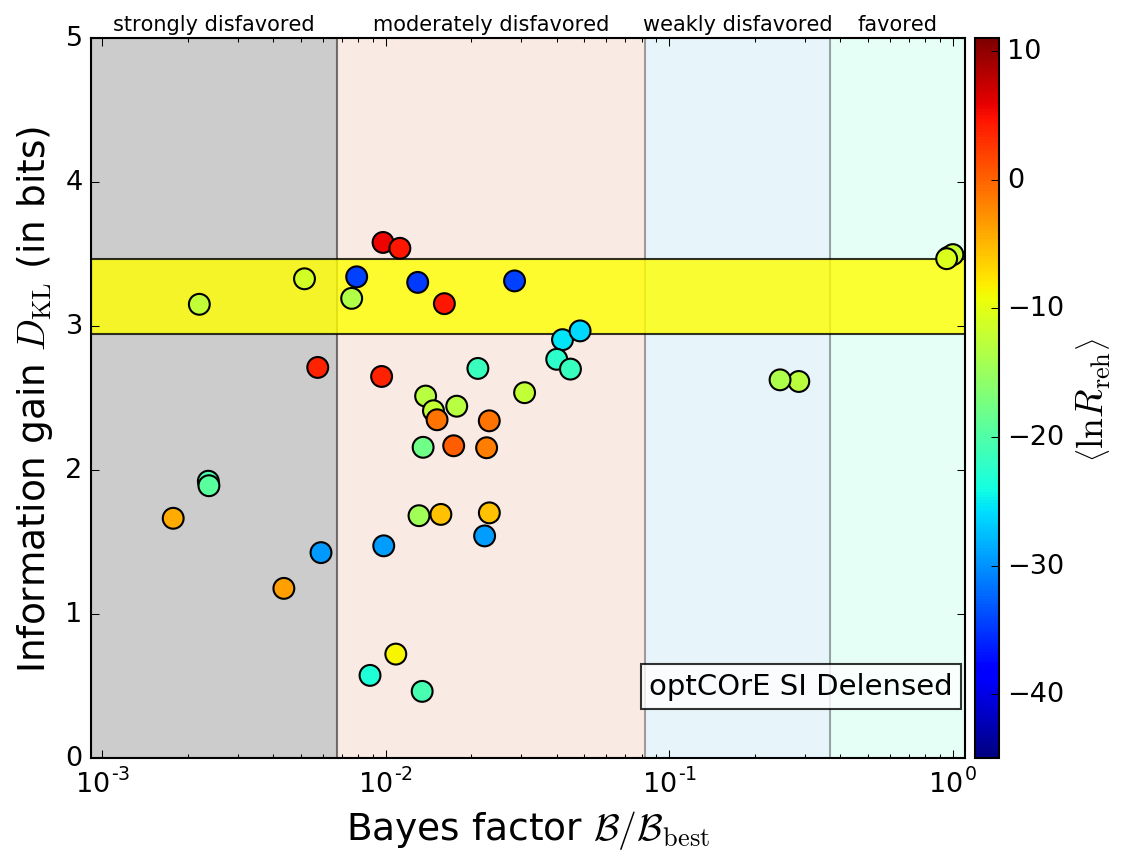}
\includegraphics[width=0.49\textwidth]{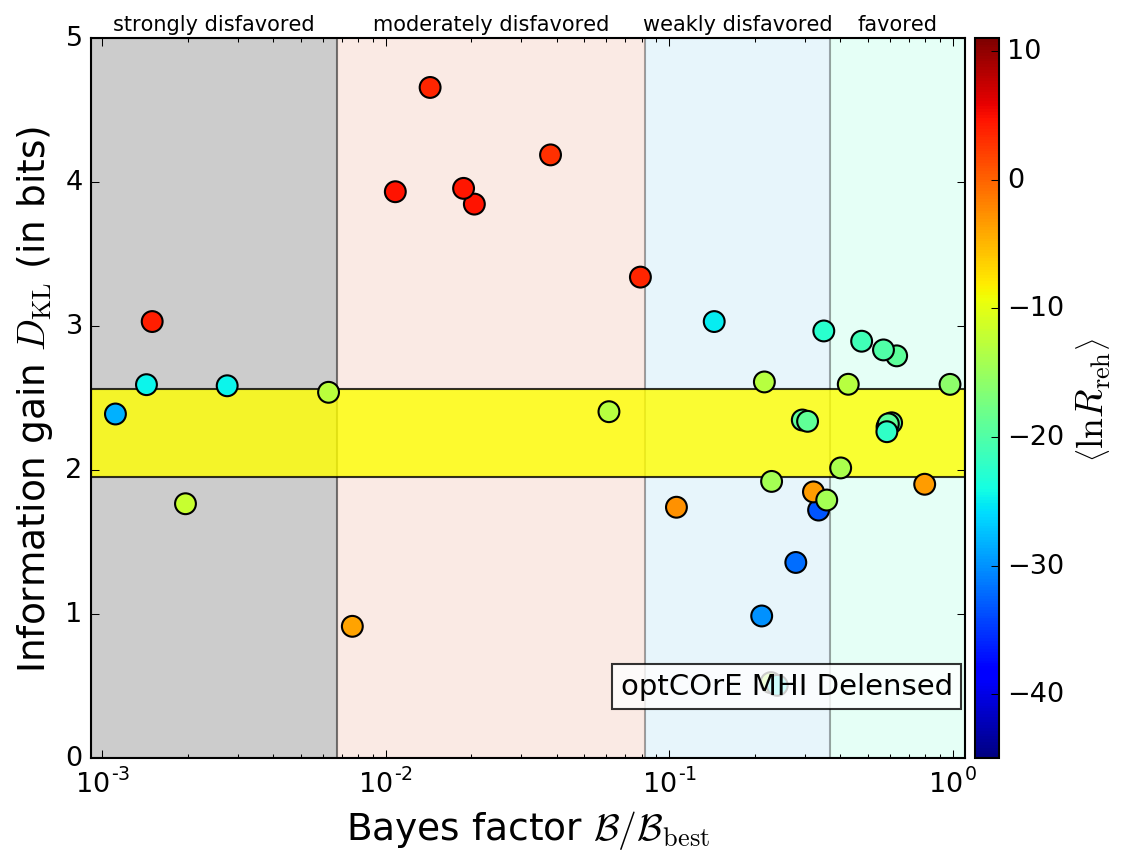}
\caption{Information gain on reheating measured by the Kullback-Leibler 
divergence $D_\mathrm{KL}$ (in bits) on the reheating parameter 
$\ln R_{\mathrm{reh}}$ for LiteCORE-120 (top panels), CORE-M5 (middle panels) 
and optCOrE+ (bottom panels), for SI (left panels) and MHI (right panels) 
as fiducial models as a function of the Bayesian evidence normalized to the best model. 
Each circle represents one model and the $x$-axis is divided into Jeffreys categories. 
The circle color traces the mean value of the reheating parameter $\ln R_{\mathrm{reh}}$. 
The yellow band represents the 1$\sigma$ 
deviation around the mean value for $D_\mathrm{KL}$. With CORE-type missions, only a few models would remain favored (compared to $52$ currently with Planck) and one would gain between $2$ and $3$ bits of information about reheating on average (compared to $0.8$ with Planck).}
\label{fig:Rreh:1}
\end{center}
\end{figure}

The amount of information gained regarding reheating 
can therefore be quantified using the Kullback-Leibler divergence~\cite{kullback1951} 
between the prior distribution $\pi$ and the posterior $P$ on $\ln R_{\mathrm{reh}}$,
\begin{eqnarray}
D_\mathrm{KL} = \int P(\ln R_\mathrm{reh}|D) \ln\left[
\dfrac{P(\ln R_\mathrm{reh}|D)}{\pi(\ln R_\mathrm{reh})}  \right] \mathrm{d} \ln R_\mathrm{reh}\, ,
\label{eq:DKL}
\end{eqnarray}
which is a measure of the amount of information
provided by the data $D$ about $\ln R_\mathrm{reh}$~\cite{Kunz:2006mc,
  Liddle:2007fy}. This quantity is also the discrepancy measure
between the posterior $P$ and the prior $\pi$ when the prior is viewed
as an approximation of the posterior. Because the Kullback-Leibler
divergence is invariant under any reparametrizations $x=f(\ln R_\mathrm{reh})$
and uses a logarithmic score function as in the Shannon's entropy, it
is a well-behaved measure of information~\cite{bernardo:2008}. In \Fig{fig:Rreh:Planck}, the Kullback-Leibler divergence of all single-field models 
within Encyclopaedia Inflationaris~\cite{Martin:2013tda} are displayed as a function 
of their Bayesian evidence for \Planck\ 2015 plus BKP joint cross-correlation~\cite{Ade:2015tva} 
and CORE-M5 if SI is the fiducial model. One can see that on average one would go from $0.8$ 
bit of information about reheating to more than $3$ bits.
Let us recall that $1$ bit is the amount of
information contained in answering ``yes'' or ``no'' to a given
question. In the present case, the question is about
the values of $\ln R_\mathrm{reh}$, and the current CMB data answer on average
whether $\ln R_\mathrm{reh}$ is large or small. In the same manner, $2$ bits would correspond to choosing between $4$ possible values for $D_\mathrm{KL}$ and $3$ bits to choosing between $8$ possible values.
Going from $1$ to $3$ bits is therefore a considerable gain that would 
open a new observational window into the physics of the end of inflation~\cite{Drewes:2015coa, Renaux-Petel:2015mga, Martin:2016iqo}.
In \Fig{fig:Rreh:1}, the same results are displayed for LiteCORE-120, CORE-M5, and optCOrE+ when 
SI and MHI are used as fiducial models. One can see that even if $r$ is not detected 
(here described by the case where the fiducial model is MHI), more than 2 bits of information would still be gained.
\FloatBarrier
\subsection{Summary}
\label{sec:discussion}
\begin{figure}%[b]
\begin{center}
\includegraphics[width=0.7\textwidth]{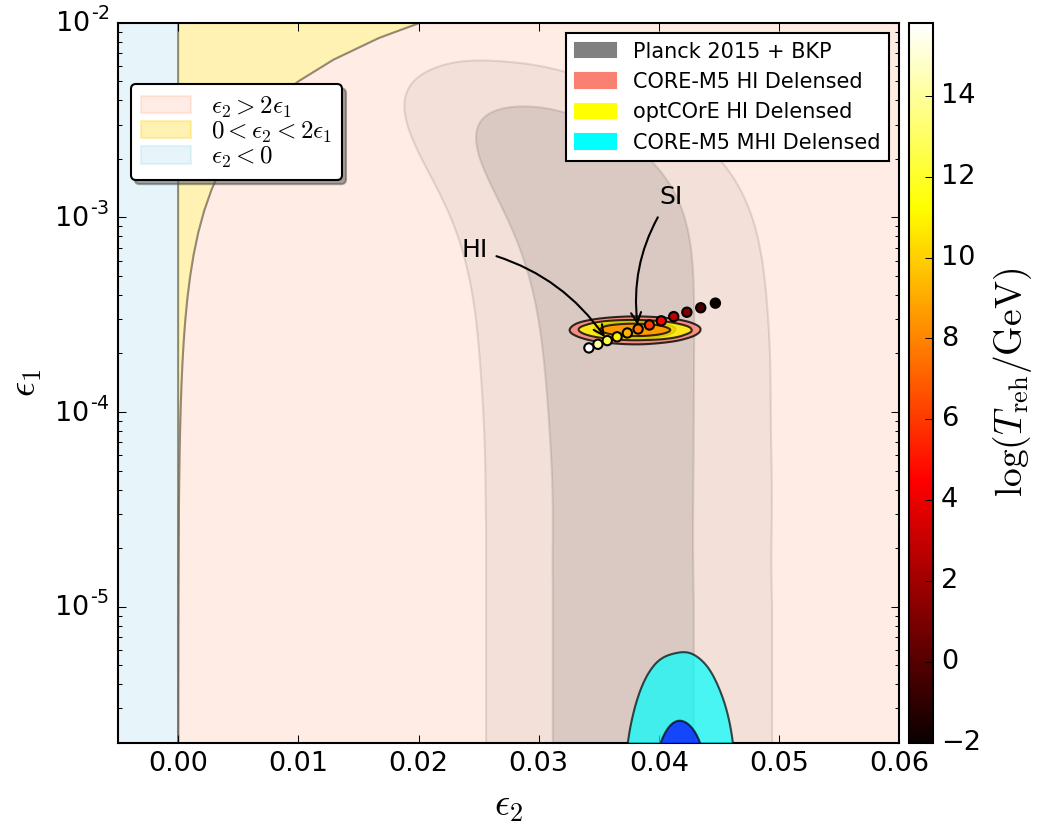}
\caption{Constraints in the $\epsilon_1$-$\epsilon_2$ plane 
(1$\sigma $ and 2$\sigma$ contours) 
for \Planck\ 2015 plus the BKP joint cross-correlation, CORE-M5  
and optCOrE+ (with SI and MHI as fiducial models). 
The predictions of the model SI have been displayed assuming 
$\bar{w}_\mathrm{reh}=0$, where the color encodes the reheating temperature 
$T_\mathrm{reh}$. The models indicated as HI (Higgs Inflation) and SI (Starobinsky Inflation) 
share the same inflationary potential, but are endowed with different reheating 
temperatures (around $10^{12}\, \mathrm{GeV}$ for HI and $10^8\,\mathrm{GeV}$ for SI), 
which a CORE-type mission could distinguish at 2$\sigma .$}
\label{fig:summary}
\end{center}
\end{figure}
 
The analysis in this Section demonstrates the potential of CORE to discriminate among different inflationary models.
Considering SI as the reference model, one ends up with only 3 favored and 2 weakly disfavored models among those listed in \Ref{Martin:2013tda}.
For the MHI reference model, which would correspond to an upper bound on primordial 
B-modes, we find $\sim$ 10 favored and $\sim$15 weakly disfavored models, always among those listed in \Ref{Martin:2013tda}. 

More than the angular resolution of the apparatus (linked to the mirror size), we have 
shown that the sensitivity of the instrument plays the most 
important role in improving the forecasts on $\epsilon_1$, and thus on $r$. 
This is particularly important for the information gain on reheating 
which increases from less than 1 bit with \Planck\ to more than 
3 bits with CORE if SI is taken as the fiducial model, and more than 2 bits if MHI is the fiducial model.

These results can be summarized by \Fig{fig:summary}, 
displaying the \Planck\, CORE-M5 and optCOrE+ posteriors in the $(\epsilon_1,\epsilon_2)$ plane, 
considering SI or MHI as the reference model. Not only would CORE allow us to bring the number of favored inflationary models 
listed in \Ref{Martin:2013tda} down to a few, 
but it would even distinguish among different reheating scenarios. Interestingly, this could allow us to distinguish models that share 
nearly the same inflationary potential such as Higgs Inflation and the Starobinsky model, but could have different reheating temperature 
under minimal assumptions on the inflaton coupling \cite{Terada:2014uia}.
%but that are derived in different frameworks and are thus endowed 
%with different reheating temperatures (around $10^{12}\, \mathrm{GeV}$ for Higgs Inflation and $10^8\,\mathrm{GeV}$ for the Starobinsky model when 
%realized in supergravity~\cite{Terada:2014uia}). 
This opens fascinating prospects to better understand the physics of the very early universe.

\section{Testing deviations from a power law spectrum}
\label{sec:features}

The fundamental contact of inflationary theory with the observations
comes from comparing the model predictions for the primordial 
scalar \(\mathcal{P}_\mathcal{R}\) and tensor \(\mathcal{P}_\mathcal{T}\) 
power spectra with the data. These are processed by cosmological transfer functions to yield 
predictions of 
CMB temperature and polarization multipoles \(C_\ell\). Comparing the theoretical 
\(C_\ell\)'s with the observed values yields constraints on the inflationary model parameters.

In this Section we reverse this procedure and reconstruct the primordial power spectra directly from the data. 
This procedure has been successfully applied to the \Planck\ 2015 data \cite{Ade:2015lrj} 
for the scalar primordial power spectrum using the methods of Gauthier and 
Bucher~\cite{Gauthier2012}, Vazquez et.~al.~\cite{Vazquez2012}, Aslanyan et~al.~\cite{Aslanyan2014} 
and the methods of Bond et al.~\cite{Ade:2015lrj}.

\subsection{Primordial power spectra reconstruction methodology}
\begin{figure}[tp]
    \centering
    \begin{tikzpicture}

        \newcommand{\movablecross}[2]{%
            \draw[->](#1) -- ++(0:#2);
            \draw[->](#1) -- ++(90:#2);
            \draw[->](#1) -- ++(180:#2);
            \draw[->](#1) -- ++(270:#2);
            \fill[red!70!black] (#1) circle (2pt);
        }

        \newcommand{\movablevert}[2]{%
            \draw[->](#1) -- ++(90:#2);
            \draw[->](#1) -- ++(270:#2);
            \fill[red!70!black] (#1) circle (2pt);
        }

    % Axes
        \draw [<->,thick] (6,0) coordinate (X) -- (0,0) coordinate (O) -- (0,4) coordinate (Y);

    % Axes labels
        \draw (X) node[right]  {\(\log k [ \mathrm{Mpc}^{-1}]\)}; 
        \draw (Y) node[above] {\(\log(10^{10}\PPS)\)};

    % Reconstruction line
        \draw (0.5,1.5) coordinate (A) -- (1.75,2.25) coordinate (B) -- (3,1.5) coordinate (C) -- (5,2.5) coordinate(D);

    % Crosses
        \movablevert{A}{0.4}
        \movablecross{B}{0.4}
        \movablecross{C}{0.4}
        \movablevert{D}{0.4}

    % Node labels
        \draw (A) node[below right] {\((k_1,\PPS_{1})\)};
        \draw (B) node[above right] {\((k_2,\PPS_{2})\)};
        \draw (C) node[below right] {\((k_3,\PPS_{3})\)};
        \draw (D) node[above left] {\((k_{\Nknots},\PPS_{\Nknots})\)};

    % Draw prior widths
        \coordinate(A1) at ($(O)!(A)!(X)$);
        \coordinate(D1) at ($(O)!(D)!(X)$);
        \draw (A1) node[below] {$10^{-4}$};
        \draw (D1) node[below] {$10^{-0.3}$};

        \coordinate(ymin) at (O |-0,0.75);
        \coordinate(ymax) at (O |-0,3.5);

        \draw[-,dashed] (ymin) -- (X |- ymin);
        \draw[-,dashed] (ymax) -- (X |- ymax);
        \draw[-,dashed] (A1) -- (A1 |- Y);
        \draw[-,dashed] (D1) -- (D1 |- Y);

        \draw (ymin) node[left] {2};
        \draw (ymax) node[left] {4};

% put some ellipses in between the start and end point

    \end{tikzpicture}
    \caption{%
    The scalar primordial power spectra are reconstructed using \(\Nknots\)-point interpolating logarithmic splines. The positions of the points in the \((k,\PPS)\) plane are treated as likelihood parameters with log-uniform priors. Further, the \(k\)-positions sorted {\em a priori} such that \({k_1<k_2<\cdots<k_{\Nknots}}\) with \(k_1\) and \(k_\Nknots\) fixed. Units of \(k\) are inverse megaparsecs.}\label{fig:linear_spline_reconstruction}
\end{figure}
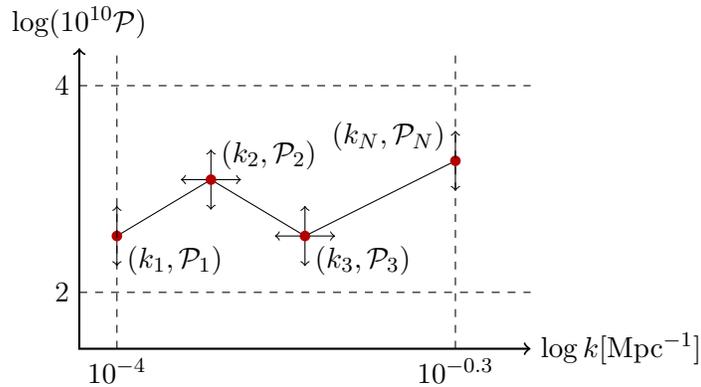

To reconstruct primordial power spectra, we follow the methodology of Section~8.2 
of the \Planck\ inflation paper~\cite{Ade:2015lrj}. Instead of using the 
traditional amplitude-tilt parameterization (\(A_\mathrm{s},n_\mathrm{s}\)), 
we instead use an \(\Nknots\)-point interpolating logarithmic 
spline (Fig.~\ref{fig:linear_spline_reconstruction}), with the positions of the knots considered as 
free parameters in the full posterior distribution. 

We compute posteriors (conditioned on \(\Nknots\)) using the PolyChord nested sampler~\cite{Handley2015a,Handley2015b}, also varying the cosmological and any nuisance parameters. We then use evidence values for each model to correctly marginalize out the number of knots \(\Nknots\).

To plot our reconstructions of the power spectra, we compute the marginalized posterior 
distribution of \(\log\PPS\) conditioned on \(k\). The iso-probability confidence intervals 
are then plotted in the $k$-$\PPS $ plane (e.g., Fig.~\ref{fig:PPSR}).

To quantify the constraining power of a given experiment, we use the conditional 
Kullback-Leibler (KL) divergence~\cite{Hee2016}, defined in Eq.~\ref{eq:DKL}. 
For our reconstructions, we compute the KL divergence of each distribution conditioned 
on \(k\) and \(\Nknots\), and then marginalize over \(\Nknots\) using evidences to produce a \(k\)-dependent number which quantifies the logarithmic compression (i.e., information) 
that each experiment provides at each value of \(k\) (e.g., Fig.~\ref{fig:DKLPPSR}).

This reconstruction approach has the advantage of being somewhat agnostic to any specific inflation model 
and constitutes a `blind' reconstruction of the power spectra. One 
could also search for parameterized features as 
discussed in Section~9 of the \Planck\ inflation paper~\cite{Ade:2015lrj}.
In this Section we shall focus on answering the 
question as to how well a blind reconstruction could pick up a variety of features using CORE-M5 
data resolution. This approach is by no means unique. Other authors have used a binned 
spectrum~\cite{Hazra:2013pps}, cubic splines~\cite{CubicSplines} and 
cross-validation~\cite{RedStraightNoBends}, as well as reconstructions motivated by 
phenomenology~\cite{PhenomenologicalPPS} or inflationary 
models~\cite{Hazra:2016pps,Hazra:2014pps}. The approach chosen for this Section is 
motivated more by its conceptual simplicity than anything else.

\subsection{Featureless scalar power spectrum}

\begin{figure}
    \centering
    \includegraphics[width=0.45\textwidth ]{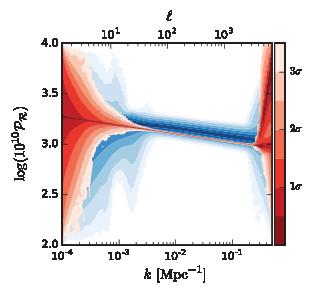}
    \includegraphics[width=0.45\textwidth ]{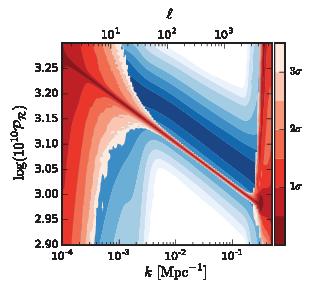}
    \caption{Left: Reconstruction of a simulated featureless scalar power spectrum for a CORE-M5 experiment (in red), 
compared to existing constraints provided by \Planck\ (in blue). Right: Zoomed-in version of the left figure to show 
the order of magnitude increase in constraining power that would be provided by CORE.}\label{fig:PPSR}
\end{figure}

\begin{figure}
\centering
\includegraphics{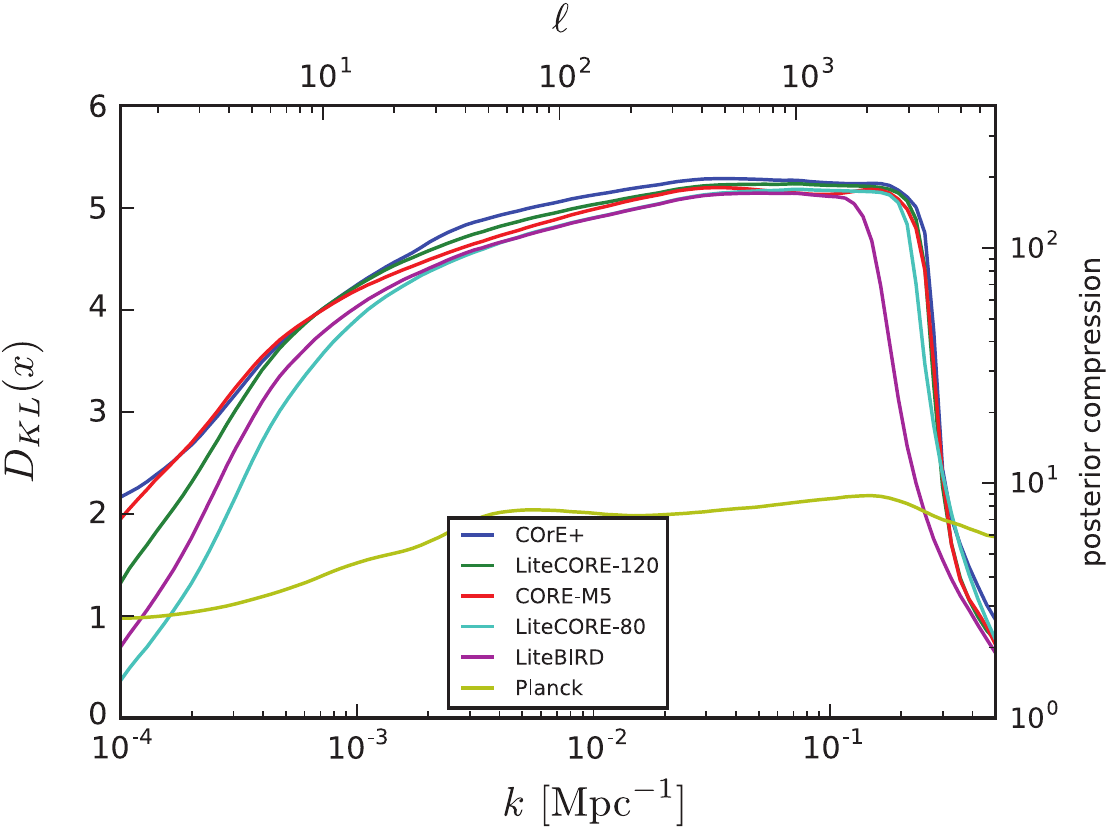}
\caption{%
The amount of information that CORE experiments would provide on the scalar primordial power spectrum.  
The Kullback-Leibler divergence shows that all configurations of the experiment provide a similar level 
of information. As expected, the prior-to-posterior compression drops off at low $\ell$ due to cosmic 
variance, and high-$\ell$ at the limits of the experiment.
All CORE configurations thus provide an order of magnitude more information than \Planck\ 2015. 
It is worth remarking that we use \Planck\ 2015 real data here. %and simulated theoretical data. 
The improvement of the CORE experiments with respect to \Planck\ 2015 real data is largely related to improved determination of $A_s$, also connected to the cosmic variance limited measurement of $\tau$.
}\label{fig:DKLPPSR}
\end{figure}

To quantify the basic constraining power of the experiments detailed in 
Tables~\ref{tab:CORE-bands} and~\ref{tab:specifications}, we begin by examining 
how well these experiments could constrain a featureless scalar power spectrum. Simulated likelihoods are 
generated from a featureless tilted power spectrum with $n_\mathrm{s}=0.96$ and $A_\mathrm{s}=2.1\times10^{-9}$ 
with \(TT\),\(TE\) , and \(EE\) data for COrE+, LiteCORE-120, LiteCORE-80, LiteBIRD, and CORE-M5 experiments. 
\begin{figure}
    \centering
    \includegraphics[width=0.45\textwidth ]{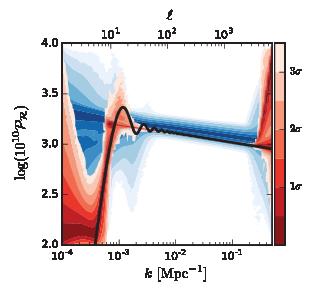}
    \includegraphics[width=0.45\textwidth ]{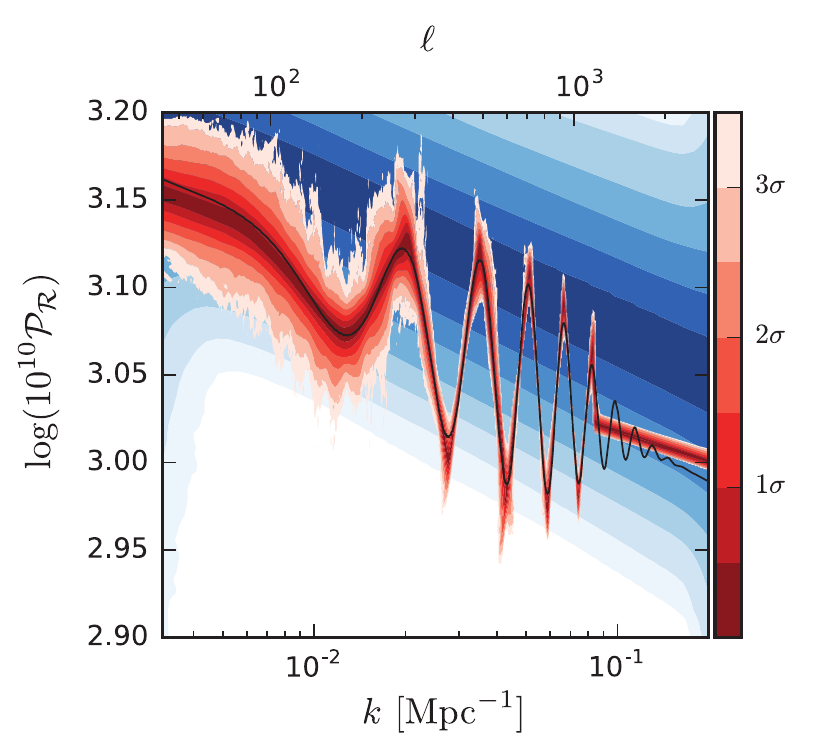}
\caption{As in Fig.~\protect\ref{fig:PPSR}, but now reconstructing a simulated power spectrum with features (black lines). 
Left: Cut-off as would be generated by a brief period of fast-roll expansion prior to slow roll inflation. The additional 
constraining power provided by CORE would allow detection of low $\ell$ features such as cutoffs and wiggles. 
Right: Reconstruction of higher-$\ell$ linearly-sinusoidal wiggles
generated by a reduction in the speed of sound of the inflaton, as
described in Section 2.3.3. In this last case, the
reconstruction only picks up the feature when the prior on the nodes' positions is restricted to the region where the feature is active; the
smaller-scale wiggles prove to be harder to reconstruct.}
\label{fig:PPSR_wiggles}
\end{figure}
\afterpage{%
\begin{figure}
    \centering
    \includegraphics[width=0.45\textwidth]{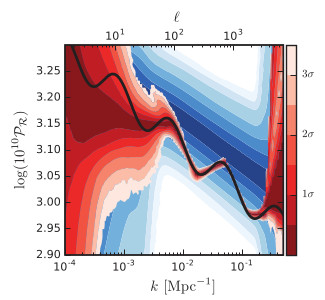}
    \includegraphics[width=0.45\textwidth]{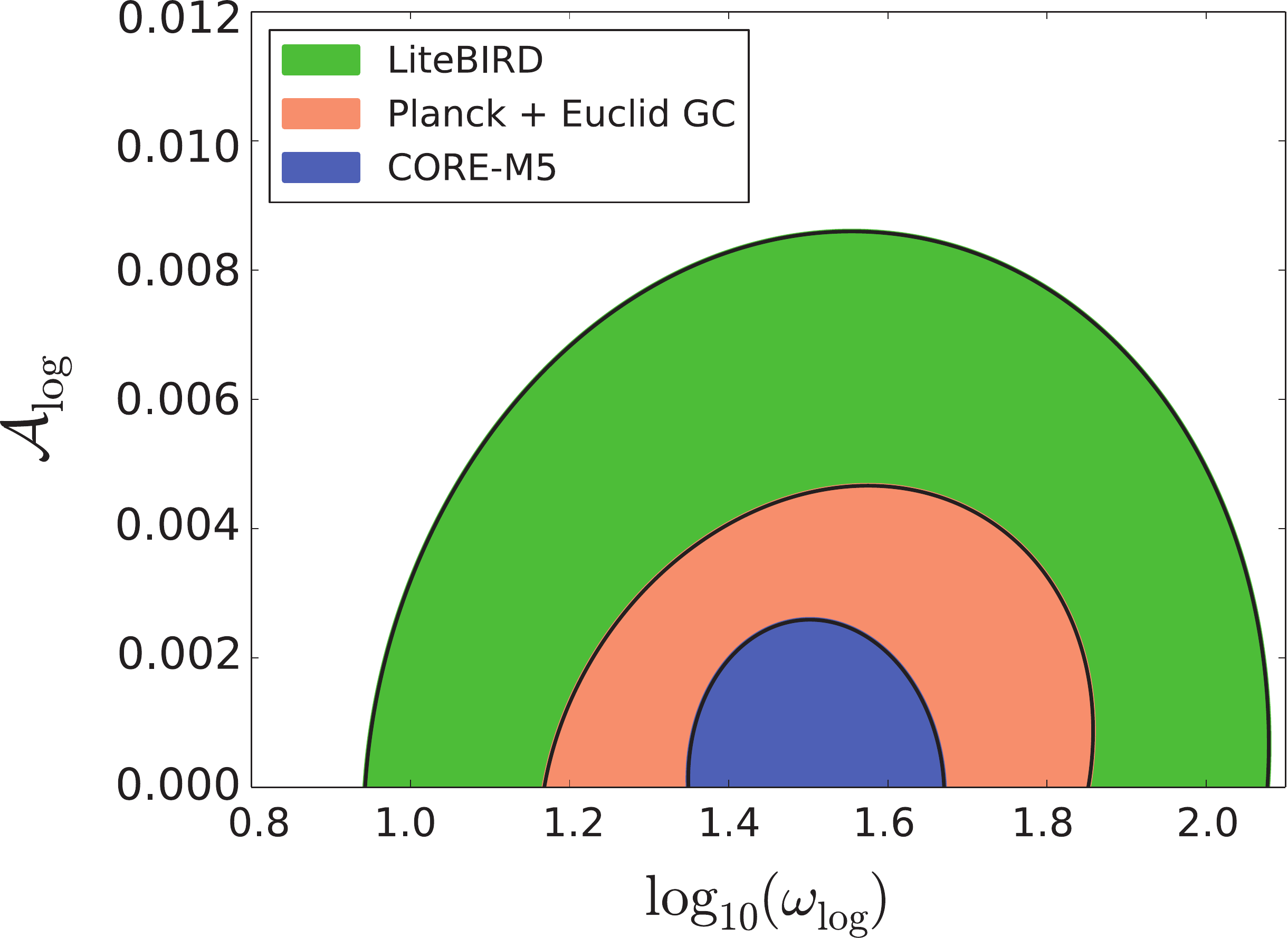}
    \caption{Left: As in Fig.~\protect\ref{fig:PPSR}, but now reconstructing a 
simulated power spectrum with 
low frequency sinusoidal logarithmic oscillations (black line), with $A_{\log}=0.03, \omega_{\log}=3$
and $\psi_{\log}=0$. Right: constraints on superimposed sinusoidal logarithmic oscillations with an higher frequency, comparable 
to those providing a best-fit to \Planck\ 2015 data \cite{Ade:2015lrj}.
Whereas a blind reconstruction technique is unsuitable for high 
frequency oscillations, CORE-M5 performs better than LiteBIRD or \Planck\ in combination with Euclid spectroscopic galaxy clustering for this type of parameterized features.
Contours indicate 68\% confidence intervals.}\label{fig:PPSR_monodromy}
\end{figure}
\begin{figure}
    \centering
    \includegraphics[width=0.79\textwidth]{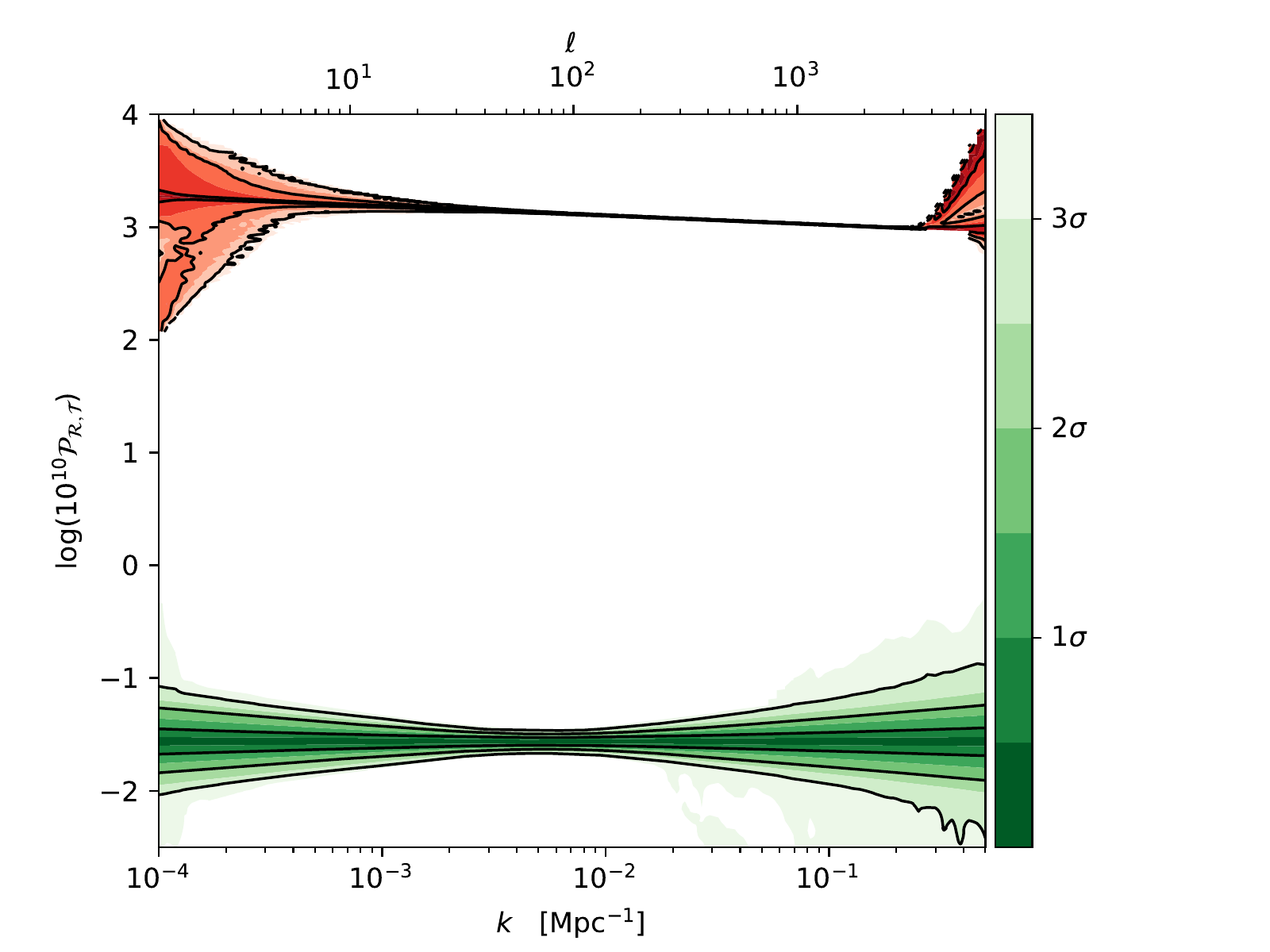}
    \caption{Simultaneous reconstructions of the tensor (lower, green) and scalar (upper, red) power spectra for CORE-M5 forecast if \(r=0.01\).\label{fig:marg_COrE}}
\end{figure}}

Figure~\ref{fig:PPSR} shows the marginalized posterior distribution for the CORE-M5 experiment. 
As usual, the reconstructions demonstrate the lack of reconstructive power at low $k$ due to 
cosmic variance and the inability to reconstruct above the resolving power of $\ell\sim3000$. 
CORE-M5 provides an order of magnitude better constraint on the amplitude of the power 
spectrum relative to the \Planck\ 2015 data (shown for reference in blue).

To quantify the level of this increase in information, Fig.~\ref{fig:DKLPPSR} shows the information 
gain for each of these experiments at each value of \(k\) in comparison with \Planck. One can see that 
all three configurations provide similar levels of information on the primordial power spectrum, and an 
order of magnitude more information than \Planck, with LiteBIRD and LiteCORE-80 failing to provide information above $\ell\sim1350$ and $2400,$ respectively.

The mock likelihoods used throughout this Section follow the methodology detailed in 
Section~\ref{sec:simplified_likelihood}, where the fiducial ``observed'' 
likelihoods $\hat{{\bf C}}_\ell$ are computed using {\tt CAMB}. A more powerful analysis has 
also been employed which incorporates cosmic variance to add ``scatter'' to these fiducial 
$\hat{{\bf C}}_\ell$. The results remain quantitatively unchanged for $\ell>10$, both 
in $D_{KL}$ divergence (Fig.~\ref{fig:DKLPPSR}) and predictive posterior plots 
(Figs.~\ref{fig:PPSR},~\ref{fig:PPSR_wiggles},~\ref{fig:PPSR_monodromy}, and \ref{fig:marg_COrE}). 
For $\ell<10$, noise features appear in higher $N$ predictive posteriors, but these are 
suppressed by correspondingly low evidence values. Quantitatively, the $D_{KL}$ 
constraining power is somewhat overestimated by using unscattered $\hat{{\bf C}}_\ell$ 
for these extremely low values of $\ell$. We have chosen to present the unscattered fiducial 
versions of our plots for simplicity and consistency with the rest of this paper.

\subsection{Scalar power spectrum with wiggles and cutoffs}

With such strong constraining power, any of these CORE experiments would be able to detect features in the primordial power 
spectrum that are currently unresolved using \Planck\ data. 
To show this, we produced a likelihood for a wiggly power spectrum 
as would be generated by an epoch of fast-roll inflation prior to slow roll~\cite{Contaldi2003}. 
The left hand plot of Fig.~\ref{fig:PPSR_wiggles} shows that CORE-M5 has the capacity to blindly 
reconstruct a cutoff and the first oscillation which could be hidden beneath \Planck\ 2015 resolution. This remains true even when cosmic variance noise is added.

CORE experiments would also be capable of reconstructing smaller localized features, 
such as those produced by reduced adiabatic sound speeds (Section~\protect\ref{sssec:features}), as demonstrated in 
Fig.~\ref{fig:PPSR_wiggles}.

\subsection{Superimposed logarithmic oscillations}

Resonant models including periodic oscillations in the potential lead to superimposed sinusoidal oscillation in the spectrum and bispectrum
\cite{Chen:2008wn}, as discussed in Section 2.
The periodic oscillations appearing in axion monodromy \cite{Silverstein:2008sg,McAllister:2008hb} belong to this general class of models \cite{Flauger:2009ab}.
We consider the phenomenological parameterization
\begin{equation}
    \mathcal{P}_\mathcal{R}(k) = A_\mathrm{s} 
{\left( \frac{k}{k_*} \right)}^{n_\mathrm{s} - 1} 
\left( \phantom{\Big|} 1 + A_{\log} \sin[\omega_{\log} k + \psi_{\log}] \right).
\label{resonant}
\end{equation}
As shown in Fig.~\ref{fig:PPSR_monodromy}, 
CORE-M5 also has the resolving power to blindly reconstruct logarithmic oscillations with an amplitude at a percent level 
for low frequencies, i.e., 
$\omega_{\log} = 3$. For higher frequencies, the blind reconstruction scheme would require prohibitively 
large numbers of knots.
Nevertheless, as is shown in the right-hand panel of Fig.~\ref{fig:PPSR_monodromy}, CORE-M5 is still capable of extracting information
about such models by directly constraining the parameters of the superimposed oscillations in Eq. (\ref{resonant}). 
Further detail on the Fisher forecast methodology for this approach is provided by~\cite{Ballardini2016}.

\subsection{Reconstructing the tensor power spectrum}

Finally, we demonstrate that if there is a detectable scalar-to-tensor ratio, 
CORE-M5 would also be able to reconstruct this spectrum. Fig.~\ref{fig:marg_COrE} 
shows a simultaneous reconstruction of both the scalar and tensor power spectrum, 
where we use two independent reconstructions on simulated data with a flat tensor power spectrum for \(r=0.01\).

%\clearpage
%\newpage

\section{Testing the adiabaticity of initial conditions: constraining isocurvature}
\label{sec:testing}

%\section{Testing the adiabaticity of initial conditions: constraining isocurvature}
%\label{sec:seven}

In the simplest single-field inflationary models only the adiabatic growing mode is excited. Indeed, the
six-parameter concordance cosmological model, which was found to provide an adequate fit to
the WMAP and \Planck\ data, includes only the growing adiabatic mode, but this does not necessarily mean
that other modes could not have been excited. Much theoretical work has been devoted to studying
multi-field inflationary models and other extended inflationary models, and many of these models predict
that isocurvature modes could have been excited. It is therefore of great interest
to test the hypothesis of adiabaticity by searching for isocurvature modes. At one time, proposals were put
forth in which isocurvature modes would offer an alternative to adiabatic modes for the formation of 
structure and for imprinting the CMB anisotropies. But it soon became apparent that such models in which the
growing adiabatic mode was not excited were not viable. As a consequence the emphasis shifted to exploring
scenarios in which both adiabatic and isocurvature modes 
were excited, possibly in a correlated manner \cite{Langlois:1999dw,Langlois:2000ar,Gordon:2000hv,Amendola:2001ni}. The
discussion below explores how CORE and other future satellite configurations will be able to improve on the
constraints already established by \Planck\ in Refs.~\cite{Planck:2013jfk,Ade:2015lrj}. See also
references therein for a more comprehensive list of literature regarding the isocurvature modes. 

We study a model where the primordial curvature perturbation, or the adiabatic mode, is  
correlated with a primordial Cold dark matter Density Isocurvature (CDI) mode.\footnote{Here we do 
not consider neutrino isocurvature modes.
In addition to the recent \Planck\ results 
\cite{Planck:2013jfk,Ade:2015lrj}, observational work on the neutrino modes include 
\cite{Beltran:2004uv,Beltran:2005xd,Savelainen:2013iwa} using the WMAP data.
Furthermore, we restrict the analysis to the power spectrum level 
where the baryon density isocurvature (BDI) mode or the total matter density isocurvature 
(simultaneously CDI and BDI) can be mapped into ${\cal I}_{\rm CDI}^{\rm effective} = {\cal 
I}_{\rm CDI} + (\Omega_\mathrm{b}/\Omega_\mathrm{c}) {\cal I}_{\rm BDI}$. The trispectrum may
be useful for distinguishing 
between CDI and BDI \cite{Grin:2013uya}, but we leave these forecasts 
to future work. In the case of an exact cancellation between  ${\cal I}_{\rm CDI}$ and
$(\Omega_\mathrm{b}/\Omega_\mathrm{c}) {\cal I}_{\rm BDI},$ we have ${\cal I}_{\rm CDI}^{\rm effective} = 0$, so
there is no isocurvature perturbation between the radiation and the matter. 
These are called compensated isocurvature perturbations (CIP), and because
baryons behave differently from dark matter on small scales, CIPs
modify the angular power spectrum compared to the pure adiabatic prediction, 
but these scales correspond to multipoles
$\ell \gtorder 10^5$--$10^6$ \cite{Grin:2011tf}. At much larger scales, CIPs
may imprint a lensing-like signal onto the angular power spectrum at second order in the CIP amplitude
$\Delta_\mathrm{rms}$ \cite{Munoz:2015fdv}.  Using adiabatic
$\Lambda$CDM fiducial TT, TE, and EE data, we find 
$\Delta^2_\mathrm{rms} < 0.0019$ at 95\% CL for CORE-M5. 
LiteBIRD gives slightly
weaker constraints than \Planck, whereas CORE-M5 gives three times
stronger constraints and is close to the cosmic variance
limit. Compared to the future ground-based instruments
\cite{Abazajian:2016yjj} CORE-M5 performs about a factor of two better. More detailed power spectrum 
based forecasts for CIP are presented in \cite{CIPinprep}.} 
In our analysis we use the \texttt{MultiNest} nested sampling algorithm and a modified version of \texttt{CosmoMC} 
and \texttt{CAMB} capable of calculating theoretical predictions for the arbitrarily correlated 
adiabatic and isocurvature modes and simultaneously including primordial tensor perturbations in a 
consistent way.

\subsection{The model and its parameterization}

The full details of the parameterization and notation including various symbols appearing in the 
Figures are explained in \cite{Planck:2013jfk,Ade:2015lrj}.
%in the \Planck\ 2015 Constraints on Inflation paper \cite{Planck:2013jfk,Ade:2015lrj}. 
Here we summarize the main assumptions and choices. The model has the usual 4 background parameters of the flat 
$\Lambda$CDM model ($\Omega_{\rm b} h^2$, $\Omega_{\rm c} h^2$, $\theta_{\mathrm{MC}}$, $\tau$) and two 
parameters describing the assumed power law primordial curvature perturbation power ${\cal P}_{\cal 
RR}(k)$. In addition we have two parameters that describe a power law primordial isocurvature 
perturbation spectrum ${\cal P}_{\cal II}(k)$ and one parameter that describes the correlation 
amplitude between the curvature and isocurvature perturbations ${\cal P}_{\cal RI}(k_0)$ at a pivot 
scale corresponding to $k_0$. The curvature perturbation has a spectral index $n_{\cal RR}$ (called $n_\mathrm{S}$ in the 
other Sections), and the isocurvature perturbation has an independent spectral index $n_{\cal II}$. 
The correlation power does not have an independent spectral index in our model, but is simply a 
power law with index $n_{\cal RI} = (n_{\cal RR} + n_{\cal II})/2$. This restriction means that the 
correlation fraction 
\begin{equation}
\cos\Delta \equiv \frac{\mathcal{P_{RI}}(k)}{\sqrt{\mathcal{P_{RR}}(k) \mathcal{P_{II}}(k)}} \in [-1,\,+1]
\label{eq:cosDelta}
\end{equation}
is constant with respect to $k$ (i.e., scale independent). If the correlation spectral index $n_{\cal RI} $ were 
allowed as an independent parameter, the correlation fraction would not remain between $-1$ and $+1$ 
over all scales.

Spectral indices are not suitable for describing this extended model, since the pure adiabatic 
$\Lambda$CDM model should be represented by a single set of values (one point) in the parameter space
of the extended model. However if we chose the isocurvature power at a pivot scale $k_0$, [i.e., 
${\cal P}_{\cal II}(k_0)$] as a primary isocurvature parameter of our model and for the other 
primary parameter the spectral index $n_{\cal II}$,  the pure adiabatic model would be 
described by a line where ${\cal P}_{\cal II}(k_0)=0$, but $n_{\cal II}$ can take any value between 
$-\infty$ and $+\infty$. As a consequence, the larger prior range we allowed for $n_{\cal II}$, the 
more biased our results would be toward the pure adiabatic model, since the marginalization over 
$n_{\cal II}$ direction with nearly zero ${\cal P}_{\cal II}(k_0)$ would artificially increase the 
weight of the nearly adiabatic models (${\cal P}_{\cal II}(k_0) \approx 0$) by a factor $\Delta 
n_{\cal II}$, (i.e., by the prior width of the $n_{\cal II}$ parameter).

A solution to this problem was proposed in Ref.~\cite{KurkiSuonio:2004mn}, and first applied to 
isocurvature analysis in Ref.~\cite{MacTavish:2005yk} by the BOOMERANG team and in 
Ref.~\cite{Keskitalo:2006qv} using in addition the WMAP 3-year and ACBAR data. Also in the \Planck\ 2013 
\cite{Planck:2013jfk} and 2015 \cite{Ade:2015lrj} papers, the same approach was adopted, in 
2015 both for isocurvature and tensor perturbations. We parameterize the primordial power 
law perturbations by specifying their amplitudes at two different scales: at a large scale corresponding 
to a small wave number $k=k_1=k_{\rm low} = 0.002\,$Mpc$^{-1},$ and at a small scale corresponding 
to a large wave number $k=k_2=k_{\rm high}=0.100\,$Mpc$^{-1}$. Thus the primary parameters 
describing the primordial perturbations are ${\cal P}_{\cal RR}(k_1)$, ${\cal P}_{\cal RR}(k_2)$, 
${\cal P}_{\cal II}(k_1)$, ${\cal P}_{\cal II}(k_2)$, and ${\cal P}_{\cal RI}(k_1)$. The sign 
convention is such that a positive ${\cal P}_{\cal RI}$ leads to extra power at low multipoles in 
the temperature angular power and a negative primordial correlation leads to a negative 
observational contribution [see also footnote \ref{foot:JVSW} on page \pageref{foot:JVSW}].

The spectral indices are derived parameters calculated from the primary parameters [e.g., $n_{\cal 
II} = \ln({\cal P}_{\cal II}(k_2)/{\cal P}_{\cal II}(k_1))\,/\,\ln(k_2/k_1) \ + 1$]. The parameter 
${\cal P}_{\cal RI}(k_2)$ is also a derived parameter. Other interesting derived parameters are the 
primordial isocurvature fraction
\begin{equation}
\beta_{\rm iso}(k) = \frac{{\cal P_{II}}(k)}{{\cal P_{RR}}(k)+{\cal P_{II}}(k)}
\label{eq:betaiso}
\end{equation}
and the correlation fraction $\cos \Delta $ defined in Eq.~\eqref{eq:cosDelta}.  

\afterpage{
\begin{figure}
\flushleft{(a)}\\
\begin{centering}
\vspace{-6mm}
\includegraphics[width=0.88\textwidth]{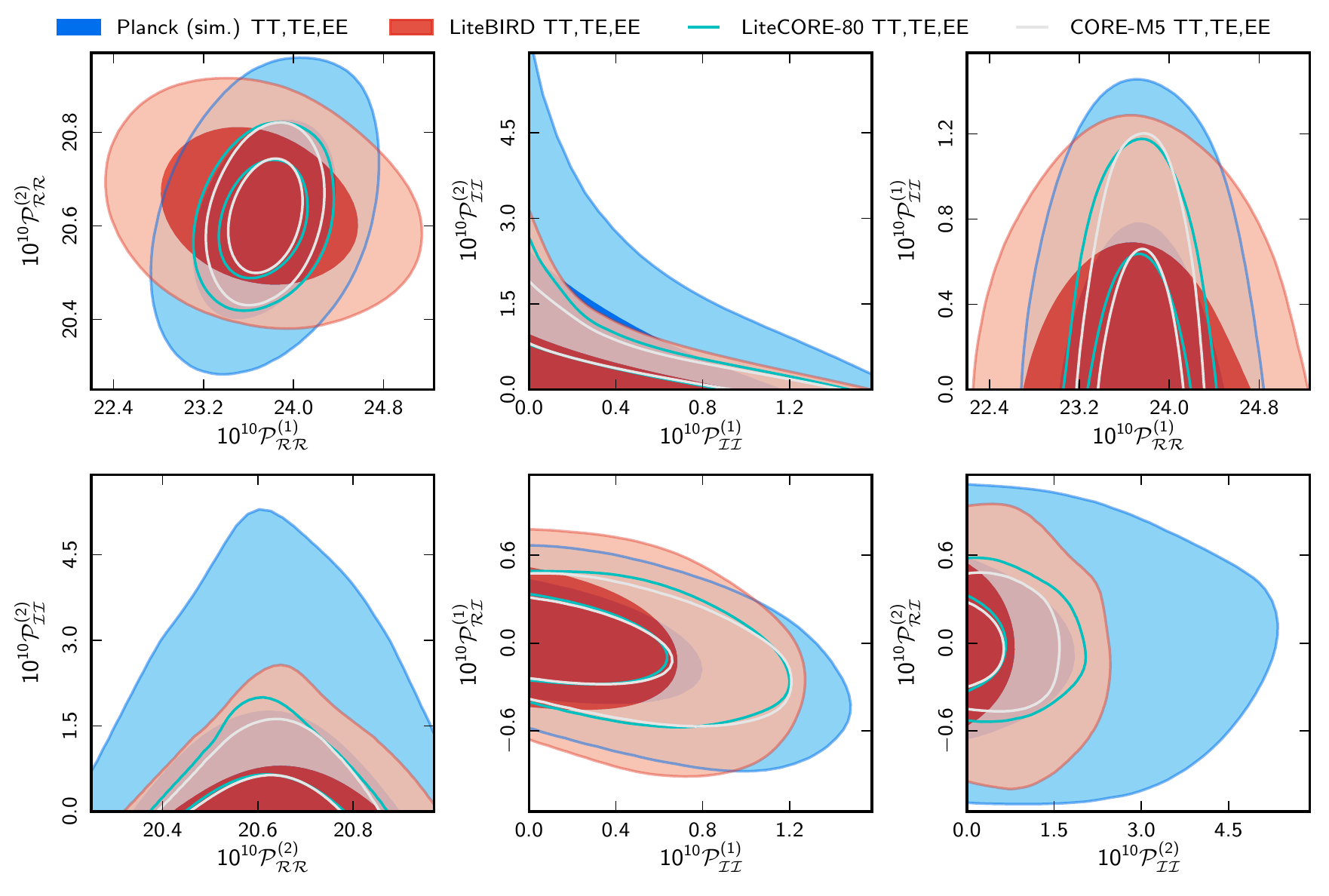}\\
\end{centering}
\vspace{-5mm}
\flushleft{(b)}\\
\begin{centering}
\vspace{-6mm}
\includegraphics[width=0.88\textwidth]{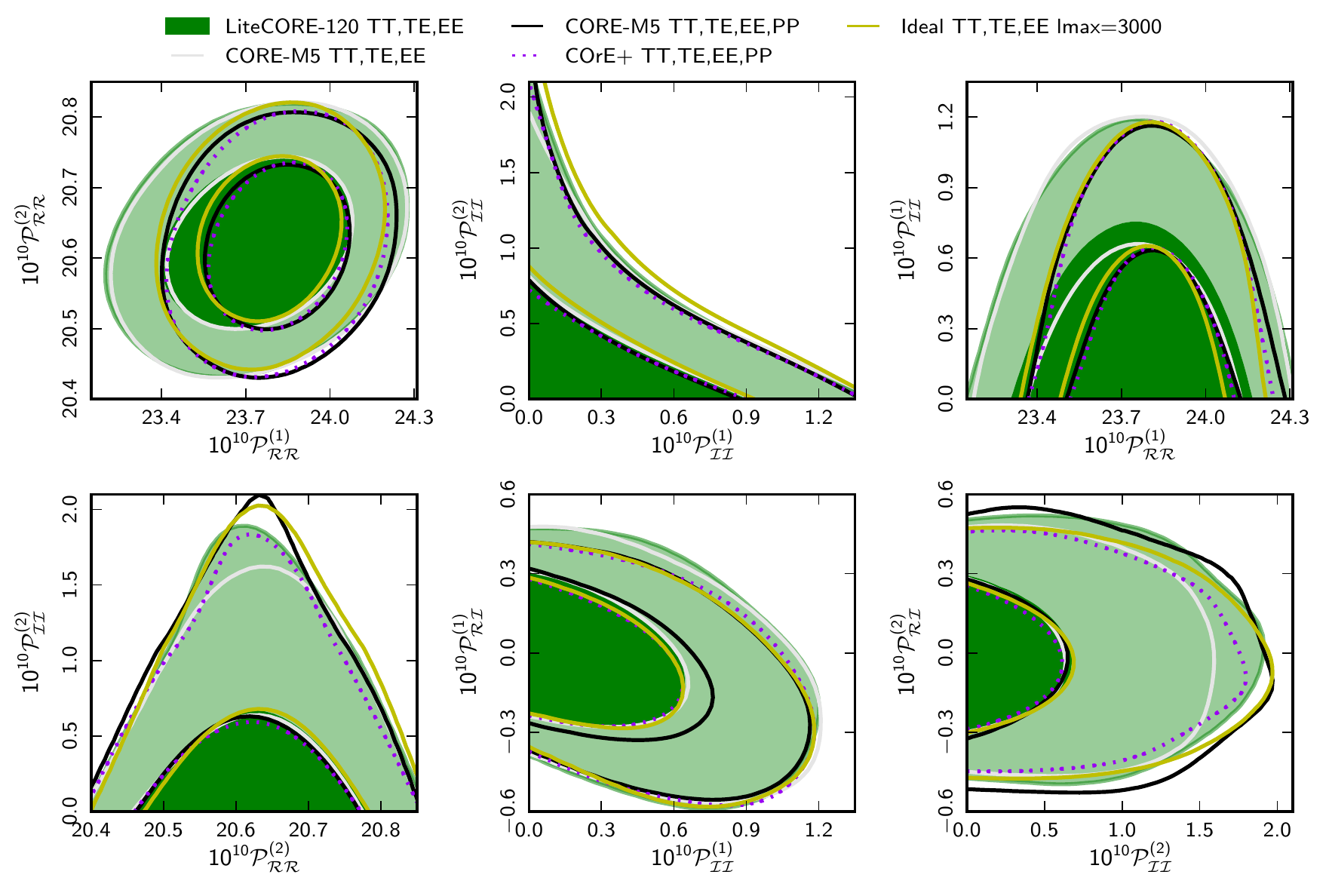}\\
\end{centering}
\vspace{-5.5mm}
\caption{ 
Constraints on primordial curvature, isocurvature and correlation power at large scales and 
small scales, denoted by the superscripts (1) and (2), respectively, 
when the fiducial data are adiabatic and have $r=0$, and the fitted model has 
generally correlated primordial adiabatic and CDI modes. (a) Simulated \Planck\ and LiteBIRD data 
lead to significantly weaker constraints than CORE-M5, but LiteCORE-80 only slightly weaker. (b) A 
zoomed version, now with LiteCORE-120, and showing CORE-M5 and COrE+ with the lensing potential PP. 
For the isocurvature and correlation powers LiteCORE-120, CORE-M5 and COrE+ are virtually 
indistinguishable and reach the cosmic variance limit (the Ideal case). PP improves the constraints 
on the curvature power.
\label{fig:JVPrimordialPowers}}
\vspace{-10mm}
\end{figure}}

\afterpage{%
\begin{table}
\setlength{\tabcolsep}{.48em}
\scriptsize
\begin{tabular}{|l|c|c|c|cc|}%|cc}
\hline
%\hline
   & $100\beta_{\rm iso}(k_{\rm low})$  &  $100\beta_{\rm iso}(k_{\rm mid})$  & $100\beta_{\rm iso}(k_{\rm high})$ & \multicolumn{2}{c|}{$100\cos\Delta$} \\ %& \multicolumn{2}{c}{$100\alpha_{\rm non-adi}$} \\
\hline
\Planck\ (sim.) TT,TE,EE &      4.70 &      8.29 &     15.31 &     -1.78 & $[   -20.81;     19.44]$ \\ % &      0.85 & $[    -0.78;      3.36]$  \\
LiteBIRD TT,TE,EE &      4.07 &      4.48 &      7.87 &     -0.91 & $[   -24.63;     25.01]$ \\ % &      0.92 & $[    -1.09;      3.86]$  \\
(LiteBIRD TT,TE,EE,BB; $r_{0.05}=10^{-3}$) &      4.01 &      5.18 &      9.89 &     -5.17 & $[   -23.04;     13.45]$ \\ % &      0.58 & $[    -0.90;      3.00]$  \\
LiteCORE-80 TT,TE,EE &      3.82 &      3.84 &      6.64 &     -0.24 & $[   -15.76;     17.50]$  \\ % &      0.94 & $[    -0.54;      3.51]$  \\
LiteCORE-120 TT,TE,EE &      3.79 &      3.76 &      6.25 &     -0.80 & $[   -15.25;     15.31]$ \\ %% &      0.93 & $[    -0.46;      3.50]$  \\
CORE-M5 TT,TE,EE &      3.91 &      3.57 &      5.67 &     -0.80 & $[   -15.51;     14.80]$ \\ %&      0.93 & $[    -0.50;      3.38]$  \\
{\bf CORE-M5 TT,TE,EE,PP} &      3.73 &      3.63 &      6.40 &     -1.26 & $[   -15.05;     14.79]$ \\ % &      0.86 & $[    -0.50;      3.27]$  \\
(CORE-M5 TT,TE,EE,BB,PP; $r_{0.05}=10^{-3}$) &      3.77 &      4.38 &      7.83 &     -3.09 & $[   -16.21;     10.50]$ \\ % &      0.73 & $[    -0.59;      3.05]$  \\
COrE+ TT,TE,EE,PP &      3.75 &      3.54 &      6.02 &     -1.56 & $[   -15.43;     13.56]$ \\ %&      0.88 & $[    -0.58;      3.30]$  \\
Ideal TT,TE,EE &      3.79 &      3.79 &      6.49 &     -1.46 & $[   -15.05;     13.51]$ \\ %&      0.87 & $[    -0.51;      3.28]$  \\
%\hline
\hline
\end{tabular}
\caption{
95\% CL upper bound on the primordial isocurvature fraction $\beta_{\rm iso}$ at three different 
scales (from large scales to small scales), and the mean posterior value of the primordial 
correlation fraction and the 95\% CL interval of its one-dimensional marginalized posterior. The fitted model 
has a general correlation between the adiabatic and CDI modes, while the fiducial data assume pure 
adiabatic $\Lambda$CDM with $r=0$, except on the lines with $r_{0.05}=10^{-3}$ (to be discussed in 
Section \ref{sec:isoTensorTensorless}).\label{tab:JVbetaisoAndAlphanonadi}}
\end{table}}

\subsection{Adiabatic fiducial data}
%\subsection{Adiabatic fiducial data with $r=0$}

\subsubsection{Fitting a generally correlated mixture of adiabatic and CDI modes}
\label{sec:JVfirstiso}

Figure \ref{fig:JVPrimordialPowers} presents the constraints on primordial curvature, isocurvature, 
and correlation powers if the true primordial perturbations were purely adiabatic without a tensor 
contribution. We fit an extension of the standard $\Lambda$CDM model with  
three isocurvature parameters to these data.
All datasets use TT, TE, and EE, but for CORE-M5 and COrE+ we  also 
give the results with the CMB lensing potential (PP) data included. 
BB and delensing analysis is not applied in 
this Subsection. In addition to forecasts for `future experiments,' for comparison 
we show the constraints for simulated \Planck\ 
TT, TE, EE data assuming the bluebook \cite{PlanckBlueBook} white noise sensitivities divided by 
$\sqrt{2}$ using the 100, 143, and 217 GHz channels in inverse noise weighting (the factor $1/\sqrt{2}$ takes into account 
that \Planck\ operated at least twice as long as the nominal mission described in the bluebook).
%$\sqrt{2}$ using the 100, 143, and 217 GHz channels. 
%We combine these channels using inverse noise 
%weighting and take into account the beam sizes in the same way as for the future experiments, as explained 
%in Section \ref{sec:three}. The factor $1/\sqrt{2}$ (i.e., $1/2$ in $N_{\ell}$) approximately takes 
%into account that \Planck\ HFI operated twice as long as the nominal mission described in the 
%bluebook. 
Finally, we make a \texttt{MultiNest} run with zero instrumental noise using the multipole range 
$\ell =2$--$3000$ (called ``Ideal TT,TE,EE $\ell _{max}$=3000'' in Fig.~\ref{fig:JVPrimordialPowers}). The 
only `noise' in this run is cosmic variance. Consequently, this represents an ideal case for TT, TE, EE.
%beyond which (using TT, TE, EE) one cannot do better by improving instrument sensitivity.

The first panel of Fig.~\ref{fig:JVPrimordialPowers} provides a convenient way to represent the 
determination accuracy of the adiabatic scalar perturbations instead of the usual ($A_\mathrm{S}$, 
$n_\mathrm{S}$) pair. We see directly the primordial curvature perturbation amplitude at 
large and small scales, denoted by the superscripts (1) and (2), respectively.  
The future missions (except LiteBIRD owing to its coarse angular resolution) 
constrain these parameters much better 
than \Planck. LiteCORE-120, CORE-M5, and COrE+ virtually reach the cosmic variance limit, and with PP 
included actually determine the curvature perturbation power better than the ideal cosmic variance 
limited experiment only with TT, TE, and EE.
\afterpage{%
\begin{figure}
\centering
\includegraphics[width=\textwidth]{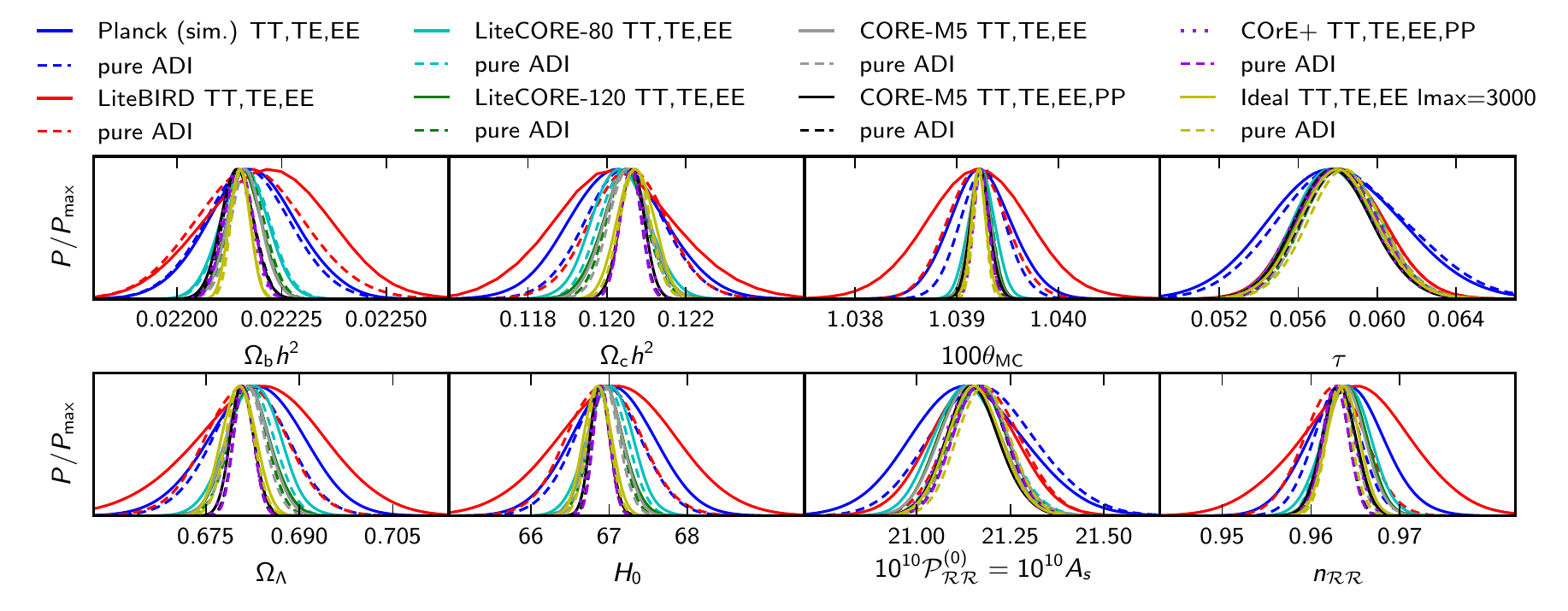}\\
\vspace{-5.5mm}
\caption{
Parameters that exist also in the standard adiabatic model---comparison of their determination 
when assuming pure adiabatic model (dashed lines) or when assuming a generally correlated mixture 
of the adiabatic and CDI mode (solid lines). The fiducial data here are pure adiabatic with $r=0$. 
\label{fig:JVadiparams}}
\end{figure}
\begin{table}[!h]
\setlength{\tabcolsep}{.320667em}
\scriptsize
\begin{tabular}{|l|cc|cc|cc|cc|cc|cc|}
\hline
%\hline
   & \multicolumn{2}{c|}{$100\Omega_{\rm b} h^2$} & \multicolumn{2}{c|}{$100\Omega_{\rm c} h^2$} & \multicolumn{2}{c|}{$10000\theta_{\rm MC}$} & \multicolumn{2}{c|}{$100\Omega_\Lambda$} & \multicolumn{2}{c|}{$H_0$} & \multicolumn{2}{c|}{$100n_{\cal RR}$} \\ \raisebox{9pt}

 $\phantom{j}$ & mean & $\sigma$ & mean & $\sigma$ & mean & $\sigma$ & mean & $\sigma$ & mean & $\sigma$ & mean & $\sigma$ \\

\hline
Fiducial data (adiab. $\Lambda$CDM) & 2.214 & & 12.06 & & 103.922 & & 68.10 & & 66.89 & & 96.25 &  \\ \raisebox{9pt}

%Planck TT,TE,EE (sim.)  & 2.218 & 0.011 & 0.1203 & 0.0012 & 1.0392 & 0.0003 & 0.6830 & 0.0076 & 67.05 & 0.53 & 0.9639 & 0.0041 \\
%LiteBIRD TT,TE,EE & 2.221 & 0.014 & 0.1202 & 0.0016 & 1.0392 & 0.0005 & 0.6834 & 0.0010 & 67.09 & 0.76 & 0.9649 & 0.0064 \\
%LiteCORE-80 TT,TE,EE & \\
%LiteCORE-120 TT,TE,EE & \\
%CORE-M5 TT,TE,EE  & \\
%{\bf CORE-M5 TT,TE,EE,PP}  & \\
%COrE+ TT,TE,EE,PP  & \\
%Ideal TT,TE,EE  &  \\ 
\Planck\ (sim.) TT,TE,EE: CDI &     2.218 &     0.011 &     12.03 &      0.12 &   103.924 &     0.031 &     68.30 &      0.76 &     67.05 &      0.53 &     96.39 &      0.41  \\
$\phantom{j}$ \hspace{3mm} pure ADI &     2.217 &     0.010 &     12.05 &      0.10 &   103.926 &     0.023 &     68.20 &      0.62 &     66.98 &      0.44 &     96.29 &      0.27  \\ \raisebox{9pt}
             
LiteBIRD TT,TE,EE: CDI &     2.221 &     0.014 &     12.02 &      0.16 &   103.921 &     0.051 &     68.34 &      1.04 &     67.09 &      0.76 &     96.49 &      0.64  \\
$\phantom{j}$ \hspace{3mm} pure ADI &     2.217 &     0.013 &     12.05 &      0.10 &   103.920 &     0.029 &     68.14 &      0.63 &     66.94 &      0.47 &     96.27 &      0.34  \\ \raisebox{9pt}
             
(LiteBIRD TT,TE,EE,BB: CDI) &     2.215 &     0.010 &     12.08 &      0.06 &   103.905 &     0.040 &     67.91 &      0.41 &     66.76 &      0.31 &     96.24 &      0.42  \\
$\phantom{j}$ \hspace{3mm} (pure ADI) &     2.213 &     0.010 &     12.07 &      0.05 &   103.921 &     0.028 &     68.03 &      0.32 &     66.84 &      0.25 &     96.23 &      0.30  \\ \raisebox{9pt}
             
LiteCORE-80 TT,TE,EE: CDI &     2.216 &     0.006 &     12.03 &      0.07 &   103.925 &     0.013 &     68.26 &      0.42 &     67.01 &      0.28 &     96.40 &      0.26  \\
$\phantom{j}$ \hspace{3mm} pure ADI &     2.217 &     0.006 &     12.04 &      0.06 &   103.924 &     0.010 &     68.23 &      0.36 &     66.99 &      0.25 &     96.31 &      0.19  \\ \raisebox{9pt}
             
LiteCORE-120 TT,TE,EE: CDI &     2.216 &     0.004 &     12.05 &      0.06 &   103.923 &     0.011 &     68.19 &      0.35 &     66.96 &      0.23 &     96.40 &      0.23  \\
$\phantom{j}$ \hspace{3mm} pure ADI &     2.217 &     0.004 &     12.05 &      0.05 &   103.923 &     0.008 &     68.18 &      0.30 &     66.96 &      0.20 &     96.32 &      0.17  \\ \raisebox{9pt}
             
CORE-M5 TT,TE,EE: CDI &     2.216 &     0.004 &     12.05 &      0.06 &   103.923 &     0.010 &     68.17 &      0.33 &     66.95 &      0.22 &     96.39 &      0.22  \\
$\phantom{j}$ \hspace{3mm} pure ADI &     2.217 &     0.004 &     12.05 &      0.05 &   103.923 &     0.008 &     68.15 &      0.29 &     66.94 &      0.19 &     96.32 &      0.17  \\ \raisebox{9pt}
             
{\bf CORE-M5 TT,TE,EE,PP}: CDI &     2.214 &     0.004 &     12.06 &      0.03 &   103.923 &     0.009 &     68.09 &      0.19 &     66.89 &      0.13 &     96.32 &      0.19  \\
$\phantom{j}$ \hspace{3mm} pure ADI &     2.215 &     0.004 &     12.06 &      0.03 &   103.924 &     0.008 &     68.11 &      0.16 &     66.90 &      0.11 &     96.28 &      0.15  \\ \raisebox{9pt}
             
(CORE-M5 TT,TE,EE,BB,PP: CDI) &     2.214 &     0.004 &     12.07 &      0.03 &   103.920 &     0.009 &     68.06 &      0.19 &     66.86 &      0.13 &     96.27 &      0.18  \\
$\phantom{j}$ \hspace{3mm} (pure ADI) &     2.214 &     0.004 &     12.06 &      0.03 &   103.922 &     0.008 &     68.11 &      0.16 &     66.90 &      0.11 &     96.25 &      0.14  \\ \raisebox{9pt}
             
COrE+ TT,TE,EE,PP: CDI &     2.214 &     0.003 &     12.06 &      0.03 &   103.923 &     0.009 &     68.08 &      0.18 &     66.88 &      0.12 &     96.31 &      0.17  \\
$\phantom{j}$ \hspace{3mm} pure ADI &     2.215 &     0.003 &     12.06 &      0.03 &   103.924 &     0.007 &     68.11 &      0.16 &     66.91 &      0.11 &     96.28 &      0.14  \\ \raisebox{9pt}
             
Ideal TT,TE,EE: CDI &     2.215 &     0.002 &     12.07 &      0.05 &   103.922 &     0.008 &     68.03 &      0.26 &     66.85 &      0.17 &     96.36 &      0.18  \\
$\phantom{j}$ \hspace{3mm} pure ADI &     2.215 &     0.002 &     12.07 &      0.04 &   103.922 &     0.006 &     68.05 &      0.23 &     66.86 &      0.15 &     96.32 &      0.15  \\ 
\hline
%\hline
\end{tabular}
\caption{Selected parameters that exist also in the standard adiabatic model---comparison 
of their recovered posterior mean value and posterior standard deviation when fitting 
the pure adiabatic model (pure ADI) or when fitting the general CDI model. The 
fiducial data here are pure adiabatic with $r=0$. (Note that even when fitting the pure 
adiabatic model, the last digits may slightly differ from the values given in the other 
Sections, since here our primary primordial perturbation parameters that have a uniform prior are 
${\cal P}^{(1)}_{\cal RR}$ and ${\cal P}^{(2)}_{\cal RR}$, not
$\ln\!\left[10^{10}A_\mathrm{S}\right]=\ln\!\left[10^{10} 
{\cal P}_{\cal RR}(k_{\rm mid})\right]$ and  $n_\mathrm{S} = n_{\cal RR}$
like in the other Sections. Furthermore, here we do not utilize BB, except on the lines 
where the experiment name is in parenthesis where the fiducial adiabatic model has 
$r_{0.05}=10^{-3}$, to be discussed in Section \ref{sec:isoTensorTensorless}.)\label{tab:JVadiparams}}
\end{table}}

For isocurvature, the second panel of Fig.~\ref{fig:JVPrimordialPowers} shows the key result.
The point $(0,0)$ represents the pure adiabatic model. We notice a region which is allowed by 
both the real \cite{Ade:2015lrj} and simulated \Planck\ data, but which could be excluded by any 
of the future missions. For example, the primordial isocurvature powers ${\cal P}_{\cal 
II}(k_1) = 0.3\times10^{-10}$ and ${\cal P}_{\cal II}(k_2) = 2.3\times10^{-10}$ (corresponding to 
$\beta_\mathrm{iso}(k_1) \approx 0.0125$ and $\beta_\mathrm{iso}(k_2) \approx 0.1$) would not 
have been detected by \Planck, but the future missions would be able to make a detection. The improvement 
compared to \Planck\ on the upper bounds on isocurvature or correlation power are 
mostly in the parameters of small scales, labelled with the superscript (2), and are about a factor of two to three better. Also for the isocurvature parameters, all 
future configurations are very nearly at the cosmic variance (ideal) limit, except for 
LiteBIRD, which is clearly worse for most parameters.

As shown in Table \ref{tab:JVbetaisoAndAlphanonadi}, CORE-M5 can improve 
the upper bound on the primordial isocurvature 
fraction by a factor of 1.3 on large scales and 
2.4 on small scales compared to the constraints from simulated \Planck\ TT, TE, EE data. The 
tightest constraints on the isocurvature fraction occur at scales in the middle of 
the $k$ or $\ell$ range probed, where $k_\mathrm{mid} = 0.05\,$Mpc$^{-1}$. Here the improvement over 
\Planck\ is by a factor of 2.3. Again, in terms of these parameters, all the future configurations
except for LiteBIRD are near to the cosmic variance limit. The last panel of Table 
\ref{tab:JVbetaisoAndAlphanonadi} shows the primordial correlation fraction. Interestingly, 
LiteBIRD constrains $\cos (\Delta )$ less well than \Planck, while the other configurations reach the 
cosmic variance limit. CORE-M5 tightens the 95\% CL interval by a factor of 1.3 compared to \Planck\ and by a 
factor of 1.7 compared to LiteBIRD.

Finally, in Fig.~\ref{fig:JVadiparams} and Table \ref{tab:JVadiparams}, we report selected 
standard cosmological parameters, which will be determined 
much better than by \Planck\ (in general with $\sigma$ two to four times smaller), virtually at the 
cosmic variance limit. In particular, $H_0$ and $\Omega_\mathrm{c}h^2$ 
will be determined with excellent (almost ideal) accuracy no matter whether or not 
isocurvature is allowed in the theoretical model. There is a well-known degeneracy (different parameter combinations 
produce almost the same observable angular power spectrum) between the isocurvature parameters and 
the standard parameters (mainly $\theta_\mathrm{MC}$ and $\Omega_\mathrm{b} h^2$, but this also is reflected in 
other parameters). With the WMAP accuracy, these degeneracies 
cause large absolute shifts and significant broadening of the posterior 
of standard parameters when isocurvature is allowed 
\cite{Crotty:2003rz,KurkiSuonio:2004mn,Valiviita:2009bp,Savelainen:2013iwa}, but with the \Planck\ data
these effects almost disappear \cite{Ade:2015lrj}. However CORE-M5 as well as both 
LiteCOREs and COrE+ would further decrease the uncertainties caused by the degeneracy and make the 
determination of the standard parameters even more robust against the assumptions made concerning the 
initial conditions. 
(In Table \ref{tab:JVadiparams}, for each experiment we quote first the results when fitting 
the general correlated three-parameter isocurvature 
CDI model and then the results when fitting the 
pure adiabatic model.) Table \ref{tab:JVadiparams} indicates only marginal broadening of the 
$1\sigma$ interval with CDI and insignificant shifts compared to the input fiducial data, 
as indicated on the first line of the Table. 

However with \Planck\ and LiteBIRD some broadening and 
shifts (compared to the pure adiabatic fit, pure ADI/dashed lines) are visible in 
Fig.~\ref{fig:JVadiparams} and confirmed in the Table. For example, with LiteBIRD the $1\sigma$ 
posterior interval of the adiabatic scalar spectral index $n_{\cal RR}$ broadens by a factor of 1.9 
when isocurvature is allowed, and the mean value of $n_{\cal RR}$ shifts by +0.25\% 
compared to the input fiducial value, or by +0.7$\sigma_{\rm ADI}$. With CORE-M5 TT, TE, EE, and PP, the 
corresponding numbers are very small: broadening by a factor of 1.3 (note that this is compared to the 
already small $\sigma_{\rm ADI}=0.0015$ of CORE-M5) and a shift by +0.07\% or +0.4$\sigma_{\rm ADI}$.
\afterpage{%
\begin{figure}
\centering
\includegraphics[width=\textwidth]{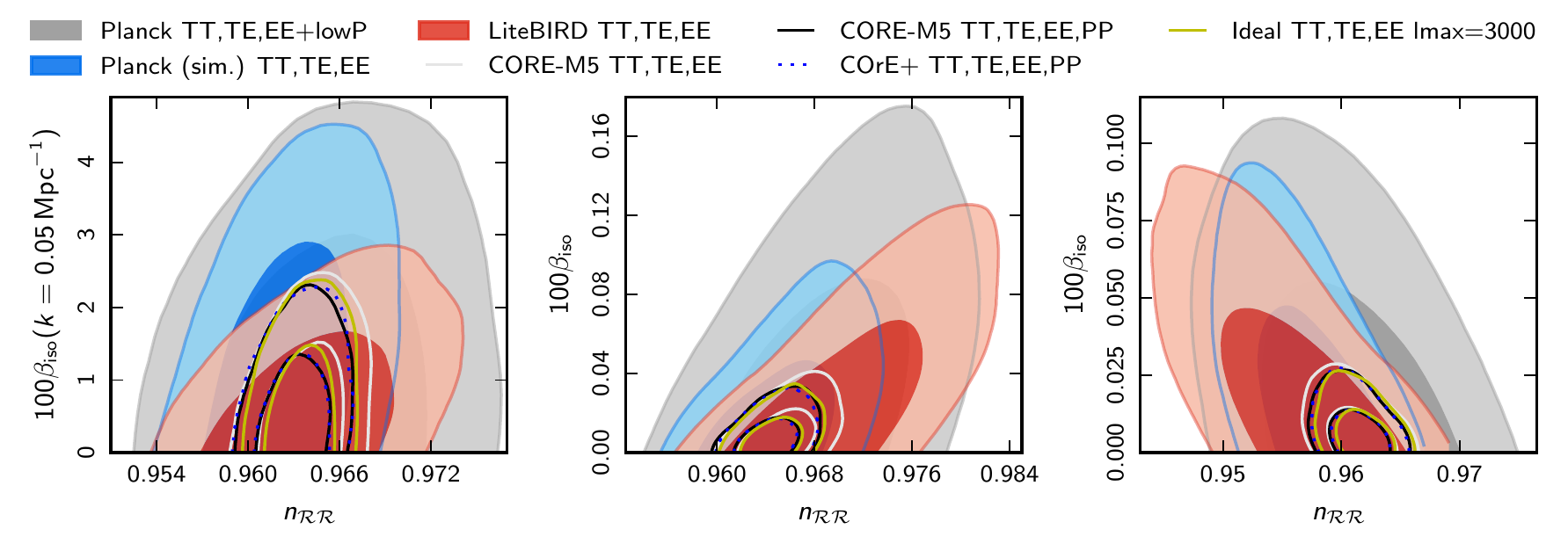}\\
\vspace{-5.5mm}
\caption{One-parameter isocurvature extensions to the adiabatic $\Lambda$CDM model. 
Left: an ``Axion'' model (uncorrelated adiabatic and isocurvature modes, 
with $n_{\cal II} = 1$, $\cos\Delta=0$). Middle: ``Curvaton I'' (fully correlated 
adiabatic and isocurvature modes, with $n_{\cal II} = n_{\cal RR}$,  
$\cos\Delta=+1$). Right: ``Curvaton II'',  (fully anti-correlated adiabatic and 
isocurvature modes, with $n_{\cal II} = n_{\cal RR}$,  $\cos\Delta=-1$). The simulated \Planck\ 
and LiteBIRD datasets contain only TT,TE,EE, but the simulated CORE-M5 is presented both 
with and without the lensing potential PP (but not with the BB nor delensing information). 
The real \Planck\ data contain also a noisy B-mode at low-$\ell$, consistent with zero. 
CORE-M5 has capability for an unprecedented constraining power in the curvaton models.
\label{fig:JVspecialcases}}
\end{figure}
\begin{table}
\setlength{\tabcolsep}{.19667em}
\scriptsize
\begin{tabular}{|l|ccc|ccc|ccc|cc|}
\hline
%\hline
    & \multicolumn{3}{c|}{Axion (NC)} & \multicolumn{3}{c|}{Curvaton I (FC)} & \multicolumn{3}{c|}{Curvaton II (AC)} &  \multicolumn{2}{c|}{Pure ADI}\\
    & $100\beta_{\rm iso}$ & $n_{\cal RR}$ & $\sigma(n_{\cal RR})$ & $100\beta_{\rm iso}$ &  $n_{\cal RR}$ & $\sigma(n_{\cal RR})$ & $100\beta_{\rm iso}$  &  $n_{\cal RR}$ & $\sigma(n_{\cal RR})$ &  $n_{\cal RR}$ & $\sigma(n_{\cal RR})$ \\
\hline
\Planck\ TT,TE,EE+lowP           & 3.842 & 0.9653 & 0.0044 & 0.129 &  0.9693 & 0.0049 & 0.082 & 0.9600 & 0.0048 & 0.9641 & 0.0044 \\
\Planck\ TT,TE,EE (sim.)           & 3.628 & 0.9628 & 0.0030 & 0.072 & 0.9655 & 0.0033 &  0.072 & 0.9572 & 0.0034 & 0.9629 & 0.0027 \\
LiteBIRD TT,TE,EE                   & 2.192 & 0.9647 & 0.0038 & 0.096 & 0.9711 & 0.0050 & 0.072 & 0.9555 & 0.0048  & 0.9627 & 0.0034 \\
CORE-M5 TT,TE,EE                      & 1.951 & 0.9639 & 0.0018 & 0.031 & 0.9656 & 0.0019 & 0.022 & 0.9612 & 0.0018 & 0.9632 & 0.0017 \\
{\bf CORE-M5 TT,TE,EE,PP}      & 1.792 & 0.9633 & 0.0015 & 0.026 & 0.9645 & 0.0016 & 0.021 & 0.9612 & 0.0016 & 0.9628 & 0.0015 \\
COrE+ TT,TE,EE,PP               & 1.823 & 0.9632 & 0.0015 & 0.025 & 0.9644 & 0.0015 & 0.021 & 0.9613 & 0.0015 & 0.9628 & 0.0014  \\
Ideal TT,TE,EE                      & 1.891 & 0.9637 & 0.0015 & 0.026 & 0.9649 & 0.0016 & 0.021 & 0.9617 & 0.0015 & 0.9632 & 0.0015\\
%\hline
\hline
\end{tabular}
\caption{95\% CL upper bounds on the primordial isocurvature fraction $\beta_{\rm iso}$, 
the posterior mean value of  the scalar spectral index $n_{\cal RR} = n_{\cal II} = n_{\cal RI}$, and $1\sigma$ uncertainty in its determination when fitting 
the one-parameter isocurvature extensions of the $\Lambda$CDM model (the first three cases) 
and when fitting the pure adiabatic model (the last case). The fiducial data are the pure 
adiabatic $\Lambda$CDM model, with $n_{\cal RR} = 0.9625$, $r=0$ (and $\beta_{\rm iso} = 0$).
\label{tab:JVspecialcases}}
\end{table}}

\subsubsection{Special one parameter extensions to adiabatic $\Lambda$CDM\label{sec:special}}

There are a number of interesting (one-parameter) isocurvature models simpler than the general model 
studied in the previous Subsection. Here we analyze the same three models as in \cite{Planck:2013jfk,Ade:2015lrj}.
%the \Planck\ 2013 and 2015 Constraints on Inflation papers \cite{Planck:2013jfk,Ade:2015xua,Ade:2015lrj}. 
These are an ``Axion'' model (no correlation between ADI and CDI, and $n_{\cal II}=1$ and $r=0$; see, e.g., 
ref.~\cite{Beltran:2006sq}), ``Curvaton I'' (100\% correlation between ADI and CDI, and $n_{\cal 
II}=n_{\cal RR}$), and ``Curvaton II'' ($-100$\% correlation between ADI and CDI, and $n_{\cal 
II}=n_{\cal RR}$). These curvaton models have $\cos\Delta=\pm1$ and do not have a primordial tensor 
contribution since the tensor contribution is associated to the curvature perturbation at the 
horizon exit during inflation, which in these models is negligible. The primordial curvature 
perturbation is generated later from the conversion of the isocurvature perturbation into curvature 
perturbations \cite{Enqvist:2001zp,Lyth:2001nq,Moroi:2001ct}, hence the full (anti)-correlation 
between them (see also \cite{Gordon:2002gv}). Indeed, as discussed in 
\cite{Bartolo:2001rt,Wands:2002bn,Byrnes:2006fr,Savelainen:2013iwa,Ade:2015lrj}, in 
slow-roll models if the tensor-to-scalar power ratio at the end of inflation was $\tilde{r}$, 
at the primordial epoch after inflation and reheating the ratio would become 
$r=(1-\cos^2\!\Delta)\tilde{r}$ --- or 
in terms of the slow-roll parameter $\epsilon$, $r=16\epsilon\sin^2\!\Delta$ to the leading order. 
Thus $r$ will be zero if $|\cos\Delta|=1$.

Constraints on these three models are presented in Fig.~\ref{fig:JVspecialcases} and Table 
\ref{tab:JVspecialcases}: ``Axion'' (no correlation, NC) in the left panel, ``Curvaton I'' in the 
middle panel (full correlation, FC), and ``Curvaton II'' in the right panel (anti-correlation, AC). 
In addition to the simulated \Planck, LiteBIRD, CORE-M5, and COrE+ data, we reproduce 
the constraint from 
the real \Planck\ TT, TE, EE+lowP data as reported in Ref.~\cite{Ade:2015lrj}.

For the axion model, the real and simulated \Planck\ data constrain the primordial isocurvature 
fraction equally well. However the simulated \Planck\ data constrain the primordial scalar spectral 
index $n_{\cal RR}$ (=$n_{\cal II}$) more tightly than the 2015 data, since the simulated data constrain 
the optical depth $\tau$ much better, at an accuracy similar to the subsequent determination using the 
\Planck\ HFI data \cite{planck2014-a10,Adam:2016hgk}. The 95\% upper bound on the isocurvature 
fraction drops from the \Planck\ level of 4\% to 2.2\% with LiteBIRD
and to 1.8\% with CORE-M5 or COrE+.

The constraints on $\beta_{\rm iso}$ are much tighter for the curvaton models, since the value of 
$\beta_{\rm iso}$ directly controls the amplitude of the correlation power spectrum, which now is 
the geometric average of the (small) isocurvature power and the (larger) curvature perturbation 
power. Hence the correlation has a much larger effect on the CMB angular power spectrum than the 
isocurvature itself, which is the only contribution in the uncorrelated axion model. {For 
curvaton models} LiteBIRD performs equally well or worse than \Planck, but {CORE-M5 gives 
unprecedented constraining power.} The upper bound on $\beta_{\rm iso}$ drops from the \Planck\ one
part per thousand by almost an order of magnitude, to 0.26 (FC) or 0.21 (AC) per thousand.

Figure \ref{fig:JVspecialcases} shows a degeneracy between the spectral index and 
isocurvature fraction in the curvaton models. Consequently, marginalizing over $\beta_\mathrm{iso}$ 
biases the one-dimensional marginalized posterior of $n_{\cal RR}$ toward larger values in the Curvaton I (FC), 
and toward smaller values in the Curvaton II (AC) case. This is evident in Table 
\ref{tab:JVspecialcases}, where for example with the simulated \Planck\ data the posterior mean value of 
$n_{\cal RR}$ is 0.9655 with FC ($0.9\sigma$ above the input fiducial value of 0.9625) or 0.9572 
with AC ($1.6\sigma$ below the fiducial value). As CORE-M5 constrains $\beta_\mathrm{iso}$ much 
better, the absolute shift in the mean value of $n_{\cal RR}$ is significantly reduced (see for example
the line in bold in the Table), but since the width ($\sigma$) of the posterior also shrinks, the 
shift naturally stays at the same $1\sigma$ level as with \Planck.
\begin{figure}
\centering
\includegraphics[width=0.88\textwidth]{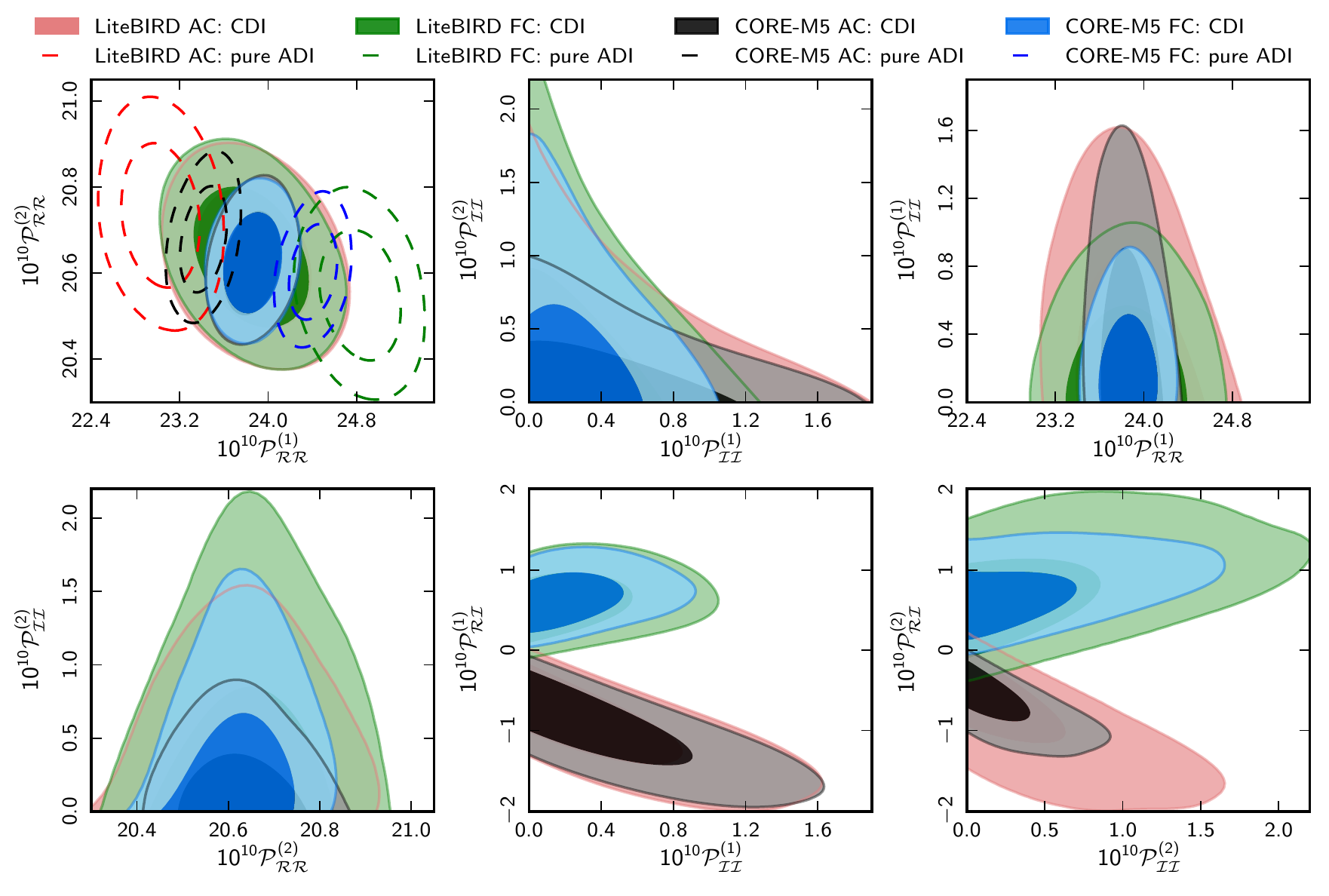}\\
\vspace{-5.5mm}
\caption{
Primordial powers when the fiducial data are produced from a curvaton model and have a 0.1\% 
anti-correlated (AC) or 0.1\% fully correlated (FC) CDI contribution, but no tensor contribution. 
The fitted model is the generally correlated model with three isocurvature parameters and 
$\tilde{r}_{0.05}$ (CDI, shaded colors) or the pure adiabatic model with $r_{0.05}$ (pure ADI, 
dashed contours). LiteBIRD uses TT,TE,EE,BB, whereas CORE-M5 uses TT,TE,EE,BB,PP and delensing 
information. (The fiducial values used are $10^{10}{\cal P}^{(1)}_{\cal RR} = 23.90$, $10^{10}{\cal 
P}^{(2)}_{\cal RR} = 20.64$, $10^{10}{\cal P}^{(1)}_{\cal II} = 0.0239$, $10^{10}{\cal 
P}^{(2)}_{\cal II} = 0.0206$, $10^{10}{\cal P}^{(1)}_{\cal RI} = \pm0.7560$, $10^{10}{\cal 
P}^{(2)}_{\cal RI} = \pm0.6528$.) For the dramatically biased posterior of ${\cal P}_{\cal RR}$ in 
the first panel in the case when we fit the wrong (pure ADI, dashed lines) model, see the main text 
and footnote \ref{foot:JVSW}.
\label{fig:JVACFCprimpowers}}
\end{figure}

\subsection{Fiducial data with a fully (anti)-correlated CDI contribution}

Isocurvature perturbations do not produce B modes except through lensing exactly as in the  
adiabatic model without tensors. Therefore BB and delensing information would not significantly help 
for the studies described in the previous Subsections. However, in this Subsection 
we utilize BB, since we fit to our fiducial data a model 
where $\tilde{r}$ is a free parameter, while $n_\mathrm{t}$ is fixed 
by the inflationary consistency condition to $n_\mathrm{t} = -2\epsilon = -r/(8(1-\cos^2\!\Delta)) = -\tilde{r}/8$.
%Hence, CORE with BB and delensing is now expected to outperform LiteBIRD.
We create two fiducial datasets, one assuming ``Curvaton I'' 
(i.e., 100\% correlated ADI and CDI) with $\beta_{\rm iso} = 0.1$\%, and another assuming 
``Curvaton II'' with the same isocurvature fraction (but setting 100\% anti-correlation between ADI 
and CDI). As explained in the previous Subsection, these fiducial data do not have any primordial 
tensor contribution. However we fit to these data our full three parameter generally correlated 
isocurvature model (CDI) and also allow for a nonzero $\tilde{r}$. In a nutshell, our fiducial data has 
only one extra parameter compared to the tensorless adiabatic $\Lambda$CDM model, but we try to fit 
to a theory with four extra parameters. For comparison, we show the dramatic effects encountered
in determining the standard cosmological parameters if we try to force fit the pure adiabatic 
model (ADI) to these data having a seemingly small one part per thousand primordial isocurvature 
fraction.

\begin{figure}[t]
\centering
\includegraphics[width=\textwidth]{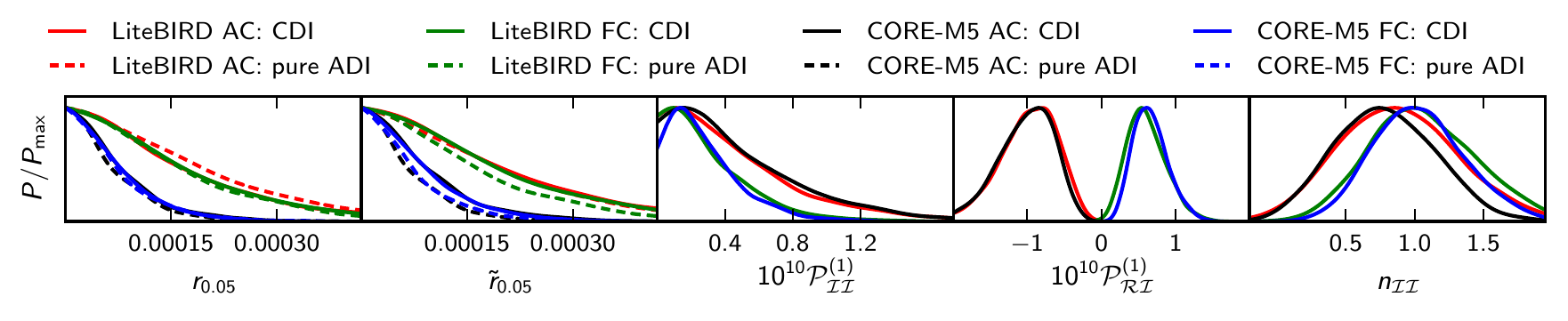}\\
\vspace{-5.5mm}
\caption{The primordial tensor-to-scalar ratio $r_{0.05}$, the horizon exit tensor-to-scalar 
ratio $\tilde r_{0.05}$, isocurvature and correlation power at large scales, and isocurvature 
spectral index, when the fiducial data have a 0.1\% anti-correlated (AC) or fully correlated 
(FC) CDI contribution and $r=0$. The fitted model is the generally correlated model with 
three isocurvature parameters and $\tilde{r}_{0.05}$ (CDI, solid lines) or the pure 
adiabatic $\Lambda$CDM model with $r_{0.05}$ (pure ADI, dashed lines). 
\label{fig:JVACFCprimfractions}}
\end{figure}

Figure \ref{fig:JVACFCprimpowers} demonstrates that both LiteBIRD and CORE-M5 easily recover the 
adiabatic and correlation input parameters of these models (though the input isocurvature power is 
too small to be detected). In particular, the bottom middle and right panels show that the nonzero 
correlation is detected at about the $2\sigma$ level. However Fig.~\ref{fig:JVACFCprimfractions} and 
Table \ref{tab:JVACFCprimfractions} reveal that the isocurvature spectral index (input fiducial 
value $n_{\cal II} = n_{\cal RR} = 0.9625$) is poorly constrained, CORE-M5 being about 15\% better 
than LiteBIRD. The difference in the capability to set an upper bound on $r$ is also evident 
from the first panels of Fig.~\ref{fig:JVACFCprimfractions} and Table 
\ref{tab:JVACFCprimfractions}. 
%When fitting the full CDI model, we obtain $r_{0.05}<1.84\times10^{-4}$ (AC) 
%or $r < 1.74\times10^{-4}$ (FC) with CORE-M5, $r_{0.05}<1.84\times10^{-4}$ (AC) or $r < 1.74\times10^{-4}$ (FC) 
%with LiteBIRD, i.e., LiteBIRD gives constraints 2.2 times weaker. 
%CORE-M5 gives 
%$r_{0.05}<1.84\times10^{-4}$ (AC) or $r < 1.74\times10^{-4}$ (FC), whereas LiteBIRD finds $r_{0.05} 
%< 4.11\times10^{-4}$ (AC) or $3.94\times10^{-4}$ (FC) at 95\% CL, i.e., LiteBIRD gives 
%constraints 2.2 times weaker.
When fitting a wrong model (i.e., the pure adiabatic model) to these data, the 
constraints on $r$ change insignificantly. %Our fiducial data have $r=0.$ 
In addition, we show in Table \ref{tab:JVACFCprimfractions} the results when fitting the full CDI model without a tensor 
contribution: the constraints on ${\cal P}^{(1)}_{\cal II}$, ${\cal P}^{(1)}_{\cal 
RI}$, $n_{\cal II}$, and ${\cal P}^{(1)}_{\cal RR}$ stay virtually unchanged. Consequently, allowing for a 
nonzero $r$ does not bias the results nor weaken the constraints. Moreover, this test proves that 
the asymmetry in the determination of ${\cal P}_{\cal RI}$ between the AC and FC cases is not 
caused by a (hypothetical) degeneracy between $r$ and the correlation contribution. Even without 
$r$, the posterior on $10^{10}{\cal P}^{(1)}_{\cal RI}$ in the AC case has a long tail toward much 
more negative values than the fiducial value $-0.76$, leading to a biased mean value $-1.0$ of the 
posterior with both LiteBIRD and CORE-M5. In the FC case the posterior on ${\cal P}^{(1)}_{\cal RI}$ 
is much more symmetric and centers near to the fiducial value $+0.76$. This asymmetry between AC 
and FC cases can be understood as follows. When the full CDI fit has a negative correlation, the 
positive contribution from the isocurvature itself can partially cancel the correlation 
contribution, thus explaining the more negative correlation to fit the data. However when the full CDI fit 
has a positive correlation, it adds to the positive contribution from isocurvature itself, and 
hence we obtain tighter constraints in the case of positive correlation for both $|{\cal 
P}^{(1)}_{\cal RI}|$ and ${\cal P}^{(1)}_{\cal II}$. This degeneracy is confirmed in Table 
\ref{tab:JVACFCprimfractions} (and in the bottom middle panel of Fig.~\ref{fig:JVACFCprimpowers}).

To summarize, fitting the three isocurvature degrees of freedom (CDI) or ignoring them (pure ADI) 
does not significantly interfere with the upper bound on $r$. Neither does allowing for a nonzero 
$r$ interfere with the determination of the isocurvature parameters in these models with a one part
per thousand fully (anti)-correlated isocurvature contribution. Furthermore, apart from the determination 
of the isocurvature spectral index, there is no significant difference in the (isocurvature) 
detection power between LiteBIRD and CORE-M5. However $r$ is constrained about 2.2 times better by 
CORE-M5 in these models.
%\afterpage{%
\begin{table}
\setlength{\tabcolsep}{.30667em}
\scriptsize
\begin{tabular}{|l|r|r|r|rl|rl|rl|}
\hline
%\hline
   & $10^{4}r_{0.05}$ & $10^{4}\tilde r_{0.05}$ &  $10^{10}{\cal P}^{(1)}_{\cal II}$  & \multicolumn{2}{c|}{$10^{10}{\cal P}^{(1)}_{\cal RI}$} & \multicolumn{2}{c|}{$100n_{\cal II}$} &   \multicolumn{2}{c|}{ $10^{10}{\cal P}^{(1)}_{\cal RR}$ } \\ \raisebox{9pt}

 $\phantom{j}$ & 95\% CL & 95\% CL & 95\%CL & mean & 68\% CL & mean & 68\% CL & mean & 68\% CL \\

\hline
Fiducial data  & $0$ & $0$ & $0.024$ & $\pm0.76$  & & $96.25$ & & $23.90$ &  \\ \raisebox{9pt}

LiteBIRD {\bf AC}: CDI & $<     4.11$ & $<     4.92$ & $<    1.275$ &     -0.99 & $[    -1.29;     -0.54]$ &     88.42 & $[    42.26;    128.28]$ &     23.88 & $[    23.54;     24.21]$  \\
$\phantom{j}$ \hspace{5mm}$r=0$ CDI &  0 &  0 & $<    1.263$ &     -0.99 & $[    -1.29;     -0.53]$ &     87.35 & $[    39.81;    125.19]$ &     23.86 & $[    23.54;     24.19]$  \\
$\phantom{j}$ \hspace{5mm} pure ADI & $<     4.43$ & $<     4.43$  & & & & & &     23.03 & $[    22.81;     23.25]$  \\ \raisebox{9pt}
             
LiteBIRD {\bf FC}: CDI & $<     3.94$ & $<     4.54$ & $<    0.818$ &      0.61 & $[     0.31;      0.82]$ &    107.04 & $[    63.71;    141.74]$ &     23.85 & $[    23.52;     24.18]$  \\
$\phantom{j}$ \hspace{5mm}$r=0$ CDI &  0 & 0 & $<    0.885$ &      0.58 & $[     0.29;      0.79]$ &    107.63 & $[    63.58;    142.59]$ &     23.88 & $[    23.52;     24.21]$  \\
$\phantom{j}$ \hspace{5mm} pure ADI & $<     3.67$ & $<     3.67$  & & & & & &     24.83 & $[    24.60;     25.07]$  \\ \raisebox{9pt}
             
CORE-M5 {\bf AC}: CDI & $<     1.84$ & $<     2.19$ & $<    1.288$ &     -1.02 & $[    -1.32;     -0.59]$ &     78.45 & $[    39.81;    113.40]$ &     23.86 & $[    23.69;     24.02]$  \\
$\phantom{j}$ \hspace{5mm}$r=0$ CDI & 0 & 0 & $<    1.306$ &     -1.04 & $[    -1.34;     -0.61]$ &     76.76 & $[    37.09;    109.63]$ &     23.85 & $[    23.68;     24.01]$  \\
$\phantom{j}$ \hspace{5mm} pure ADI & $<     1.68$ & $<     1.68$  & & & & & &     23.42 & $[    23.29;     23.55]$  \\ \raisebox{9pt}
             
CORE-M5 {\bf FC}: CDI & $<     1.74$ & $<     2.02$ & $<    0.717$ &      0.66 & $[     0.40;      0.85]$ &    103.21 & $[    67.26;    133.28]$ &     23.86 & $[    23.69;     24.03]$  \\
$\phantom{j}$ \hspace{5mm}$r=0$ CDI & 0 & 0 & $<    0.737$ &      0.66 & $[     0.40;      0.85]$ &    102.15 & $[    66.18;    132.75]$ &     23.87 & $[    23.70;     24.04]$  \\
$\phantom{j}$ \hspace{5mm} pure ADI & $<     1.82$ & $<     1.82$  & & & & & &     24.41 & $[    24.27;     24.54]$  \\
%\hline
\hline
\end{tabular}
\caption{
The same information in numbers as in Fig.~\ref{fig:JVACFCprimfractions}, with the curvature 
perturbation amplitude at large scales added as a last column, and we also compare the full CDI 
with and without $r$.
\label{tab:JVACFCprimfractions}}
\end{table}

\afterpage{%
\begin{figure}
\centering
\includegraphics[width=\textwidth]{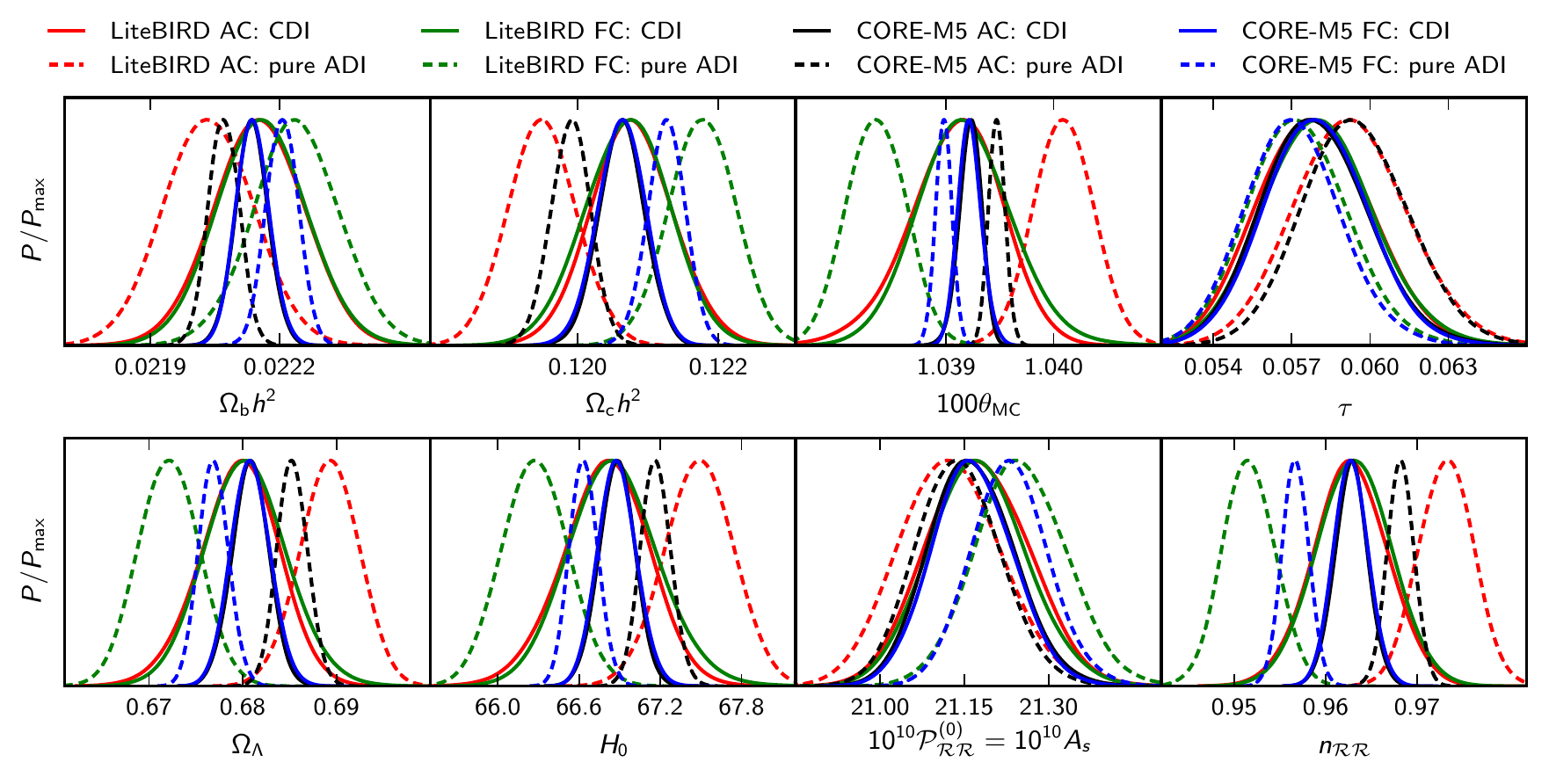}\\
\vspace{-5.5mm}
\caption{Determination of the standard parameters when the fiducial data have a 
0.1\% anti-correlated (AC) or fully correlated (FC) CDI contribution and zero $r$. The solid 
lines (CDI) represent the fit of the generally correlated three isocurvature parameter model 
plus $\tilde{r}_{0.05}$ to these one isocurvature parameter models. The input fiducial data are 
recovered extremely well in this case. However, despite of the smallness of the isocurvature 
contribution, fitting a ``wrong'' model, i.e., the pure adiabatic model (pure ADI, dashed lines),
to these data leads to a large bias (error) in the determination of the standard parameters. 
\label{fig:JVACFCstandardparams}}
\end{figure}
\begin{table}
\setlength{\tabcolsep}{.12667em}
\scriptsize
\begin{tabular}{|l|crr|crr|crr|crr|crr|crr|}
\hline
%\hline
   & \multicolumn{3}{c|}{$100\Omega_{\rm b} h^2$} & \multicolumn{3}{c|}{$100\Omega_{\rm c} h^2$} & \multicolumn{3}{c|}{$10000\theta_{\rm MC}$} & \multicolumn{3}{c|}{$100\Omega_\Lambda$} & \multicolumn{3}{c|}{$H_0$} & \multicolumn{3}{c|}{$100n_{\cal RR}$} \\ \raisebox{7pt}

 $\phantom{j}$ & mean & {\tiny $\Delta(\sigma)$} & {\tiny $\Delta$(\%)} & mean & {\tiny $\Delta(\sigma)$} &  {\tiny $\Delta$(\%)} & mean & {\tiny $\Delta(\sigma)$} &  {\tiny $\Delta$(\%)} & mean & {\tiny $\Delta(\sigma)$} &  {\tiny $\Delta$(\%)} & mean & {\tiny $\Delta(\sigma)$} &  {\tiny $\Delta$(\%)} & mean & {\tiny $\Delta(\sigma)$}  & {\tiny $\Delta$(\%)} \\

\hline
Fiducial data  & 2.21 & & & 12.06 & & & 103.92 & & & 68.10 & & & 66.89 & & & 96.25 & &  \\ \raisebox{7pt}

LB {\bf AC}: CDI &      2.22 &    0.1 &    0.1 &     12.08 &    0.3 &    0.1 &    103.91 &   -0.3 &   -0.0 &     67.96 &   -0.3 &   -0.2 &     66.80 &   -0.3 &   -0.1 &     96.27 &    0.1 &    0.0  \\
$\phantom{j}$ \hspace{3mm} pure ADI &      2.20 &   -1.0 &   -0.5 &     11.95 &   -1.9 &   -0.9 &    104.01 &    2.0 &    0.1 &     68.92 &    2.0 &    1.2 &     67.49 &    1.9 &    0.9 &     97.32 &    2.6 &    1.1  \\ \raisebox{7pt}
             
LB {\bf FC}: CDI &      2.22 &    0.2 &    0.1 &     12.07 &    0.2 &    0.1 &    103.92 &   -0.1 &   -0.0 &     68.03 &   -0.2 &   -0.1 &     66.85 &   -0.1 &   -0.1 &     96.31 &    0.1 &    0.1  \\
$\phantom{j}$ \hspace{3mm} pure ADI &      2.22 &    0.9 &    0.4 &     12.18 &    1.9 &    1.0 &    103.83 &   -2.0 &   -0.1 &     67.21 &   -2.0 &   -1.3 &     66.27 &   -1.8 &   -0.9 &     95.16 &   -2.6 &   -1.1  \\ \raisebox{7pt}
             
CORE-M5 {\bf AC}: CDI &      2.21 &   -0.1 &   -0.0 &     12.06 &    0.1 &    0.0 &    103.92 &    0.0 &    0.0 &     68.08 &   -0.1 &   -0.0 &     66.88 &   -0.1 &   -0.0 &     96.28 &    0.2 &    0.0  \\
$\phantom{j}$ \hspace{3mm} pure ADI &      2.21 &   -1.8 &   -0.3 &     11.99 &   -2.2 &   -0.6 &    103.95 &    2.4 &    0.0 &     68.53 &    2.2 &    0.6 &     67.17 &    2.1 &    0.4 &     96.82 &    3.1 &    0.6  \\ \raisebox{7pt}
             
CORE-M5 {\bf FC}: CDI &      2.21 &   -0.1 &   -0.0 &     12.06 &    0.1 &    0.0 &    103.92 &    0.0 &    0.0 &     68.08 &   -0.1 &   -0.0 &     66.88 &   -0.1 &   -0.0 &     96.28 &    0.1 &    0.0  \\
$\phantom{j}$ \hspace{3mm} pure ADI &      2.22 &    1.7 &    0.3 &     12.13 &    1.9 &    0.5 &    103.90 &   -2.2 &   -0.0 &     67.69 &   -2.0 &   -0.6 &     66.63 &   -1.8 &   -0.4 &     95.67 &   -3.0 &   -0.6  \\
%\hline
\hline
\end{tabular}
\caption{Determination of the standard parameters when the fiducial data have a 0.1\% anti-correlated 
(AC) or fully correlated (FC) CDI contribution and zero $r$. $\Delta$ indicates the shift (bias) 
with respect to the input fiducial values in units of the standard deviation ($\sigma$) of the 
posterior of the full three-parameter CDI model plus $\tilde{r}_{0.05}$ or in percentage (\%). 
Fitting a wrong model, the pure adiabatic model (pure ADI), leads to seemingly large biases 
(as large as $|\Delta|=3\sigma$) in terms of $\sigma$, but all shifts are below 
$|\Delta| = 1.3\%$.\label{tab:JVACFCstandardparams}}
\end{table}}

In Fig.~\ref{fig:JVACFCstandardparams} and Table \ref{tab:JVACFCstandardparams} we study a 
different question: What happens to the determination of the standard cosmological parameters if 
we (unaware of the one part per thousand AC/FC isocurvature contribution in the data) fit the pure 
adiabatic model (pure ADI) to these data? When fitting the full CDI model, we recover the input 
fiducial model with  very high accuracy. On the contrary, force fitting a wrong model (pure 
ADI) leads to a huge bias in the standard parameters, as large as $3\sigma$. However, in particular 
with CORE-M5, the shifts are small in terms of percentage of the input fiducial values. With this 
measure, the largest shifts with CORE-M5 are for $\Omega_\mathrm{c}h^2$, $\Omega_\Lambda$, and 
$n_{\cal RR}$ whose values can be corrupted by $0.6\%$. In case of LiteBIRD, the value of 
$\Omega_\Lambda$ can be misestimated by fitting the pure adiabatic model by as much as $1.3\%$ .

Most of the shifts seen in Fig.~\ref{fig:JVACFCstandardparams} and Table 
\ref{tab:JVACFCstandardparams} can be traced back to the determination of the sound horizon angle 
$\theta_\mathrm{MC}$. As first pointed out in \cite{Keskitalo:2006qv} and later in 
\cite{Valiviita:2009bp,Valiviita:2012ub} (when fitting isocurvature models to the WMAP data), the 
positive correlation effectively shifts the acoustic peaks toward the right, leading to a 
smaller estimated $\theta_\mathrm{MC}$, and the negative correlation leads to a larger estimated $\theta_\mathrm{MC}$ if 
the other parameters are kept fixed. By reversing the argument, we realize that having an AC 
component in the data but fitting the pure adiabatic model gives us a too large $\theta_\mathrm{MC}$, while 
with the FC data fitting the pure adiabatic model gives us a too small $\theta_\mathrm{MC}$. These changes are 
then reflected in $\Omega_\mathrm{m}h^2$, and further in $H_0$, $\Omega_\Lambda$, and the other 
parameters. The adiabatic spectral index $n_{\cal RR}$ is further affected by the ratio of the 
large scale to small scale (temperature) angular power. Since the AC/FC contribution with $n_{\cal 
II}=n_{\cal RR}$ only affects the low multipoles (large scales), the ratio is reduced (increased) 
in the AC (FC) case, leading to a too large (small) $n_{\cal RR}$ if the pure adiabatic model is 
fitted to these data.\footnote{%It should be noted that 
The large-scale Sachs-Wolfe effect is such 
that $\delta T / T \approx -(1/5)({\cal R} + 2f_c {\cal S})$, where $f_c = 
\Omega_\mathrm{c}/(\Omega_\mathrm{c}+\Omega_\mathrm{b})$ and ${\cal P_{RR}} \propto \langle |{\cal 
R}|^2 \rangle$, ${\cal P_{II}} \propto \langle |{\cal S}|^2 \rangle$, and ${\cal P_{RI}} \propto 
\langle {\cal R}^\ast {\cal S} \rangle$ at the primordial time \cite{KurkiSuonio:2004mn}. If $b^2 
\equiv {\cal P_{II}} / {\cal P_{RR}}$, then
%\begin{equation}
$C_\ell^{TT} \propto \langle (\Delta T / T)^2 \rangle \propto (1 \pm 2\times2|b|f_c + 4b^2f_c^2){\cal P_{RR}}\,.$
%\label{eq:JVCell}
%\end{equation}
Now we have $f_c=0.8451$ and $b\approx0.0316$, which leads to $C_{\ell}^{TT} \propto 0.90 {\cal 
P_{RR}}$ (AC, the minus sign) and $1.11 {\cal P_{RR}}$ (FC, the plus sign). Thus, our one part per thousand 
primordial isocurvature contribution translates to a $\pm10\%$ observational non-adiabatic 
contribution at large scales. (In addition to the ordinary Sachs-Wolfe effect, the large-scale angular power is affected by the integrated Sachs-Wolfe effect. A precise numerical examination of the temperature angular power spectrum of our fiducial model shows that the maximum non-adiabatic contribution, coming mainly from correlation, is actually $\pm8\%$ at multipole $\ell = 5$, and after $\ell \approx 65$ the non-adiabatic contribution falls below 1\%, being $\pm0.6\%$ to the first acoustic peak at $\ell \approx 220$.) This together with the change in $\theta_\mathrm{MC}$ explains the dramatic 
shifts of ${\cal P}^{(1)}_{\cal RR}$ in the upper left panel of Fig.~\ref{fig:JVACFCprimpowers} 
when incorrectly fitting the pure ADI model.
%However, with CORE-M5 even the ``heavily'' biased ADI values are within the proper CDI curves of LiteBIRD.
\label{foot:JVSW}}

In the case of these simulated AC/FC data, it is naturally easy to tell that the pure adiabatic 
model is the wrong model, since with LiteBIRD the best fit $\chi^2$ of pure ADI is about 30 points 
worse and with CORE-M5 about 60 points worse than the one of CDI. However, with real data the question 
is more subtle. For example, is 12060 an acceptable best fit $\chi^2$ with 12000 ``data points'', or 
should one seriously look for another model beyond the pure adiabatic $\Lambda$CDM model? Indeed, the 
first panel of Fig.~\ref{fig:JVACFCprimpowers} helps answer this question, at least with respect to
the isocurvature extensions. Large shifts in the ``ADI contours'' when 
a generally correlated CDI mode is allowed constitute a clear hint that the CDI model may be the 
correct one. Conversely, if the contours do not shift, there is very little or no room for 
isocurvature.
\afterpage{%
\begin{figure}
\centering
\includegraphics[width=0.88\textwidth]{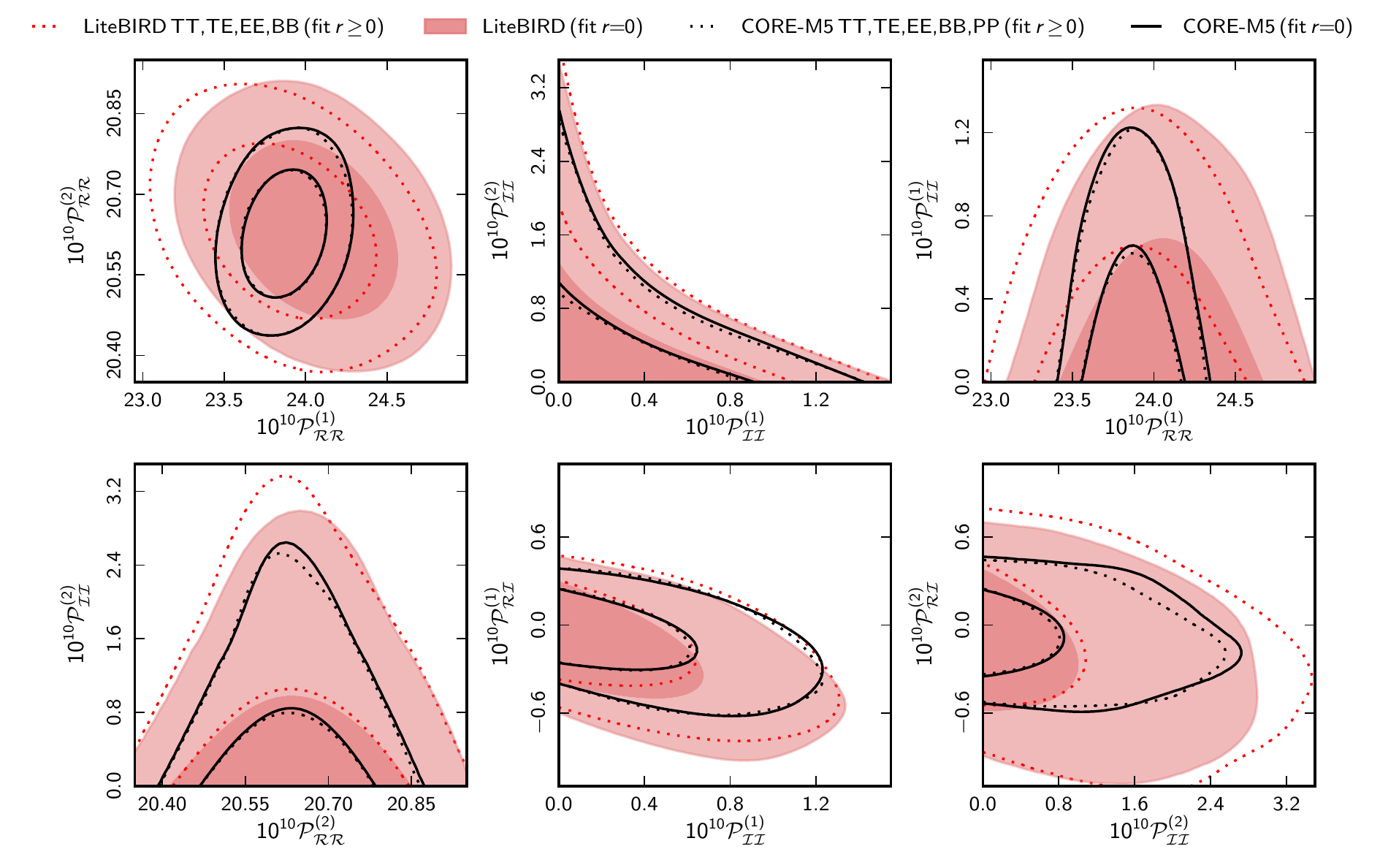}\\
\vspace{-5.5mm}
\caption{Primordial powers when the fiducial data are adiabatic with $r_{0.05}=10^{-3}$. The fitted model 
is the generally correlated CDI model with three isocurvature parameters and a free $\tilde{r}_{0.05}$ 
or a fixed $r=0$. CORE-M5 utilizes delensing and its results with a free $\tilde{r}_{0.05}$ or 
with $r=0$ are indistinguishable.
\label{fig:JVtensorprimpowers}}
\end{figure}}

%\subsection{Adiabatic fiducial data with $r_{0.05}=10^{-3}$}
\subsection{Adiabatic fiducial data plus a tensor contribution}
\label{sec:isoTensorTensorless}

Our final test of isocurvature is with pure adiabatic fiducial data plus a tensor contribution with $r_{0.05}=10^{-3}$ 
and $n_\mathrm{t} = -r_{0.05}/8$. We fit to these data the full three-parameter generally correlated 
isocurvature model, and a free $\tilde{r}$ (with $n_\mathrm{t} = -\tilde{r}/8$). LiteBIRD with TT, TE, EE, BB 
and CORE-M5 with TT, TE, EE, BB, PP, and delensing constrains the isocurvature power equally well to near 
zero (see Fig.~\ref{fig:JVtensorprimpowers}). However CORE-M5 constrains the correlation power both 
at large scales and small scales (last two panels of the Figure) about two times closer to zero than 
LiteBIRD. Figure \ref{fig:JVtensorprimpowers} shows also the results of fitting a wrong model (i.e, 
the $r=0$ model) to these data that in reality have $r_{0.05}=10^{-3}$. Overall, this does not 
significantly affect the determination of isocurvature parameters. In particular, CORE-M5 does not 
misidentify the tensor contribution as an additional isocurvature or correlation contribution, but 
with LiteBIRD the correlation power becomes slightly more biased toward negative values.

The input value of $r_{0.05}=10^{-3}$ is recovered with CORE-M5 much better than with LiteBIRD, see also Section 4.
%detected by CORE-M5 at $5.4\sigma$ and by LiteBIRD at $3.2\sigma$. 
%When fitting the CDI model, the mean values and $1\sigma$ uncertainties are 
%$r_{0.05}=(1.03\pm0.19)\times10^{-3}$ with CORE-M5, and $(1.12\pm0.35)\times10^{-3}$ with LiteBIRD. 
%When fitting the pure adiabatic model, these change 
%These results can be compared to those obtained purely adiabatic case, i.e. 
%$r_{0.05}=(1.01\pm0.19)\times10^{-3}$ with CORE-M5, and 
%$r_{0.05}=(1.09\pm0.34)\times10^{-3}$ with LiteBIRD. 
When fitting the CDI model, a slightly larger $r$ is 
favored than by fitting the pure adiabatic model. This can be understood by looking at the bottom middle panel of 
Fig.~\ref{fig:JVtensorprimpowers}. The negative correlation power at large scales, which is more 
pronounced than the positive correlation power, leaves more room for a positive tensor contribution 
to the temperature angular power spectrum and hence allows for a larger $r$ to fit the data. 
Likewise, to make more room for the positive TT contribution from $r$, $\beta_\mathrm{iso}$ 
at large scales becomes smaller (with LiteBIRD) and $\cos\Delta$ more
negative, as reported in Table \ref{tab:JVbetaisoAndAlphanonadi} in Section \ref{sec:JVfirstiso}.
%Compare the lines 
%where LiteBIRD and CORE-M5 are in parenthesis (fiducial $r_{0.05}=10^{-3}$) to the lines without 
%parenthesis (fiducial $r=0$) in Table \ref{tab:JVbetaisoAndAlphanonadi} in Section 
%\ref{sec:JVfirstiso}. Finally, we note from Table \ref{tab:JVadiparams} that the standard 
%parameters are recovered with LiteBIRD much better and with CORE-M5 slightly better than without $r$ 
%in the fiducial model, which may actually be explained by the fact that here we also used the BB 
%(and for CORE-M5 delensing) information, unlike in Section \ref{sec:JVfirstiso}, where TT, TE, EE, (PP) 
%were used.

%\subsection{Summary of probing the adiabaticity of the primordial perturbations}
\subsection{Summary}

Depending on  the details of the model and the measure used, \emph{CORE-M5 determines the nature} (the pure 
adiabatic mode or possibly a correlated mixture of adiabatic and isocurvature modes) \emph{of 
primordial perturbations 2--5 times better than \Planck.} LiteBIRD, on the other hand, cannot improve 
much on \Planck\ because of its coarse angular resolution. Downgrading CORE-M5 to 
LiteCORE-120 does not affect the constraining power what comes to isocurvature. Downgrading to 
LiteCORE-80 shows insignificant weakening of constraints. On the other hand, upgrading to COrE+ 
does not improve the results, since already CORE-M5 is virtually cosmic variance limited what comes to 
probing the isocurvature perturbations. In cases where both the primordial tensor contribution and 
an isocurvature mode are present, CORE-M5 distinguishes between these better than LiteBIRD.

\section{Primordial non-Gaussianity}
\label{sec:NG}

This Section investigates the implications of the different configurations
proposed for the new CORE satellite for studying primordial non-Gaussianity
(NG). In most cases, we also compare the forecasts to the current state-of-the-art 
{\it Planck} results \cite{Ade:2015ava} as well as to forecasts for the proposed
LiteBIRD satellite and a hypothetical ideal noiseless experiment. 

Results are presented as predicted $1\sigma$ error bars 
of the so-called $f_\mathrm{NL}$ and $g_{\rm NL}$ parameters, 
measuring the amplitudes of various theoretical bispectrum and 
trispectrum templates. The bispectrum and connected trispectrum are the 
Fourier or spherical harmonic transform of the three-point and four-point 
correlation function, respectively, and would be zero in the purely Gaussian 
case. Bispectrum (trispectrum) templates arising from different inflationary scenarios 
display a specific functional dependence on 
 triangular (quadrilateral) configurations in Fourier space. 
Such a dependence defines the bispectrum {\em shape}.
Unless otherwise stated, all bispectrum and trispectrum shapes analyzed here are 
considered independent.

So far the best constraints on primordial NG have been obtained from direct measurements of the 
angular bispectrum and trispectrum of CMB temperature and polarization fluctuations. However, new 
promising techniques have recently been proposed. Such techniques rely on measurements of either 
CMB spectral distortion anisotropies---tiny departures of the CMB energy spectrum from that of a 
blackbody across the sky---or the cosmic infrared background (CIB), and they can lead, for 
specific shapes and using future surveys, to better constraints than those achievable through 
standard bispectrum and trispectrum measurements. We will analyze bispectrum and trispectrum 
constraints in Section~\ref{sec:CMBbisp}, and consider these new approaches in 
Section~\ref{sec:CIB_dist}. We conclude in Section~\ref{sec:conclbisp}.

\subsection{CMB temperature and polarization bispectrum and trispectrum}
\label{sec:CMBbisp}
\subsubsection{Standard bispectrum shapes}

We start by analyzing the so-called standard bispectrum shapes: local, 
equilateral, and orthogonal (sometimes collectively called LEO), as well 
as the foreground lensing-ISW bispectrum 
due to correlations between the integrated Sachs-Wolfe (ISW) effect and 
gravitational lensing. More information about these shapes can be found in 
\cite{Ade:2015ava,binned2} and references therein. 
Our $f_{\rm NL}$ forecasts are obtained via a standard bispectrum Fisher matrix analysis. 
The signal-to-noise ratio is given by 
(see e.g., \cite{2001PhRvD..63f3002K, 2010PhRvD..82b3502F} for details)
\begin{equation}
\left(\frac{S}{N}\right)^2 = \frac{1}{6} \sum_{\ell_1 \ell_2 \ell_3} 
\frac{B_{\ell_1 \ell_2 \ell_3}^2}  {C_{\ell_1} C_{\ell_2} C_{\ell_3}} 
\end{equation}
where $B_{\ell_1 \ell_2 \ell_3}$ is the CMB angular averaged bispectrum, computed for a given shape 
assuming $f_{\rm NL}=1$. Both the bispectrum and power spectrum include beam and noise, computed 
by coadding cosmological channels for the different configurations considered
(LiteCORE, COrE$+$, CORE-M$5$, 
together with {\it Planck} and LiteBIRD for comparison).
Details of these configurations have been presented in Section~\ref{sec:three}. 
The forecasts are shown in Table~\ref{table_NGleo}. We remind the
reader that the CORE configurations assume $\ell_\mathrm{max}=3000$
(except the LiteCORE-80 configuration that has $\ell_\mathrm{max}=2400$),
while the much lower resolution LiteBIRD has $\ell_\mathrm{max}=1350$.
A sky masked to leave 70\% available has been assumed for all results.
%, as well
%as maps pixelized with a Healpix resolution parameter 
%$n_\mathrm{side}=2048$ (except for LiteBIRD where $n_\mathrm{side}=1024$ was
%assumed). 
For comparison we also present the error bars
of the {\it Planck} 2015 release. 
We stress that these are real error
bars, not forecasts. Moreover, as far as polarization is concerned they 
are not yet the ultimate {\it Planck} error bars, since the 2015 polarization
analysis was restricted to $\ell_\mathrm{min}=40$, which has an impact
in particular on the local shape. The {\it Planck} analysis used 
$\ell_\mathrm{max}=2500$ for temperature and $2000$ for $E$-polarization.
Finally we also show for comparison the forecasts for an ideal full-sky
and noiseless experiment with $\ell_\mathrm{max}=3000$.

Compared to {\it Planck} we see that the CORE configurations provide just a modest
improvement in temperature-only (as expected since {\it Planck} already is nearly cosmic 
variance limited in temperature), but a very significant improvement in
polarization-only, for a final improvement in full $T+E$ of about a factor of 2.
There is very little difference between the different CORE configurations.
Because of its lower resolution, LiteBIRD performs significantly worse than
CORE, with final $T+E$ error bars comparable to the current ones from {\it Planck}
and hence about a factor of 2 worse than CORE. We also see that CORE
performs within 50\% of the ultimate error bars of an ideal noiseless 
full-sky experiment.
It is also very interesting to note that CORE should provide a more than 
20$\sigma$ detection of the predicted $f_\mathrm{NL}=1$ lensing-ISW bispectrum,
which was for the first time observed by {\it Planck} but at only $3\sigma$ 
significance. At this level of detection, it might be interesting to start 
using the ISW-lensing signal to estimate cosmological parameters.

\begin{table}[t!]
\begin{tabular}{l|ccccccc}
& LiteCORE & LiteCORE & {\bf CORE} & COrE+ & Planck & LiteBIRD & ideal\\
& 80 & 120 & {\bf M5} && 2015 && 3000\\
\hline
T local & 4.5 & 3.7 & {\bf 3.6} & 3.4 & (5.7) & 9.4 & 2.7\\
T equilat & 65 & 59 & {\bf 58} & 56 & (70) & 92 & 46\\
T orthog & 31 & 27 & {\bf 26} & 25 & (33) & 58 & 20\\
T lens-isw & 0.15 & 0.11 & {\bf 0.10} & 0.09 & (0.28) & 0.44 & 0.07\\
\hline
E local & 5.4 & 4.5 & {\bf 4.2} & 3.9 & (32) & 11 & 2.4\\
E equilat & 51 & 46 & {\bf 45} & 43 & (141) & 76 & 31\\
E orthog & 24 & 21 & {\bf 20} & 19 & (72) & 42 & 13\\
E lens-isw & 0.37 & 0.29 & {\bf 0.27} & 0.24 &  & 1.1 & 0.14\\
\hline
T+E local & 2.7 & 2.2 & {\bf 2.1} & 1.9 & (5.0) & 5.6 & 1.4\\
T+E equilat & 25 & 22 & {\bf 21} & 20 & (43) & 40 & 15\\
T+E orthog & 12 & 10.0 & {\bf 9.6} & 9.1 & (21) & 23 & 6.7\\
T+E lens-isw & 0.062 & 0.048 & {\bf 0.045} & 0.041 &  & 0.18 & 0.027\\
\end{tabular}
\caption{Forecasts for the $1\sigma$ $f_\mathrm{NL}$ error bars for the standard
primordial shapes as well as for the lensing-ISW shape for the indicated 
configurations. Results are given for $T$-only, $E$-only and full $T+E$. The
results for {\it Planck} have been put in parentheses because they are not
forecasts but real measured error bars. See the main text for further details.}
\label{table_NGleo}
\end{table}

\begin{figure}[tbh]
\subfloat{\includegraphics[width=0.45\textwidth]{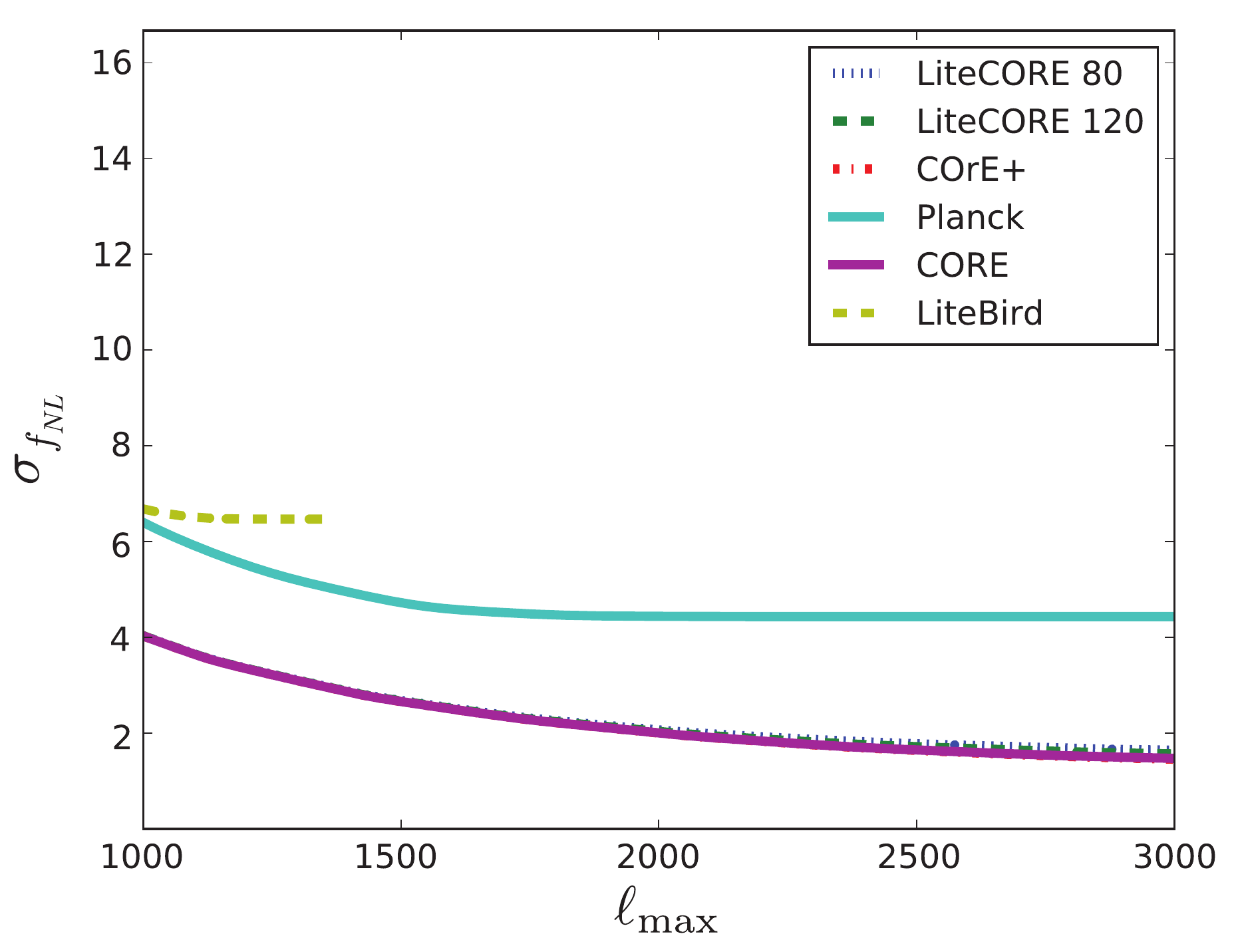}} 
\subfloat{\includegraphics[width=0.45\textwidth]{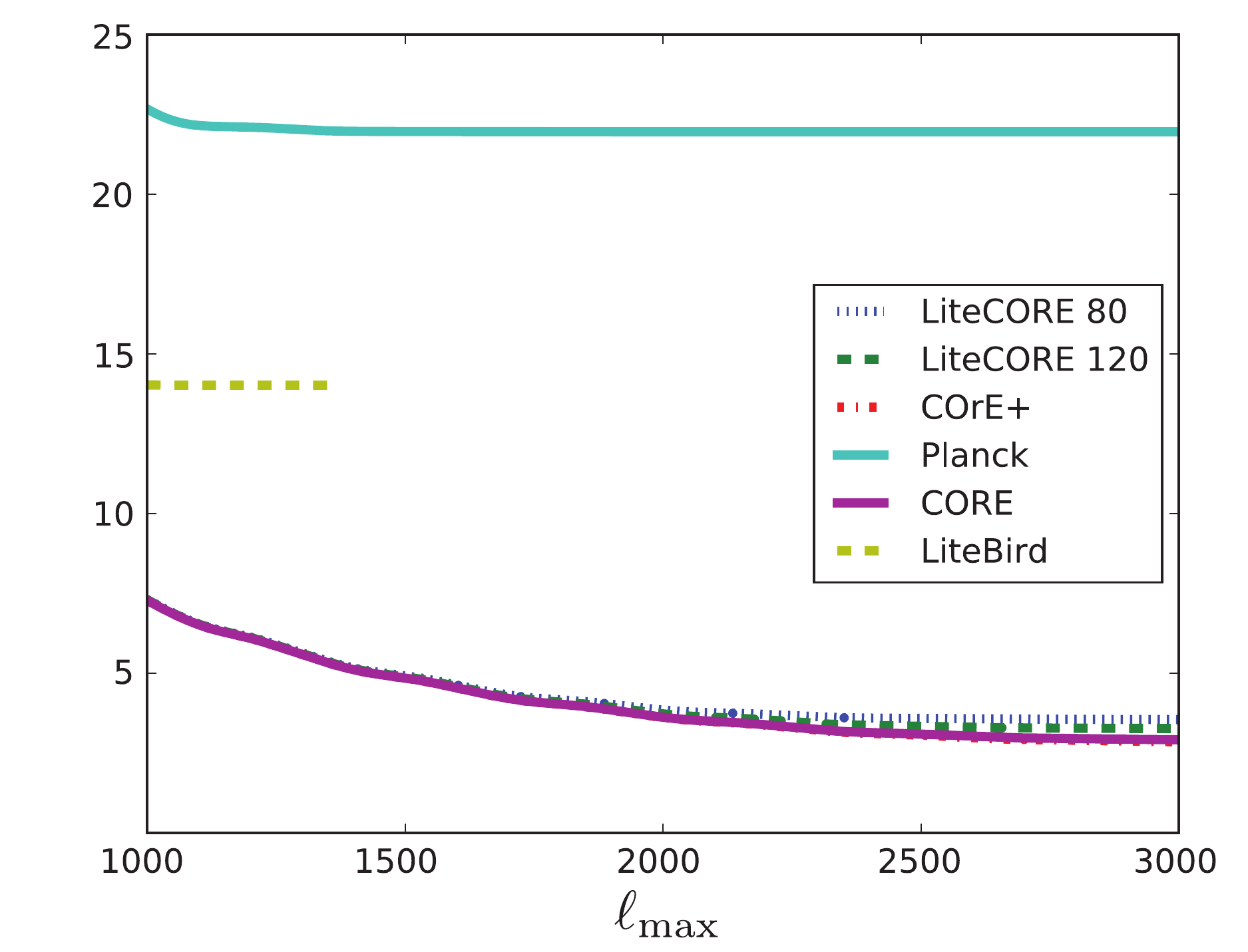}} \\
\subfloat{\includegraphics[width=0.45\textwidth]{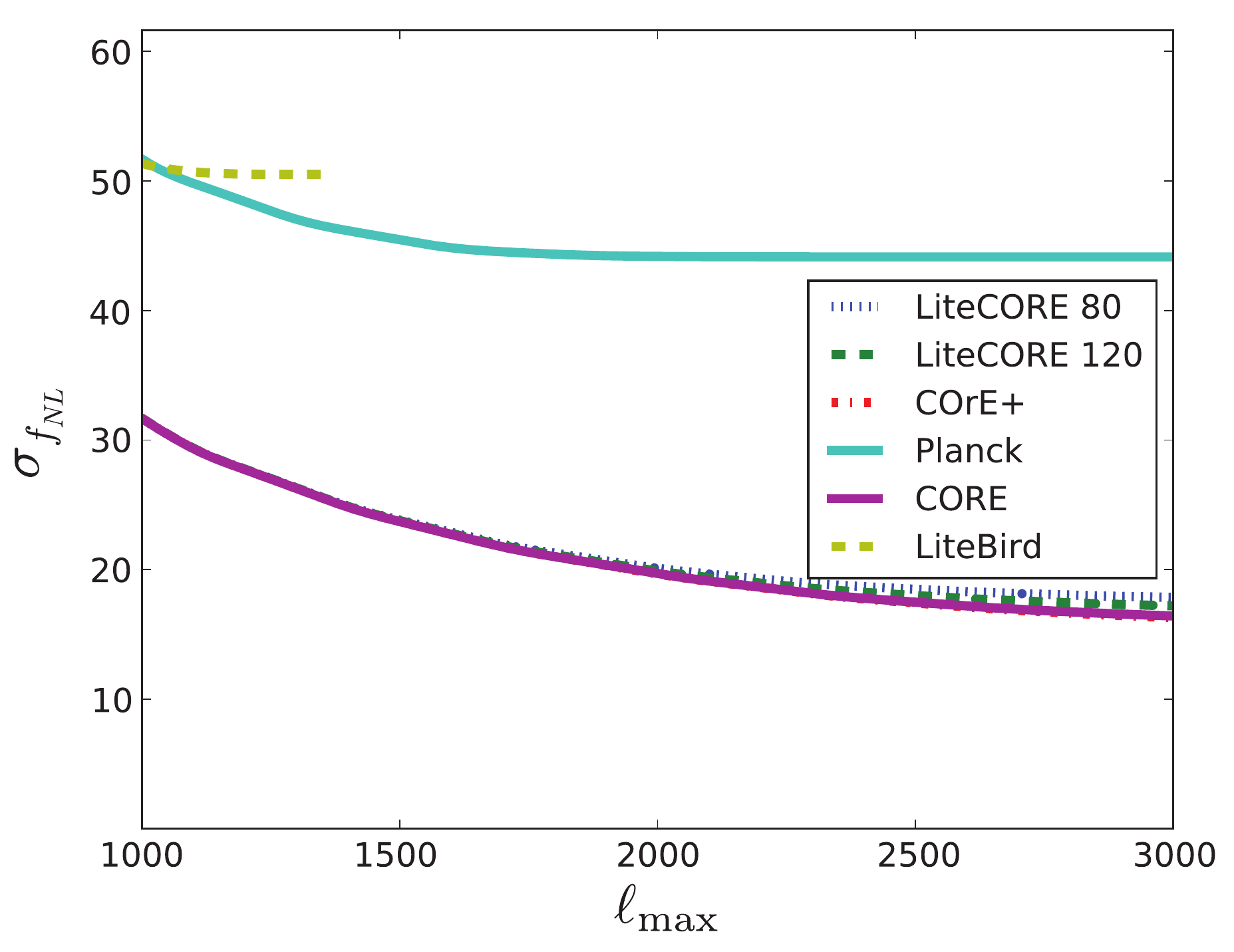}}
\subfloat{\includegraphics[width=0.45\textwidth]{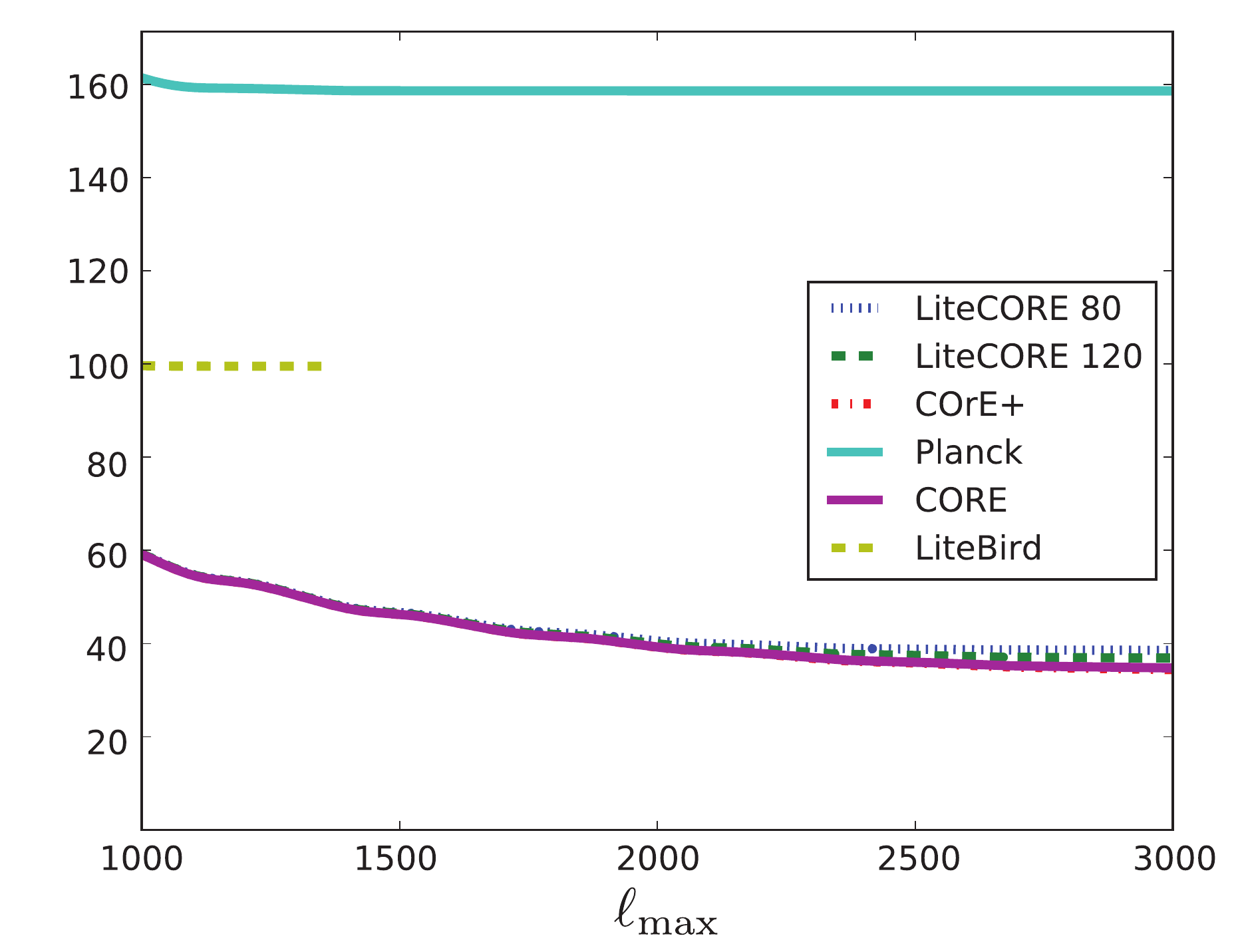}} \\
\subfloat{\includegraphics[width=0.45\textwidth]{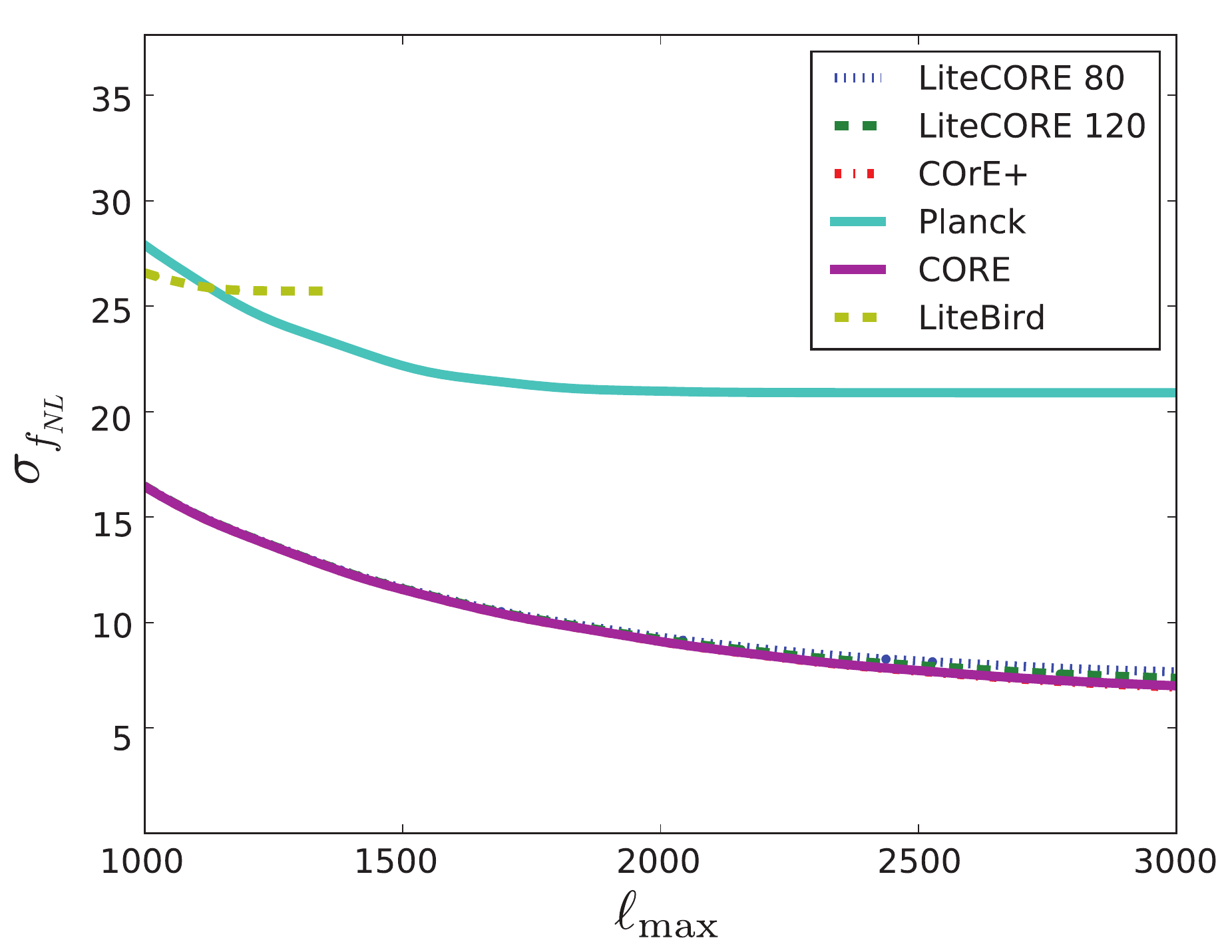}}
\subfloat{\includegraphics[width=0.45\textwidth]{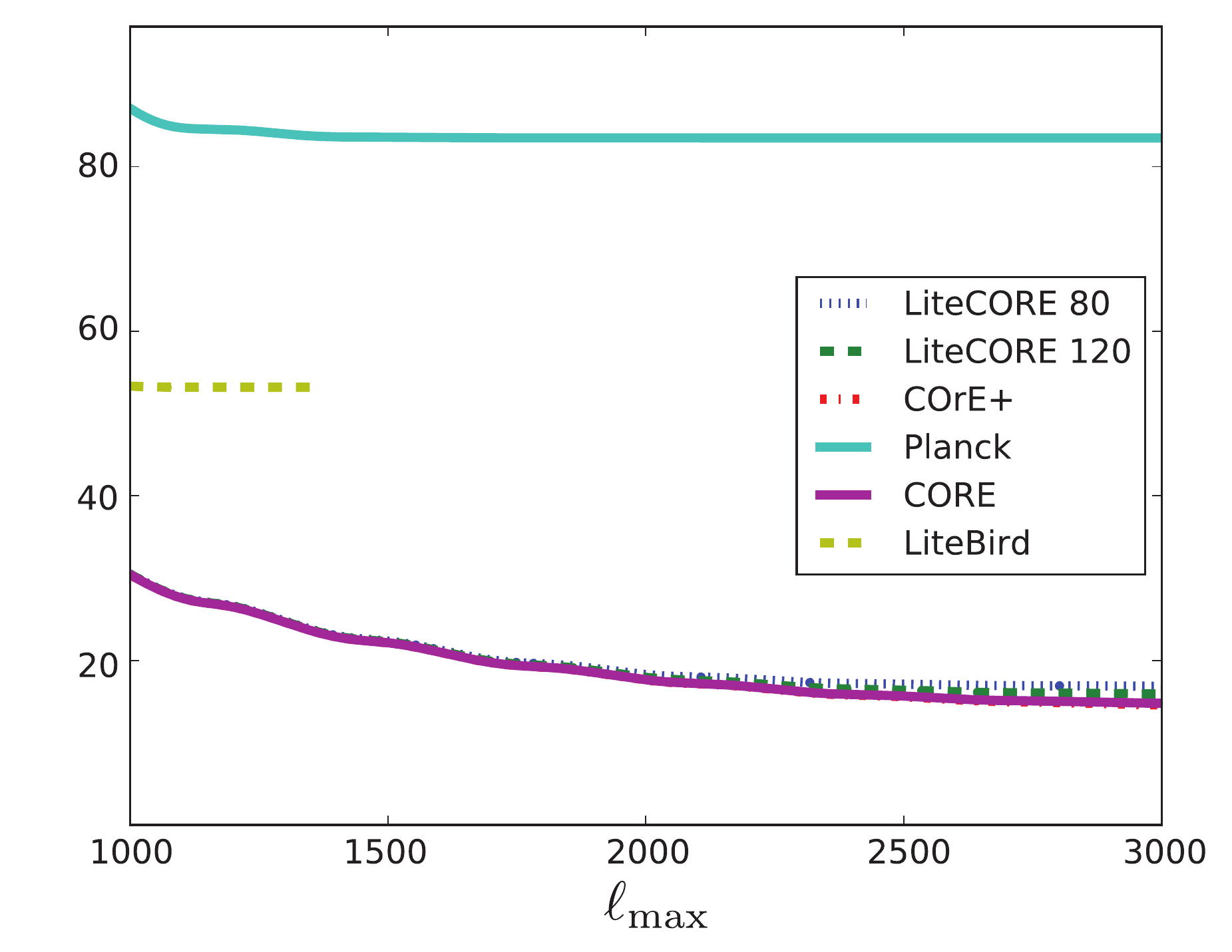}}
\caption{Expected local (first row), equilateral (second row), orthogonal (third row) $f_{\rm NL}$ error bars, obtained 
by combining temperature and polarization, on the left side, and using only polarization data, on the right side, 
for the different CORE configurations. The forecasts are compared to current constraints from {\it Planck}.}
\label{fig:forecasts_leo}
\end{figure}

A graphical comparison between forecasts for different experiments and
configurations is shown in Fig.~\ref{fig:forecasts_leo}, where we consider the $E$-only 
and $T+E$ case.  As a figure of merit, in order to summarize the
overall improvement achievable with a CORE-like experiment for the
three standard shapes, we also computed the overall constrained volume
in the three-dimensional local-equilateral-orthogonal bispectrum space
(accounting for correlations between them and using the shape
correlator to define a scalar product between bispectra). We find
that going from {\it Planck} to CORE shrinks the volume of the allowed
$f_{\rm NL}$ parameter region by a factor $\approx 20$, using $T+E$, with
little difference between the LiteCORE and COrE$+$
configurations. The improvement reaches a level of $\approx 200$ if we
consider polarization data only. Besides the improvements in error bars, 
we stress that having $EEE$ measurements at the same
level of sensitivity as $TTT$ is important because it allows a much
tighter control of systematics and foreground contamination, via
internal cross-validation of $T$-only and $E$-only results.

\begin{figure}[t!]
\centering
\includegraphics[width=0.5\textwidth]{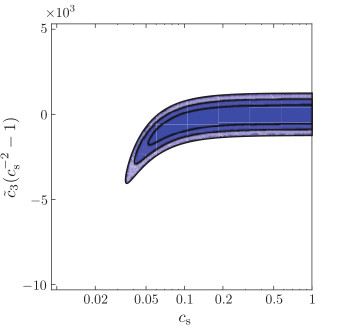}
\caption{
CORE-M5 forecasts 
of typical parameters of the effective field theory of inflation for general single-field models as obtained from the equilateral and orthogonal $\fnl$ predictions ($68\%$, $95\%$ and $99.7\%$ confidence regions are shown). 
Vanishing central values for $f_{\rm NL}^{\rm equil}$ and $f_{\rm NL}^{\rm ortho}$ have been assumed.
There are two ``microscopic'' non-Gaussianity parameters of interest. $c_{\rm s}$ is the sound speed of the inflaton field, with $(1-c^{-2}_{\rm s})$ being the amplitude of the inflaton self-interaction $\dot{\pi} (\nabla \pi)^2$, while $\tilde{c}_3$ is related to the amplitude of the inflaton self-interaction $(\dot{\pi})^3$.}
\label{fig:cs}
\end{figure}

Even in the absence of a detection, tight bounds on $\fnl$ parameters are of course very 
important to constrain parameters in different inflationary scenarios. As an example of this, in 
Fig.~\ref{fig:cs} we forecast constraints on the inflaton speed of sound in the effective field 
theory of single-field inflation, derived from our equilateral and orthogonal $\fnl$ predictions. 
In this case (see~\cite{Cheung:2007st} and also~\cite{Chen:2006xjb}) a lower bound $c_{\rm s} > 
0.045$ ($95\%$ CL) can be achieved, improving by almost $50\%$ the present $\it Planck$ 
constraints.

\subsubsection{Isocurvature non-Gaussianity}

In addition to these standard shapes, it is also interesting to investigate other shapes. One 
class of shapes where one would expect a significant improvement compared to {\it Planck} is 
isocurvature non-Gaussianity. If there was more than one degree of freedom during inflation, it 
is possible for one or more isocurvature modes to have survived in addition to the standard 
adiabatic mode. Such an additional mode will not only potentially produce a signal in the power 
spectrum (see Section~\ref{sec:testing}), but also in the bispectrum. It should be noted that some 
inflation-curvaton models (e.g.~\cite{Langlois:2011zz}) predict an even larger isocurvature than 
adiabatic bispectrum and at the same time a negligible isocurvature power spectrum.

\begin{table}[t!]
\begin{tabular}{l|ccccc|ccccc}
& \multicolumn{5}{c|}{independent analysis} & \multicolumn{5}{c}{joint analysis}\\
& LC120 & {\bf C} & C+ & Planck & ideal & LC120 & {\bf C} & C+ & Planck & ideal\\
\hline
T a,aa   & 3.7 & \bf{3.6} & 3.4 & (5.4) & 2.7  & 9.7 & \bf{9.4} & 8.8 & (13) & 7.1\\    
T a,ai   & 7.1 & \bf{6.9} & 6.4 & (10) & 5.2   & 19 & \bf{18} & 17 & (26) & 14\\      
T a,ii   & 990 & \bf{990} & 990 & (910) & 830  & 7800 & \bf{7800} & 7800 & (8200) & 6500\\
T i,aa   & 54 & \bf{54} & 54 & (50) & 45       & 120 & \bf{120} & 120 & (120) & 100\\   
T i,ai   & 70 & \bf{70} & 70 & (66) & 58       & 950 & \bf{950} & 940 & (1000) & 790\\  
T i,ii   & 290 & \bf{290} & 290 & (280) & 240  & 1900 & \bf{1800} & 1800 & (2000) & 1500\\
\hline                                                
E a,aa   & 4.5 & \bf{4.2} & 3.9 & (34) & 2.4    & 7.9 & \bf{7.3} & 6.7 & (50) & 4.1\\
E a,ai   & 16 & \bf{15} & 13 & (200) & 8.2      & 27 & \bf{26} & 23 & (310) & 14\\    
E a,ii   & 610 & \bf{610} & 610 & (4000) & 500  & 1100 & \bf{1100} & 1100 & (6100) & 930\\
E i,aa   & 20 & \bf{20} & 20 & (87) & 17        & 39 & \bf{38} & 38 & (180) & 31\\     
E i,ai   & 41 & \bf{41} & 41 & (250) & 34       & 130 & \bf{130} & 130 & (770) & 100\\   
E i,ii   & 220 & \bf{210} & 210 & (2200) & 180  & 440 & \bf{440} & 440 & (5300) & 360\\  
\hline                                                
T+E a,aa & 2.2 & \bf{2.1} & 1.9 & (4.9) & 1.4   & 3.3 & \bf{3.1} & 2.9 & (10) & 2.0\\
T+E a,ai & 5.3 & \bf{5.1} & 4.7 & (9.7) & 3.3   & 8.0 & \bf{7.6} & 7.0 & (20) & 4.9\\   
T+E a,ii & 200 & \bf{200} & 200 & (450) & 170   & 480 & \bf{470} & 470 & (1300) & 390\\ 
T+E i,aa & 10 & \bf{10} & 10 & (26) & 8.6       & 19 & \bf{19} & 19 & (47) & 16\\     
T+E i,ai & 16 & \bf{16} & 16 & (38) & 13        & 54 & \bf{54} & 53 & (170) & 44\\    
T+E i,ii & 76 & \bf{76} & 76 & (170) & 63       & 140 & \bf{140} & 140 & (390) & 110\\  
\end{tabular}
\caption{Forecasts for the $1\sigma$ local $f_\mathrm{NL}$ error bars 
with 
both an adiabatic and a CDM density isocurvature mode
for the indicated configurations and showing both the independent and joint 
analysis. Results are given for $T$-only, $E$-only and full $T+E$. The
results for {\it Planck} have been put in parentheses because they are not
forecasts but real measured error bars. See the main text for further details.}
\label{table_NGiso_CDI}
\end{table}

\begin{table}[h!]
\begin{tabular}{l|ccccc|ccccc}
& \multicolumn{5}{c|}{independent analysis} & \multicolumn{5}{c}{joint analysis}\\
& LC120 & {\bf C} & C+ & Planck & ideal & LC120 & {\bf C} & C+ & Planck & ideal\\
\hline
T a,aa   & 3.7 & \bf{3.6} & 3.4 & (5.4) & 2.7   & 41 & \bf{40} & 38 & (52) & 31\\       
T a,ai   & 10.4 & \bf{10.1} & 9.5 & (15) & 7.6  & 160 & \bf{160} & 160 & (210) & 130\\    
T a,ii   & 190 & \bf{190} & 170 & (280) & 140   & 3700 & \bf{3600} & 3500 & (4500) & 2900\\ 
T i,aa   & 6.1 & \bf{6.0} & 5.6 & (9.0) & 4.5   & 84 & \bf{83} & 81 & (99) & 66\\       
T i,ai   & 15 & \bf{15} & 14 & (22) & 11        & 540 & \bf{530} & 520 & (630) & 430\\    
T i,ii   & 170 & \bf{170} & 160 & (250) & 130   & 2200 & \bf{2100} & 2100 & (2400) & 1700\\ 
\hline                                               
E a,aa   & 4.5 & \bf{4.2} & 3.9 & (34) & 2.4   & 22 & \bf{21} & 20 & (120) & 14\\      
E a,ai   & 17 & \bf{16} & 15 & (140) & 8.9     & 107 & \bf{102} & 96 & (640) & 68\\     
E a,ii   & 300 & \bf{290} & 270 & (2300) & 170 & 1300 & \bf{1200} & 1200 & (6200) & 900\\ 
E i,aa   & 6.7 & \bf{6.3} & 5.9 & (42) & 3.7   & 35 & \bf{34} & 33 & (170) & 24\\      
E i,ai   & 19 & \bf{18} & 17 & (130) & 11      & 180 & \bf{170} & 170 & (850) & 130\\ 
E i,ii   & 200 & \bf{190} & 180 & (1400) & 130 & 910 & \bf{880} & 860 & (5300) & 650\\
\hline                                               
T+E a,aa & 2.2 & \bf{2.1} & 1.9 & (4.9) & 1.4  & 8.9 & \bf{8.6} & 8.2 & (27) & 6.1\\     
T+E a,ai & 7.0 & \bf{6.6} & 6.2 & (14) & 4.4   & 35 & \bf{34} & 33 & (94) & 24\\
T+E a,ii & 120 & \bf{120} & 110 & (240) & 77   & 550 & \bf{540} & 520 & (1400) & 410\\  
T+E i,aa & 3.4 & \bf{3.2} & 3.0 & (7.7) & 2.2  & 15 & \bf{15} & 14 & (45) & 11\\
T+E i,ai & 8.7 & \bf{8.4} & 7.9 & (19) & 5.7   & 70 & \bf{69} & 67 & (210) & 53\\ 
T+E i,ii & 82 & \bf{79} & 76 & (180) & 57      & 300 & \bf{290} & 290 & (860) & 230\\
\end{tabular}
\caption{Same as Table~\ref{table_NGiso_CDI}, but for the neutrino density
isocurvature mode.}
\label{table_NGiso_NDI}
\end{table}

As explained in \cite{LvT1, LvT2}, in the case of a local bispectrum produced by two modes, one 
adiabatic and one isocurvature, there will be six different $f_\mathrm{NL}$ parameters: one 
purely adiabatic ($a,aa$; which is the normal adiabatic local shape), one purely isocurvature 
($i,ii$) and four mixed ($a,ai$; $a,ii$; $i,aa$; $i,ai$). Assuming a cosmology with a standard 
particle content, the isocurvature mode can be cold dark matter (CDM) density isocurvature, 
neutrino density isocurvature, or neutrino velocity isocurvature.\footnote{There could also be a 
baryon density isocurvature mode, but its shape is identical to the CDM density isocurvature 
mode, just with a rescaled amplitude.} As explained in \cite{LvT1, LvT2}, several of the 
isocurvature NG modes profit significantly from the addition of polarization data to the 
analysis, much more than for the adiabatic mode. Already for {\it Planck} the improvement due to 
including polarization was up to a factor of six.

The forecasts for the isocurvature NG modes are given in Tables~\ref{table_NGiso_CDI}, 
\ref{table_NGiso_NDI}, and \ref{table_NGiso_NVI}, for the CDM density, neutrino 
density, and neutrino velocity isocurvature modes, respectively. We present both the case where all six modes 
are considered independent and the more realistic case when they are analyzed in a fully joint 
way.\footnote{The independent $a,aa$ result in these Tables corresponds exactly with the local 
result in Table~\ref{table_NGleo}. The small difference in the {\it Planck} column is due to a 
difference between estimators: the {\it Planck} errors cited in Table~\ref{table_NGleo} are those 
computed by the KSW estimator, while the errors in the isocurvature Tables are those computed 
by the binned estimator.} Like for the standard shapes we see that the difference between the 
different CORE configurations is small. However, here the improvement compared to {\it 
Planck} in $T+E$ is much more impressive: for the joint analysis the improvement varies from a factor of 3 up 
to almost a factor of 10 for the neutrino velocity isocurvature mode. We also see the huge 
importance of polarization: for some shapes the error bars in the joint analysis decrease by a 
factor of almost 30 when going from $T$-only to $T+E$. Like for the standard shapes, we see that 
for all isocurvature modes CORE approaches to within 50\% the ultimate error bars of an ideal 
noiseless and full-sky experiment.

\begin{table}
\begin{tabular}{l|ccccc|ccccc}
& \multicolumn{5}{c|}{independent analysis} & \multicolumn{5}{c}{joint analysis}\\
& LC120 & {\bf C} & C+ & Planck & ideal & LC120 & {\bf C} & C+ & Planck & ideal\\
\hline
T a,aa   & 3.7 & \bf{3.6} & 3.4 & (5.4) & 2.7   & 40 & \bf{39} & 38 & (48) & 32\\     
T a,ai   & 20 & \bf{20} & 19 & (29) & 15        & 290 & \bf{280} & 280 & (350) & 230\\  
T a,ii   & 260 & \bf{250} & 240 & (360) & 190   & 3300 & \bf{3200} & 3200 & (3800) & 2600\\
T i,aa   & 3.3 & \bf{3.2} & 3.0 & (4.7) & 2.4   & 47 & \bf{46} & 45 & (51) & 38\\     
T i,ai   & 15 & \bf{15} & 14 & (21) & 11        & 150 & \bf{150} & 150 & (170) & 130\\  
T i,ii   & 170 & \bf{170} & 160 & (230) & 130   & 1300 & \bf{1300} & 1300 & (1400) & 1100\\
\hline                                                    
E a,aa   & 4.5 & \bf{4.2} & 3.9 & (34) & 2.4    & 28 & \bf{27} & 25 & (150) & 18\\    
E a,ai   & 6.9 & \bf{6.4} & 5.9 & (93) & 3.6    & 55 & \bf{53} & 51 & (620) & 37\\    
E a,ii   & 61 & \bf{58} & 53 & (940) & 34       & 530 & \bf{520} & 510 & (3900) & 390\\ 
E i,aa   & 3.7 & \bf{3.5} & 3.2 & (27) & 2.0    & 27 & \bf{26} & 25 & (120) & 19\\    
E i,ai   & 5.0 & \bf{4.8} & 4.4 & (62) & 2.8    & 74 & \bf{73} & 72 & (420) & 56\\    
E i,ii   & 30 & \bf{28} & 27 & (460) & 19       & 140 & \bf{140} & 140 & (1600) & 110\\ 
\hline                                                    
T+E a,aa & 2.2 & \bf{2.1} & 1.9 & (4.9) & 1.4   & 9.8 & \bf{9.5} & 9.2 & (24) & 7.0\\   
T+E a,ai & 3.6 & \bf{3.4} & 3.2 & (22) & 2.3    & 19 & \bf{19} & 18 & (130) & 14\\    
T+E a,ii & 31 & \bf{29} & 28 & (230) & 20       & 220 & \bf{220} & 220 & (1200) & 170\\ 
T+E i,aa & 1.8 & \bf{1.8} & 1.6 & (4.1) & 1.2   & 9.9 & \bf{9.7} & 9.5 & (24) & 7.4\\   
T+E i,ai & 2.8 & \bf{2.7} & 2.5 & (15) & 1.8    & 24 & \bf{24} & 23 & (74) & 19\\     
T+E i,ii & 16 & \bf{15} & 15 & (130) & 11       & 47 & \bf{46} & 46 & (430) & 37\\    
\end{tabular}
\caption{Same as Table~\ref{table_NGiso_CDI}, but for the neutrino velocity
isocurvature mode.}
\label{table_NGiso_NVI}
\end{table}

\subsubsection{Spectral index of the bispectrum}

\begin{figure}[t!]
\centering
\subfloat{\includegraphics[width=0.5\textwidth]{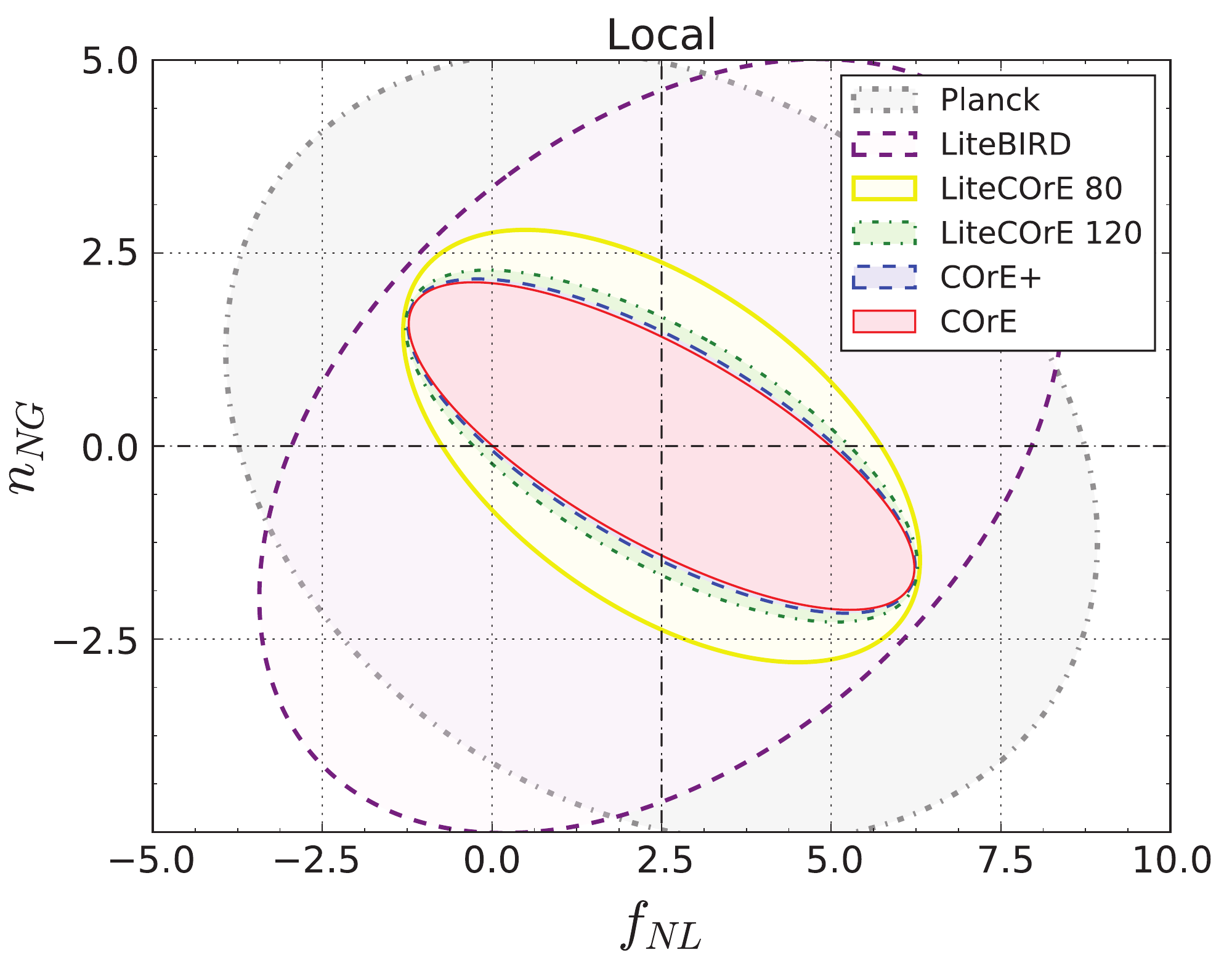}}
\subfloat{\includegraphics[width=0.5\textwidth]{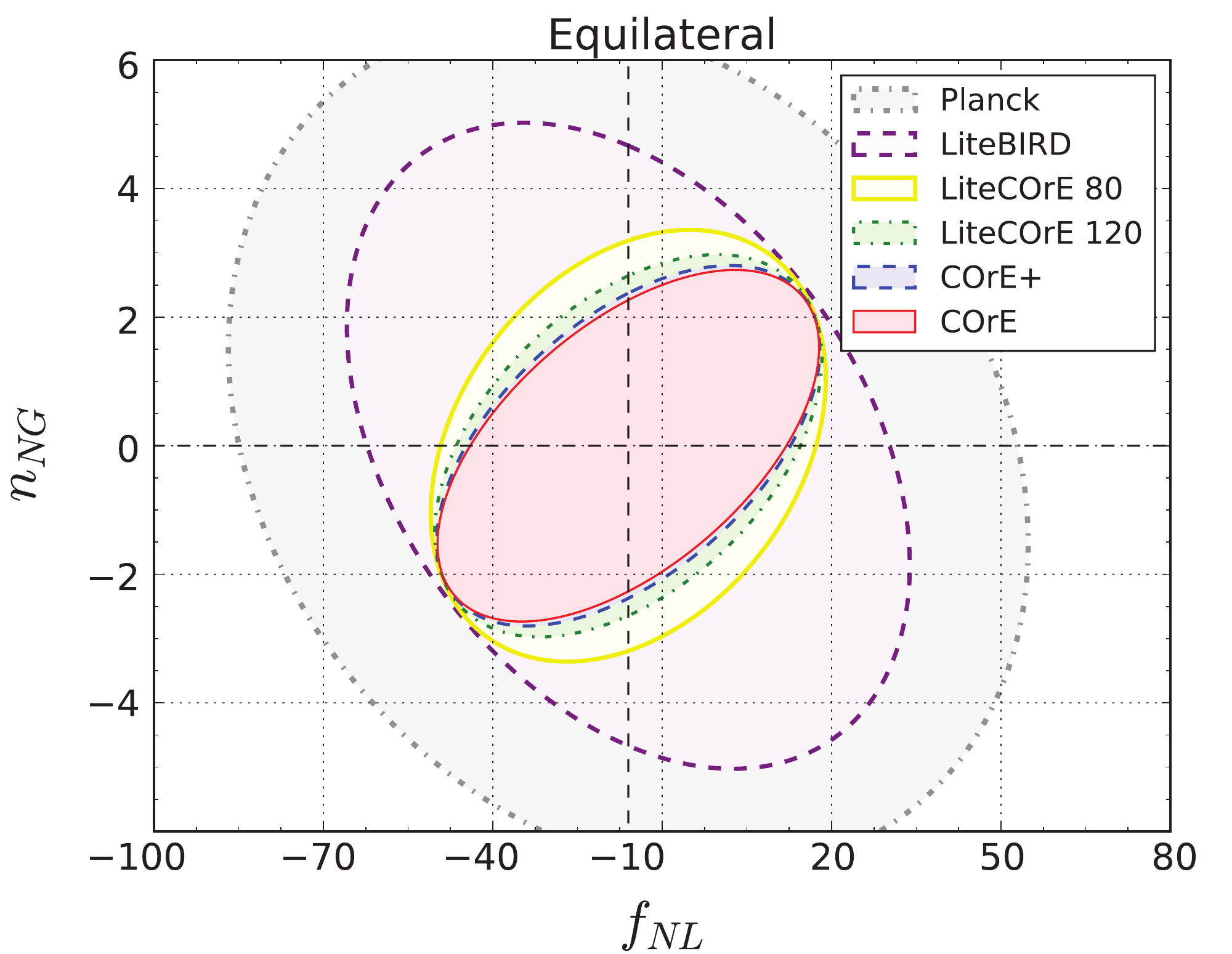}}
\caption{1-sigma contours in the $f_{\rm NL}$--$n_{\rm NG}$ plane, for local (left panel) and equilateral (right panel) scale-dependent bispectra. Pivot 
scale $k_P = 0.055 \, {\rm Mpc}^{-1}$.}\label{fig:scaledep}
\end{figure}

We also considered models with a mild running of the NG parameter $f_{\rm NL}$, parameterized by 
a spectral index $n_\mathrm{NG}$. Scale dependence of this type can arise, for example in single-field 
models with non-standard kinetic terms or in specific multi-field scenarios. The bispectrum in 
our forecast is written as a scale-independent part (typically, one of the standard local, 
equilateral, orthogonal shapes) multiplying a scale-dependent term, proportional to $(k_1 + k_2 + 
k_3)^{n_{\rm NG}}$, as in \cite{loverde/miller/etal:2008, 2008PhRvD..78l3534T}. Results in the 
$f_{\rm NL}$--$n_{\rm NG}$ plane from a full $T+E$ analysis are plotted in 
Fig.~\ref{fig:scaledep}, while marginalized error bars are reported in Table~\ref{tab:scaledep}. 
CORE allows an improvement of a factor of $> 3$ when constraining the NG spectral index, with 
respect to {\it Planck}. We note that these models are not constrained yet using {\em Planck} 
data; constraints for local scale-dependent NG exist from WMAP data \cite{2012PhRvL.109l1302B}. 
Therefore the {\em Planck} constraints in Table~\ref{tab:scaledep} also refer to a Fisher matrix 
analysis in this case.

\begin{table}[t!]
\centering
\begin{tabular}{l|cccccc}
& LiteCORE & LiteCORE & {\bf CORE} & COrE+ & Planck & LiteBIRD \\
& 80 & 120 & {\bf M5} &&& \\
\hline
$f_\mathrm{NL}$ local & 2.5 & 2.5 & {\bf 2.3}  & 2.3 & 4.3 & 3.9 \\
$f_\mathrm{NL}$ equil. & 23 & 22 & {\bf 21} & 21 & 47 & 33 \\
\hline
$n_{\rm NG}$ local & 1.9 & 1.5 & {\bf 1.4} &  1.4 & 3.4 & 3.3 \\
$n_{\rm NG}$ equil. & 2.2 & 2.0 & {\bf 1.8} & 1.8 & 4.4 & 3.9 \\
\end{tabular}
\caption{Forecasts for the marginalized $1\sigma$ $f_\mathrm{NL}$ and $n_{\rm NG}$ error bars in scale-dependent NG
models. We assumed $n_{\rm NG} = 0$ as fiducial value for the NG spectral index (for both shapes), corresponding to scale invariance.
For the amplitudes we took instead 
$f_{\rm NL}^{\rm fid, local} = 2.5$, $f_{\rm NL}^{\rm fid, equil.} = -16$, which represent the current {\it Planck} central values.}
\label{tab:scaledep}
\end{table}

\subsubsection{Oscillatory bispectra}

A strong scale-dependent running and sinusoidal oscillations in the CMB bispectrum can be 
produced in a variety of inflationary scenarios, generally characterized by temporary violations 
of the slow-roll conditions. These can arise in single-field inflation from the presence of 
sharp features in the inflaton potential or from changes in the inflaton sound speed. Large field 
models based on string theory, such as axion monodromy, can also produce this behavior, as well 
as multi-field scenarios with sharp turns in field space. See for example \cite{Ade:2015ava} and 
references therein for a more detailed survey and explanations. It is interesting to note that
in all the scenarios above, we not only have oscillatory bispectra, but also model-dependent 
oscillatory counterparts in the power spectrum and trispectrum. An observation of matching 
features in the power spectrum and bispectrum would of course enhance the statistical evidence 
in favor of specific models. Searches of this type have been conducted in 
\cite{2015PhRvD..91l3506F,2016PhRvD..93d3536M}.

We consider here two specific primordial oscillatory shapes. One of them associated with sharp 
features in the inflaton potential has the form 
\begin{equation}\label{eqn:feat}
B^\mathrm{feat}(k_1,k_2,k_3) = 6A f^\mathrm{feat}_{\rm NL} \cos \left[\omega (k_1 + k_2 + k_3) + \phi \right] \; ,
\end{equation}
where $A$ is the power spectrum amplitude, $\phi$ is a phase factor, and $\omega$ defines the 
oscillation frequency. Our forecasts for this shape are shown in Table~\ref{tab:feat}. Also in 
this case as for standard LEO shapes, the final $S/N$ improvement with respect to {\em Planck} 
amounts to a factor $\approx 2$, is essentially independent of the specific CORE configuration 
considered and is mostly driven by improved polarization sensitivity.

The other type of feature considered has the form described in Section~\ref{sssec:features} from 
a transient reduction in the speed of sound away from $c_s = 1$, with uninterrupted slow roll. 
This is part of a general class of effectively single-field models where the leading contribution 
to the bispectrum is given by the power spectrum feature and its first and second derivatives, 
with coefficients that are independent of scale, see Eq.~(\ref{DeltaBispectrum}).

The results in Table~\ref{tab:table_transient_cs} are for the simplest case of a reduction in 
$c_s$ given by a Gaussian in $e$-folds.\footnote{More precisely, a Gaussian in $u = 1 - c_s^{-2}$ 
away from zero, $u = - u_\mathrm{max} \exp(-\beta (N-N_0)^2) = -u_\mathrm{max} 
\exp(-\beta(\log(\eta/\eta_0))^2)$, where $\eta$ is the conformal time \cite{Achucarro:2013cva}. 
The rate of change of $c_s$ is given by $s \equiv \dot{c}_s/(c_s H)$.} This leads to enveloped 
oscillations in the bispectrum (correlated with similar features in the power spectrum and higher 
correlation functions). The oscillations are linear in $k$ with a frequency determined by the 
instant ($\eta_0$) of maximum reduction in $c_s$. An important difference with the previous 
oscillating shape is that here the phase depends on the $k$-triangle: oscillations in the 
equilateral shape are $\pi/2$ out of phase with respect to the ones in the squeezed limit. The 
oscillations are modulated and the envelope is largest away from $k = 0$ \cite{Torrado:2016sls}.

\begin{table}[t!]
\centering
\begin{tabular}{l|cccccc}
& LiteCORE & LiteCORE & {\bf CORE} & COrE+ & Planck  \\
& 80 & 120 & {\bf M5} & & \\
\hline
T, $\omega=50$ & 60 &  53 & {\bf 52} & 52 & 73\\ 
T, $\omega=100$ &  87 & 86 &  {\bf 85} & 85 & 102\\
\hline
T$+$E, $\omega=50$ & 21 & 21 & {\bf 20} & 20 & 42\\  
T$+$E, $\omega=100$ & 26 & 25 & {\bf 24} & 24 & 51\\
\end{tabular}       
\caption{
Forecasts for $1 \sigma$ error bars on the $\fnl$ parameter for the feature model of Eq.~(\ref{eqn:feat}). 
We consider two different oscillation frequencies and we take $0$ as fiducial 
value for $\phi$.}\label{tab:feat}
\end{table}

\begin{table}[t!]
\centering
%\scriptsize
\begin{tabular}{c|ccccc}
\multicolumn{6}{c}{\textbf{T}, $|s|_\mathrm{max} = 0.1$}\\[1mm]
$-\eta_0$ & Planck & LiteCORE-80 & LiteCORE-120 & {\bf CORE} & COrE+ \\
\hline\rule{0pt}{0.9\normalbaselineskip}
30 & \sci{2.5}{-6} & \sci{7.1}{-7} & \sci{3.6}{-7} & \scib{3.5}{-7} & \sci{3.5}{-7} \\
60 & \sci{1.7}{-5} & \sci{1.2}{-5} & \sci{8.8}{-6} & \scib{8.8}{-6} & \sci{8.8}{-6} \\
\multicolumn{6}{c}{ }\\[-2mm]
\multicolumn{6}{c}{\textbf{T+E}, $|s|_\mathrm{max} = 0.1$}\\[1mm]
$-\eta_0$ & Planck & LiteCORE-80 & LiteCORE-120 & {\bf CORE} & COrE+ \\
\hline\rule{0pt}{0.9\normalbaselineskip}
30 & \sci{2.0}{-6} & \sci{2.6}{-7} & \sci{1.6}{-7} & \scib{1.4}{-7} & \sci{1.4}{-7} \\
60 & \sci{1.2}{-5} & \sci{4.0}{-6} & \sci{3.6}{-6} & \scib{3.3}{-6} & \sci{3.3}{-6} \\
\multicolumn{6}{c}{ }\\[-2mm]
\multicolumn{6}{c}{\textbf{T}, $|s|_\mathrm{max} = 0.2$}\\[1mm]
$-\eta_0$ & Planck & LiteCORE-80 & LiteCORE-120 & {\bf CORE} & COrE+ \\
\hline\rule{0pt}{0.9\normalbaselineskip}
30 & \sci{4.4}{-7} & \sci{9.3}{-8} & \sci{2.5}{-8} & \scib{2.5}{-8} & \sci{2.5}{-8} \\
60 & \sci{8.0}{-7} & \sci{3.6}{-7} & \sci{2.4}{-7} & \scib{2.4}{-7} & \sci{2.4}{-7} \\
\multicolumn{6}{c}{ }\\[-2mm]
\multicolumn{6}{c}{\textbf{T+E}, $|s|_\mathrm{max} = 0.2$}\\[1mm]
$-\eta_0$ & Planck & LiteCORE-80 & LiteCORE-120 & {\bf CORE} & COrE+ \\
\hline\rule{0pt}{0.9\normalbaselineskip}
30 & \sci{3.9}{-7} & \sci{3.6}{-8} & \sci{1.3}{-8} & \scib{1.1}{-8} & \sci{1.1}{-8} \\
60 & \sci{6.4}{-7} & \sci{1.3}{-7} & \sci{1.2}{-7} & \scib{1.1}{-7} & \sci{1.1}{-7} \\
\multicolumn{6}{c}{}\\[-2mm]
\end{tabular}
\caption{\label{tab:table_transient_cs}
Forecasts for the 1$\sigma$ error bars of the intensity $|u|_\mathrm{max}$ of a Gaussian-shaped 
transient reduction in the speed of sound of the inflaton, for different combinations of the 
maximum change rate $|s|_\mathrm{max}$ and the instant of maximum reduction $\eta_0$, which is 
also the oscillation frequency of the feature, equal to $\omega$ in Eq.\ (\ref{eqn:feat}). 
Typical expected values for $|u|_\mathrm{max}$ are $\mathcal{O}(10^{-2}$--$10^{-1})$. In general, 
CORE's high signal-to-noise polarization data shrinks the error bars by a few units. For low 
values of $-\eta_0$ (e.g.\ $-\eta_0=30$), the feature peaks in intensity at high $\ell$; this 
makes the shrinkage of the error bars even more dramatic, since it is also driven by the 
increased effective $\ell_\mathrm{max}$ of CORE, especially for LiteCORE-120 and higher 
configurations.}
\end{table}

\subsubsection{Trispectrum}

So far we have discussed only the bispectrum parameters $f_{\rm NL}$, but 
interesting information is also contained in the angular trispectrum (4-point 
function in harmonic space; see \cite{Sekiguchi:2013hza,Ade:2015ava,2015arXiv150200635S} and 
references therein). 

To get an idea of future improvements in this case, we forecasted the expected 
performance of CORE to constrain the $g_{\rm NL}$ local trispectrum parameter
using T-only and E-only trispectra, as shown in Table~\ref{gnl_tab}.
If we consider temperature-only data ($TTTT$ trispectrum), we find 
that CORE can improve on current error bars from ${\it Planck}$ by a factor 
$\sim 3$ from the current $\sigma_{g_{\rm NL}} = 7.7 \times 10^{4}$ 
({\it {Planck}}) constraint \cite{Ade:2015ava} to 
$\sigma_{g_{\rm NL}} = 2.8 \times 10^{4}$ (CORE). However, this mostly comes from 
the fact that the error bar for  {\it {Planck}} refers to the actual data 
analysis, with $\ell_{\rm max} = 1600$ (imperfect knowledge of the noise model 
did not allow to go to higher multipoles with {\it Planck} so far for the 
trispectrum), while the CORE Fisher matrix forecast has $\ell_{\rm max} = 3000$.
If we consider the polarization-only $EEEE$ trispectrum, we see, in line with 
the $EEE$ bispectrum forecasts, that CORE will allow for massive improvements 
compared to {\it Planck}. The trispectrum $E$-only error bars shrink by a 
factor $\sim 20$, from $\sigma_{g_{\rm NL}} \sim 5 \times 10^{5}$ ({\it Planck} at 
$\ell_{\rm max} = 1600$; note that only the $TTTT$ trispectrum was actually 
analyzed with {\it Planck} data, so this is a forecast) to 
$\sigma_{g_{\rm NL}} \sim 2.5 \times 10^{4}$ (CORE, $\ell_{\rm max} = 3000$). 
We find that different CORE configurations (LiteCORE-120, CORE, COrE$+$) perform
nearly identically.

\begin{table}[t!]
  \centering
  \begin{tabular}{l|cccc}
    $\sigma_{\rm g_{NL}}\times10^4$ & LiteCORE-120 & {\bf CORE} & COrE+ & Planck \\
  \hline
    TTTT & $2.8$ & {\bf 2.8} & $2.8$ & (7.7) \\
    EEEE & $2.8$ & {\bf 2.7} & $2.6$ & 50 \\
  \end{tabular}
\caption{Forecast $g_\mathrm{NL}$ standard deviations for different experiments.}
  \label{gnl_tab}
\end{table}

\subsection{Other methods}\label{sec:CIB_dist}
 
Two interesting and potentially powerful new approaches have been proposed over the past few 
years to extract NG information: one using observations of the cosmic infrared background (CIB), 
and 
the other based on spectral distortions created through energy release in the early universe 
\cite{Zeldovich1969, Sunyaev1970mu, Burigana1991, Hu1993, Chluba2011therm}. Both these methods work 
only for a specific class of bispectrum (trispectrum) shapes, namely those peaking in the 
so-called squeezed limit (i.e., bispectrum or trispectrum configurations in which one wavenumber 
is much smaller than the others). This happens notably for the local shape, which is of 
particular interest since it allows one to discriminate between single-field and multiple-field 
inflation. On the other hand, a complete study of primordial NG must include a large number of 
shapes, the vast majority of which does not peak in the squeezed limit (e.g., equilateral and 
folded shapes, as well as many other cases characterized by breaking of scale-invariance or 
isotropy; see also previous Section). For those scenarios, a direct estimate of temperature and 
polarization angular bispectra (and trispectra) is the only way forward. 

The first method was recently considered in \cite{Tucci:2016aa}. It is based on exploiting 
primordial NG signatures (i.e.\ scale-dependent bias on very large scales) in the CIB power 
spectrum. In this case, the main obstacle is represented by dust contamination. However, dust 
contamination is also the main issue to overcome for the primordial $B$-mode analysis. Therefore 
many high-frequency channels are planned in future surveys, and the achievable level of 
foreground subtraction should make CIB-based local $\fnl$ measurements very promising, as 
originally pointed out in \cite{Tucci:2016aa}. This is shown here explicitly for {\em CORE, 
which, according to our forecasts, will be able to achieve $f_{\rm NL}^{\rm local} < 1$ 
sensitivity with this probe}.

The second method consists of extracting $f_{\rm NL}$ via measurements of correlations between 
CMB temperature and $\mu$-type spectral distortion\footnote{A $\mu$-type distortion is created 
by energy release at redshifts $5\times 10^4\lesssim z\lesssim 2\times10^6$, when Comptonization 
is very efficient \cite{Sunyaev1970mu, Burigana1991, Hu1993}.}  {\em anisotropies} generated by 
dissipation of primordial acoustic waves 
\cite{Pajer:2012vz,2013PhRvD..87f3521B,Ganc:2012ae,2015PhRvD..91l3531E,2015PhRvD..92h3502S,2016JCAP...03..029B}. 
Interestingly, such measurements do not require absolute calibration, as they do not make direct 
use of the $\mu$-distortion monopole. However, an interpretation of the data in terms of $f_{\rm 
NL}^{\rm local}$ relies on knowledge of the average dissipation-induced $\mu$-distortion, such 
that a combination with an absolute spectrometer (e.g., PIXIE) is necessary 
\citep{2016arXiv161008711C}.

\subsubsection{CIB power spectrum}

The angular power spectrum of the cosmic infrared background (CIB) is another sensitive probe of 
the {\em local} primordial bispectrum (and potentially of other bispectra peaking in the 
squeezed limit). CIB measurements are integrated over a large volume so that the scale-dependent 
bias from the primordial non-Gaussianity leaves a strong signal in the CIB power spectrum. 
Although galactic dust dominates over the non-Gaussian CIB signal, it is possible to mitigate the 
dust contamination with enough frequency channels, especially if high frequencies such as the 
\planck\ 857 GHz channel are available.

We adopt here a Fisher matrix approach to investigate the sensitivity of CORE and COrE+ to $\fnl$ 
through measurements of the CIB power spectrum. The Fisher matrix elements can be written as % 
\begin{equation} 
F_{ij}=\sum_{\ell=\lmin}^{\lmax}\,\frac{2\ell+1}{2}f_\text{sky}\, {\rm 
Tr}\Bigg({\bf C}_{\ell}^{-1}\frac{\partial{\bf C}_{\ell}}{\partial\theta_i} {\bf 
C}_{\ell}^{-1}\frac{\partial{\bf C}_{\ell}}{\partial\theta_j} \Bigg)\,, 
\label{eq:fisher} 
\end{equation} 
where the elements of the covariance matrix $\bf{C}_{\ell}$ are defined as the 
auto-- and cross--power spectra of data at $N_{\nu}$ different observational frequencies, 
computed on a fraction of the sky $f_\text{sky}$. Model parameters ${\bf \theta}$ include $\fnl$, 
the CIB model parameters (see \citep{planck_xxx_2013,Tucci:2016aa}) plus the parameters related 
to the Galactic dust emission. We refer the reader to \cite{Tucci:2016aa} for all technical 
details and for the full calculation of the CIB power spectrum in the presence of primordial NG.

The Galactic dust emission is the main contaminant for the detection of CIB anisotropies on large 
angular scales, where it is orders of magnitude brighter than the CIB. Distinguishing Galactic 
from extragalactic dust emission is especially difficult because of their fairly similar spectral 
energy distribution (SED) which approximately scales in both cases like a modified blackbody law. 
Extracting the CIB signal from Galactic contamination thus requires a very accurate component 
separation. Here we assume that CIB maps can be reconstructed by linearly combining the set of 
frequency maps (at $\nu>200$\,GHz and after subtracting CMB) on the basis of the ``mixing'' 
matrix ${\bf A}$ that describes the frequency dependence of the sky signal components 
\citep{Errard:2015cxa}. If observations are available at $n$ frequencies, we assume that CIB maps 
can be reconstructed only in $N_{\nu}\sim n/2$ frequencies, while the other channels are 
dedicated to the estimation of the mixing matrix and the Galactic dust template used in the 
subtraction. This can be seen as a general approach, independent of the specific component 
separation method employed. The auto-- and cross--power spectra in clean CIB maps then read as
\begin{equation}
C_{\ell}^{\nu_i\nu_j}=C_{\ell}^{(\CIB)}(\nu_i,\nu_j)+\varepsilon 
C_{\ell}^d(\nu_i,\nu_j)+\Sigma_{\CIB}^2(\nu_i)\delta_{ij}\,,
\label{eq:remov4}
\end{equation}
where $\Sigma_{\CIB}$ is the noise variance and $\varepsilon$ is the fraction of the total 
Galactic dust power spectrum left over in the reconstructed CIB maps (we use as reference value 
$\varepsilon=10^{-2}$). The noise variance in the maps will be degraded after the component 
separation according to the frequency spectrum of the sky signals and the noise in the channels 
involved in the foreground subtraction \citep{Tegmark:2000,Errard:2015cxa}. In our case, the 
noise degradation can be very severe due to the similar SED of the CIB and the dust emission, and 
frequencies much larger than 300~GHz are mandatory to accurately separate them. Regarding the 
residual dust emission, we model the power spectrum $C_{\ell}^d$ as a power law, whose amplitude 
and slope are free parameters determined directly from the data. The SED of the dust emission is 
instead assumed to be perfectly known.

%%%%%%%%%%%%%%%%%%%%%%%%%%%%%%%%%%%%%%%
\begin{table*}[t]
\begin{center}
\begin{tabular}{ccccccccccccccccccc}
\hline
& \multicolumn{16}{c}{ \bf COrE+}  & & ${\bf \sigma(\fnl)}$ \\
\hline
$\nu$ [GHz] & & \multicolumn{2}{c}{220} & \multicolumn{2}{c}{255} &
\multicolumn{2}{c}{295} & \multicolumn{3}{c}{340} &
\multicolumn{2}{c}{390} & 450 & 520 & 600 &  & & \\
fwhm [arcmin] & & \multicolumn{2}{c}{3.82} & \multicolumn{2}{c}{3.29} &
\multicolumn{2}{c}{2.85} & \multicolumn{3}{c}{2.47} &
\multicolumn{2}{c}{2.15} & 1.87 & 1.62 & 1.40 & & & \\
w$^{-1}$ [Jy$^2$\,sr$^{-1}$] & & \multicolumn{2}{c}{0.654} & 
\multicolumn{2}{c}{1.43} &
\multicolumn{2}{c}{5.20} & \multicolumn{3}{c}{8.31} &
\multicolumn{2}{c}{13.50} &  22.98 & 39.88 & 69.26 & & & \\
\hline
$\Sigma_{\CIB}^2$ [Jy$^2$\,sr$^{-1}$] & & \multicolumn{2}{c}{--} & 
\multicolumn{2}{c}{8.04} &
\multicolumn{2}{c}{--} & \multicolumn{3}{c}{47.3} &
\multicolumn{2}{c}{--} & 212. & 382. & -- &  & & {\bf 1.6} \\
\hline
& \multicolumn{16}{c}{\bf COrE+ with {\bf{\em Planck}}} & & \\
\hline
$\Sigma_{\CIB}^2$ [Jy$^2$\,sr$^{-1}$] & & \multicolumn{2}{c}{--} & 
\multicolumn{2}{c}{1.97} &
\multicolumn{2}{c}{--} & \multicolumn{3}{c}{10.8} &
\multicolumn{2}{c}{--} & 45.6 & 90.2 & 163.6 &  & & {\bf 0.6} \\
\hline
\hline
& \multicolumn{16}{c}{\bf CORE}  & & ${\mathbf {\sigma(\fnl)}}$ \\
\hline
$\nu$ [GHz] & & \multicolumn{2}{c}{220} & \multicolumn{2}{c}{255} &
\multicolumn{2}{c}{295} & \multicolumn{3}{c}{340} &
\multicolumn{2}{c}{390} & 450 & 520 & 600 &  & & \\
fwhm [arcmin] & & \multicolumn{2}{c}{5.23} & \multicolumn{2}{c}{4.57} &
\multicolumn{2}{c}{3.99} & \multicolumn{3}{c}{3.49} &
\multicolumn{2}{c}{3.06} & 2.65 & 2.29 & 1.98 & & & \\
w$^{-1}$ [Jy$^2$\,sr$^{-1}$] & & \multicolumn{2}{c}{0.29} & 
\multicolumn{2}{c}{0.57} &
\multicolumn{2}{c}{0.77} & \multicolumn{3}{c}{1.08} &
\multicolumn{2}{c}{2.16} & 3.55 & 6.2 & 11.0 & & & \\
\hline
$\Sigma_{\CIB}^2$ [Jy$^2$\,sr$^{-1}$] & & \multicolumn{2}{c}{--} & 
\multicolumn{2}{c}{1.8} &
\multicolumn{2}{c}{--} & \multicolumn{3}{c}{8.8} &
\multicolumn{2}{c}{--} & 42.9 & 80.1 & -- &  & & {\bf 0.7} \\
\hline
& \multicolumn{16}{c}{\bf CORE with {\bf {\em Planck}}} & & \\
\hline
$\Sigma_{\CIB}^2$ [Jy$^2$\,sr$^{-1}$] & & \multicolumn{2}{c}{--} & 
\multicolumn{2}{c}{1.03} &
\multicolumn{2}{c}{--} & \multicolumn{3}{c}{5.2} &
\multicolumn{2}{c}{--} & 23.3 & 43.9 & 68.9 &  & & {\bf 0.34} \\
\hline
\hline
\end{tabular}
\caption{Instrumental specifications of COrE+ and CORE, the noise variance in
    the reconstructed CIB maps ($\Sigma_{\CIB}^2$) and the uncertainty on
    $\fnl$ (assuming $\fnl=0$). \label{tab:1}}
\end{center}
\end{table*}
%%%%%%%%%%%%%%%%%%%%%%%%%%%%%%%%%%%%%%

In Table~\ref{tab:1} we report the uncertainty on $\fnl$, $\sigma(\fnl)$, estimated by the Fisher 
analysis for the COrE+ and CORE configurations, assuming $\fnl=0$ and a usable sky fraction of 
40\%. $\sigma(\fnl)$ is computed after marginalizing over all the other model parameters. Only 
frequencies higher than 200~GHz are considered in the analysis. Channels at lower frequencies are 
in fact dominated by the CMB and should be dedicated to removing CMB fluctuations from the 
signal. Given the 8 frequency channels of CORE and COrE+ at $\nu>200$~GHz, we assume that clean 
CIB maps can be obtained in 4 of them. We can note in Table~\ref{tab:1} that, after the component 
separation, the noise variance in these maps ($\Sigma_{\CIB}$) increases by a factor of 3--12 
with respect to the original noise. This significant degradation in sensitivity is due to the 
frequency coverage of CORE and COrE+, which does not include the wavelengths at which the SEDs of 
the dust and CIB mostly differ (i.e.\ at $\nu\bsim\ 800$~GHz).

We see that the results for CORE here are actually better than those for COrE+. In the analysis 
we have assumed a maximum multipole of $\ell_{max}=1000$. This choice guarantees that the 
shot--noise contribution from star--forming dusty galaxies is negligible. The resolution of the 
CORE experiment therefore is not a critical point to constrain $\fnl$ using the CIB. While COrE+ 
has a smaller beam, CORE has a higher sensitivity, and that is more important in this analysis. 
We find that $|\fnl|=5$ would be measured by CORE at a $\sim7\sigma$ level ($\sim3\sigma$ for 
COrE+), and $|\fnl|$ of 1--2 would also be detectable with CORE with a significance of about 
1--\,3\,$\sigma$. Achieving $\sigma(\fnl)\sim1$ for COrE+ would require a more challenging dust 
removal at the order of 0.1\,percent. We also find that the best results are for sky fractions 
around 0.4--0.6. This is already a significant improvement with respect to {\it Planck} that is 
expected to provide an upper limit on $\fnl$ of $\sim3.5$ at 1$\sigma$ \citep{Tucci:2016aa}.

However, the sensitivity of CORE and COrE+ to the $\fnl$ parameter can strongly improve when the 
highest \planck\ channels are taken into account and combined with CORE/COrE+ data. Extending the 
frequency coverage to 857~GHz is key to better separate CIB and dust emission. The noise 
degradation in the reconstructed CIB maps is significantly reduced when \planck\ channels are 
included in the mixing matrix, even more for COrE+ than for CORE. Under the hypothesis of 5 
CORE/COrE+ channels dedicated to the CIB, we find $\sigma(\fnl)\approx 0.6$ for COrE+ and 
$\sigma(\fnl)\approx 0.3$ for CORE, the latter allowing a 3$\sigma$ detection of $\fnl$ down to 
values of 1. In this configuration the residual dust emission is efficiently separated from the 
CIB fluctuations, and reducing the level of the dust residual in the CIB maps does not give 
significant improvements in the $\fnl$ detection.

In summary, as anticipated at the beginning of this Section, the important conclusion of 
this analysis is that the combination of CORE and Planck should be able to reach a sensitivity 
$\sigma(\fnl)<1$ with this probe, for the local shape.

\subsubsection{Spectral distortions}

The dissipation of acoustic modes set up by inflation in the early universe creates a small 
spectral distortion of the CMB \cite{Sunyaev1970diss, Daly1991, Hu1994, Chluba2012}. Spatial 
fluctuations in the dissipation rate caused by primordial non-Gaussianity can enhance the 
amplitude of distortion anisotropies. It has been shown that this effect can, in principle, be 
used to estimate local $\fnl$ from measurements of the $T$-$\mu$ power spectrum, potentially 
leading to improvements up to many orders of magnitude over current constraints 
\cite{Pajer:2012vz}. However, generally this requires futuristic levels of sensitivity (due to 
the necessity of measuring $\mu$ {\em anisotropies} with high accuracy), well beyond those 
achievable with a CORE-like survey. On the other hand, a strongly enhanced signal in the squeezed 
limit is expected for a specific class of models, characterized by excited initial states 
\cite{2011PhRvD..83f3526A,2011PhRvD..84f3514G}. It is interesting to consider such models using 
this technique with the next generation surveys like CORE. Therefore, we forecast here the 
detectability of a $C_{\ell}^{\mu T}$ signal in these non-Bunch-Davies (NBD) scenarios. Our 
results are obtained by closely following the methodology originally proposed in 
\cite{Ganc:2012ae}, to which we refer the reader for details.

The signal-to-noise ratio now reads
\begin{equation}
\left(\frac{S}{N}\right)^2 = \sum_\ell \left( 2 \ell + 1 \right) \frac{\left( C_{\ell}^{\mu T} \right)^2}{C_\ell^{TT} C_\ell^{\mu \mu}} 
\end{equation}
where large-scale temperature data are completely signal dominated, while the $\mu \mu$ spectrum 
is dominated by noise, for the experiments under study. Maps of $\mu$-anisotropies can be created 
from differences of temperature maps at different frequencies, $\nu_1$ and $\nu_2$, and the 
corresponding noise is computed as:
\begin{equation}\label{eqn:munoise} 
C_{\ell}^{\mu \mu, N} = \Omega_{\rm pix} b_{\ell}^{-2} \left[ \frac{   \frac{\nu_1 \nu_2}{\nu_1 - \nu_2}   }{56.80 \, {\rm Ghz}} \right]^2 \left[(\sigma_{\rm pix}^{\nu_1})^2 + (\sigma_{\rm pix}^{\nu_2})^2 \right] \; .
\end{equation}
In this formula, $\Omega_{\rm pix}$ represents the solid angle (in steradians) subtended by a 
given pixel, $b_{\ell}$ is the beam and $\sigma^X_{\rm pix}$ is the noise per pixel in channel X.

\begin{figure}[t!]
\centering
  \includegraphics[width=\textwidth]{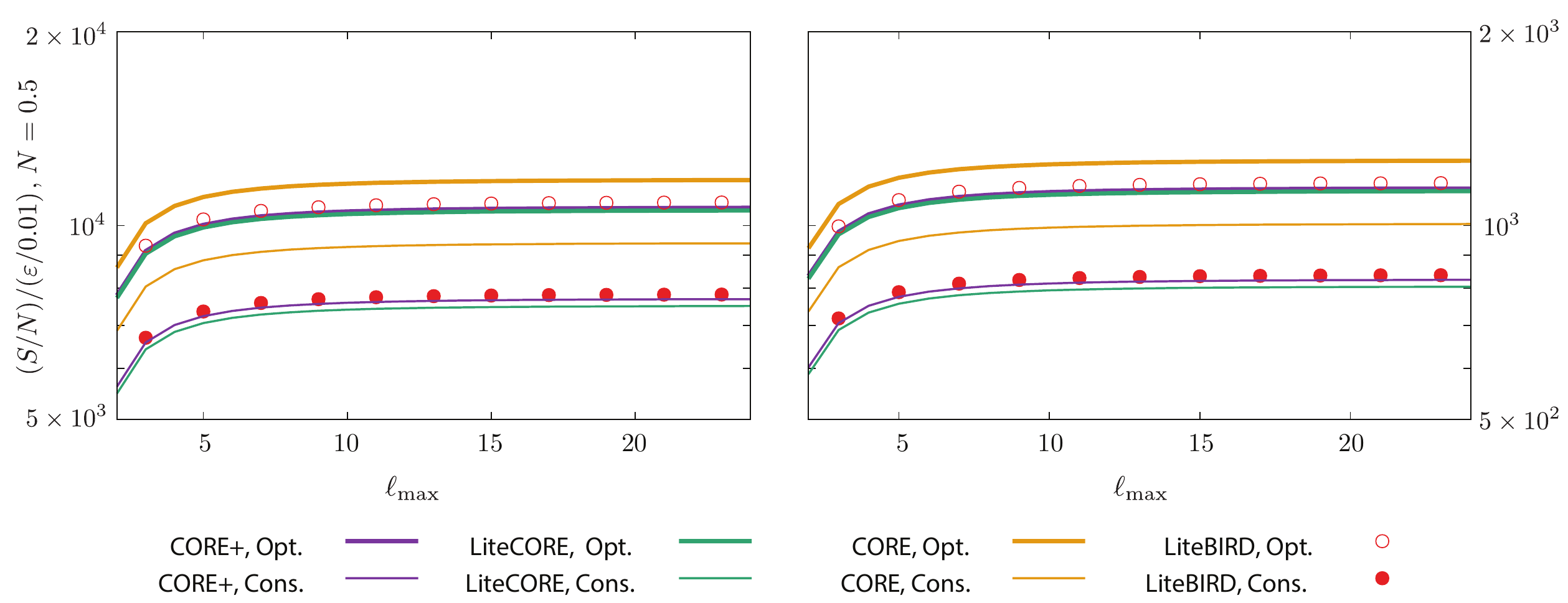}
  \caption{Signal-to-noise ratio of $C_\ell ^{\mu T}$ from a modified initial state as a 
function of $\ell_{max}$ for a fixed occupation number $N=0.5$. In our forecast we considered 
a `conservative' and an `optimistic' case. In the optimistic case we computed the 
signal-to-noise ratio by coadding the $4$ lowest noise combinations of frequencies 
in the range $[80,200]$ GHz that can be obtained from the LiteCORE, COrE$+$ and CORE 
configurations. For COrE$+$ we use $\Delta T = 9.1$ $\mu$K arcmin, ${\rm FWHM} = 10.5$ 
arcmin at $80$ GHz and $\Delta T = 6.5$ $\mu$K arcmin, ${\rm FWHM} = 9.3$ arcmin 
at $90$ GHz. All other channels and experimental configurations are shown in the 
Tables of Section~\ref{sec:three}.
In the conservative case we instead take only the best couple of frequencies for 
each configuration. This choice is justified by the fact that the issues of component 
separation 
and relative calibration of channels are not accounted for 
in this type of analysis. Therefore it is not obvious that all the channels 
which are assumed clean for standard temperature analysis 
will also be available for $T$-$\mu$ measurements. We see that the signal saturates in 
the first few multipoles. For each configuration of the satellite are shown models with 
both $\theta_k =$ const (left) and $\theta_k \approx k\eta_0$ (right). LiteCORE-80 and 
LiteCORE-120 perform essentially in the same way, due to the saturation of the signal 
after the first few multipoles, and are described by a single line. $S/N$ for 
$\it Planck$ is a factor $\approx 100$ lower than for CORE.}
  \label{fig:U-squeezedModelForecast}
\end{figure}

\begin{figure}[t!]

  \centering
  \includegraphics[width=0.75\textwidth]{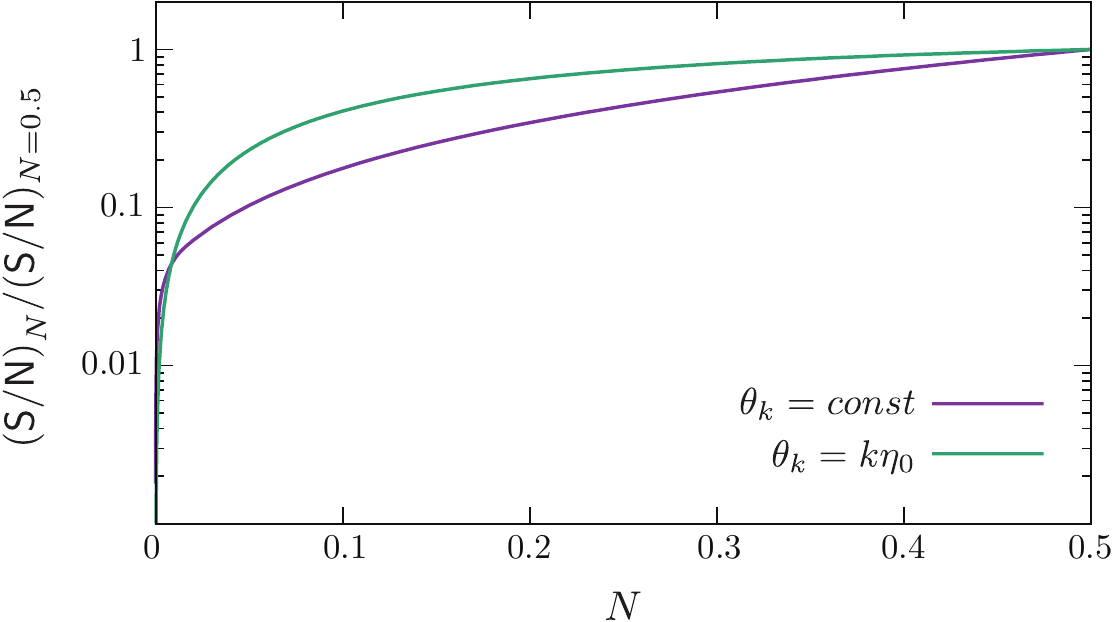}
  \caption{Dependence of the signal-to-noise ratio on the value of the occupation number 
$N$. Signal-to-noise ratio for different $N$ is here 
normalized to the values obtained 
for $N=0.5$ as displayed in Fig.~\ref{fig:U-squeezedModelForecast}.}
  \label{fig:U-squeezedModelNDependence}
\end{figure}

The initial conditions, set at a finite conformal time $\eta_0$, can be effectively parameterized 
through the occupation number of excited states, $N$, and a phase, $\theta_k$. The $C_\ell ^{\mu 
T}$ are then computed for two different parameterizations of the phase: $\theta_k=$ constant or 
$\theta_k \approx k\eta_0$. The final signal-to-noise results, displayed in 
Figs.~\ref{fig:U-squeezedModelForecast} and \ref{fig:U-squeezedModelNDependence}, show (in 
agreement with similar findings in \cite{Ganc:2012ae}, which referred to a PIXIE-like experiment) 
that {\em NBD models can be probed by CORE with high statistical significance} using this test. 
More specifically, in Fig.~\ref{fig:U-squeezedModelForecast} we show the signal-to-noise ratio of 
$C_\ell ^{\mu T}$ from a modified initial state as a function of $\ell_{max}$ for a fixed 
occupation number $N=0.5$, slow-roll parameter $\epsilon = 0.01$, and for both parameterizations 
of $\theta_k$.  We see that all experimental configurations achieve a very high signal-to-noise. 
In all the considered cases we obtain $S/N > 500$. By comparison, direct $f_{\rm NL}$ 
measurements for these models, based on bispectrum template fitting, achieve a sensitivity 
$\sigma_{\rm fNL} \approx 5$, which would make the case plotted in 
Fig.~\ref{fig:U-squeezedModelForecast} undetectable. We also see that for this test, CORE 
outperforms the other configurations, including COrE+, due to better sensitivity at $80$ GHz.
Fig.~\ref{fig:U-squeezedModelNDependence} shows how the signal-to-noise changes when varying the 
occupation number $N$. Results are normalized to the values shown in 
Fig.~\ref{fig:U-squeezedModelForecast} for $N=0.5$. Unless occupation numbers become very low, 
the signal should remain detectable.

Interestingly, {\it Planck} is already expected to have a strong discriminating power for these 
models, although with a signal-to-noise ratio a factor roughly $100$ below CORE. On the other 
hand, we have to stress that systematic sources of error are not taken into account here. The 
expected two orders of magnitude improvement of CORE over {\it Planck} looks therefore even more 
important and reassuring, for the purpose of retaining a high signal-to-noise after relative 
calibration error and foreground subtraction are taken into account.

\subsection{Summary}
\label{sec:conclbisp}

In this Section we considered NG signatures produced by many different inflationary scenarios. We 
forecasted the detectability of these signatures with CORE-type surveys in terms of future 
bounds on the NG parameters $\fnl$. For standard local, equilateral, and orthogonal (LEO) shapes, we 
find that CORE will allow for improvements of a factor between $2$ and $3$ over current {\it 
Planck} bounds, giving $1\sigma$ $\fnl$ error forecasts of about 2, 20, and 10 for the local, 
equilateral, and orthogonal shapes, respectively (see Table~\ref{table_NGleo}). These 
forecasts are obtained using 
direct bispectrum measurements with the same methods used to analyze the {\it Planck} data. 
Similar improvements are also expected for the $g_{\rm NL}$ trispectrum constraints, for 
the $n_{\rm NG}$ constraints in models with a mild running of $\fnl$, and for the $\fnl$ constraints in 
the case of oscillatory bispectra. This improvement shrinks the available volume in LEO 
bispectrum space by a factor of $\approx 20$ with respect to {\it Planck}. Our CMB bispectrum-based 
constraints are slightly worse than recent $\fnl$ forecasts using the bispectrum of forthcoming 
galaxy surveys, like Euclid, at least for the local shape (see e.g. 
\cite{2016JCAP...06..014T,Agarwal:2013qta,dePutter:2014lna,Camera:2014bwa}). However, LSS 
bispectrum measurements must correctly and reliably account for
complex non-primordial NG 
contamination arising from non-linear gravitational evolution of structure (such as bias), 
making the CMB a much cleaner probe. Non-primordial LSS non-linearities seem moreover more 
problematic in the equilateral limit. Consequently, the CMB might provide the only way to make progress 
for many nonlocal shapes (see e.g.~\cite{2016arXiv160200674B}) if we exclude futuristic 21-cm 
measurements \citep{2015PhRvD..92h3508M}.
 
Looking at other NG tests not based on direct bispectrum estimation, we find a very interesting 
result: CORE should be able to achieve $\fnl \approx 1$ sensitivity for the local shape 
by measuring the scale-dependent bias in the CIB power spectrum on large scales. (Note that 
these $\fnl$ CIB forecasts account for the dust subtraction explicitly.) This is a 
crucial threshold for discriminating between single and multi-field inflation. Such a  
sensitivity would be comparable or better than what is achievable by estimating $\fnl^{\rm 
local}$ from scale-dependent bias in future galaxy surveys.

If we move beyond `standard' LEO bispectra, we find that CORE can achieve massive improvements 
over current constraints for specific targeted models such as 
isocurvature NG, where $S/N$ improvements by an order of magnitude are expected via bispectrum 
estimation, and for some NBD models, where of order $\approx 100$ improvements in sensitivity can be 
achieved via measurements of cross-correlations between temperature and $\mu$-distortion 
anisotropies. The combination of CORE with 
an absolute spectrometer such as  PIXIE \citep{2016arXiv161008711C}
would allow
a model-independent interpretation of the $C_\ell ^{\mu T}$ 
constraints.
Moreover, CORE should detect the 
lensing-ISW bispectrum with almost an order of magnitude better signal-to-noise than {\it 
Planck}.
For most of the performed bispectrum tests, we find that the LiteCORE, CORE, and COrE+ 
configurations are nearly equivalent in terms of final $\fnl$ sensitivity, with CORE falling in 
between LiteCORE and COrE+. The CIB and T$\mu$ tests, on the other hand, favor CORE over the 
other configurations due to its higher sensitivity.

\section{Topological defects}
\label{sec:defects}

In this Section we will take Abelian Higgs gauge cosmic strings (as are present for example in D-term hybrid inflation) as a representative topological defect.  The power spectra coming from gauge strings and other defects are relatively similar, though the precise shape and normalization are different  \cite{Mukherjee:2010ve}. 
However, the procedure for searching for strings and other defects is in principle the same. In the next 
Subsection we will briefly introduce how the CMB power spectra from defects is obtained
before we describe the statistical analysis performed to assess the sensitivity of the 
different missions to either detect defects or to distinguish them from primordial gravitational waves.

\subsection{Calculation of CMB from defects}

A general method for calculating CMB power spectra from defects starts with the numerical solution of the classical field theory in an expanding background, from which the unequal time correlators (UETCs) of the energy-momentum tensor are extracted \cite{Pen:1997ae,Durrer:2001cg}.  One can also model the defect sources in various computationally less expensive 
ways \cite{Pogosian:1999np,Charnock:2016nzm,Achucarro:2013mga}, 
but these effective models must be checked against large-scale numerical simulations.

Using rotational symmetry, it can be shown that there are only five  independent correlators: three scalar, one vector, and one tensor. The three scalar correlators can be thought of as sources for the auto- and cross-correlation of the two Bardeen potentials $\Phi$ and $\Psi$. Unlike in the case of inflationary perturbations, the vector perturbations coming from defects do not die out, as they are constantly being seeded, and they contribute (together with the tensor modes) to create a B-mode polarization signal.

The correlation functions can then be diagonalized 
and their eigenfunctions used as sources for an Einstein-Boltzmann solver.  
The CMB power spectra from each eigenfunction can finally be summed. 

\begin{figure}[htbp]  
   \centering
   \includegraphics[width=0.6\textwidth]{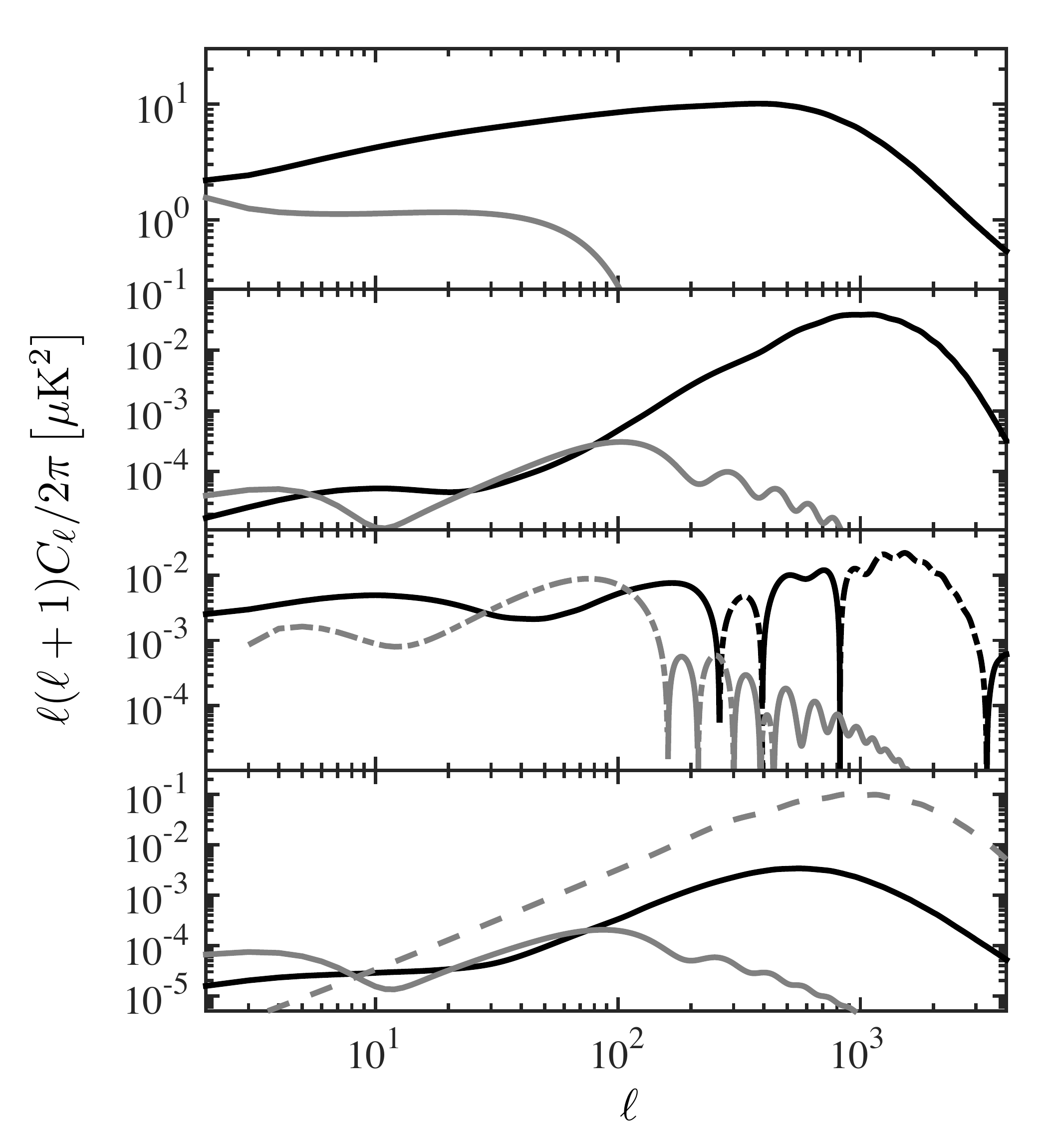}
   \caption{ CMB perturbation power spectra from cosmic strings \cite{Lizarraga:2016onn} with $\fd = 5\times10^{-3}$ (solid black), compared to inflationary tensor modes with $r = 3\times 10^{-3}$ (solid grey) and lensing B-modes (grey dashed). Top to bottom are TT, EE, TE, and BB. Negative values in the TE power spectrum are shown as dotted 
lines.  The signal from inflationary scalar perturbations in the TT, EE, and TE channels are not shown, as they are much larger.}
   \label{f:DefectCMB}
\end{figure}

For parameterizing defect power spectra, it is convenient to define the dimensionless parameter
\be
G\mu = 2\pi G \phi_0^2
\ee
where 
$\phi_0$ is the expectation value of the symmetry-breaking field (assumed complex) and 
$\mu$ is the cosmic string tension in models with strings. Power spectra are proportional to $(G\mu)^2$. It is also convenient to parameterize the fractional contribution of the defects to the CMB temperature power spectrum $C_{\ell}$ by 
\be
f_{10} =\frac{C_{10}^{\rm def}}{C_{10}^{\rm inf}+C_{10}^{\rm def}}
\ee
where $C_{\ell}^{\rm def}$ is the power spectrum produced by defects and $C_{\ell}^{\rm inf}$ the power spectrum from the underlying inflationary model. The two sets of fluctuations are statistically independent. Note that all different power spectra (temperature and polarization) are linked in the sense that 
for a given defect model, the normalization of one forces the normalization of the others. 
There is no freedom to normalize them independently. Therefore, even though $f_{10} $ is defined for the temperature power spectrum, it also gives the normalization of the polarization power spectra.
Note also that the constant of proportionality relating $f_{10}$ and $(G\mu)^2$ varies between kinds of 
topological defects, but not by more than a factor of a few. Constraints expressed in terms 
of $f_{10}$ are similar for different kinds of defects, reflecting the similarity of the shape of the power spectra.
The power spectra from the most recent computations for cosmic strings 
\cite{Daverio:2015nva,Lizarraga:2016onn} are shown in Fig.~\ref{f:DefectCMB}.

\subsection{MCMC Fits}

In order to quantify the ability of future satellite CMB missions to constrain the abundance
of cosmic strings in the universe, we adopt a MCMC approach as in previous
works \cite{Urrestilla:2011gr,Ade:2013xla}. 
Our parameters are 
the standard six $\Lambda$CDM parameters 
$\{\omega_b, \omega_c, \theta_{\rm MC}, \tau, \ln 10^{10} A_s, n_s\}$
with the addition
of the scalar to tensor ratio $r$ and/or the string amplitude $10^{12} (G\mu)^2$. 
We use flat priors in these parameters that are always wide enough to encompass 
the posteriors, and impose the condition $r\geq0$ and $G\mu\geq0$ when those
parameters are varied (else they are set to zero). The flat prior on $(G\mu)^2$
translates (for small admixtures of cosmic strings) to a flat prior on the fractional
contribution at $\ell=10$ $\fd$.

For Planck temperature and polarization data, the 95\% upper limit on $\fd$ is about 1\% \cite{Lizarraga:2016onn};
strings are thus a small contribution. We are therefore justified in computing the string spectra only for a reference cosmological model, varying only the amplitude of the contribution as described above.
The small errors in the shape of the power spectra will be insignificant.

Strings have an important B-mode signal, which is obscured by the lensing of the E-mode polarization of the dominant inflationary fluctuations. 
In the analysis we consider two different approaches to analyzing the B-mode polarization data, which span the range of success in disentangling the lensed B-modes from the signal from primordial gravitational waves ($r$) or defects ($G\mu$):

\begin{enumerate}
	\item Standard, where the full lensing	signal is present;
%	\item delensed, by an amount depending on the specification of the mission; 
	\item Fully delensed, where the B-mode is entirely due to gravitational waves or defects.
\end{enumerate}

%The fully delensed case represents an ideal and is unlikely to be realized in practice.  
 
\begin{figure}[htbp]  
   \centering

	\includegraphics[width=0.5\textwidth]{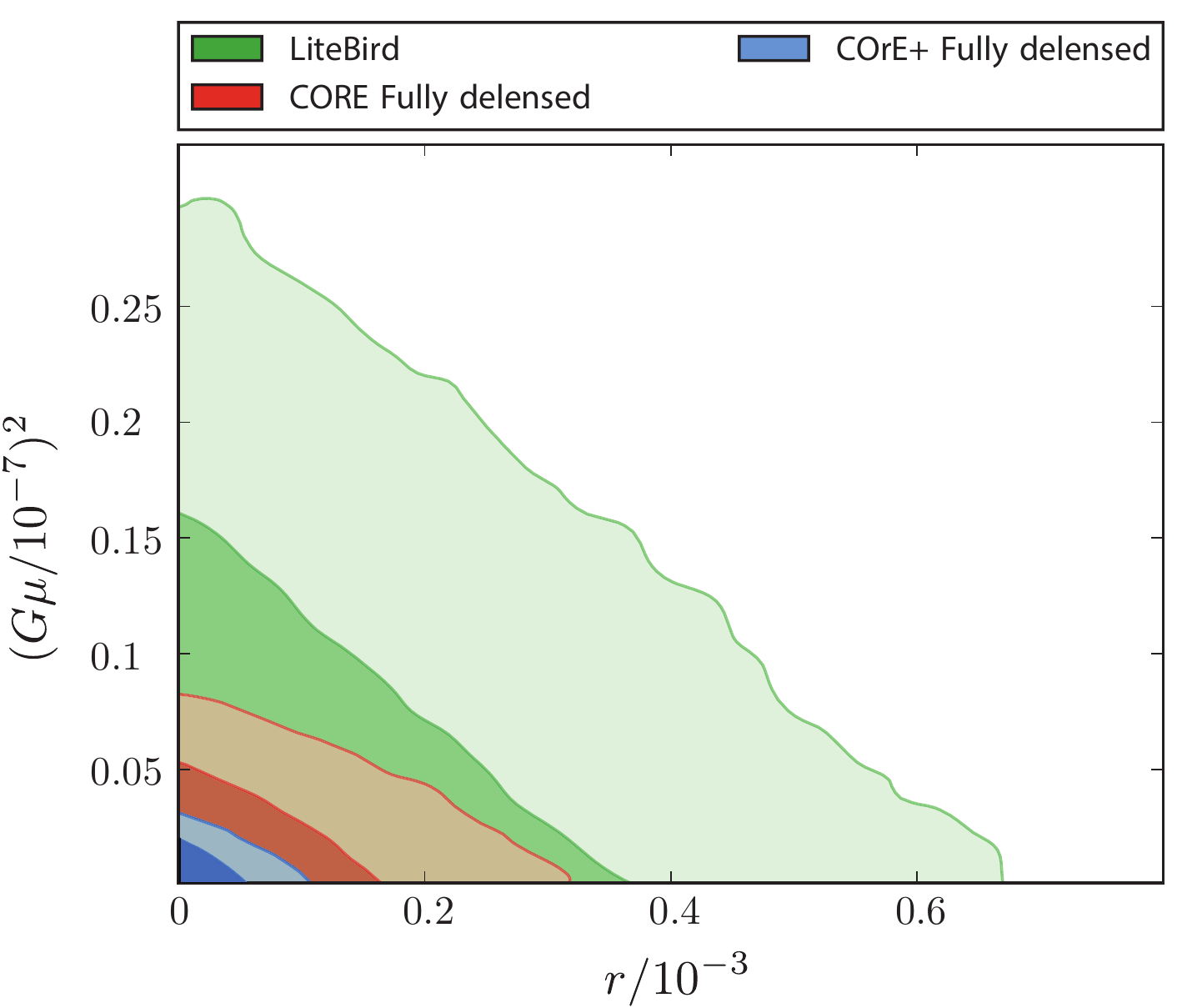}

   \caption{ Likelihood contours of the $r$-$(G\mu)^2$ posterior for models with both tensors and defects (cosmic strings), 
   marginalized over all other parameters, for LiteBIRD, CORE, and COrE+. 
The CORE and COrE+ contours are given for the fully delensed case, 
representing the best possible constraints attainable, while 
 LiteBIRD contours are given for the standard case. In all cases light shading represents 1$\sigma$ and dark shading 2$\sigma$. 
 For comparison, the current \Planck\ marginalized 95\% confidence limits on these models are 
 $(G\mu/10^{-7})^2 < 3.4$  ($\fd < 0.011$) 
 and $r < 0.11$.
}
   \label{f:theBusiness}
\end{figure}

In Fig.~\ref{f:theBusiness} we show 1$\sigma $ and 2$\sigma$ likelihood contours for 
the $r$-$(G\mu)^2$ posterior, marginalized over
all other parameters, for the fully delensed case.  The 95\% upper limits for all cases considered here are listed in Table \ref{tab:missions}.

\begin{table}[!h]
\centering
\begin{tabular}{lcccl}
\hline
B-mode analysis         & $f_{10}$                  & $10^{12}(G\mu)^2$ & $r$                 \\ \hline\hline
  &  \multicolumn{3}{l}{{\bf COrE+}} \\\hline
standard         & $5.0\times10^{-4}$  & $1.5\times10^{-3}$  & $3.2\times10^{-4}$   \\
standard         & $2.3\times10^{-4}$  & $7.0\times10^{-4}$  & $-$                             \\
%standard         & $-$                          & $-$                           & $3.5\times10^{-4}$  \\
%delensed         & $1.2\times10^{-4}$  & $3.5\times10^{-4}$  & $-$                            \\
fully delensed & $7.1\times10^{-5}$  & $2.2\times10^{-4}$  & $5.8\times10^{-5}$                            \\
fully delensed & $6.3\times10^{-5}$  & $1.9\times10^{-4}$  & $-$                            \\
\hline\hline
  &  \multicolumn{3}{l}{{\bf CORE}} \\\hline
standard         & $5.4\times10^{-4}$  & $1.6\times10^{-3}$  & $4.3\times10^{-4}$   \\
standard         & $3.0\times10^{-4}$  & $9.0\times 10^{-4}$  & $-$                           \\
%standard         & $-$                          & $-$                           & $4.5\times10^{-4}$    \\
fully delensed  & $1.9\times10^{-4}$  & $5.8\times10^{-4}$  & $1.7\times10^{-4}$    \\
fully delensed & $1.5\times10^{-4}$  & $4.5\times10^{-4}$  & $-$                            \\
%fully delensed  & $-$                          & $-$                           & $1.8\times10^{-4}$    \\
\hline\hline
   &  \multicolumn{3}{l}{{\bf LiteBIRD}} \\\hline
standard         & $6.5\times10^{-4}$  & $2.0\times10^{-3}$  & $3.3\times10^{-4}$   \\
standard         & $7.1\times10^{-4}$  & $2.1\times10^{-3}$  & $-$                            \\
%fully delensed & $2.0\times10^{-4}$  & $5.9\times10^{-4}$  & $-$                            \\
\hline\end{tabular}
\caption{
	Upper limits at 95\% confidence on the parameters in $\Lambda$CDM models with cosmic strings and/or primordial gravitational waves tested against an input spectrum 
with neither strings nor gravitational waves. Note that $\fd$ (the 
proportion of the temperature power spectrum due to strings) is proportional to 
$(G\mu)^2$ where $\mu$ is the string tension and $G$ is Newton's constant.
\label{tab:missions}
}
\end{table}

%\subsection{Conclusions}
\subsection{Summary}

We find that COrE+ places the tightest 
constraints on the abundance of topological defects, followed by CORE, 
and finally by LiteBIRD. 
%LiteCORE-150 and LiteCORE-120 fall between CORE and LiteBIRD.
%The absolute difference between these missions is however not very large. 
When looking only for defects 
and without delensing, the limit on $\fd$ achievable by CORE when fitting against a strings-only model  ($3.0\times 10^{-4}$) 
is approximately twice as strong as that by LiteBIRD, with COrE+ about 20\% stronger still.
%When only looking for defects, and without delensing, the limits on $f_{10}$ range from $2.2\times10^{-4}$ (COrE+) to $6.1\times 10^{-4}$ for LiteBIRD, with the LiteCORE configurations not far from the COrE+ numbers. But as $G\mu$ scales like $\sqrt{f_{10}}$, this corresponds to a bound on $G\mu$ for COrE+ that is only a factor of about 1.7 better than the one by LiteBIRD. 

All these future missions will perform
much better than Planck, however, for which $f_{10}<0.011$ at 95\% CL \cite{Lizarraga:2016onn} in models with strings and tensors.  
COrE+ could improve the constraints on $f_{10}$ by up to two orders of magnitude, and thus by about one order of magnitude in $G\mu$. 
%A somewhat curious fact is that for COrE+ and LiteCORE the bounds on defects 
%weaken significantly when fitting simultaneously for defects and primordial 
%gravitational waves, while for LiteBIRD the opposite is true {\color{red} [THIS EFFECT PROBABLY NEEDS TO BE CLARIFIED]}. 
%Because of this, all missions perform at a similar level when fitting for both.

Delensing leads to further improvements in the constraints for COrE+ and CORE. 
If the B-mode map can be fully delensed, the improvement is a factor of 2 in $f_{10}$  for CORE and nearly  
a factor of 4 for COrE+. 
%The improvement for LiteCORE-150 is smaller, and LiteCORE-120 does not profit from delensing. 
LiteBIRD will not be able to delens its data on its own. 
%If ever a complete removal of all lensing B-modes became possible then COrE+ would 
%profit the most, gaining another factor of 2 over delensed case.
% The gain for the LiteCORE missions  on the other hand would be more limited. 
%Interestingly, LiteBIRD would also profit significantly, and reach a sensitivity on par with LiteCORE. {\color{blue}[EXPLAIN OR DROP?]}

%This order of magnitude improvement in the constraints on $G\mu$ achievable by CORE and COrE+  is important as it will push 
%the maximally allowed string tension $G \mu$ decisively below the GUT scale
%(or detecting cosmic strings if any are formed at GUT energies).
%Similar gains in sensitivity are expected for all types of defect.

With full delensing, the upper bound on the string tension parameter can be brought down to $G\mu < 2.1 \times 10^{-8}$, which corresponds to a bound on the symmetry-breaking scale of $\phi_0 < 6.9 \times 10^{14}$, well below the GUT scale.

%\bibliography{CoreStrings}
%
%\end{document}

\section{Conclusions}
\label{sec:conclusions}

A broad consensus has emerged that the discovery of primordial gravitational waves generated 
during the epoch of inflation will likely be one of 
the next breakthrough discoveries of observational cosmology. 
These would be detected through their imprint on the B mode of the CMB polarization anisotropy. 
The experimental requirements for a definitive search for the B mode signal 
from inflationary gravitational waves 
would lead to far reaching improvements in 
the measurement of T and E and would allow for a tremendous 
advance in our understanding of cosmic inflation,  
beyond measuring the energy scale of inflation from 
the primordial B-mode polarization amplitude. 

In this paper we have explored the potential of an experiment like CORE, a proposed CMB space satellite 
submitted in response to the ESA fifth call for a medium-size mission opportunity, 
for improving our understanding of cosmic inflation. CORE would also enable a broad 
range of science objectives to be achieved in cosmology beyond inflation science, 
as reported in the companion papers of this series 
for cosmological parameters \cite{ecoParams}, extragalactic sources \cite{DeZotti:2016qfg}, clusters \cite{ecoCluster}, 
and Doppler/aberration effects \cite{ecoPeculiar}.  
We have quantified how CORE compares to other concepts for a next generation CMB space mission dedicated 
to the ultimate measurement of the polarization anisotropies in improving our knowledge of 
the physics of inflation, at present
largely based on \Planck\ and ground experiments such as BICEP 2/Keck Array.

This paper uses idealized
mock likelihoods that assume that the lowest and highest frequency channels suffice 
to remove foreground contaminations from the 
inverse noise weighted combination of the six central frequency `cosmological'
channels used for our forecasts. Although this is an idealized case, CMB space 
missions such as CORE (denoted in this paper as CORE-M5) are experimental concepts that come
close to achieving this limit as
a result of the large number of frequency channels present and the possibility to 
measure the Galactic dust polarization signal at high frequencies
accurately. We refer the reader to the companion paper \cite{ecoCompSep} for a 
the CORE capability to measure primordial B-mode polarization by 
a component separation approach on maps which include contributions from of all the relevant foregrounds.
%\Planck\ has demonstrated how the measurement of dust polarization at high frequency has been essential 
%for the B-mode joint BKP cross-correlation study \cite{Ade:2015tva} and for the improved
%determination of the optical depth from E-modes on large angular scales 
%\cite{Aghanim:2015xee,Aghanim:2016yuo,Adam:2016hgk}.  
Table~\ref{tab:key} summarizes the expected CORE uncertainties in the 
inflationary parameters compared to the current measurements based on 
the \Planck\ 2015 data. 
 
We included the JAXA LiteBIRD proposal \cite{2014JLTP..176..733M,Matsumura:2016sri} among 
the different configurations to be compared with CORE by adopting specifications of an 
extended LiteBIRD mission \cite{Errard:2015cxa}.
We have shown that CORE can achieve an uncertainty in $r$ at 
least a factor of 2 better than the LiteBIRD configuration. 
Because of its coarse angular resolution (owing to its smaller aperture),
LiteBIRD needs to be complemented with external CMB data to reach the level of the 
current \Planck\ uncertainties for the scale dependence of the spectral index, primordial 
non-Gaussianities and isocurvature perturbations, which are invaluable for
probing the physics of inflation beyond searching for the energy scale at which inflation occurred.

Highlights of the new information regarding 
the physics of inflation that can be achieved with CORE include
the following:

\vspace{0.4cm}
\noindent
{\em Primordial B-modes and fundamental physics}

\vspace{.2cm}
With its sensitivity to the primordial B-mode polarization, CORE will test new physics at energy
scales approximately a trillion times higher than those probed by the Large Hadron Collider (LHC).
Such scales are of key relevance to fundamental physics. If gravitational waves from inflation are
detected by CORE, this discovery will constitute the first experimental signature of
quantum gravity and of physics at the Planck scale, implying large-field inflation with a super-Planckian field
excursion. If $r$ is  measured at high statistical significance, we have shown how CORE
will be superior to LiteBIRD for constraining the shape of the spectrum of primordial
gravitational waves. The CORE window on energy scales probed by B-mode polarization is
unique. The target sensitivity of the ESA-L3 gravitational wave mission is not
sufficient to detect inflationary gravitational waves for the simplest inflationary models.
However, synergies between CORE and the ESA-L3 gravitational wave mission will be able to
constrain alternatives to the simplest models of inflation.

\begin{table}[t]
\begin{center}
\small{%
\hspace{- 2.1 cm}
 \begin{tabular}{llccc}
  \toprule
Parameter & Results from \Planck\ 2015 release & CORE & Improvement \\
& & expected uncertainties & factor \\
\midrule[0.065em]
\hline
$\Lambda$CDM model & & & \\
\hline
$A_\mathrm{s}$ & $ A_\mathrm{s} = ( 2.130 \pm 0.053 ) \times 10^{-9} \,$ (68\% CL) \cite{Ade:2015xua} & $\sigma (A_\mathrm{s}) = 0.0073$ & $7.3$ \\
$n_{\rm s}$ & $n_{\rm s} = 0.9653 \pm 0.0048 \,$ (68\% CL) \cite{Ade:2015xua} & $\sigma (n_{\rm s}) = 0.0014$ & $3.4$ \\
$\Omega_\mathrm{b} h^2$ & $\Omega_\mathrm{b} h^2=0.02226 \pm 0.00016 \,$ (68\% CL) \cite{Ade:2015xua}& $\sigma (\Omega_bh^2) = 0.000037$ & $4.3$ \\
$\Omega_\mathrm{c} h^2$ & $\Omega_\mathrm{c} h^2=0.1193 \pm 0.0014 \,$ (68\% CL) \cite{Ade:2015xua} & $\sigma (\Omega_ch^2) = 0.00026$ & $5.4$ \\
$\tau$ & $\tau = 0.063 \pm 0.014\,$ (68\% CL) \cite{Ade:2015xua} & $\sigma (\tau) = 0.002$ &  $7.0$ \\
$H_0$ [km/s/Mpc] & $H_0=67.51 \pm 0.64\,$ (68\% CL) \cite{Ade:2015xua} & $\sigma (H_0) = 0.11$ & $5.8$ \\
\hline
                        $\mathrm{d} n_{\rm s}/\mathrm{d} \ln k$ & $\mathrm{d} n_{\rm s}/\mathrm{d} \ln k = -0.0023 \pm 0.0067 \,$ (68\% CL) \cite{Ade:2015xua,Ade:2015lrj}
& $\sigma (\mathrm{d} n_{\rm s}/\mathrm{d} \ln k) = 0.0023$ & $2.9$ \\
                        $\mathrm{d}^2 n_{\rm s}/\mathrm{d} \ln k^2$ & $\mathrm{d}^2 n_{\rm s}/\mathrm{d} \ln k^2 = 0.025 \pm 0.013 \,$ (68\% CL) \cite{Ade:2015lrj}
& $\sigma (\mathrm{d}^2 n_{\rm s}/\mathrm{d} \ln k^2) = 0.0046$ & $2.8$ \\
%                      $r$ & $r < 0.07$ (95 \% CL) \cite{Ade:2015xua} & $\sigma (r=10^{-3}) = {\bf 4 \cdot 10^{-4}}$ &  {\cal O} (10^2) \\
$\Omega_{\rm k}$ & $\Omega_{\rm k} = -0.0037^{+0.0083}_{-0.0069} \,$ (68\% CL) \cite{Ade:2015xua} & $\sigma (\Omega_{\rm k})
= 0.0019$ & 4 \\
%                       $r$ & $r < 0.08$ (95\% CL) \cite{Ade:2015tva,Ade:2015lrj} & $\sigma (r) = 4 \cdot 10^{-4}$ & ${\cal O} (10^2)$ \\ % \, \, (r_{\mathrm{fid}} = 0.01 )$ & \\
%& & $(r_{\mathrm{fid}} = 0.01 )$ &  \\
%                       $n_{\rm t}$ & $-0.38 < n_{\rm t} < 2.6$ (95\% CL) \cite{Ade:2015lrj} 
%& $\sigma (n_{\rm t}) = 0.08$ & $10$ \\ % \, \, (r_{\mathrm{fid}} = 0.01 \,, n_{\mathrm{fid} \, {\mathrm t}} = - r_{\mathrm{fid}}/8 )$ \\
%& & $(r_{\mathrm{fid}} = 0.01 \,, n_{\mathrm{fid} \, {\mathrm t}} = - r_{\mathrm{fid}}/8 )$ & \\
                       $r$ & $r < 0.08$ (95\% CL) \cite{Ade:2015tva,Ade:2015lrj} & $\sigma (r) = 4 \cdot 10^{-4}$ & $10^2$ \\ % \, \, (r_{\mathrm{fid}} = 0.01 )$ & \\
& & $(r_{\mathrm{fid}} = 0.01 )$ &  \\
                       $n_{\rm t}$ & $-0.38 < n_{\rm t} < 2.6$ (95\% CL) \cite{Ade:2015lrj}
& $\sigma (n_{\rm t}) = 0.08$ & $10$ \\ % \, \, (r_{\mathrm{fid}} = 0.01 \,, n_{\mathrm{fid} \, {\mathrm t}} = - r_{\mathrm{fid}}/8 )$ \\
& & $(r_{\mathrm{fid}} = 0.01 \,, n_{\mathrm{fid} \, {\mathrm t}} = - r_{\mathrm{fid}}/8 )$ & \\
                        $\beta_\mathrm{iso}$ & $\beta_\mathrm{iso}^\mathrm{curvaton} < 0.0013 \,$ (95\% CL) \cite{Ade:2015lrj} 
& $\beta_\mathrm{iso}^\mathrm{curvaton} < 0.00026 \,$ (95\% CL) & 5.0 \\
                                             & $\beta_\mathrm{iso}^\mathrm{axion} < 0.038 \,$ (95\% CL) \cite{Ade:2015lrj} 
& $\beta_\mathrm{iso}^\mathrm{axion} < 0.018 \,$ (95\% CL) & 2.1 \\
%                        $\beta_\mathrm{iso}^\mathrm{curvaton}$ & $\beta_\mathrm{iso} < 0.0013 \,$ (95\% CL) \cite{Ade:2015lrj} & $\beta_\mathrm{iso} < 0.00026 \,$ (95\% CL) & \\
%                        $\beta_\mathrm{iso}^\mathrm{axion}$ & $\beta_\mathrm{iso} < 0.037 \,$ (95\% CL) \cite{Ade:2015lrj} & $\beta_\mathrm{iso} < 0.018 \,$ (95\% CL) & \\
                        $f_{\rm NL}$ & $f_\mathrm{NL}^\mathrm{local}=0.8 \pm 5.0 \,$ (68\% CL) \cite{Ade:2015ava} & $\sigma (f_\mathrm{NL}^\mathrm{local}) = 2.1$ & 2.4 \\
                                     & $f_\mathrm{NL}^\mathrm{equil} = -4 \pm 43 \,$ (68\% CL) \cite{Ade:2015ava} &  $\sigma (f_\mathrm{NL}^\mathrm{equil}) = 21$ & 2.0 \\
                                     & $f_\mathrm{NL}^\mathrm{ortho} =-26 \pm 21 \, $ (68\% CL) \cite{Ade:2015ava}
&  $\sigma \left( f_\mathrm{NL}^\mathrm{ortho} \right) = 9.6$ & 2.2 \\
                                     & $f_\mathrm{NL}^\mathrm{ISW-lens} = 0.79 \pm 0.28 \, $ (68\% CL) \cite{Ade:2015ava}
 &  $\sigma \left( f_\mathrm{NL}^\mathrm{ISW-lens} \right) = 0.045$ & 6.2 \\
                        $c_{\rm s}$ & $c_\mathrm{s} > 0.023 \,$ (95\% CL) \cite{Ade:2015ava} & $c_\mathrm{s} > 0.045 \,$ (95\% CL) & 2.0 \\
                        $G \mu$ & $G \mu < 2.0 \times 10^{-7} \,$ (95\% CL) \cite{Lizarraga:2016onn} & $G \mu < 2.1 \times 10^{-8} \,$ (95 \% CL) & 9.5 \\
%                        $f_{10}$ & $f_{10} < 2.0 \times 10^{-7} \,$ (95\% CL) \cite{Lizarraga:2016onn} & $f_{10}  < 1.4 \times 10^{-4} \,$ (95\% CL) & 140 \\
          \bottomrule
 \end{tabular}
}
\caption{\footnotesize Summary of the current results based on the latest \Planck\ 2015 release and CORE forecasts presented in this paper.
The third column gives the figure of merit of the improvement expected with CORE.
%In the third column in parenthesis we quote the improvement on the current measurement.
}
\label{tab:key}
\end{center}
\end{table}

\vspace{0.4cm}
\noindent
{\em Boosting the precision in the measurements of cosmological parameters}

\vspace{.2cm}
The precision of the determination of cosmological parameters will be greatly improved
by CORE. We have shown that the specifications of CORE suffice to provide
nearly ideal (i.e., cosmic-variance limited) measurement
of the CMB temperature and polarization power spectra up to high multipoles. 
%and thus determine the cosmological parameters of the $\Lambda$CDM model. 
With the addition of the CMB lensing, whose information will be exploited
up to the scales where linear theory is reliable, CORE will improve the uncertainties on key
inflationary parameters such as $n_\mathrm{s}$, $\mathrm{d}n_\mathrm{s}/\mathrm{d}\ln k,$ and 
$\Omega_{\mathrm k}$ by approximatively factors of 3.4, 2.9, and 4, respectively.

\vspace{.4cm}
\noindent
{\em Slow-roll inflationary models}

\vspace{.2cm}
CORE will explore values of $r$ approximately two orders of magnitude below the current 
limit. %and will chart the landscape of slow-roll inflationary models down to the Lyth bound. 
We have shown how CORE specifications are sufficient to test the predictions of the $R^2$ model, 
which is the simplest among the inflationary models favored by the \Planck\ data. 
CORE could also target smaller values of $r$, as predicted for instance in 
maximally supersymmetric realizations of inflation  
with the largest possible value of the moduli space curvature. CORE 
will also help provide information about the {\em reheating} stage for a given inflationary 
model and could allow us to distinguish models that share the same inflationary potential but 
have different reheating mechanisms.

\vspace{.4cm}

\noindent
{\em Testing deviations from a simple power-law}

\vspace{.2cm}
By providing a cosmic variance limited measurement of the \(EE\) power spectrum up to high 
multipoles, any of the proposed CORE configurations will increase the amount of information 
available on the scalar primordial power spectrum by an order of magnitude with respect to the existing 
\Planck\ 2015 data. We have shown that if there are any features in the scalar primordial 
power spectrum hidden beneath \Planck's resolution, CORE would reliably reconstruct them. This 
reconstruction can be performed in either in a model independent or a non-parametric manner,
or by using more 
traditional parametric methods. We demonstrate this success for both low-$\ell$ and 
high-$\ell$ features. Additionally, CORE-like experiments would be able to establish constraints on 
the tensor primordial power spectra even for relatively low values of the tensor-to-scalar 
ratio $r$. For the purposes of primordial power spectrum reconstruction, CORE is the 
natural successor to \Planck\ to provide answers to the many questions surrounding these 
critical predictions of inflationary models.

\vspace{.4cm}

\noindent
{\em Beyond the adiabatic initial condition}

\vspace{.2cm}
CORE will determine the nature of the initial conditions of primordial fluctuations 2--5 times 
better than \Planck\ (and also LiteBIRD because of its coarse angular resolution) 
by providing nearly cosmic variance limited upper bounds on the 
allowed isocurvature fraction. By including cosmological information contained in the CMB 
lensing power spectrum, the isocurvature bounds will improve compared to the bounds obtained 
using an ideal cosmic variance limited experiment with only the temperature and polarization 
anisotropies. We have shown that CORE can recover $10^{-3}$ level isocurvature fractions 
(one order of magnitude below the current constraints) without biasing the cosmological 
parameters, even in the presence of tensor modes. We finally have shown that the uncertainties in 
the cosmological parameters of the $\Lambda$CDM model will be increased at most by 25\% by 
allowing generally correlated isocurvature fluctuations with respect to the adiabatic case. 
All these analyses have been carried for a correlated mixture of adiabatic and isocurvature 
CDI modes, but similar results are expected for the neutrino density and velocity modes as well.

\vspace{.4cm}

\noindent
{\em Primordial non-Gaussianity}

\vspace{.2cm}

As a figure of merit for primordial non-Gaussianities, the direct bispectrum measurements by CORE 
will shrink the allowed $f_{\rm NL}$ volume in the three-dimensional 
Local-Equilateral-Orthogonal (LEO) shape-function space by a factor of approximately $20$ with respect to 
the current {\em Planck} results. This corresponds to signal-to-noise ratio improvements by a 
factor $\approx 2-3$ for each shape, giving $1\sigma$ $\fnl$ error forecasts of about 2, 20, and 
10 for local, equilateral, and orthogonal shapes, respectively. Similar levels of improvement are 
also found for other models such as oscillatory bispectra or scenarios with running of 
$f_{\rm NL}$, and for $g_\mathrm{NL}$ trispectrum constraints. Even larger improvements can be 
found for isocurvature NG models, where the improved polarization sensitivity is crucial. 
In this case $f_{\rm NL}$ error bars shrink by a factor of up to $\sim 10$.

Very interesting results are also expected if we consider alternative ways of estimating
local NG not based on direct temperature and polarization bispectrum estimation. Estimating 
the power spectrum of the correlation between temperature and $\mu$-distortion anisotropies 
can for example lead to better $f_{\rm NL}$ constraints by a factor of up to $\sim 100$ for 
specific models with excited initial states. Finally, a crucial finding is that through 
scale-dependent bias measurements of the CIB power spectrum on large scales, {\em CORE will be 
able to achieve $\sigma (f_{\rm NL}) \lesssim 1$ for the local shape}. This is a crucial 
threshold for discriminating between single- and multi-field inflation, and such a level of 
sensitivity would be comparable to or better than the sensitivity
expected from future galaxy surveys.

\vspace{.4cm} 

\noindent
{\em Topological Defects}

\vspace{.2cm}

CORE will search for primordial B-modes generated by vector and tensor perturbations induced 
by topological defects, including cosmic strings. The resulting B-mode power spectrum produced 
by topological defects is quite different from the shape produced by inflationary gravitational 
waves. CORE improves the limits from the CMB anisotropy power spectrum on Abelian Higgs gauge 
cosmic strings by nearly an order of magnitude compared to \Planck , pushing the maximally 
allowed string tension $G \mu$ decisively below the GUT scale (or detecting cosmic strings if 
any are formed at GUT energy scales). Similar gains in sensitivity are expected for all types of 
defects. Previous work \cite{Mukherjee:2010ve} indicates that a mission with CORE's 
capabilities can not only detect topological defects, but also distinguish them from primordial 
tensors and even from each other. CORE can therefore probe different physical mechanisms 
generating a B-mode signal at the GUT energy scale.

\vspace{0.4cm}

\section*{Acknowledgements}

%The following grants are aknowledged: 
%Work supported in part by Academy of Finland grants 257989 and 295113. 

FA is supported by the National Taiwan University (NTU) under Project No. 103R4000 and by the
NTU Leung Center for Cosmology and Particle Astrophysics (LeCosPA) under Project No. FI121.
KK acknowledges support by the Magnus Ehrnrooth Foundation.
JV acknowledges support by the Finnish Cultural Foundation. 
KK and JV work is supported in part by Academy of Finland grants 257989 and 295113. 
CJM is supported by an FCT Research Professorship, contract reference
IF/00064/2012, funded by FCT/MCTES (Portugal) and POPH/FSE (EC).
%KK is supported by the Magnus
%Ehrnrooth Foundation and JV by the Finnish Cultural Foundation.
%We thank the following centers for computational resources: 
Partial support by ASI/INAF Agreement 2014-024-R.1 for the
{\it Planck} LFI Activity of Phase E2.
The calculations in this work were performed in 
CSC -- the IT Center for Science Ltd. (Finland), IN2P3 Computer Center 
(France), HPC-cluster facilities of RWTH Aachen University (Germany), INAF-IASF Bologna HPC cluster (Italy), the Cambridge COSMOS SMP system (UK), part of the 
STFC DiRAC HPC Facility supported by BIS NeI capital grant ST/J005673/1 and STFC grants ST/H008586/1, ST/K00333X/1.

\bibliographystyle{JHEP}
\bibliography{biblio}

\end{document}